\documentclass[11pt,a4paper,english,twoside]{article}

\usepackage{a4wide}
\usepackage{amssymb, amsmath, amsthm}
\usepackage{graphicx}
\usepackage{subcaption}
\usepackage[all]{xy}
\usepackage[pdftex,hyperref,svgnames]{xcolor}
\usepackage[pdftex,colorlinks=true,
pdfstartview=FitV,
pdfnewwindow=true,
linktoc = page,
linkcolor= blue,
citecolor= red,
urlcolor= blue,
hyperindex=true,
hyperfigures=false]{hyperref}
\hypersetup{linktocpage}
\usepackage{dsfont}
\usepackage{empheq}
\usepackage[nosort, nocompress]{cite}
\usepackage{float}
\usepackage{cancel}
\usepackage{relsize}
\usepackage{soul}
\usepackage{enumerate}
\usepackage{enumitem}
\usepackage{hhline}
\usepackage{multirow}
\usepackage{xspace}
\usepackage{bbm}
\usepackage{pdfpages}
\usepackage{setspace}
\usepackage{comment}

\usepackage{bbold} 
\usepackage{stmaryrd} 

\usepackage{tocloft}

\newcommand{\nn}{\nonumber}

\def\beq{\begin{equation}}
\def\eeq{\end{equation}}
\def\bea{\begin{eqnarray}}
\def\eea{\end{eqnarray}}

\def\del {\partial}
\def\d {{\rm d}}
\def\w {\wedge}

\begin{document}
\numberwithin{equation}{section}

\begin{titlepage}
\begin{center}

\phantom{DRAFT}

\vspace{1.8 cm}

{\LARGE \bf{Dark energy from string theory: \\ \vspace{0.5cm} an introductory review}}\\

\vspace{2.0 cm} {\Large David Andriot}\\

\vspace{0.9 cm} {\small\slshape Laboratoire d'Annecy de Physique Th\'eorique (LAPTh),\\[0.15cm]
CNRS, USMB, UMR 5108,\\[0.08cm]
9 Chemin de Bellevue, 74940 Annecy, France}\\

\vspace{0.9cm} {\upshape\ttfamily andriot@lapth.cnrs.fr}\\

\vspace{2.2cm}
{\bf Abstract}
\vspace{0.1cm}
\end{center}
\begin{quotation}
Dark energy, the main constituent in our expanding universe, responsible for its acceleration, is currently being observed with unprecedented precision through various experiments. While several cosmological models can fit this latest data, deriving some of them from string theory would provide a valuable theoretical prior, with information on the nature of dark energy. This article reviews the efforts towards such a derivation, namely the options from string theory to get a cosmological constant (a de Sitter solution) or a dynamical dark energy (via a quintessence model).

After providing a brief historical perspective, we first review proven or conjectured constraints on obtaining dark energy from string theory, in classical or asymptotic regimes. Circumventing such obstructions, by changing regime or ansatz, one can try to construct a de Sitter solution: we present a long list of such attempts, and the difficulties encountered. Among them, we discuss in detail efforts towards classical de Sitter solutions. Then, we review quintessence from string theory, focusing on single-field exponential models. Related topics are discussed, including the coupling to matter, the comparison to observational data, and the absence of a cosmological event horizon.
\end{quotation}

\end{titlepage}

\tableofcontents

\newpage

\section{Introduction}\label{sec:intro}

For about one century, our universe has been observed to be in expansion. Matter (non-relativistic massive content) and radiation (relativistic one) as we know them make this expansion slow down. However, in 1998, the universe has been observed to be accelerating during its recent history \cite{SupernovaSearchTeam:1998fmf, SupernovaCosmologyProject:1998vns}. The non-standard constituent responsible for this behaviour has been named {\sl dark energy}. Today, it represents about $69\%$ of the total energy in the universe, while matter represents most of the remaining $31\%$. {\sl What is dark energy?} This simple question is the main motivation behind this review article: the nature of dark energy is not understood. While several cosmological models exist to describe it, additional insights could come from a more fundamental theory of Nature. String theory is candidate to the latter, as being a quantum gravity theory, able to unify all known interactions. The question is then whether dark energy, as seen in observations of the late universe, can be derived from string theory, and what form it then takes. This article reviews some tentative answers to this question.

\subsection{Scientific context}

\subsubsection{From string theory to models and observations of dark energy}

A first step in this program builds on known cosmological models that describe dark energy in our 4-dimensional (4d) universe. Many of them take the simple form of a theory of scalar fields $\{\varphi^i\}$, evolving in a scalar potential $V(\varphi)$, and minimally coupled to gravity, as given by the following action
\beq
{\cal S}_{4d} = \int \d^4 x \sqrt{|g_{4}|}   \left(\frac{M_p^2}{2}  {\cal R}_4 - \frac{1}{2} g_{ij}\del_{\mu} \varphi^i \del^{\mu} \varphi^j - V \right) \ ,\label{S4dintro}
\eeq
where $M_p=1/\sqrt{8\pi G}$ is the 4d reduced Planck mass, related to Newton's constant $G$, in natural units $\hbar=c=1$, and $g_{ij}$ is the field space metric governing kinetic terms. To the above, one should add a Lagrangian to describe matter, while non-minimal couplings could also be considered. The formalism of \eqref{S4dintro} can be phrased in terms of perfect fluids, each with an energy density $\rho$ and a pressure $p$, as reviewed in Appendix \ref{ap:cosmo}. In this framework, the simplest realisation of dark energy (DE) is then a cosmological constant $\Lambda >0$, namely $\rho_{{\rm DE}}=\Lambda \, M_p^2$ is constant in time. In terms of \eqref{S4dintro}, this corresponds to having a constant value of the potential, with $\Lambda = V/M_p^2$, and constant scalar fields. In general, this is realised at an extremum of the scalar potential, $\del_{\varphi^i}V=0$, where (classical) scalar fields without kinetic energy stand and stay constant. Given the possibility of (quantum) fluctuations, having a minimum of $V$ is usually preferred for stability to realise a cosmological constant; the relation of $\Lambda$ to a ``vacuum energy'' is then manifest. On the gravity side, the solution corresponding to an extremum of the potential gives ${\cal R}_4 = 4 \Lambda$, which is realised by a de Sitter spacetime for $\Lambda >0$. The latter is a maximally symmetric spacetime, but it can also be described through a Friedmann-Lema\^itre-Robertson-Walker (FLRW) metric. Then, it corresponds to a homogeneous and isotropic universe whose only constituent is dark energy in the form of $\Lambda$. This can be viewed as an approximation of our universe today, or as its future if indeed $\rho_{{\rm DE}}=\Lambda \, M_p^2$ is constant while matter gets diluted by the expansion. In the following, we refer to extrema of the potential (maxima or minima) as de Sitter solutions.

Another option is that the dark energy density $\rho_{{\rm DE}}$ is not constant in time. In that case, one talks of a dynamical dark energy. A simple realisation through \eqref{S4dintro} is to have the scalar fields rolling in the scalar potential, meaning being on a slope instead of an extremum. In those {\sl rolling solutions}, the kinetic energy ``kin'' is non-zero. The fields are varying with time and so is the value of $V(\varphi)$. As a result, the combination $\rho_{{\rm DE}} = {\rm kin} + V(\varphi)$ is most likely varying with time under the fields dynamics. Such models, realising dynamical dark energy, are known as {\sl quintessence} models \cite{Ratra:1987rm, Peebles:1987ek, Wetterich:1994bg, Caldwell:1997ii}.

Whether dark energy is described by a cosmological constant or through quintessence, the relevant 4d cosmological models are of the form \eqref{S4dintro} (see \cite{Copeland:2006wr} for an early review). Up to now, the vast majority of 4d effective theories explicitly derived from string theory turn out to take precisely the form \eqref{S4dintro}. There are two reasons for this. First, at low energy and weak string coupling, (super)string theory gives a 10d spacetime governed by Einstein gravity. This starting point eventually gives Einstein gravity in 4d as well, as in \eqref{S4dintro}. To go from a 10d theory to a 4d one, the 6 extra space dimensions should be taken care of. A common procedure is compactification: one gathers them as a 6d compact manifold ${\cal M}$, whose typical size $L$ is small and currently beyond experimental reach. The corresponding energy scale, $1/L$ in natural units, is sometimes called the Kaluza--Klein scale. The latter may then be large and serve as a cut-off to a 4d effective theory. Nevertheless, degrees of freedom from the extra dimensions may still contribute to 4d physics if they are massless or light: among those, one typically finds scalar fields. This is the second reason why \eqref{S4dintro} is ubiquitous among string effective theories: any theory with compact (and geometric) extra dimensions gives rise to massless or light scalar fields in 4d. The scalar potential is also dictated by the 6d manifold ${\cal M}$: if it is curved, if it carries some physical content such electromagnetic fluxes, or their charged sources, then a non-trivial scalar potential is generated. In short, {\sl 4d effective theories of the form \eqref{S4dintro} are easily obtained from string theory, and the 4d model characteristics, namely $\varphi^i,\, g_{ij},\, V$, are dictated by the 6d compact manifold ${\cal M}$ formed by the extra dimensions} (for the common, though not mandatory, setting of a compactification). From this perspective, {\sl dark energy gets naturally explained from string theory: it is due to extra dimensions.}

This situation calls for two important comments. First, we see that, at least within the framework described, string theory cannot give rise to any 4d model. Even though one can choose ${\cal M}$ and its physical content, the dimensional reduction, a.k.a.~the procedure giving rise to the 4d effective theory, is not arbitrary and does not allow to get any set of $\varphi^i,\, g_{ij},\, V$. As a consequence, one should be able to discriminate among 4d cosmological models of the form \eqref{S4dintro}, whether they can or cannot be obtained from string theory, at least as a compactification. This is an example of the swampland program \cite{Vafa:2005ui}, which aims at characterising outcomes of string theory, or more generally quantum gravity \cite{Palti:2019pca, vanBeest:2021lhn, Grana:2021zvf, Agmon:2022thq}. Models that cannot be obtained from such UV-complete gravity theories populate the so-called swampland, in contrast to those in the landscape of string theory. This perspective is interesting for cosmology, as it offers a theoretical prior on the various cosmological models at hand. The second comment is that even though models of the form \eqref{S4dintro} can be obtained, they do not necessarily fit the observational data. Precisely because the string theory models are not arbitrary, there is at this stage no guarantee for them to be in agreement with observations. On the optimistic side, one may hope for guidance from string theory models to discriminate among cosmological models that do fit the observational data. In this way, one may learn about the nature of dark energy from a fundamental theory.

This brings us to observations, which play a prominent role in the understanding of dark energy. Soon after the discovery of dark energy \cite{SupernovaSearchTeam:1998fmf, SupernovaCosmologyProject:1998vns}, both options of having a cosmological constant or a quintessence scenario were discussed. In the following years however, a concordance or standard model of cosmology has been established, under the name of $\Lambda$CDM. It assumes a cosmological constant and cold dark matter. For decades, this model has agreed with most observations. Meanwhile, precision in measurements increased, as experiments improved and more data got collected and treated. Nowadays, several surveys are probing to unprecedented precision level the nature of dark energy, focusing on whether dark energy is constant or dynamical. Through observations and some input model describing the expansion of the late universe, these experiments can provide as an output the recent evolution of the dark energy equation of state parameter: $w_{{\rm DE}}= p_{{\rm DE}}/\rho_{{\rm DE}}$. In the case of a cosmological constant (e.g.~using $\Lambda$CDM), one obtains the constant value $w_{{\rm DE}}=-1$. In the case of a rolling solution in quintessence, one gets a varying $w_{{\rm DE}}$. To be detected, such a variation should be large enough in amplitude, and over the scanned time range; a priori, there is no guarantee of this. In other words, the constant average of $\rho_{{\rm DE}}$ will always be observed, and viewed as a cosmological constant; the deviation from it should be strong enough to detect a dynamical dark energy. Interestingly though, this may have happened in the latest observations. DES \cite{DES:2024jxu} and DESI \cite{DESI:2024mwx, DESI:2025zgx} have reported in the last 2 years that both $w_{{\rm DE}}=-1$ and a varying $w_{{\rm DE}}$ are compatible with their data, depending which input model is used. Importantly, in the dynamical case, $\Lambda$CDM is excluded at more than 3$\sigma$. While these observations are still on-going, and their results are under heavy scrutiny, this is an interesting hint at the possibility of having dynamical dark energy. In addition, Euclid and LSST, two other independent experiments, will provide in the coming years new observational data constraining dark energy. {\sl In view of this new, and coming, observational data, studying cosmological models of dark energy, and testing their compatibility with string theory, becomes a very timely task.} We hope this review article will offer some guidelines for it.

\subsubsection{Dark energy and string theory: a troubled though constructive history}\label{sec:history}

Before reviewing dark energy models derived from string theory, we provide a brief ``historical'' perspective on this field. Hopefully, this should help navigating through a large literature.

In the mid 1990's, (super)string theory was the promise of a (possibly unique) fundamental theory of Nature, unifying all known interactions at the quantum level. From it, particle physics could be recovered, at least qualitatively in terms of its spectrum and interactions. This was essentially achieved in a Minkowski spacetime, with the help of supersymmetry, which was hoped to be soon discovered at the LHC. Our universe was known to be expanding, but the expansion could be due to its content in matter and radiation, particle physics remaining therefore the primary focus. The discovery of dark energy in 1998 was an important change in this paradigm. Considering dark energy to be realised as a cosmological constant, one is led to focus on a de Sitter spacetime. The latter is not easily obtained from string theory: it breaks supersymmetry, and requires the 6d manifold ${\cal M}$ to carry several non-trivial ingredients, increasing complexity of the setting. One ingredient is e.g.~an orientifold plane, as was soon pointed-out in \cite{Maldacena:2000mw}. Another major topic emerged in the early 2000's: the moduli problem, that is, the presence of many (unobserved) massless scalar fields in 4d models. A scalar potential, possibly stabilising (some of) them, can be generated, at the cost of adding again ingredients on ${\cal M}$, such as electromagnetic fluxes. These two phenomenological points illustrate that the realm of Ricci flat, supersymmetric and simple string constructions became somewhat outdated in the 2000's. When trying to reproduce a positive dark energy and solve the moduli problem, the path from string theory to our universe appeared more intricate. String phenomenology, whose purpose is to connect string theory to observations or constrain such a relation, became a field on its own, with a first annual conference in 2002.

Over the last quarter of a century, much progress has been made in constructing 4d models with positive scalar potentials, of interest for dark energy and the moduli problem. Important efforts have been devoted to exploring the stringy options, sharpening the methods, and understanding the limitations of what could be done or derived. The work related to particle physics went on as well, but mostly taking a parallel, independent path to that of getting cosmology; of course the two should eventually be done together, and we will come back to this point. This article is meant to review the efforts towards dark energy, focusing on the constructive approaches, i.e.~the attempts to get an explicit and viable 4d model.\\ 

\begin{figure}[ht!]
\begin{center}
\includegraphics[width=0.6\textwidth]{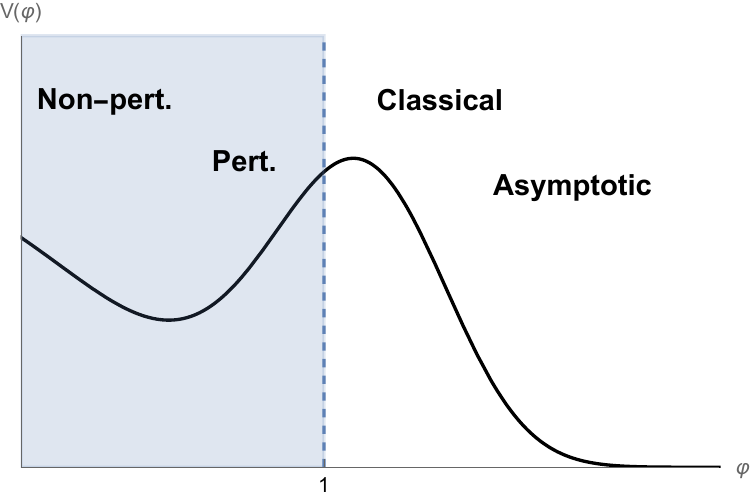}
\caption{Schematic representation of a typical scalar potential $V(\varphi)$ obtained in constructed examples of 4d effective string theories. It is common that the 4d scalar field value determines the string regime, namely with non-negligible perturbative or non-perturbative contributions ($\varphi< 1$), or without in the classical ($\varphi> 1$) or asymptotic ($\varphi \rightarrow \infty$) regimes. In turn, constructions of de Sitter minima were proposed with perturbative and non-perturbative terms, while candidate de Sitter maxima, strongly unstable, were found classically. All these attempts were then subject to scrutiny, regarding the corrections to these constructions. Asymptotically, the typical expectation for a positive potential is a runaway, forbidding any de Sitter extremum.}\label{fig:potintro}
\end{center}
\end{figure}

In more detail, most of the efforts were at first dedicated to obtain a de Sitter minimum, in order to get a positive cosmological constant. Among the main players were the KKLT construction \cite{Kachru:2003aw} and the Large Volume Scenario (LVS) \cite{Balasubramanian:2005zx, Conlon:2005ki}. Including perturbative and/or non-perturbative stringy corrections, these constructions proposed a 4d scalar potential with a positive minimum. Having in mind the cosmological constant problem, in short the smallness of the observed $\Lambda$ (see below), part of these constructions were also motivated by having a small value for $V$. Recent reviews on them can be found e.g.~in \cite{Cicoli:2023opf, McAllister:2023vgy, McAllister:2025qwq}. Another popular approach was one looking for ``classical de Sitter solutions''. Those are found in 10d supergravity, corresponding in principle to the low energy and perturbative regime of string theory, referred to as {\sl classical}; no string correction is allowed there, since they are supposed to be negligible. The advantage of this regime is that one has a clear control on corrections and approximations. The drawback is that the 4d potential gets less contributions than in other string regimes, thus being possibly poorer physics-wise. De Sitter solutions were nevertheless found, starting with \cite{Caviezel:2008tf}, but those turned out to be maxima of the 4d potential, with a strong perturbative instability. These various approaches were among the main ones before 2018, as reviewed back then in \cite{Bena:2017uuz, Danielsson:2018ztv}. We illustrate the situation in Figure \ref{fig:potintro}. 

An important point is that coupling constants in 4d effective string theories are determined by the value of scalar fields. For example, the dilaton $\phi$ gives, through its background value, the string coupling constant $g_s = e^{\phi}$, that governs e.g.~string loops. Another example is given by a 6d length $L$. The latter appears through a derivative acting on a background field (e.g.~the 6d Riemann tensor). The ratio $l_s/L$, with the string length $l_s$, then governs the size of higher derivative terms, a.k.a.~$\alpha'$-corrections; recent (10d) evaluations of those can be found e.g.~in \cite{Cicoli:2021rub, Castellano:2023aum, Wulff:2024mgu, Garousi:2025wfk, Garousi:2025dil}. As $L$ is related to a component of the 6d metric, it also corresponds to a 4d scalar field (a radius, a volume or a K\"ahler modulus), so again the value of a 4d scalar field $\varphi$ determines the size of corrections and perturbative expansion constants. As a consequence, the value of $\varphi$ also determines the string regime, as illustrated in Figure \ref{fig:potintro}. Different string theory contributions can be important to a 4d scalar potential depending on the value of $\varphi$; equivalently, a 4d theory is valid only in a certain field range. We distinguish the regimes where {\sl perturbative} and {\sl non-perturbative} string corrections become relevant (schematically $\varphi < 1$), from the regimes where they are not ($\varphi > 1$). Among the latter, the {\sl classical} regime discussed above (usually large 6d lengths or volume, small string coupling) can be found at (possibly) large but finite field values, while the {\sl asymptotic} regime corresponds to dominant contributions in the limit $\varphi \rightarrow \infty$. As recently discussed in \cite{Kaufmann:2026fli, Kaufmann:2026mha}, classical and asymptotic regimes should be distinguished: examples of the latter without the former can be found e.g.~in F-theory.

Before 2018, discussions emerged regarding corrections to the proposed de Sitter constructions, with a particular focus on KKLT. Without detailing them here, this raised the question of {\sl control on corrections}, which became central then, and later. Essentially, one asks whether all corrections to a construction are known, and for each of them, whether it can effectively be verified to be small, i.e.~negligible to the desired order, compared to contributions to the de Sitter solution. We will come back in more detail to this point, starting with Section \ref{sec:nodSQG}, but the bottom line is that control is not easily achieved: in most attempts to construct de Sitter, some stringy corrections are either unknown, or not computed, or too large. This leads to the statement that {\sl up to now, there exists no well-controlled construction of a 4d de Sitter solution}, as reviewed in \cite{Bena:2023sks}. As we will stress though, many constructions have been proposed, and a lot of progress has been made in investigating the corrections.

In 2018 was proposed the swampland de Sitter conjecture (dSC) \cite{Obied:2018sgi}. We highlight this work, not so much because of the conjecture itself, that we will discuss in \eqref{dSconj}, but because it was a turning point in the focus on dark energy from string theory. As explained, the swampland program discusses what can or cannot be obtained from string theory. It is then a natural question to ask whether a 4d theory of the form \eqref{S4dintro} can be derived, which admits a de Sitter solution, namely $\del_{\varphi^i} V =0$ and $V>0$. The dSC answered negatively to this question, giving in addition a minimal bound to the potential slope. If true, this conjecture had two immediate consequences: all proposed de Sitter constructions would actually be non-existent, e.g.~due to misevaluated corrections, and potentials would be runaways, suggesting a dynamical form of dark energy. This proposed conjecture led to many reactions: let us mention three of them, justifying the terminology of a turning point. First, the conjecture was quickly argued to be too strong, and refined in various fashions. In particular, it was proposed that the conjecture should be valid only in the asymptotics: the asymptotic runaway depicted in Figure \ref{fig:potintro} was then given a minimal bound to its slope. This stands under the name of the Strong de Sitter conjecture (SdSC) \cite{Bedroya:2019snp, Rudelius:2021oaz, Rudelius:2021azq}, given in \eqref{SdSC}, and serves today as a well-tested constraint on asymptotic potentials from string theory. Second, there was an important revived activity in attempts to construct de Sitter, with a renewed attention to the control on corrections. Third, having quintessence to realise dark energy became as well a revived topic. This gained even more interest with the latest observational data on dynamical dark energy reported above.\\

In this brief history, we have not mentioned other motivations in getting de Sitter or dark energy from string theory. A main one is holography, namely the realisation of a correspondence between a gauge and a gravity theory, the latter being here for a universe in accelerated expansion. More generally, this is related to the question of defining quantum gravity over de Sitter. We come back to this topic in Section \ref{sec:nodSQG}. While the motivation is here different than previous phenomenological ones, obtaining an explicit and controlled example of a de Sitter solution in a quantum gravity theory would still be useful to these questions. Our review of explicit attempts at such constructions may then be relevant to this topic.

\subsection{This review article}

\subsubsection{Content and structure}

The purpose of this review article is to discuss constructive and explicit approaches to realise dark energy from string theory. Special focus is given to the latest developments: the constraints worked-out as no-go theorems, or those conjectured (including from the swampland program), the most recent attempts at getting either de Sitter or quintessence, with possible comparison to the latest observational data, and the difficulties encountered. A special interest is given in reviewing attempts in the classical or asymptotic regimes of string theory. This is where this article can be found complementary to the others in the literature, such as the recent and comprehensive string cosmology review \cite{Cicoli:2023opf}, to which we refer on several occasions. As mentioned, the classical and asymptotic regimes are interesting because the control on corrections is a priori easier. The number of contributions to the model is less than in other regimes, but this also allows to face simpler and well-defined mathematical problems. In addition, more constraints are known there (e.g.~the SdSC), making the options for dark energy clearer. This article still discusses attempts at dark energy in different regimes, as they may offer richer physics.

The structure of the review article is simple. We start in Section \ref{sec:constraints} with the constraints to dark energy constructions, mostly in classical or asymptotic regimes. Those include obstructions to the existence of de Sitter solutions, to their perturbative stability, and constraints on the slope of the 4d scalar potential. These are either proven in the form of no-go theorems, or conjectured and tested, or simply argued for. In Section \ref{sec:circumvent}, we then discuss dark energy constructions, that necessarily circumvent these constraints. To do so, one option is to change regime, even though remnants of the classical and asymptotic constraints may still play a role. We list in Section \ref{sec:listattempts} attempted constructions of de Sitter solutions from string theory at large, together with the various difficulties faced. We then present quintessence in Section \ref{sec:quint}, with a focus on the case of a single field with an exponential potential, as a first standard example easily connected to string theory. Finally, we dedicate Section \ref{sec:dSsol} to the search for de Sitter solutions in the classical regime of string theory, the study of their properties, and the difficulties encountered. While a review on such solutions can be found in \cite{Andriot:2019wrs}, many recent developments motivate this update. Summaries are provided at the end of each section, in Section \ref{sec:dSconjsum}, \ref{sec:sumquint} and \ref{sec:sumclassdS}, with various outlook comments. {\sl We invite the reader to primarily use these summaries, together with the detailed table of contents}, in order to find information on a specific topic of interest.

A few points deserve to be emphasized. A detailed discussion is given in Section \ref{sec:nodSQG} on the question of control and corrections, together with a list of arguments on the difficulties to have a de Sitter solution in quantum gravity. This can be linked to the list of attempts at de Sitter in Section \ref{sec:listattempts} and their difficulties, which may also be found useful. These discussions are also related to the {\sl conjectured absence of a cosmological event horizon} in solutions from quantum gravity (NCHC) \eqref{NCHC}. The relation of that statement to the question of (no) asymptotic acceleration is investigated in some detail in Section \ref{sec:accnohor}, the results of which can be considered as new. While we have not discussed the matter content of the universe so far, we argue in Section \ref{sec:sumquint} and \ref{sec:sumclassdS} that incorporating matter from the start could play a role in the string construction realising dark energy. This is particularly obvious when considering coupling of a quintessence field to matter in Section \ref{sec:phantom}, while having specific intersecting brane configurations is another example in Section \ref{sec:dSsolexistence}. Finally, the 10d supergravity de Sitter solution $s_{55}^+ 29$ of \cite{Andriot:2024cct}, which possesses a (partial) parametric control on classicality is discussed in some more detail in Section \ref{sec:dSsolclassical}. The use of this parametric scaling in the backreaction of its $O_5$ sources is discussed in Section \ref{sec:backreact}. Those can also be viewed as new results. On the mathematical side, the review of group manifolds and relevant Lie algebras (e.g.~solvable) provided in Section \ref{sec:groupman} and Appendix \ref{ap:groupman} may be of broader interest.

\subsubsection{Topics not covered}\label{sec:topicsnotcovered}

Several important topics, related to above questions, are not covered in this article; we briefly discuss them here. Let us start with the {\sl cosmological constant problem} \cite{Weinberg:1988cp}, reviewed e.g.~in \cite{Martin:2012bt, Burgess:2013ara}. Assuming that dark energy is given by a cosmological constant, we obtain $\Lambda = V_0/M_p^2 = 3H_0^2\, \Omega_{{\rm DE}}$ (see Appendix \ref{ap:cosmo}). Using the data from \cite{Planck:2018vyg}, namely $H_0 = 67.4\, km.s^{-1}.Mpc^{-1} = 5.87 \cdot 10^{-61}\, M_p$ and $\Omega_{{\rm DE}}=0.685$, one gets $\Lambda = 7.08 \cdot 10^{-121}\, M_p^2 $ or $V_0 = 24.9\, meV^4$. One part of the problem is the huge hierarchy between this observed $\Lambda$ and the ``natural'' fundamental scale $M_p$, which is not easy to engineer in concrete models. Another aspect is that matter contributions to a vacuum energy, e.g.~in the form of quantum contributions from particle physics, should be found extremely small in order to get such a value for $\Lambda$. However, their estimates are not usually found small, in addition to being divergent with the cut-off or UV-dependent. Without a mechanism such as supersymmetry, allowing for cancelations among quantum corrections, it appears difficult to eventually obtain such a small scale (see however \cite{Dudas:2025yqm} for a recent non-supersymmetric stringy realisation). This article barely discusses the matter content, up to the points mentioned above, so it does not tackle matter contributions to the cosmological constant. But even when considering gravity alone, one should face radiative quantum corrections from gravitons, that could also contribute non-trivially to the vacuum energy; we do not discuss either such quantum contributions. Independently of those quantum effects, we are still left with the question of the value of $V_0$, i.e.~the extremum value of the scalar potential. Some attempts at constructing de Sitter minima try to reach a small and realistic value for $V_0$; we will not comment on this here, focusing on the less ambitious question of getting a de Sitter extremum (preferably a well-controlled one).

We mention in passing a famous proposal to ``solve'' the cosmological constant problem. It makes use of the string landscape, viewed as the set of solutions or even vacua that can be obtained. Naive counts of those (not necessarily taking into account tadpole cancelation or moduli stabilisation) provide huge numbers: an often quoted one is $10^{500}$ \cite{Douglas:2006es}, while later work suggests $10^{272\, 000}$ \cite{Taylor:2015xtz}. Imagining a homogenously distributed (in terms of $\Lambda/M_p^2$) set of de Sitter minima among this landscape, one may then argue that a vacuum with the observed $\Lambda$ value can be found \cite{Bousso:2000xa}. A more detailed review of such arguments can be found in \cite[Sec.6]{Hebecker:2020aqr} and \cite[Sec.6.1]{Cicoli:2023opf}. One should note that the solutions counted above are not de Sitter ones, rather Minkowski ones where the 6d ${\cal M}$ is a Calabi-Yau manifold with specific fluxes, but one may hope to reach de Sitter following e.g.~the KKLT construction. However, the validity of the latter is subject to various discussions. More generally, this early proposal solves the cosmological constant problem by referring to a large set of de Sitter minima, without explicitly constructing any of them, in particular the one corresponding to our universe. One may be satisfied by assuming that they exist, maybe even in a very quantum or strongly coupled regime, where no clear description is available. In this article, we rather focus on constructive approaches, which try to get at least one well-controlled and explicit de Sitter extremum.

Another non-constructive approach, related to the observed value of $\Lambda$, is the dark dimension scenario \cite{Montero:2022prj}. It builds on extrapolations of swampland conjectures, especially \cite{Lust:2019zwm}, to relate the tiny value of $\Lambda$ to the mass of neutrinos and to the size of extra dimensions. On these general grounds, it then proposes that one extra dimension should be large, compared to others. While this chain of arguments is not constructive, in the sense of not being built on concrete models, some attempts to get explicit realisations of this scenario have been made, and we will mention some.\\

Other relevant topics, not covered in the article, include {\sl supersymmetry breaking}, {\sl scale separation}, {\sl matter}, {\sl Hubble tension} and {\sl inflation}; we now briefly comment on them. De Sitter solutions, and typical cosmological rolling solutions, break supersymmetry. But most string theory starting points (though not all) are supersymmetric: the 10d (or 11d) theories, superstring or supergravity, are supersymmetric, and supersymmetry can sometimes be preserved through the compactification to the 4d effective theory. To reach our observed world, a supersymmetry breaking must occur at a certain energy scale, and through a certain mechanism, which can have an important impact on the resulting 4d physics. We do not cover explicitly this question; a recent review can be found in \cite{Dudas:2025ubq}.

Scale separation is, in short, the fact of having a gap between the 4d energy scale given by $\Lambda$, and the 6d energy scale given by $1/L$, the Kaluza--Klein scale. Such a gap can justify a low energy truncation to 4d physics. This topic is briefly mentioned for quintessence in Section \ref{sec:quint}, and in relation to the classicality of de Sitter solutions in Section \ref{sec:dSsolclassical}. But this question is important for any 4d cosmological model obtained from string theory, and would deserve a more thorough study. A review on scale separation in anti-de Sitter solutions, by far more investigated, can be found in \cite{Coudarchet:2023mfs}.

Semi-realistic models of particle physics can be derived from string theory \cite{Marchesano:2024gul}, but this is rarely done in a cosmological context (see however \cite{Cicoli:2012vw, Cicoli:2013mpa, Cicoli:2013cha, Cicoli:2017shd, Cicoli:2021dhg}). For cosmology, most constructions focus on an action of the form \eqref{S4dintro}, from which matter is (classically) decoupled. In this article, we do not discuss much the inclusion of matter, even though it could have important impacts on the cosmology. The notable exception is Section \ref{sec:phantom}, where we discuss the coupling of a quintessence scalar field to matter with crucial impact on the dynamics, and Section \ref{sec:sumclassdS} where we connect intersecting branes needed for de Sitter solutions to their role in particle physics models. We also recalled above the role of matter in the cosmological constant problem through quantum contributions. Last but not least, having at least $30\%$ of matter in the universe seriously contributes to the Hubble friction, which plays an important role in classical cosmological dynamics. In thawing quintessence models (e.g.~exponential), this friction is even crucial, since it freezes the scalar fields on a slope, as reviewed in Section \ref{sec:matter}. It is sometimes believed that one may obtain as a first step the correct particle physics model in Minkowski, and then have a tiny and harmless stringy correction that would provide a small positive cosmological constant; nothing guarantees that such a mechanism can be achieved with adequate control. In addition, this point of view neglects the impacts and interactions just described between matter and dark energy, which would deserve more joint studies. 

The Hubble tension is a discrepancy on the value of $H_0$, the Hubble parameter today, as inferred from early universe observations (e.g.~CMB) and from late ones (e.g.~supernovae), at least when using $\Lambda$CDM. Models of Early Dark Energy, that generate a temporary rise of $\Omega_{{\rm DE}}$ at some point in the past, have been proposed to solve the tension \cite{Kamionkowski:2022pkx, Poulin:2023lkg}. In the context of this article, obtaining a de Sitter solution alone does not offer an answer to this puzzle. Dynamical dark energy models may on the contrary provide options for this, including a realisation of Early Dark Energy. We only mention one such possibility in Section \ref{sec:phantom}, when considering quintessence coupled to matter, as illustrated in Figure \ref{fig:coupling}. 

Finally, we do not discuss much inflation, this phase of accelerated expansion that is believed to have happened in the early universe. Reviews on inflation and its string theory realisations can be found in \cite{McAllister:2007bg, Baumann:2014nda, Cicoli:2023opf}. A standard formalism to describe inflation uses the same action as \eqref{S4dintro}. Therefore, the constraints discussed in Section \ref{sec:constraints} also apply to inflation, and we discuss their consequences in Section \ref{sec:dSconjsum}. We however do not discuss attempts to construct inflation in spite of these constraints. The reason is that inflation solutions are rather different than those realising the acceleration of the late universe. A typical potential for a single-field inflation model, giving a solution agreeing with observation, is a long, slightly concave, plateau. The latter can be viewed as being almost a mildly unstable de Sitter maximum, which we however do not obtain easily from string theory. Viewed as a dynamical scenario, one should have a scalar field rolling for about 60 e-folds, while the recent acceleration has happened since less than 2 e-folds, making a quintessence realisation very different than an inflation one. We may still mention the fact that inflation is often proposed to end in a de Sitter minimum, where reheating occurs through oscillations. The search for such a de Sitter minimum could still be connected to the attempts for dark energy described in this article.

\subsubsection{Acknowledgements}

We warmly thank C.~Bock, P.~Brax, M.~Cicoli, C.~Delaunay, B.~Freivogel, A.~Guarino, A.~Hebecker, A.~Hees, D.~Junghans, J.P.~van der Schaar, R.~Scherrer, G.~Shiu, D.~Tsimpis, J.-P.~Uzan and E.~Verlinde, for helpful discussions on various points in this article while writing it.

\newpage

\section{No-go theorems and swampland conjectures}\label{sec:constraints}

In this section, we discuss constraints on realisations of dark energy from string theory in the classical or asymptotic regimes. Dark energy would be obtained as a cosmological constant in a 4d de Sitter spacetime, or through a quintessence model, where scalar field(s) are rolling along a 4d scalar potential; we thus look respectively for de Sitter or rolling solutions. To that end, we first focus on 10d supergravity theories (introduced in Appendix \ref{ap:sugra}), as effective theories for string theories in the classical regime, and introduce an adequate ansatz for their solutions. This ansatz, with 6 compact extra dimensions, will also serve as a compactification ansatz, eventually providing a 4d effective theory with scalar fields and a scalar potential. Having set the stage in Section \ref{sec:compactif}, we derive constraints on the existence and the stability of de Sitter solutions in Section \ref{sec:nogo}, in the form of no-go theorems. The former also provide constraints on the slope of 4d scalar potentials, hence on properties of rolling solutions. In Section \ref{sec:swamp}, we turn to conjectured constraints, inspired in part by the no-go theorems, known as the swampland de Sitter conjectures. Those are discussed in detail and tested against examples, to eventually turn into trusted constraints. We summarize all constraints in Section \ref{sec:dSconjsum}, together with cosmological implications on realisations of dark energy (as well as inflation).

This will lead us in Section \ref{sec:circumvent} to find ways around such obstructions, in order to obtain suitable realisations of dark energy, with cosmological solutions in accelerated expansion. In particular, any ingredient not considered in the framework of this section may serve the purpose of circumventing the constraints.

\subsection{Compactification of 10d type II supergravities to 4d}\label{sec:compactif}

\subsubsection{10d ansatz and equations}\label{sec:10dansatz}

We start with 10d type IIA and IIB supergravities, presented in Appendix \ref{ap:sugra}. We will look for solutions where the 10d spacetime, of signature $(-,+,\dots,+)$, is split into a 4d spacetime and a 6d compact space ${\cal M}$. The ansatz for the 10d solution metric in string frame is given by
\beq
\d s_{10}^2 = g_{MN} \d x^M \d x^N  = e^{2A(y)} \, \tilde{g}_{\mu \nu}(x) \d x^{\mu} \d x^{\nu} + g_{mn}(y) \d y^m \d y^n \label{10dmetricmaxsym}
\eeq
where in the last terms, the coordinates $x$, resp.~$y$, are 4d or external, resp.~6d or internal. The 10d spacetime is then almost a direct product, up to the warp factor $e^{A(y)}$. The 4d metric $\tilde{g}_{\mu \nu}$ is that of a maximally symmetric spacetime, that is anti-de Sitter (AdS), Minkowski (Mink) or de Sitter (dS); we will see later how to extend the ansatz to allow more generally for a 4d Friedmann-Lema{\^i}tre-Robertson-Walker (FLRW) metric describing a cosmological spacetime. The compact space ${\cal M}$ is a manifold without boundary, a property which will allow us to integrate certain total derivatives on ${\cal M}$ to zero.

Preserving 4d maximal symmetry (or more strictly 4d Lorentz invariance) forbids distinguishing a direction in the 4d spacetime. As a consequence, in our solution ansatz, the supergravity fluxes $F_0,\, F_1,\, F_2,\, F_3,\, H$ are restricted to be purely internal forms. $F_4^{10}$ and $F_5^{10}$ are allowed to have a 4d component $F_{q=4,5}^4$ proportional to ${\rm vol}_4 = e^{4A}\, \tilde{{\rm vol}}_4$, and a purely internal one $F_{q=4,5}$. The 4d components are given by
\beq
F_4^4 = {\rm vol}_4 \w *_6 F_6 \ ,\quad F_5^4 = - {\rm vol}_4 \w *_6 F_5 \ . \label{Fq4}
\eeq
For $F_4^4$, we introduce an ad-hoc internal $6$-form $F_6$, whose 6d Hodge star captures the free proportionality factor. $F_5^4$ is fixed by the anti-self dual condition \eqref{F510ASD} on $F_5^{10}$. From now on, $F_{q=0, \dots, 6}, H$ only denote internal forms. Also, $F_q^{10}= F_q + F_q^4$ and the signature gives
\beq
\hspace{-0.15in} |F_4^{10}|^2= |F_4|^2 - |F_6|^2 \ ,\ |F_5|^2= |*_6 F_5|^2 \ ,\ \ \frac{1}{(q-1)!} F^{10}_{q\, \mu M_2 \dots M_q} F^{10\ M_2 \dots M_q}_{q\ \nu} = - g_{\mu \nu} |F_{10-q}|^2 
\eeq
Finally, for the same reason, the dilaton only has an internal dependence: $\phi(y)$.

In the same way, maximal symmetry imposes extended objects whose world-volume fills the 4d spacetime. Considering $D_p$-branes or orientifold $O_p$-planes, we thus restrict to space-filling ones, i.e.~they are along the 3d external space, leading to $p\geq 3$ and $p-3$ internal directions. Furthermore, we assume that the set of directions parallel to these sources, or transverse to them, can be globally identified in the 10d spacetime. In practice, this means that the embedding, or the pull-back, is not too convoluted, and the 10d version of the sources action presented in Appendix \ref{ap:sugra} makes sense. The projection to 10d parallel or transverse dimensions is then well-defined, for example
\beq
T_{\mu\nu} = e^{2A} \tilde{g}_{\mu\nu} \sum_{{\rm sources}} \frac{T_{10}^s}{p+1} \ ,\ T_{mn} =  \sum_{{\rm sources}} \delta_m^{m_{||}} \delta_n^{n_{||}} g_{m_{||} n_{||}} \frac{T_{10}^s}{p+1} \ .\label{Tmn}
\eeq
For each source $s$, the transverse volume form ${\rm vol}_{\bot_s}$ is then well-defined, and is now understood as an internal form. Due to the space-filling property, we decompose the world-volume form into ${\rm vol}_4$ and the volume form of internal parallel (or wrapped) directions ${\rm vol}_{||_s}$, such that
\beq
{\rm vol}_{10} = {\rm vol}_4 \w {\rm vol}_{||_s} \w {\rm vol}_{\bot_s} \ , \ {\rm vol}_{6} = {\rm vol}_{||_s} \w {\rm vol}_{\bot_s} \ ,\ {\rm vol}_{\bot_s} =*_6 {\rm vol}_{||_s}  \ . \label{volumeforms}
\eeq\\

With the above solution ansatz, the 10d supergravity equations, listed in Appendix \ref{ap:sugraeom}, simplify: the resulting versions of these equations are given in Appendix \ref{ap:warpeq}. Here, we make one more assumption: we work in the ``smeared approximation''. Technically, this amounts to replace the warp factor, the dilaton and the source contribution as follows
\begin{gather}
\label{smearing}
\begin{aligned}
&\text{Smeared approximation:}\\
& e^A={\rm constant} \ ,\quad e^{\phi}=g_s = {\rm constant} \ ,\quad \delta(\bot) \rightarrow \int_{\bot} \d^{\bot} y\, \delta(\bot) = 1 \ .
\end{aligned}
\end{gather}
In practice, this drops all derivatives of $A$ and $\phi$ from equations, and trades source contributions $T_{10}^s$ for constants, with one unit per source; this makes the equations easier to handle. The idea behind this approximation is the following. The warp factor $e^A$ usually captures the backreaction of $D_p$ and $O_p$ sources as a Green's function: it localizes them at a point in their transverse directions (see e.g.~\cite{Andriot:2019hay}). In addition, in typical asymptotically Minkowski solutions, the dilaton is related to the warp factor: $e^{\phi} =g_s\, e^{A (p-3)}$. Considering $e^A$ constant thus amounts to average the backreaction over ${\cal M}$. The same effect is produced when trading the localizing $\delta$-function for its integral: this amounts to ``smear'' the source over its transverse directions, and provides an averaged source contribution instead of a localized one. Given this interpretation, the approximation may be understood as solving an integral (or averaged) version of the equations, where $\delta$-functions become $1$ and derivatives become total ones and drop out.\footnote{The interpretation of solving ``integral versions'' of the equations, and obtaining an averaged, non-localized solution, is especially valuable in view of orientifolds. Indeed, the latter correspond to fixed points of the geometry that cannot be moved, contrary to $D_p$-branes which are dynamical objects. ``Smearing an orientifold'' is therefore a priori a delicate operation, but interpreting this as averaging through an integral can make sense.} In any case, the smeared solution obtained through this approximation is a first step towards a complete solution. The hope is that the smeared solution can later be ``localized'', into a solution where backreaction is properly taken into account, with an adequate warp factor, dilaton and source contribution. Whether this second step can be achieved requires a longer discussion that we will present in Section \ref{sec:backreact}; let us mention that examples exist where this localization can be done. For now, we will assume it is the case, giving value to results on smeared solutions.

In the next subsection, we will rather consider no-go theorems against smeared solutions. A localized backreaction, e.g.~through gradients of $A,\phi$, may then be considered as an ingredient, not included in our ansatz, that could be added to circumvent the no-go theorems. As for other ingredients, we will discuss this option in the next section. It will be natural to ask whether the no-go can then be generalized to include the warp factor and the dilaton: we will mention some cases where this is known to be possible.\\

The solution ansatz together with the smeared approximation simplify equations to solve as follows. For the 4d metric, given the smeared approximation, we use from now on the 4d component of the 10d metric in string frame, $g_{\mu\nu}$, instead of $e^{2A} \tilde{g}_{\mu\nu}$. The 4d Ricci scalar ${\cal R}_4$ and the volume form ${\rm vol}_4$ are defined with respect to it. To start with, the 10d trace-reversed Einstein equation gets the following 6d components in type IIA and IIB
\begin{gather}
\label{Einstein6d}
\begin{aligned}
{\cal R}_{mn} & = \frac{g_s^2}{2} \left( F_{2\, mp}F_{2\, n}{}^p + \frac{1}{3!} F_{4\, mpqr} F_{4\, n}^{\ \ pqr} \right) + \frac{1}{4} H_{mpq}H_n{}^{pq} + \frac{g_s}{2} T_{mn}  \\
& +\frac{g_{mn}}{16} \left(- 2 |H|^2 + g_s^2 (|F_0|^2 - |F_2|^2 - 3 |F_4|^2 + 3 |F_6|^2 ) - g_s T_{10}  \right) \ ,\\
{\cal R}_{mn} &= \frac{g_s^2}{2} \left( F_{1\, m}F_{1\, n} + \frac{1}{2} F_{3\, mpq}F_{3\, n}{}^{pq} + \frac{1}{2\cdot 4!} F_{5\, mpqrs}F_{5\, n}^{\ \ pqrs} - \frac{1}{2} (*_6 F_{5})_m (*_6 F_{5})_n  \right) \\
& +\frac{1}{4} H_{mpq}H_n{}^{pq} + \frac{g_s}{2} T_{mn} + \frac{g_{mn}}{16} \left(- 2 |H|^2 -2 g_s^2 |F_3|^2 - g_s T_{10}  \right) \ .
\end{aligned}
\end{gather}
The 4d components indicate that the 4d Ricci tensor is proportional to the metric, i.e.~the 4d spacetime is an Einstein manifold, as expected for a maximally symmetric spacetime. In that case, the 4d Einstein equation is equivalent to its trace. We give it in the following, together with the simplified 10d trace of the Einstein equation and finally the dilaton e.o.m.
\begin{subequations}\label{3eq}
\begin{align}
({\rm 4d})\quad \quad & 2 {\cal R}_4 + |H|^2 + \frac{g_s^2}{2} \sum_{q=0}^6 (q-1) |F_q|^2  + \frac{g_s}{2} \sum_{{\rm sources}} \frac{p-7}{p+1} T_{10}^s = 0\ ,\label{Einstein4d}\\
({\rm 10d})\quad \quad & 4 {\cal R}_{4} + 4 {\cal R}_{6} - |H|^2 - \frac{g_s^2}{2} \sum_{q=0}^6 (5-q) |F_q|^2  + \frac{g_s}{2} T_{10} = 0 \ ,\label{10dtraceEsmear}\\
({\rm dil.})\quad \quad & 2 {\cal R}_{4} + 2 {\cal R}_{6} - |H|^2  + g_s \sum_{{\rm sources}} \frac{T_{10}^s}{p+1} = 0 \ ,\label{dileom}
\end{align}
\end{subequations}
where one should select even/odd RR fluxes for IIA/B. We also recall the trace notation $T_{10} = \sum_{{\rm sources}} T_{10}^s$. The gauge potential e.o.m.~simplify towards
\begin{gather}
\label{eomfluxes}
\begin{aligned}
& g_s^{-2}\ \d (*_{6} H) - \sum_{q=0}^4 F_q \w *_{6} F_{q+2} = 0 \ ,\\
&\quad \d( *_{6} F_q) + H\w *_{6} F_{q+2} = 0 \ , \quad \text{(for $1\leq q\leq 6$)}\ ,\\
\end{aligned}
\end{gather}
while the flux BI become
\begin{gather}
\label{BI}
\begin{aligned}
&\quad \d H = 0 \ , \phantom{\frac12} \\
&\quad \d F_{q} - H \w F_{q-2} = (-1)^{q+1+[(q+1)/2]} \sum_{\text{$(8-q)$-sources}} \frac{T_{10}^s}{9-q}\, {\rm vol}_{\bot_s} \ , \quad \text{(for $0\leq q\leq 5$)} \ ,\\
&\quad 0 = \sum_{\text{$9$-sources}} T_{10}^s \ ,
\end{aligned}
\end{gather}
with even/odd $p$-sources for IIA/B. $[\cdot]$ stands for the integer part; to be more explicit, the sign in the BI right-hand side (r.h.s.) is $+$ only for $q=2,3$. In all these equations, we recall that the only possibly non-zero (internal) RR fluxes to be considered are $F_{0,1,\dots,6}$.

Finally, the target space involution of orientifolds needs to be respected by the solution, as detailed in Appendix \ref{ap:Op}. Assuming constant field components would turn these extra constraints into projection conditions, which would further simplify the above equations. Also, the Riemann BI \eqref{BIRiemann} needs to be satisfied.\\

The 10d solution ansatz detailed above allows for a dimensional reduction of the 10d theory towards a 4d one. Provided adequate truncations, one can then look or constrain the solutions of interest directly from a 4d perspective, as we now explain.

\subsubsection{Dimensional reduction, 4d effective theories, extremum and stability}\label{sec:dimred}

The 10d solution ansatz just presented considers a 10d spacetime that is a direct product between a 4d spacetime and a 6d compact space ${\cal M}$. The latter having a finite volume, given by $\int \d^6 y \sqrt{|g_6|}$, this configuration allows a dimensional reduction from the 10d theory to a 4d one, sometimes also referred to as a compactification. Indeed, 6d-dependent quantities can be integrated over ${\cal M}$ to finite values, resulting effectively in a purely 4d theory. This will be shown explicitly in the following.

To get started, we face the central question in dimensional reductions, that is: what are the fields of the 4d theory? Starting from the 10d theory, one needs to truncate the initial 10d fields to a finite set of purely 4d fields. Physically, an interesting choice is to perform a truncation that keeps {\sl only} light 4d fields, and keeps {\sl all} of them: in that case, the 4d theory is a low energy effective theory. The cut-off energy scale is then often (but not always) related to a typical length scale of ${\cal M}$, called the Kaluza--Klein scale: this is the scale at which the 6 extra dimensions cannot be ignored anymore. In practice, getting such a 4d theory is not always easy. What is often done is to truncate to a finite set of 4d fields of interest, and discuss their 4d physics; we will do the same here, and come back to the question of the low energy. This approach can still be well-defined in the case where the finite set of fields is decoupled from any other field: indeed, a more complete description carrying all degrees of freedom would then not alter the physics of this first finite set of fields, in absence of coupling (note that we work in a classical regime; indirect coupling by gravitational mediation is thus neglected). Obtaining in this way an ``effective'' theory for such an independent finite set of fields is sometimes called a consistent truncation. Whether a 4d theory obtained by consistent truncation can also be a low energy one is not clear; in particular some modes kept by the consistent truncation can be heavy, while other light ones can have been truncated. It can still happen that the two match; for related discussions, see e.g.~\cite[Sec.5]{Andriot:2018tmb} with ${\cal M}$ a nilmanifold, and \cite{Tsimpis:2020ysl, Lin:2024eqq} for a Calabi-Yau.\\

As a first illustration, we consider our starting theory to be the restriction of 10d supergravities to their 10d Einstein-Hilbert term with dilaton factor: see \eqref{SNSNS}. For the truncation, we choose only the 4d metric to be a dynamical 4d field. The 6d metric is then set to a background value, $g_{mn}^0(y)$, and similarly for the dilaton that we take constant: $e^{\phi^0} = g_s$. We now perform the dimensional reduction by rewriting the 10d theory as an effective 4d one:
\begin{gather}
\label{dimred1}
\begin{aligned}
{\cal S} & = \frac{1}{2 {\kappa}_{10}^2} \int \d^{10} x \sqrt{|g_{10}|}\, e^{-2\phi}\,  {\cal R}_{10} \\
& = \frac{1}{2 {\kappa}_{10}^2\, g_s^2} \int \d^4 x \sqrt{|g_{4}|} \int \d^6 y \sqrt{|g_{6}^0|}\,  \left( {\cal R}_4 + {\cal R}_6^0 \right) \\
& = \int \d^4 x \sqrt{|g_{4}|}   \left(\frac{M_p^2}{2}  {\cal R}_4 - V_0 \right) \ ,\\
{\rm where} &\quad M_p^2 = \frac{\int \d^6 y \sqrt{|g_{6}^0|}}{{\kappa}_{10}^2\, g_s^2} \ ,\quad V_0 = -\frac{M_p^2}{2} \frac{\int \d^6 y \sqrt{|g_{6}^0|}\, {\cal R}_6^0}{\int \d^6 y \sqrt{|g_{6}^0|}} \ .
\end{aligned}
\end{gather}
As mentioned, the internal volume with the 6d background metric is a finite constant, allowing us to define a 4d reduced Planck mass $M_p$. Similarly, assuming the integral of the internal background curvature ${\cal R}_6^0$ finite, we get a constant $V_0$. The effective 4d theory resulting from this dimensional reduction is gravity with a cosmological constant, $V_0/ M_p^2$. The latter originates from a 6d quantity, namely the curvature.

In the following, we will consider truncations giving in addition scalar fields $\varphi^i$, where $i$ labels them. The previous $V_0$ will get traded for a scalar potential $V(\varphi^i)$, whose origin will again be internal quantities. We will obtain generically
\beq
{\cal S}_{4d} = \int \d^4 x \sqrt{|g_{4}|}   \left(\frac{M_p^2}{2}  {\cal R}_4 - \frac{1}{2} g_{ij}\del_{\mu} \varphi^i \del^{\mu} \varphi^j - V \right) \ ,\label{S4d}
\eeq
i.e.~a gravity theory minimally coupled to scalar fields. The kinetic terms depend on $g_{ij}(\varphi^k)$, the field space metric, that will be positive-definite for us. For a canonically normalised field $\hat{\varphi}^i$, one has $g_{ij}=\delta_{ij}$.\\ 

In such a 4d theory, having scalar fields placed at an extremum of the scalar potential, without any kinetic energy, is a solution to the e.o.m.:
\beq
\text{Extremum solution:}\quad \del_{\mu} \varphi^i=0 \ ,\ \del_{\varphi^i} V = 0 \ ,\ {\cal R}_4 = 4\frac{V}{M_p^2} \ . \label{extremumsol}
\eeq
In this solution, $\frac{V}{M_p^2}=\Lambda$ is constant (value of $V$ at the extremum) and plays the role of a cosmological constant; in other words this provides a solution with a maximally symmetric 4d spacetime. Different solutions can be found when the fields have kinetic energy, i.e.~are evolving by e.g.~rolling along a slope of the scalar potential with $\del_{\varphi^i} V \neq 0$. Such ``rolling solutions'' are relevant for cosmology, as we will see.

Given an extremum solution, it is interesting to determine its (perturbative) stability: for $V\geq0$, a stable solution is one that is a minimum in all field directions, while an unstable one is a maximum for at least one field. A distinction is sometimes made between metastability, which refers to a local minimum, and (full) stability for a global minimum. In case of a single field, the (in)stability is determined by the sign of the second derivative of the potential $\del_{\varphi}^2 V$. In a multifield situation, we use the mass matrix $M$. $\del_{\varphi}^2 V$ gets generalized to the Hessian $\nabla_{i}\del_j V= \del_i \del_j - \Gamma^k_{ij} \del_k V $, where the index $i$ stands for $\varphi^i$ and $\Gamma^k_{ij}$ is the Christoffel symbol for $g_{ij}$. Note that at an extremum, the covariant field space derivative reduces to a simple one: $\nabla_{i}\del_j V |_0 = \del_{i}\del_j V |_0 $. The mass matrix is defined as $M^i{}_j = g^{ik} \nabla_{k}\del_j V $: its eigenvalues correspond to the square of scalar masses, $m^2$. Evaluated at an extremum, the stability is then determined by the sign of these eigenvalues. A negative eigenvalue indicates an unstable solution, and equivalently, the latter admits a tachyon, $m^2<0$.

For an anti-de Sitter extremum, $V<0$, the previous definition is slightly modified to tolerate ``light tachyons'': a perturbatively stable scalar field of mass $m$ is one that obeys the Breitenlohner-Freedmann (BF) bound, given in 4d by
\beq
m^2 > - \frac{9}{4} \frac{1}{l^2} = \frac{3}{4} \Lambda   \ \Rightarrow \ \eta_V < \frac{3}{4} \label{BFbound}
\eeq
where $l$ is the AdS radius, and $\eta_V$ is defined as follows.\footnote{In $d$ dimensions, one has at an extremum $\frac{d-2}{2d}{\cal R}_d=\frac{V}{M_p^2}=\Lambda_d$. For AdS, one defines $\frac{1}{l^2}= \frac{-2\Lambda_d}{(d-1)(d-2)}$, and the BF bound is $m^2 l^2 > - \frac{(d-1)^2}{4}$. We also refer to footnote \ref{foot:Delta} about this bound.}

For $V \neq0$, it is useful to introduce the standard cosmological quantities
\beq
\epsilon_V = \frac{M_p^2}{2} \, \left( \frac{\nabla V}{V} \right)^2 \ ,\quad  \eta_V = M_p^2\ \frac{{\rm min}\, M}{V} \ , \label{epseta}
\eeq
where $\nabla V = \sqrt{g^{ij} \del_{\varphi^i} V \del_{\varphi^j} V}$ is the norm of the gradient, and $\ {\rm min}\, M =$ minimal eigenvalue of the mass matrix $M$. Indeed, an (anti-)de Sitter extremum is found when $\epsilon_V=0$, and its stability can be expressed through $\eta_V$, with $\eta_V <0$ for an unstable de Sitter.

Since the scalar fields result from a truncation of 10d fields, and as we tend to consider at first a small finite set of them, one may wonder what happens to the solution and its stability when including more fields. Let us consider that allowing for more fields does not change the potential dependence on the first ones: this amounts to say that the new fields had simply been fixed to their (constant) value at the extremum. In that case, by having the dependence on the new fields, the mass matrix becomes larger and incorporates the first mass matrix as a diagonal block (at least in a canonical basis). In the Lemma of \cite[Sec.3.3]{Andriot:2020wpp} (see \eqref{Lemma}), we showed that the minimal eigenvalue of the new larger mass matrix can then only be smaller that any eigenvalue of the first mass matrix. In other words, considering the dependence of more fields can only make the solution {\sl more unstable} than it was at first. To conclude, if the solution is already unstable with few fields, it will remain unstable by including more fields; this conclusion can also be reached using the Sylvester criterion \cite{Shiu:2011zt} (see \eqref{Sylvester}). On the contrary, declaring stability is only valid given a certain set of modes, and can a priori not be ensured in a more complete set-up with more fields (unless more arguments are brought, like symmetries). These considerations will be crucial in the following.\\

Obtaining 4d effective theories of string theory of the form \eqref{S4d} has been at the heart of string phenomenology for decades. It would take another review to summarize the options there. In the context of compactifications and classical regime, let us briefly mention two main avenues. A first one is the compactification on manifolds which possess certain geometric structures, such as Calabi-Yau manifolds, and generalisations thereof (SU(3)$\times$SU(3) structures, and more). These structures are such that background supersymmetry can be preserved. For instance, Minkowski and AdS solutions in such compactifications can preserve some supersymmetry, and the formalism makes this possibility manifest. Note that the background supersymmetry should be distinguished from the theory supersymmetry, the amount of which is typically low, but not zero. For example, in an ${\cal N}=1$ theory, the 4d scalar potential is written in terms of the K{\"a}hler potential $K$ and the superpotential $W$, while the field space metric is also given by $K$. In the aforementioned compactifications, the quantities $K, W$ can further be written in terms of the existing geometric structures on the manifold ${\cal M}$ (related e.g.~to the K{\"a}hler form $J$ and the holomorphic $(3,0)$-form $\Omega_3$): these are the central objects of the formulation. Due to supersymmetry and to the Ricci-flatness of a Calabi-Yau manifold, a 4d theory obtained on it is often a low energy effective theory, which highlights its physical relevance. Whether it is the case for SU(3)$\times$SU(3) structures (or more advanced ones) is a more difficult question, see however \cite{Smith:2024ejf}; these structures may still provide consistent truncations. Relevant reviews here include \cite{Grana:2005jc, Koerber:2010bx}.

A second avenue is that of gauged supergravities \cite{Samtleben:2008pe, Trigiante:2016mnt}. There, the amount of supersymmetries of the theory can be high, even maximal. The central object there, generating the scalar potential, is the embedding tensor, related to (gauged) symmetries, whose components can be interpreted as internal fluxes, curvature contributions, and more. The set of allowed manifolds ${\cal M}$ is broader than before (e.g.~spheres), since they are not required to carry the geometric structures previously mentioned. As a consequence, it is a priori less obvious whether the solutions preserve or not supersymmetry. This formalism typically leads to a consistent truncation in 4d (see examples in Section \ref{sec:groupmandimred}); whether the theory can be a low energy effective theory is however far from guaranteed. The two approaches still have overlaps, for instance when ${\cal M}$ is a torus or certain group manifolds (e.g.~nilmanifolds).

We will refer to one approach or the other when requiring a ``complete'' 4d theory. In contrast, we turn in the following to a 4d effective theory for 10d type II supergravities of the form \eqref{S4d}, that captures only few scalar fields. But those are universal to all compactifications.

\subsubsection{4d theory for $(\rho,\tau,\sigma)$}\label{sec:rhotau}

Having explained the first principles of dimensional reduction on a compact space, and properties of the resulting 4d theories with scalar fields, we now focus on a concrete and universal example. The truncation of interest will keep as 4d fields the 4d metric, and scalar fields denoted $(\rho,\tau,\sigma)$. To get there, we start with a compactification ansatz, which is nothing but the 10d solution ansatz of Section \ref{sec:10dansatz}. This is interpreted as the background, around which we will study fluctuation modes. The 4d scalar fields will come as scalar fluctuations, depending on 4d coordinates $x^{\mu}$, around background valued (internal) fields. Let focus at first on two 4d scalar fluctuations: $\rho$ is a diagonal fluctuation of the 6d metric, also called the volume field, and $\delta \phi$ is the fluctuation of the dilaton. We combine the latter with $\rho$ into the so-called 4d dilaton $\tau$, namely
\beq
g_{mn} = \rho(x) \, g_{mn}^0 (y) \ ,\ e^{\phi} = g_s \, e^{\delta \phi (x)} \ ,\ \tau(x) = e^{-\delta \phi}\, \rho^{\frac{3}{2}} \ , \quad \rho,\tau > 0 \ .
\eeq
From these definitions, the background is recovered by setting the 4d fields to the following values
\beq
\text{Background values:}\quad \rho =\tau = 1 \ .\label{backgroundvalues}
\eeq

Let us derive, as explained in Section \ref{sec:dimred}, the 4d gravity theory for $(\rho,\tau)$. This theory was first obtained in 4d in \cite{Hertzberg:2007wc, Silverstein:2007ac}, and for a spacetime of arbitrary dimension $d \geq 3$ in \cite{VanRiet:2011yc, Andriot:2022xjh}. To derive the 4d theory, we start with the following terms of 10d type II supergravities (see \eqref{SNSNS})
\beq
{\cal S}= \frac{1}{2 {\kappa}_{10}^2} \int \d^{10} x \sqrt{|g_{10}|}\, e^{-2\phi} \left( {\cal R}_{10} + 4 (\del \phi)^2  \right)  \ . \label{S1}
\eeq
There are two differences with \eqref{dimred1}. First, one has
\beq
\sqrt{|g_{10}|}\, e^{-2\phi} = g_s^{-2} \sqrt{|g_4|} \sqrt{|g_6^0|}\, \tau^2 \ ,
\eeq
which gives a $\tau(x)$ factor which was not there previously. Together with ${\cal R}_4$, the result is different than the usual 4d Einstein-Hilbert term. Reaching the latter requires to remove the $\tau$ factor: this is done by redefining the 4d metric towards $g_{\mu\nu E}$, going to the so-called 4d Einstein frame. The adequate redefinition is $g_{\mu\nu} = \tau^{-2} g_{\mu\nu E}$. Reaching the 4d theory \eqref{S4d}, given in Einstein frame, now requires to consider the 10d metric
\beq
\d s_{10}^2 = \tau^{-2}(x)\, g_{\mu\nu E}(x) \d x^{\mu} \d x^{\nu} + \rho(x)\, g_{mn}^0(y) \d y^m \d y^n \ . \label{10dmetric}
\eeq
In passing, note the difference with the maximally symmetric spacetime ansatz considered in \eqref{10dmetricmaxsym}, or its smeared version: here, dynamics of $\rho, \tau$ could allow to have $g_{\mu\nu E}(x)$ as an FLRW metric; we will come back to this point.

The second difference with \eqref{dimred1} is ${\cal R}_{10}$: evaluated on this metric, it will not be simply the sum ${\cal R}_4 + {\cal R}_6^0$. There will be extra factors, and more importantly, derivatives of the 4d fields. The term $(\del \phi)^2$ will also generate such derivatives. Remarkably, those derivatives combine to provide diagonal kinetic terms for the two 4d fields: we refer to \cite[App.D]{Andriot:2020wpp} for the detailed computation in 4d, and to \cite{Andriot:2022xjh} for the $d$-dimensional version. The result, starting from \eqref{S1}, is
\begin{gather}
\label{dimred2}
\begin{aligned}
{\cal S} & = \int \d^4 x \sqrt{|g_{4E}|}   \left(\frac{M_p^2}{2}  {\cal R}_{4E} - M_p^2\, \tau^{-2}(\del \tau)^2 - \frac{3}{4} M_p^2\, \rho^{-2}(\del \rho)^2  - V(\rho,\tau) \right) \ ,\\
{\rm where} &\quad V(\rho,\tau) = \frac{M_p^2}{2} \frac{\int \d^6 y \sqrt{|g_{6}^0|}\, \left( - \tau^{-2} \rho^{-1} {\cal R}_6^0\right)}{\int \d^6 y \sqrt{|g_{6}^0|}} \ ,
\end{aligned}
\end{gather}
and the kinetic terms are squared with the Einstein frame metric. Note this theory is of the form \eqref{S4d}. One is led to introduce the following canonical scalar fields
\beq
\hat{\tau} = \sqrt{2} M_p \ln \tau \ , \quad \hat{\rho} = \sqrt{\frac{3}{2}} M_p \ln \rho \ . \label{hattaurho}
\eeq
In the following, we will simplify notations. First, we will drop the subscript ${}_E$ on the 4d metric, at the risk of confusing with the initial 10d string frame 4d component; on the background ($\tau=1$) though, the two match. Second, we will drop the index ${}^0$ for background quantities in $V$, with a similar possible confusion. Finally, we will drop the integrals in the scalar potential: strictly speaking this would only be possible if integrated quantities (e.g.~${\cal R}_6^0$) were constant; here we do it all the time to simplify notations. In summary, we adopt the following notations
\beq
\text{Simplified notations:}\quad g_{\mu\nu E} \rightarrow g_{\mu\nu} \ ,\quad \frac{\int \d^6 y \sqrt{|g_{6}^0|}\, {\cal R}_6^0}{\int \d^6 y \sqrt{|g_{6}^0|}} \rightarrow {\cal R}_6 \ , \label{simplenotations}
\eeq
giving for now the potential to be expressed as $\frac{2}{M_p^2} V(\rho,\tau) =  - \tau^{-2} \rho^{-1} {\cal R}_6 $.

We now turn to the other terms of 10d type II supergravities action (see Appendix \ref{ap:actionsugra}), that will give further terms to $V$. The 6d metric and the dilaton appear in the flux square terms, and in the DBI term. For all those, this is only a multiplicative dependence, i.e.~these terms will not generate derivatives of the fields but only powers. We determine those by power counting, considering the overall Lagrangian as $\sqrt{|g_{10}|} e^{-2\phi} ( \dots )$, and fluctuating within the parentheses
\beq
|H|^2 \rightarrow \rho^{-3}\, |H|^2 \ ,\quad e^{2\phi}|F_q|^2 \rightarrow \tau^{-2}\rho^{3-q}\, g_s^2|F_q|^2 \ ,\quad e^{\phi} T_{10}^s \rightarrow \tau^{-1} \rho^{\frac{p-6}{2}}\, g_s T_{10}^s \ ,\label{powerrhotau}
\eeq
while the overall Lagrangian factor gives an extra $\tau^{-2}$. We recall that we use here the compactification ansatz described as the 10d solution in Section \ref{sec:10dansatz}, considering in particular in \eqref{powerrhotau} only internal fluxes, with $q\leq 5$.

A last contribution is that of 4d fluxes, $F_4^4$ and $F_5^4$, or $F_6$ and $*_6 F_5$, as given in \eqref{Fq4}. Proceeding as above, one naively gets terms of opposite sign and opposite field powers with respect to \eqref{powerrhotau}. But these contributions will eventually take the same form as above. This non-trivial fact is because these contributions involve a 4d $4$-form U(1) flux, say generically $f_4$, and this situation together with gauge invariance require an extra boundary term in the 4d action. The latter eventually contributes as well to the scalar potential, as explained in detail in \cite[App.A]{Andriot:2020lea}. The net effect is that a naive term $\varphi |f_4|^2$ becomes $-\varphi^{-1} |f_4|^2$, where $f_4$ is taken on-shell (or background-valued here) and $\varphi$ is a 4d scalar field. The generalisation to the $d$-dimensional case can be found in \cite{Andriot:2022xjh}. This results in having the same dependence as in \eqref{powerrhotau} for $F_6$ in IIA. In IIB, the 4d $F_5^4$ term contributes exactly as the internal term $1/2\, |F_5|^2$, leading to remove the $1/2$ and giving exactly the same contribution as for any other $F_q$.

We conclude with the following 4d theory for $(\rho,\tau)$, obtained with the compactification ansatz of Section \ref{sec:10dansatz}, and using the simplified notations \eqref{simplenotations}
\begin{eqnarray}
&& \hspace{-0.3in} {\cal S} = \int \d^4 x \sqrt{|g_{4}|}   \left(\frac{M_p^2}{2}  {\cal R}_{4} - \frac{1}{2} \left( \frac{2 M_p^2}{\tau^{2}} (\del \tau)^2 + \frac{3M_p^2}{2 \rho^{2}} (\del \rho)^2 \right)  - V(\rho,\tau) \right) \ ,\label{rhotautheory}\\
&& \hspace{-0.3in} \frac{2}{M_p^2} V(\rho,\tau) =  \tau^{-2} \left(-\rho^{-1} {\cal R}_6 + \frac{1}{2} \rho^{-3}\, |H|^2 \right) - \tau^{-3} \sum_{{\rm sources}} \rho^{\frac{p-6}{2}}\, g_s \frac{T_{10}^s}{p+1} + \frac{1}{2} \tau^{-4} \sum_{q=0}^6 \rho^{3-q}\, g_s^2|F_q|^2  \ .\nn
\end{eqnarray}
Let us emphasize that this theory is of the form \eqref{S4d}, the discussion in Section \ref{sec:dimred} therefore applies.\\ 

A first question is whether this is a low energy effective theory of 10d type II supergravities: we recall from that section that this requires {\sl only} light fields, and {\sl all} of them, to be present in the 4d theory. Let us first point-out that $\rho,\tau$ are ``universal'' scalar fields, in the sense that they would appear as fluctuations in any compactification of type II supergravities; this is the main interest of this dimensional reduction. On a more technical note, if one would perform a Kaluza--Klein reduction and keep only the zero-modes (i.e.~the field components independent of internal coordinates), thereby having a chance (although not strictly guaranteed) to keep only the lightest modes, the volume and the dilaton would be part of the 4d fields. It is therefore very likely that they are always present in a low energy effective theory obtained from a compactification of type II supergravities. A definite answer can only be obtained by a detailed case-by-case study of the mass spectrum and the energy scales. However, the other important point remains, namely whether other (light) 4d fields should be included. A compactification that would only lead to $\rho,\tau$ is fairly specific (e.g.~on a compact Einstein manifold); in many known examples, more fields also occur, and we will give an example in the following with $\sigma$, or also in \eqref{potIIAB}. We refer to \cite{Andriot:2025gyr} for a discussion on the minimal number of scalar fields in effective theories from string theory. To conclude, we can say that in most (known) compactifications, the 4d theory \eqref{rhotautheory} is not the complete low energy effective theory of type II supergravities, but it is still part of it, offering a glimpse of the physics through the two universal fields $\rho,\tau$. The missing fields can be interpreted as ``frozen'', since we have not considered the corresponding fluctuations in the truncation. We will also come back to $\rho,\tau$ as a consistent truncation.

The ``low energy'' approximation we are referring to is with respect to 10d supergravity, one of the main energy scales in between being therefore related to the typical length of the 6 extra dimensions, say a radius $R$. There is however another low energy approximation that has been considered previously: the one making 10d supergravities effective theories for string theories. This one requires the string length $l_s$ to be negligible compared to the typical lengths encountered, here the relevant one being $R$, a priori smaller than the 4d lengths. Having a small $l_s/R$ allows to neglect the massive string modes, as well as the $\alpha'$-corrections to 10d supergravities: this is part of the classical regime approximations. Having a large volume $\rho$ is consistent with a large $R/l_s$, since $g_{mn} = \rho\, g_{mn}^0$. Similarly, neglecting the string loops as required in the classical regime amounts to have a small value for $e^{\phi}$, which eventually amounts to a large $\tau$. In short, it is important to note the adequate 4d field direction, namely the large volume and weak coupling limits, to respect the approximation and regime
\beq
\text{(Asymptotic) classical regime:}\quad \rho \rightarrow \infty \ ,\quad \tau \rightarrow \infty \ . \label{asymptoticclass}
\eeq
A consequence for the 4d theory \eqref{rhotautheory} is that it cannot be trusted, as a valuable string effective theory, in one of the opposite limits where one field vanishes.\\

As discussed in Section \ref{sec:dimred}, the extrema of the 4d potential correspond to solutions with maximally symmetric spacetimes. Here, this is obtained with ${\cal R}_4 = 4 V / M_p^2$ and $\del_{\rho} V = \del_{\tau} V = 0$. Setting the background values \eqref{backgroundvalues}, i.e.~$\rho=\tau=1$, to be those at such a critical point of $V$, we obtain the following equations on the background fields
\begin{subequations}\label{CriticalpointV}
\begin{align}
(V)\quad \quad & {\cal R}_4 + 2{\cal R}_6 - |H|^2 + 2 \sum_{{\rm sources}}  g_s \frac{T_{10}^s}{p+1} - \sum_{q=0}^6  g_s^2|F_q|^2 = 0  \label{eqV} \\
(\del_{\rho} V=0) \quad \quad & {\cal R}_6 - \frac{3}{2}  |H|^2  - \sum_{{\rm sources}} \frac{p-6}{2}\, g_s \frac{T_{10}^s}{p+1} + \frac{1}{2} \sum_{q=0}^6 (3-q)\, g_s^2|F_q|^2 = 0\label{eqdelrho}  \\
(\del_{\tau} V=0) \quad \quad &  2{\cal R}_6 -  |H|^2  +3 \sum_{{\rm sources}} \, g_s \frac{T_{10}^s}{p+1} -2 \sum_{q=0}^6  g_s^2|F_q|^2 = 0\label{eqdeltau}
\end{align}
\end{subequations}
It is straightforward to verify that these are equivalent to certain 10d equations of motion, namely the 4d Einstein trace, the 10d Einstein trace (or equivalently the 6d trace), and the dilaton e.o.m. Indeed, one can show the following relations
\begin{gather}
\label{equiv}
\begin{aligned}
\eqref{Einstein4d} &= 2\, \eqref{eqV} - \eqref{eqdelrho} - \frac{3}{2}\, \eqref{eqdeltau} \\
\eqref{10dtraceEsmear} &= 4\, \eqref{eqV} - \eqref{eqdelrho} - \frac{3}{2}\, \eqref{eqdeltau} \\
\eqref{dileom} &= 2\, \eqref{eqV} - \eqref{eqdeltau}
\end{aligned}
\end{gather}
Given the relation of $\rho, \tau$ to the 10d fields, this equivalence is not surprising. This is however an explicit verification that the background fields here indeed correspond to a 10d solution of supergravity. This is also because the 10d solution ansatz was tailored for a maximally symmetric spacetime. We conclude that {\sl we can learn on maximally symmetric spacetime solutions of 10d supergravities by studying extrema of this 4d theory}. We will make use of this important relation to prove non-existence of solutions (no-go theorems), or instability of the solution (as explained in Section \ref{sec:dimred}).

In the case where all other 10d equations were satisfied, e.g.~thanks to an adequate, more restrictive, ansatz, then a solution to the 4d theory would automatically be a solution of the 10d theory: this is the actual definition of a {\sl consistent truncation}, a concept introduced in Section \ref{sec:dimred}. Indeed, if the 4d field and their equations are sufficient to have complete 10d solution, then they can be considered as an independent set of modes, as introduced previously. We see here that this is a possibility for maximally symmetric spacetime.

One may wonder whether such a relation could be possible as well for a rolling solution, i.e.~along a slope of the potential. In such a 4d solution, $\rho,\tau$ cannot be set to the fixed ``background values'' \eqref{backgroundvalues}, since they are dynamical. Getting the same solution in 10d would require to deviate from the 10d solution ansatz for a maximally symmetric spacetime \eqref{10dmetricmaxsym}, to one with e.g.~the 10d metric considered here \eqref{10dmetric}, and derive the corresponding 10d e.o.m. The 4d metric should also be promoted to a more dynamical one, such as FLRW. Without more detail, let us mention that such a consistent truncation of 10d type II supergravities (on time-dependent compactifications) towards 4d rolling solutions can be realised, for example recently in \cite{Marconnet:2022fmx}, reviewed in \cite[Sec.4.1]{Andriot:2023wvg}.\\

We now include one more field, $\sigma$, analogously to $\rho,\tau$. The 4d potential for this field was first proposed in \cite{Danielsson:2012et}, and revisited more completely in \cite{Andriot:2018ept}; the derivation in arbitrary dimension $d\geq3$ can be found in \cite{Andriot:2022xjh}. As we will see in Section \ref{sec:nogo}, in the considered compactification setting, orientifolds are a necessary ingredient to obtain a de Sitter or a Minkowski solution. In addition, the compactification ansatz of Section \ref{sec:10dansatz} assumes that directions parallel and transverse to an $O_p$ can be globally identified in the 10d spacetime. As a consequence, we can split the 6d metric into parallel and transverse dimensions for a given $O_p$. It is then natural to introduce a scalar field capturing the fluctuation of the parallel or transverse dimensions. The (almost) genericity of having an $O_p$ makes $\sigma(x)$ a (semi-)universal field. For a given $O_p$, the field $\sigma >0$ is defined through the following 10d metric that builds on \eqref{10dmetric}
\beq
\d s_{10}^2 = \tau^{-2}\, \d s_{4E}^2 + \rho \left(\sigma^A (\d s_{||}^2)^0 + \sigma^B (\d s_{\bot}^2)^0  \right) \ , \quad A= p-9\ ,\ B= p-3 \ . \label{10dmetricsigma}
\eeq
This definition only makes sense for $4\leq p \leq 8$, giving in turn $AB \neq 0$. $A$ and $B$ are chosen in such a way that $\sigma$ does not appear in the 6d volume, $\sqrt{|g_6|} = \rho^3 \sqrt{|g_6^0|}$. This convenient choice makes $\rho$ still the volume (fluctuation), while $\sigma$ captures off-diagonal fluctuations parallel and transverse to the $O_p$. The definition of $\tau$ is then the same. As before, the background value for this new field is $\sigma =1$.

The kinetic term was computed in 4d in \cite[App.B]{Andriot:2019wrs} (see also \cite[App.D]{Andriot:2020wpp}), and in arbitrary dimension $d\geq 3$ in \cite{Andriot:2022xjh}, giving the following 4d canonically normalised field
\beq
\hat{\sigma} = \sqrt{\frac{3(9-p)(p-3)}{2}} M_p \ln \sigma \ .
\eeq
Deriving the scalar potential requires to identify parallel and transverse directions in the field components. For internal fluxes, we use the notation $F_q^{(n)}$ defined in \eqref{F(n)}: it denotes the form with components have $n$ parallel legs and $q-n$ transverse ones. One also has $|F_q|^2 = \sum_n |F_q^{(n)}|^2$; this allows to extract the multiplicative dependence in $\sigma$ as above. For the 4d fluxes, one has to proceed as described above, and we refer again to \cite[App.A]{Andriot:2020lea}. Getting the sources contributions is straightforward and multiplicative, provided all sources have the same dimensionality $p$ and are parallel to each other; we will come back to different situations. Eventually, the complete 4d theory for $(\rho,\tau,\sigma)$ is given by
\begin{eqnarray}
&& \hspace{-0.3in} {\cal S} = \int \d^4 x \sqrt{|g_{4}|}   \left(\frac{M_p^2}{2}  {\cal R}_{4} - \frac{1}{2} \left( \frac{2 M_p^2}{\tau^{2}} (\del \tau)^2 + \frac{3 M_p^2}{2 \rho^{2}} (\del \rho)^2 + \frac{3(-AB) M_p^2}{2 \sigma^{2}} (\del \sigma)^2 \right)  - V(\rho,\tau,\sigma) \right) \ ,\nn\\
&& \hspace{-0.3in} \frac{2}{M_p^2} V(\rho,\tau,\sigma) =  \tau^{-2} \left(-\rho^{-1} {\cal R}_6(\sigma) + \frac{1}{2} \rho^{-3}\sum_n \sigma^{-An-B(3-n)} |H^{(n)}|^2 \right) \label{rhotausigmatheory}\\
&& \hspace{-0.3in} \phantom{\frac{2}{M_p^2} V(\rho,\tau,\sigma)} - \tau^{-3} \rho^{\frac{p-6}{2}} \sigma^{\frac{AB}{2}}\, g_s \frac{T_{10}}{p+1} + \frac{1}{2} \tau^{-4} \sum_{q=0}^6 \rho^{3-q} \sum_n \sigma^{-An-B(q-n)} g_s^2|F_q^{(n)}|^2  \ ,\nn
\end{eqnarray}
where we used again simplified notations \eqref{simplenotations}. We also took into account the restriction on sources (parallel of same dimensionality $p$), and recall the trace $T_{10} =\sum_s T_{10}^s$. The curvature dependence ${\cal R}_6(\sigma)$ is also multiplicative, but can be involved: we refer to \cite[(2.32)]{Andriot:2018ept} for a generic expression, which simplifies to \cite[(2.11)]{Andriot:2020lea} on group manifolds. Finally, when restricting to constant flux components subject to orientifold projection conditions, as in \eqref{fluxlistconstant}, the $\sigma$ dependence of the flux terms simplifies: see e.g.~\cite[(3.2)]{Andriot:2018ept}.

As can be guessed from the definition of $\sigma$, the equation $\del_{\sigma} V = 0$ is related to the trace of the internal Einstein equation \eqref{Einstein6d} along parallel (or transverse) directions. More precisely, as in \eqref{equiv}, a linear combination of $\del_{\sigma} V = 0$ with the other 3 equations can be shown to be identical to this partial internal Einstein trace, provided (the integral of) one curvature term vanishes \cite[(2.49)]{Andriot:2018ept}; the latter condition holds on group manifolds, and is likely to happen more generally. This result extends the equivalence between the 10d and 4d extremum solutions, that we will use in the following.

In general, one may have sources of different dimensionalities, and/or non-parallel ones, that we sometimes refer to as intersecting \cite{Andriot:2017jhf}. This situation leads to defining multiple fields $\sigma_I$. For the kinetic terms, we then refer to \cite[App.D]{Andriot:2020wpp} and \cite{Andriot:2022yyj}. For the potential, we refer to \cite[(4.11)]{Andriot:2019wrs}, and to \cite{Andriot:2020wpp, Andriot:2022yyj} for explicit examples. The same equivalence of $\del_{\sigma_I} V =0$ to the internal trace along parallel directions, as for the single $\sigma$ case, is shown for intersecting sources in \cite[App.C]{Andriot:2019wrs}.\\

We conclude this section by observing that the scalar potential is made of {\sl exponentials}. This can be seen by going to the canonically normalised basis of fields, which is the only one that allows a proper comparison of theories. To illustrate this point we rewrite the $(\rho,\tau)$ theory \eqref{rhotautheory} as follows
\begin{eqnarray}
&& {\cal S} = \int \d^4 x \sqrt{|g_{4}|}   \left(\frac{M_p^2}{2}  {\cal R}_{4} - \frac{1}{2} \left(  (\del \hat{\tau})^2 + (\del \hat{\rho})^2 \right)  - V(\rho,\tau) \right) \ ,\label{rhotautheoryexp}\\
&& \frac{2}{M_p^2} V(\rho,\tau) =  e^{-\sqrt{2} \frac{\hat{\tau}}{M_p}} \left(- e^{-\sqrt{\frac{2}{3}} \frac{\hat{\rho}}{M_p}} {\cal R}_6 + \frac{1}{2} e^{-\sqrt{6} \frac{\hat{\rho}}{M_p}}\, |H|^2 \right) - e^{-\frac{3}{\sqrt{2}} \frac{\hat{\tau}}{M_p}} \sum_{{\rm sources}}  e^{\frac{p-6}{\sqrt{6}} \frac{\hat{\rho}}{M_p}} \, g_s \frac{T_{10}^s}{p+1} \nn\\
&& \phantom{\frac{2}{M_p^2} V(\rho,\tau) } + \frac{1}{2} e^{-2\sqrt{2} \frac{\hat{\tau}}{M_p}} \sum_{q=0}^6 e^{\sqrt{\frac{2}{3}}(3-q) \frac{\hat{\rho}}{M_p}}\, g_s^2|F_q|^2  \ ,\nn
\end{eqnarray}
and the same could be done with $\sigma$ (see e.g.~\cite[(5.3)]{Andriot:2022brg} for a $d$-dimensional version). Such an exponential behaviour is typical of a classical or asymptotic string effective theory. In addition, exponentials of $\hat{\tau}$ are all decreasing (in absolute value) in the classical asymptotic direction \eqref{asymptoticclass} $\hat{\tau} \rightarrow \infty$, bringing the potential to vanish. Decreasing exponentials in positive potentials will be important in the context of cosmology.

Since the volume and the dilaton are universal fields in 4d theories obtained from a string compactification, exponentials will always be present. Including more scalar fields can bring different behaviours in the potential: for instance, axions appear through polynomials. But exponentials then remain as overall factors. Another option to change the potential is to go away from the classical regime, which we do not consider for now.

We will now make use of all the material introduced so far, to get constraints on the existence and stability of de Sitter solutions, as well as on rolling solutions.

\subsection{No-go theorems}\label{sec:nogo}

We introduced in Section \ref{sec:10dansatz} a 10d solution ansatz with 4d maximally symmetric spacetimes. We start here by presenting in Section \ref{sec:nogo10d} constraints on their existence, mostly focusing on de Sitter. These constraints take the form of no-go theorems, in the sense that they emphasize the impossibility of getting a solution without a certain ingredient. The outcome will not be a complete exclusion of classical de Sitter solutions, rather an identification of all required ingredients in the compactification (i.e.~curvature, fluxes, sources); explicit solutions will then be discussed in Section \ref{sec:dSsolexistence}.

We explained in Section \ref{sec:dimred} how a 4d effective theory can also exhibit solutions with maximally symmetric spacetimes, as extrema of the potential, and how such a theory allows to study the perturbative stability of the solution. In addition, we gave in Section \ref{sec:rhotau} an explicit and universal 4d theory whose equations are equivalent to some of the 10d ones. We then use in Section \ref{sec:nogo4d} this 4d theory to rederive the same no-go theorems on the existence of extrema. Doing so will have the advantage to get constraints beyond extrema, on the slope of the scalar potential. We finally discuss in Section \ref{sec:nogostab} no-go theorems on perturbative stability. Those will not allow, at this stage, to completely exclude a classical (meta)stable de Sitter solution.\\

The prime example of an existence no-go theorem against de Sitter, derived with 10d equations, is the so-called Maldacena-Nu\~nez no-go theorem \cite{Gibbons:1984kp, deWit:1986mwo, Maldacena:2000mw, Townsend:2003qv}. It will be circumvented by including orientifolds. A first no-go theorem explicitly formulated in 4d with a scalar potential was obtained for a type IIA compactification on a Calabi-Yau \cite{Hertzberg:2007wc}; in the following, it will be extended to require a negative internal curvature of ${\cal M}$. A more systematic treatment, including stability, was first discussed in \cite{Silverstein:2007ac}. Since then, many papers have investigated no-go theorems. A recent list in 4d can be found in \cite[Sec.2.3]{Andriot:2020lea}. A similar list for $d$-dimensional maximally symmetric spacetimes, $d\geq3$, obtained in 10d or in the $d$-dimensional effective theory, can be found in \cite{Andriot:2022xjh}.

\subsubsection{Existence no-go in 10d}\label{sec:nogo10d}

A no-go theorem, against the existence of a solution with a certain 4d maximally symmetric spacetime, is obtained as follows in 10d. One considers a combination of 10d equations (having to be satisfied on the solution), that gives a specific expression of ${\cal R}_4$. Provided an assumption, the sign of ${\cal R}_4$ may then get fixed, in which case one can conclude. For instance, if ${\cal R}_4 \leq 0$, this proves there is no de Sitter solution.\footnote{For cosmological observations, it would be important to have ${\cal R}_4$ of the 4d Einstein frame metric, while the ${\cal R}_4$ used here corresponds at first sight to the 4d component of the 10d string frame metric. However, the smeared approximation makes $e^A$ irrelevant, and the focus on extremum solutions sets $\tau=1$. These two points erase any (on-shell) distinction between these frames.} We start by giving several examples of such no-go theorems, for both type IIA and IIB.

\begin{enumerate}[label=\textbf{\roman*.},resume]
\item \label{nogo1}{\bf Maldacena-Nu\~nez} \cite{Maldacena:2000mw}

Considering the combination of 10d equations $\frac12 \eqref{Einstein4d} -\frac12 \eqref{10dtraceEsmear} + \eqref{dileom}$, we obtain the following expression
  \beq
  {\cal R}_4 = g_s \sum_{{\rm sources}} \frac{T_{10}^s}{p+1} - g_s^2 \sum_{q=0}^{6} |F_q|^2 \ , \label{R4MaldNun}
  \eeq
  where here and in the following, one should restrict to appropriate RR fluxes in type IIA/IIB. The only positive terms in this expression of ${\cal R}_4 $ are, in our compactification ansatz, the orientifold contributions. Then, {\sl to obtain a de Sitter solution, we must have an $O_p$}, giving one $T_{10}^s >0$. Note that a stronger conclusion can be reached, since the whole sum on source contributions needs to be positive. In the particular case of sources of a single dimensionality $p$, we deduce the requirement $T_{10} =\sum_{{\rm sources}} T_{10}^s >0$: in other words, orientifold contributions must dominate those of $D_p$-branes. Actually, {\sl the same conclusion applies to a Minkowski solution}, in case one requires at least one non-zero RR flux, or some $D_p$.
  
  Unsmeared versions of this no-go theorem exist, starting with \cite{Maldacena:2000mw}. This underlines the necessity, at least in such a compactification framework, to have singularities in ${\cal M}$ due to localized sources \cite{Denef:2008wq}, i.e.~move away from a smooth geometry, if one wants to have a positive cosmological constant. As discussed in Section \ref{sec:10dansatz} and \ref{sec:backreact}, the treatment of the corresponding backreaction is sometimes delicate. It should be noted that accelerated expansion via (classical) rolling solutions can circumvent this no-go theorem, thus avoiding such matters (see e.g.~\cite{Townsend:2003qv}, or more recently \cite[Sec.4.3]{Andriot:2023wvg}). Other options, discussed in Section \ref{sec:listattempts}, amount to leave the present framework and include other contributions, in particular corrections.
  
\end{enumerate} 
  
\begin{enumerate}[label=\textbf{\roman*.},resume]
\item \label{nogo2}{\bf No-go for $p=9,8,7$, or $p=6,5,4$ with $F_{6-p}=0$} \cite{Wrase:2010ew, Andriot:2016xvq}
  
  This no-go theorem was first derived in 4d in \cite{Wrase:2010ew}, and in 10d in \cite{Andriot:2016xvq}. It considers sources of a single dimensionality $p$. The combination $\frac{p-3}{2} \eqref{Einstein4d} + \frac{7-p}{2} \eqref{10dtraceEsmear} + (p-7)\eqref{dileom}$ gives the following expression
  \beq
  (p-3)\, {\cal R}_4 = g_s^2 \sum_{q=0}^{6} (8-p-q) |F_q|^2 - 2|H|^2 \ .
  \eeq
  This allows to conclude on a {\sl no-go against de Sitter for $p=9$, $8$ or $7$}. In addition, there is {\sl no de Sitter solution for $p=6$, $5$ or $4$ if $F_{6-p} =0$}. For $p=4$, it is necessary to recall the point made in \cite{Andriot:2018ept}, concluding that $F_0=0$ and providing the no-go. Indeed, the single dimensionality $p=4$ means here an absence of $p=8$ sources. In that case, the BI $\d F_0=0$ gives a constant $F_0$, which does not survive the $O_4$ projection conditions \eqref{Opflux}.
  
  We refer to \cite[Sec.2.2.2]{Andriot:2022xjh} for more details. An extension to sources of multiple dimensionalities can also be found there. Finally, an unsmeared version of this no-go theorem can be found in \cite[Sec.3]{Andriot:2016xvq}, and a more general one for $p=8$ in \cite{Cribiori:2019clo}.
  
\end{enumerate}

\begin{enumerate}[label=\textbf{\roman*.},resume]
\item \label{nogo3}{\bf No-go for positive or vanishing ${\cal R}_6$} \cite{Wrase:2010ew, Andriot:2016xvq}  
  
  This no-go theorem was first derived in 4d in \cite{Wrase:2010ew} (see also \cite{Hertzberg:2007wc} for $p=6$), and in 10d in \cite{Andriot:2016xvq}. It considers sources of a single dimensionality $p$. The combination $\frac{p-1}{8}\eqref{Einstein4d} +\frac{5-p}{8} \eqref{10dtraceEsmear} +\frac{p-3}{4} \eqref{dileom}$ gives the following expression
  \beq
  \frac{(p+3)}{4}\, {\cal R}_4 = \frac{g_s^2}{4} \sum_{q=0}^{6} (6-p-q) |F_q|^2 - {\cal R}_6 \ . \label{R4nogoR6}
  \eeq
  Using again that $F_0=0$ for $p=4$ \cite{Andriot:2018ept}, one concludes on a {\sl no-go against de Sitter for $p\geq4$ if ${\cal R}_6 \geq 0$}. An extension to general dimension $d\geq 3$ can be found in \cite[Sec.2.2.3]{Andriot:2022xjh} (see also \cite{VanRiet:2011yc}). Finally, the no-go theorem also works for $p=3$ with $F_1=0$.
  
  It is straightforward to extend the above to multiple dimensionalities: we consider the same combination of equations without restricting to a single source dimensionality. We obtain
  \beq
  \frac{(p+3)}{4}\, {\cal R}_4 = \frac{g_s^2}{4} \sum_{q=0}^{6} (6-p-q) |F_q|^2 - {\cal R}_6 + \frac{g_s}{4} \sum_{p'{\rm -sources}} \frac{T_{10}^s}{p'+1} (p-p')  \ . 
  \eeq
  At first sight, it seems the new terms have the wrong sign to circumvent the requirement ${\cal R}_6 <0$.\footnote{Consider for instance a flat 6-torus in IIA. The BI sourced by 6-sources requires a non-zero $F_0$. This implies having no $O_4$. The $D_4$ give precisely the wrong sign in the above with $p=6, p'=4$. The same reasoning can be made in IIB with 5- and 3-sources, having a non-zero $F_1$. A manifold allowing for non-closed fluxes would deserve more investigation.}
  
\end{enumerate}
  
\begin{enumerate}[label=\textbf{\roman*.},resume]
\item \label{nogo4}{\bf No-go for $p=3$} \cite{Blaback:2010sj, Andriot:2016xvq}  
  
   An interest of this no-go theorem is that beyond the three equations \eqref{3eq}, it requires a sourced flux BI which is a purely 10d equation, as well as an integration. We start with the combination $\frac{p+1}{2} \eqref{Einstein4d} + \frac{3-p}{2} \eqref{10dtraceEsmear} + (p-3) \eqref{dileom}$, considering again sources of single dimensionality $p$, giving
  \beq
  {\cal R}_4 = \frac{1}{p+1} \left(4 g_s \frac{T_{10}}{p+1} - 2 |H|^2 - g_s^2 \sum_{q=0}^6 (p+q-4) |F_q|^2 \right) \ .
  \eeq
  We now focus on $p=3$ (a first extension to higher $p$ can be found in \cite[(4.5)]{Andriot:2016xvq} in 10d and \cite[(2.47)]{Andriot:2020lea} in 4d). In that case, ${\rm vol}_{\bot} = {\rm vol}_6$. The sourced BI \eqref{BI} can be rewritten as follows
  \begin{gather}
  \label{BIrewriting}
  \begin{aligned}
  & \d F_5 - H \w F_3 = - \frac{T_{10}}{4}\, {\rm vol}_6 \\
  \Leftrightarrow \ \ & 2 g_s \frac{T_{10}}{4} = - |*_6 H - g_s F_3 |^2 + |H|^2 + g_s^2 |F_3|^2 - 2 g_s (\d F_5) \ ,
  \end{aligned}
  \end{gather}
  by projecting on ${\rm vol}_6$ with the notation $\d F_5 = (\d F_5) {\rm vol}_6$. We deduce the expression
  \beq
  {\cal R}_4 = - g_s (\d F_5) - \frac{1}{2} \left( |*_6 H - g_s F_3 |^2  + 2g_s^2|F_5|^2  \right) \ .
  \eeq
  We are left to integrate this expression on ${\cal M}$ which has no boundary
  \beq
  \int_{{\cal M}}  {\rm vol}_6 (\d F_5) = \int_{{\cal M}}  \d F_5 = \int_{\del {\cal M} =0}  F_5 = 0 \ .
  \eeq
  Since $\int_{{\cal M}}  {\rm vol}_6 {\cal R}_4 = {\cal R}_4 \int_{{\cal M}}  {\rm vol}_6 > 0 $ for a de Sitter solution, we conclude on a {\sl no-go against de Sitter for $p=3$}.
  
  As for the other no-go theorems, this result can be generalized to an external spacetime of arbitrary dimension $d\geq4$ for $p=d-1$ (and a refinement for $d=3$). We refer to \cite[Sec. 2.2.4]{Andriot:2022xjh} about it, where in addition, a version for sources of multiple dimensionalities can be found. An unsmeared version of this no-go theorem can also be found in \cite{Blaback:2010sj, Andriot:2016xvq}.  
  
\end{enumerate}
  
\begin{enumerate}[label=]
\item {\bf Summary (so far)}
 
  The four no-go theorems against de Sitter obtained so far indicate necessary ingredients to a classical de Sitter solution, that respects the ansatz of Section \ref{sec:10dansatz}. Restricting to sources of single dimensionality $p$, the ingredients are
  \beq
  \text{$T_{10}>0$ (implying having $O_p$), $\ {\cal R}_6 <0$, $\ p=4$, $5$ or $6$,$\ $ and $F_{6-p} \neq0$}\ . \label{summarydS}
  \eeq
  To reach these claims, the manifold has not been specified, and sources can be parallel or intersecting (a distinction that will be clarified below). The information \eqref{summarydS} will be useful in Section \ref{sec:dSsolexistence} to actually find de Sitter solutions. In Section \ref{sec:nogo4d}, we will give 4d versions of the above four no-go theorems.
  
  These results have been derived along the years, going from case by case no-go theorems to the more general formulations here above. Early works discussing existence no-go theorems, mostly in type IIA with $O_6$, but also in the T-dual setting in type IIB with $O_5, O_7$, include \cite{Hertzberg:2007wc, Silverstein:2007ac, Haque:2008jz, Caviezel:2008tf, Flauger:2008ad, Danielsson:2009ff, Caviezel:2009tu}. In 2010, the status among the various compactification options was summarized in Table 1 of \cite{Wrase:2010ew}, which became for some time in the literature a standard in presenting the constraints (see e.g.~\cite[Tab.1]{Obied:2018sgi}). The table, finally updated to the above results, appeared in 2019 in\cite[Tab.1]{Andriot:2019wrs}, and we give it for completeness here as Table \ref{tab:nogosummary}.
\begin{table}[H]
\begin{center}
\begin{tabular}{|c||c|c|}
    \hline
     & \multicolumn{2}{|c|}{} \\[-8pt]
     & \multicolumn{2}{|c|}{$\quad$ A de Sitter solution requires $T_{10}>0$ (\ref{nogo1}) and $\quad$}\\[5pt]
    \hhline{~||~~}
    $p$ & \hspace{0.52in} ${\cal R}_6 \geq 0$ \hspace{0.52in} & ${\cal R}_6 <0$ \\[3pt]
    \hhline{===}
     & \multicolumn{2}{|c|}{} \\[-8pt]
    3 & \multicolumn{2}{|c|}{(\ref{nogo4})} \\[3pt]
    \hhline{-||--}
     & & \\[-8pt]
    4 &  &  \\[3pt]
    \hhline{-||~~}
     & & \\[-8pt]
    5 & (\ref{nogo3}) & $F_{6-p}\ (\ref{nogo2})$ \\[3pt]
    \hhline{-||~~}
     & & \\[-8pt]
    6 &  &  \\[3pt]
    \hhline{-||--}
     & & \\[-8pt]
    7 &  &  \\[3pt]
    \hhline{-||~~}
     & & \\[-8pt]
    8 & (\ref{nogo2}) or (\ref{nogo3}) & (\ref{nogo2}) \\[3pt]
    \hhline{-||~~}
     & & \\[-8pt]
    9 &  &  \\[3pt]
    \hline
\end{tabular} \caption{Each ({\bf number.}) refers to a no-go theorem listed above. Their presence in an empty cell implies the absence of de Sitter solutions in the corresponding compactification, while together with an entry (flux, $T_{10}$), they indicate a necessary ingredient. The only options for a de Sitter solution (with sources of single dimensionality $p$) are left in only one cell, in agreement with \eqref{summarydS}.}\label{tab:nogosummary}
\end{center}
\end{table} 

\end{enumerate}

In the remainder of this section, we mention other no-go theorems against the existence of de Sitter solutions. Those are found by using more technicalities (e.g.~focusing on directions parallel to a source), or when going to more specific situations (e.g.~restricting to group manifolds) or by going away from above restrictions (e.g.~considering multiple dimensionalities). In the following, we list such no-go theorems, staying within the ansatz of Section \ref{sec:10dansatz}. We will eventually discuss others beyond this ansatz in Section \ref{sec:circumvent}.

\begin{enumerate}[label=\textbf{\roman*.}]
\setcounter{enumi}{4}
\item \label{nogo5}{\bf Heterotic at order ${\alpha'}^0$}
  
Many constraints exist in heterotic string against de Sitter solution (see Section \ref{sec:circumvent}). Here we focus on its simplest version: tree-level and ${\alpha'}^0$. This boils down heterotic string to the NSNS sector of type II, and therefore fits within our ansatz. It is then easy to show that the only solutions are Minkowski for an external spacetime of dimension $d\geq 4$: see for instance \eqref{R4MaldNun} restricted to NSNS. For $d=3$ however, there is an option for AdS: see e.g.~\cite[Sec.2.2.5]{Andriot:2022xjh}. Here in $d=4$, we rather propose, for later convenience, the obstruction against de Sitter in the following form: we consider $\frac{1}{2} \eqref{Einstein4d} + \frac{3}{2} \eqref{10dtraceEsmear} -3 \eqref{dileom}$ with $T_{10}^s=0, F_q=0$, giving
\beq
{\cal R}_4 = -2 |H|^2 \ .
\eeq
{\sl This forbids any de Sitter solution}.

\end{enumerate}  

\begin{enumerate}[label=\textbf{\roman*.}]
\setcounter{enumi}{5}
\item {\bf Other no-go theorems within our solution ansatz}

\begin{enumerate}[label=\textbf{\alph*.}]
\item {\bf Parallel directions}

Other no-go theorems against de Sitter can be found when splitting internal dimensions into those parallel or transverse to a source. In 10d, the internal Einstein equation \eqref{Einstein6d} can be split in this way, as done e.g.~in \cite[Sec.3.2]{Andriot:2019wrs} for ${\cal M}$ being a group manifold and for parallel sources only, giving a no-go theorem on the parallel components of the equation, in the form of a condition on the Ricci tensor: ${\cal R}_{a_{||}a_{||}}>0$. A 4d version was obtained in \cite[Sec.2.3]{Andriot:2020lea} (no-go theorem 9), using an effective theory with a radion field. It got generalized to arbitrary dimension $d$ in \cite[Sec.4.2.7]{Andriot:2022xjh}.

One can further trace the internal Einstein equation along parallel dimensions, giving e.g.~\cite[(6)]{Andriot:2016xvq} in 10d, related in 4d to $\del_{\sigma} V$ as explained in \cite[(2.48)]{Andriot:2018ept} and mentioned in Section \ref{sec:rhotau}. At this stage it becomes important to distinguish the situation of parallel sources (and therefore of same dimensionality $p$), or intersecting ones (and possibly different dimensionalities $p$), since the trace of the energy momentum tensor for sources $T_{MN}$ contributes differently. Focusing on parallel sources, one deduces from this trace several no-go theorems against de Sitter, or conditions. A first one is obtained on curvature terms \cite[(12)]{Andriot:2016xvq}, whose 4d version is discussed in \cite[Sec.2.3]{Andriot:2020lea} (no-go theorem 5). For completeness, let us briefly state these conditions without entering details of the notations: de Sitter requires $|f^{||}{}_{\bot \bot}|^2 \neq0$ and $2{\cal R}_{||} + 2{\cal R}_{||}^{\bot} +|f^{||}{}_{\bot \bot}|^2 > 0$ (or integrals thereof).\footnote{The former condition was used in \cite{Andriot:2016xvq} to conclude on the absence of a de Sitter solution with an $O_4$ and $f^{a_{||}}{}_{b_{\bot}c_{\bot}}=0$. That setting is required to embed the monodromy inflation mechanism of \cite{Silverstein:2008sg}; this situation challenges the realisation of this mechanism, as first discussed in \cite{Andriot:2015aza}. The no-go theorem to be mentioned against de Sitter on nilmanifolds with parallel sources also plays against the embedding of this monodromy inflation mechanism.} A second one is on a (possibly T-duality invariant) combination of curvature terms and $H$-flux \cite[(1)]{Andriot:2016xvq} that must be non-zero for de Sitter \cite[(12)]{Andriot:2016xvq}: $2{\cal R}_{||} + 2{\cal R}_{||}^{\bot}  - |H^{(2)}|^2 - 2|H^{(3)}|^2 <0$ (or integrals thereof); its 4d version is discussed in \cite[Sec.2.3]{Andriot:2020lea} (no-go theorem 6). This last no-go theorem is the generalization to $p\geq4$ of the above no-go (\ref{nogo4}). These two no-gos were extended to intersecting sources (with single dimensionality $p$) in \cite[Sec.4]{Andriot:2017jhf}, considering various types of intersections, and different manifolds.

\item {\bf Multiple dimensionalities}

Considering several sources of different dimensionalities $p$ goes beyond the constraints summarized in Table \ref{tab:nogosummary}. The Maldacena-Nu\~nez no-go theorem, as presented in (\ref{nogo1}) still holds. As mentioned, no-go theorems (\ref{nogo2}) and (\ref{nogo4}) admit versions for multiple dimensionalities, given in \cite{Andriot:2022xjh}, while we gave above the extension of (\ref{nogo3}). In particular, a result of \cite{Andriot:2017jhf} is reproduced in these extensions, namely that there is no de Sitter solution with a combination of $p=3 \& 7$ sources. Another constraint is obtained for IIA sources in \cite[(6.28)]{Andriot:2017jhf}. On group manifold, a more systematic approach to identify the relevant dimensionalities of sources was followed in \cite{Andriot:2022way}, and will be presented in Section \ref{sec:dSsolgroupman}.

\item {\bf Group manifolds}

Restricting to group manifolds, for which spin connection coefficients are expressed as structure constants $f^a{}_{bc}$ (see Section \ref{sec:groupman}), and to parallel sources of single dimensionality $p$, constraints were obtained on a parameter $\lambda$ given by a specific ratio of $f^a{}_{bc}$: $\lambda= -\delta^{cd} f^{a_{||}}{}_{b_{\bot} c_{\bot}} f^{b_{\bot}}{}_{a_{||} d_{\bot}} /|f^{||}{}_{\bot \bot}|^2 $. Indeed, outside a window $0 < \lambda <1$, one proves the absence of de Sitter solution \cite{Andriot:2018ept}. This is shown using the $(\rho,\tau,\sigma)$ 4d theory of Section \ref{sec:rhotau}, combined with the 10d BI of RR fluxes. A 10d version of this no-go theorem has not been established; it would involve the trace of Einstein equation along internal parallel directions. Off-shell 4d versions are given in \cite[Sec.2.3]{Andriot:2020lea} (no-go theorems 7 and 8), while the constraint $\lambda>0$ was extended to arbitrary dimension $d\geq3$ in \cite[4.2.6]{Andriot:2022xjh}. Extension of these constraints to the case of intersecting sources is also proposed in \cite[(4.30)]{Andriot:2019wrs}.

As argued in \cite{Andriot:2018ept}, these constraints on $\lambda$ restrict the algebras underlying the group manifolds, up to a change of basis. In addition, we note here that $\delta^{cd} f^{a_{||}}{}_{b_{\bot} c_{\bot}} f^{b_{\bot}}{}_{a_{||} d_{\bot}}$ is the trace, along transverse directions, of the Killing form of the algebra.\footnote{We refer to \cite[(2.6)]{Andriot:2022yyj} about the Killing form, and to \cite[(2.9)]{Andriot:2019wrs} for admissible structure constants.} The Killing form signature is basis independent. Furthermore, if the algebra is nilpotent, then $B$ is identically zero. This gives here $\lambda=0$, from which we conclude, independently of the basis, that there is {\sl no de Sitter solution on nilmanifolds with parallel sources}. Having intersecting sources might change this result, as pointed out around \cite[(3.20)]{Andriot:2018ept}, even though no solution is known on nilmanifolds \cite{Andriot:2022yyj}; see also \cite[Sec.4.4]{Andriot:2019wrs}.

As we will discuss in Section \ref{sec:dSsolgroupman}, a more systematic approach can be followed on group manifolds, especially when allowing intersecting sources and/or multiple dimensionalities. Then considering specific classes of compactification, namely $s_{555}$ (three intersecting $O_5$) and the T-dual one $m_{466}$ (one $O_4$ and two intersecting $O_6$), one can prove an {\sl absence of de Sitter solution}, and interestingly, an {\sl absence of anti-de Sitter solution}. We refer to \cite[Sec.3.3]{Andriot:2022way} for the 10d version, and to \cite[Sec. 5.1.2]{Andriot:2022brg} for the 4d version. Further {\sl no-go theorems against anti-de Sitter} in specific group manifold compactifications can be found in \cite[Sec.4.2]{Andriot:2022yyj}.

\item {\bf Conjectured existence no-go theorems}

All no-go theorems listed above have been proven. Analytic experience on this topic, as well as explicit searches for de Sitter solutions, have led us to conjecture more no-go theorems, without being able to prove them. While this will be the topic of Section \ref{sec:dSsolexistence}, let us already mention the main ones. It is believed that {\sl de Sitter solutions with only parallel sources, also referred to as one set of sources, do not exist} (Conjecture 1, \cite{Andriot:2019wrs}). More than this, it is also believed that {\sl de Sitter solutions do not exist with only two sets of intersecting sources} (Conjecture 4, \cite{Andriot:2022way}); at least three sets are therefore needed. In other words, this implies that {\sl a corresponding 4d effective theory, having a de Sitter solution, is at most ${\cal N}=1$ supersymmetric}, i.e.~has at most 4 supercharges, since each source set breaks supersymmetry by half. This also implies that {\sl there is no de Sitter solution in dimensions $d \geq 5$}, where supersymmetry requires more than 4 supercharges, as further argued in \cite[Sec.3.3]{Andriot:2022xjh}, {\sl unless the $d$-dimensional effective theory there is not supersymmetric}. We will provide more arguments in Section \ref{sec:listclassdim} and \ref{sec:dSsolexistence}, relating in particular to gauged supergravities.

\end{enumerate}
\end{enumerate}

\subsubsection{Existence no-go in 4d, and slope constraints}\label{sec:nogo4d}

In Section \ref{sec:nogo10d}, we have presented no-go theorems against 10d solutions with 4d maximally symmetric spacetimes, focusing mostly on de Sitter. These no-go theorems were derived using combinations of 10d equations, concluding on an impossibility through ${\cal R}_4 \leq 0$ (no de Sitter). We discuss here the derivation of the same no-go theorems directly in a 4d effective theory of the form \eqref{S4d}. A no-go against a de Sitter solution is derived through the following type of inequality (as first proposed in \cite{Hertzberg:2007wc})
\beq
\del_{\hat{\varphi}} V + \frac{c}{M_p}\, V \leq 0 \ , \quad c >0 \ , \label{nogoinequality4dv0}
\eeq
with $\hat{\varphi}$ a canonically normalised field. Such inequalities will be reached under some assumptions. Then, considering \eqref{nogoinequality4dv0} at a critical point or extremum \eqref{extremumsol}, one deduces $V|_0 \leq 0$: this forbids de Sitter. The derivation of the {\sl same} no-go theorems as in 10d is guaranteed by the equivalence \eqref{equiv} between the 10d equations \eqref{3eq} used in Section \ref{sec:nogo10d}, and the 4d critical point equations \eqref{CriticalpointV} of the $(\rho,\tau)$ theory. These re-derivations will be given explicitly.

Beyond reproducing known 10d no-go theorems, or finding new ones against de Sitter solutions (some examples of the latter were mentioned at the end of Section \ref{sec:nogo10d}), there is an important advantage to the 4d derivation. One accesses in 4d information away from the extrema of the potential, constraining in this way the slope. Indeed, the inequality \eqref{nogoinequality4dv0} can be reformulated as follows: using that $-|\del_{\hat{\varphi}} V| \leq \del_{\hat{\varphi}} V$, one obtains for $V>0$
\beq
\frac{|\del_{\hat{\varphi}} V|}{V} \geq \frac{c}{M_p} \ . \label{nogoinequality4d}
\eeq
This constrains the (absolute value of the logarithmic) slope of a positive potential, giving it a lower bound. In other words, a positive potential verifying this inequality cannot be too flat, forbidding in passing a de Sitter critical point, as well as a quasi-de Sitter solution. This can be further reformulated in terms of $\epsilon_V$ defined in \eqref{epseta}. Using that $\nabla V \geq |\del_{\hat{\varphi}} V|$, the slope constraint gets rephrased \cite{Hertzberg:2007wc, Wrase:2010ew} as
\beq
\frac{\nabla V}{V} \geq \frac{c}{M_p} \ \ \Leftrightarrow \ \ \epsilon_V \geq \frac{c^2}{2} \ , \label{nogoinequality4dv2}
\eeq
which has phenomenological consequences, e.g.~possibly forbidding slow-roll. This formulation is also the one used in de Sitter swampland conjectures, starting with \cite{Obied:2018sgi}, as discussed in Section \ref{sec:swamp}.

Instead of an inequality of the form \eqref{nogoinequality4dv0}, what one typically obtains is a linear combination of field derivatives. Provided those can be expressed in terms of canonically normalised fields, the inequality \eqref{nogoinequality4dv0} can be recovered
\begin{gather}
\label{combifields}
\begin{aligned}
& \frac{a}{M_p}\, V + \sum_i b_i\, \del_{\hat{\varphi}^i} V  \leq 0 \ , \quad {\rm with}\ a>0 \ ,\ \exists\ b_i \neq 0 \\
\Leftrightarrow \ \ & \del_{\hat{\varphi}} V + \frac{c}{M_p}\, V \leq 0 \ , \quad {\rm with}\ \ c = \frac{a}{\sqrt{\sum_i b_i^2}} >0 \ ,\ {\rm and} \ \sqrt{\sum_i b_i^2}\ \del_{\hat{\varphi}} = \sum_i b_i\, \del_{\hat{\varphi}^i} \ .
\end{aligned}
\end{gather}
The last equality corresponds to an orthonormal transformation from one canonical field basis to another one, the coefficients of the transformation being $b_i/\sqrt{\sum_j b_j^2}$ \cite{Andriot:2021rdy}. The linear combination of field derivatives should therefore really be understood as a single field derivative, along which the slope constraint is found.\\

We now give explicitly no-go theorems in 4d, focusing on their $c$ value. We use the $(\rho,\tau)$ theory \eqref{rhotautheory} with canonical fields \eqref{hattaurho}. Upon restriction to extremum values ($\rho=\tau=1$), these no-go theorems match their 10d counterparts (\ref{nogo1}) - (\ref{nogo5}); we refer to the latter for discussion, extensions, and references. Let us mention that the following 4d list, and more no-go theorems, can be found in \cite[Sec.2.3]{Andriot:2020lea}, while $d$-dimensional extensions are given in \cite[Sec.4.2]{Andriot:2022xjh}.

\begin{enumerate}[label=\textbf{\roman*.}]
\item {\bf Maldacena-Nu\~nez}

Assuming a compactification where all source contributions are such that $T_{10}^s \leq0$, we get the following inequality and reproduce the 10d version (\ref{nogo1})
\begin{gather}
\label{nogo14d}
\begin{aligned}
& \frac{2}{M_p^2} \left( 2 V+ \tau \del_{\tau} V \right)\\
=\ & \tau^{-3} \sum_{{\rm sources}} \rho^{\frac{p-6}{2}} \, g_s \frac{T_{10}^s}{p+1} - \tau^{-4} \sum_{q=0}^6 \rho^{3-q}\, g_s^2 |F_q|^2 \leq 0\\
\Leftrightarrow \ \ & \frac{2}{M_p} V + \sqrt{2} \, \del_{\hat{\tau}} V \leq 0 \quad \Rightarrow \quad  c =\sqrt{2} \ .
\end{aligned}
\end{gather}
  
\end{enumerate} 

\begin{enumerate}[label=\textbf{\roman*.},resume]
\item {\bf No-go for $p=9,8,7$, or $p=6,5,4$ with $F_{6-p}=0$}
  
We assume a compactification with sources of a single dimensionality $p$. We consider either $p=9$, $8$, $7$, or $p=6$, $5$, $4$ with $F_{6-p} =0$. We also recall from (\ref{nogo2}) that for $p=4$, one has $F_0=0$. With these assumptions, we get the following inequality and reproduce the 10d version (\ref{nogo2})  
\begin{gather}
\label{nogo24d}
\begin{aligned}
& \frac{2}{M_p^2} \left( 2 (p-3) V+ 2\rho\del_{\rho} V + (p-4) \tau \del_{\tau} V \right)\\
=\ & \tau^{-4} \sum_{q=0}^{6} (8-p-q)\, \rho^{3-q}\, g_s^2 |F_q|^2 - 2 \tau^{-2} \rho^{-3}\,|H|^2 \leq 0\\
\Leftrightarrow \ \ & \frac{2(p-3)}{M_p} V +\sqrt{6}\, \del_{\hat{\rho}} V+ (p-4)\sqrt{2} \, \del_{\hat{\tau}} V \leq 0 \quad \Rightarrow \quad  c =\frac{\sqrt{2}(p-3)}{\sqrt{3+(p-4)^2}} \ .
\end{aligned}
\end{gather}
Note here that with $p\geq 4$, the value of $c$ gets minimized for $p=4$ with $c_{p=4}= \sqrt{\frac{2}{3}}$.    
  
\end{enumerate}

\begin{enumerate}[label=\textbf{\roman*.},resume]
\item {\bf No-go for positive or vanishing ${\cal R}_6$}  
  
We assume a compactification with sources of a single dimensionality $p\geq 4$, and ${\cal R}_6 \geq 0$. We also recall that for $p=4$, one has $F_0=0$. With these assumptions, we get the following inequality and reproduce the 10d version (\ref{nogo3})  
\begin{gather}
\label{nogo34d}
\begin{aligned}
& \frac{2}{M_p^2} \left( \frac{p+3}{2}\, V+ \frac12\,\rho\del_{\rho} V + \frac{p}{4}\, \tau \del_{\tau} V \right)\\
=\ & \frac{1}{4} \tau^{-4} \sum_{q=0}^{6} (6-p-q)\, \rho^{3-q}\, g_s^2 |F_q|^2 - \tau^{-2}\rho^{-1}\, {\cal R}_6 \leq 0\\
\Leftrightarrow \ \ & \frac{2(p+3)}{M_p} V +\sqrt{6}\, \del_{\hat{\rho}} V+ p\sqrt{2} \, \del_{\hat{\tau}} V \leq 0 \quad \Rightarrow \quad  c =\frac{\sqrt{2}(p+3)}{\sqrt{3+p^2}}  \ .
\end{aligned}
\end{gather}
We note that $c>1$.  
 
\end{enumerate}

\begin{enumerate}[label=\textbf{\roman*.},resume]
\item {\bf No-go for $p=3$}

To reproduce (\ref{nogo4}), we start again by assuming a compactification with sources of single dimensionality $p$. We obtain the following equality
\begin{gather}
\label{nogo44dv0}
\begin{aligned}
\hspace{-0.2in} & \frac{2}{M_p^2} \left( 2(p+1)\, V+2\,\rho\del_{\rho} V + p\, \tau \del_{\tau} V \right)\\
\hspace{-0.2in} =\ &  - 2\, \tau^{-2} \rho^{-3} |H|^2  +4\, \tau^{-3} \sum_{{\rm sources}} \rho^{\frac{p-6}{2}} \, g_s \frac{T_{10}^s}{p+1} + \tau^{-4} \sum_{q=0}^{6} (4-p-q)\, \rho^{3-q}\, g_s^2 |F_q|^2 \ .
\end{aligned}
\end{gather}  
This no-go theorem requires a sourced flux BI. From a 4d perspective, the BI is an extra constraint that comes from 10d consistency, i.e.~it is due to the fact that the 4d theory comes from a more fundamental one. The rewriting of this BI \eqref{BIrewriting} can be extended to include scalar fields. Focusing on $p=3$, we obtain
\begin{gather}
\label{BIrewritingscalars}
\begin{aligned}
\hspace{-0.2in}& 2 \tau^{-3} \rho^{-\frac{3}{2}} g_s \frac{T_{10}}{4} \\
\hspace{-0.2in}=\ & - |\tau^{-1} \rho^{-\frac{3}{2}} *_6 H - \tau^{-2} g_s F_3 |^2 + \tau^{-2} \rho^{-3} |H|^2 + \tau^{-4} g_s^2 |F_3|^2 - 2 \tau^{-3} \rho^{-\frac{3}{2}} g_s (\d F_5) \ .
\end{aligned}
\end{gather} 
Combined with \eqref{nogo44dv0}, we get the expression
\begin{gather}
\label{nogo44dv1}
\begin{aligned}
 & \frac{2}{M_p^2} \left( 8\, V+2\,\rho\del_{\rho} V + 3\, \tau \del_{\tau} V \right)\\
 =\ & - 4 \tau^{-3} \rho^{-\frac{3}{2}} g_s (\d F_5) - 2 |\tau^{-1} \rho^{-\frac{3}{2}} *_6 H - \tau^{-2} g_s F_3 |^2 - 4 \tau^{-4} \rho^{-2}\, g_s^2 |F_5|^2 \ .
\end{aligned}
\end{gather}  
In 10d, we were left with integrating the expression over ${\cal M}$. In 4d, the integration has already been performed: we recall the simplified notation of the potential terms \eqref{simplenotations}, indicating here that $(\d F_5)$ actually stands for
\beq
(\d F_5) \ \leftrightarrow \ \frac{\int \d^6 y \sqrt{|g_{6}^0|}\, (\d F_5)^0}{\int \d^6 y \sqrt{|g_{6}^0|}} = \frac{\int \d F_5^0}{\int {\rm vol}_6} = 0 \ .
\eeq
We conclude that this term is simply vanishing here, and obtain the following inequality reproducing the 10d version (\ref{nogo4})  
\begin{gather}
\label{nogo44d}
\begin{aligned}
 & \frac{2}{M_p^2} \left( 8\, V+2\,\rho\del_{\rho} V + 3\, \tau \del_{\tau} V \right)\\
 =\ & - 2 |\tau^{-1} \rho^{-\frac{3}{2}} *_6 H - \tau^{-2} g_s F_3 |^2 - 4 \tau^{-4} \rho^{-2}\, g_s^2 |F_5|^2 \leq 0 \\
\Leftrightarrow \ \ & \frac{8}{M_p} V +\sqrt{6}\, \del_{\hat{\rho}} V+ 3\sqrt{2} \, \del_{\hat{\tau}} V \leq 0 \quad \Rightarrow \quad  c =2 \sqrt{\frac{2}{3}} \ . 
\end{aligned}
\end{gather}  
  
\end{enumerate}

\begin{enumerate}[label=\textbf{\roman*.},resume]
\item {\bf Heterotic at order ${\alpha'}^0$}
  
We consider a heterotic compactification at order ${\alpha'}^0$, namely setting $T_{10}^s=0, F_q=0$ in the $(\rho,\tau)$ theory. We then get the following inequality and reproduce the 10d version (\ref{nogo5})  
\begin{gather}
\label{nogo54d}
\begin{aligned}
& \frac{2}{M_p^2} \left( 2\, V+ 2\,\rho\del_{\rho} V \right)\\
=\ & - 2 \, \tau^{-2}\rho^{-3}\, |H|^2 \leq 0\\
\Leftrightarrow \ \ & \frac{2}{M_p} V +\sqrt{6}\, \del_{\hat{\rho}} V \leq 0 \quad \Rightarrow \quad  c =\sqrt{\frac{2}{3}} \ .
\end{aligned}
\end{gather}  

\end{enumerate}  

More no-go theorems are mentioned at the end of Section \ref{sec:nogo10d}, including some in 4d, sometimes involving the third scalar $\sigma$ introduced in Section \ref{sec:rhotau}. In references given there, the value of $c$ is typically calculated. One remarkable example is the no-go theorem obtained for ${\cal M}$ being a group manifold, with parallel sources of single dimensionality $p$, and with a parameter $\lambda \leq 0$: we mentioned above that this forbids de Sitter solutions on nilmanifolds (with parallel sources). The derivation uses a sourced RR flux BI. For any $p\geq4$, one obtains for this one $c=\sqrt{\frac{2}{3}}$ \cite{Andriot:2019wrs, Andriot:2020lea}.\\

In all examples so far, one always gets as a value $c \sim {\cal O}(1)$, the lowest value in previous 4d examples being $c \geq \sqrt{\frac{2}{3}}$. This corresponds in Planckian units to lower bounds on the logarithmic slope of the potential, as discussed around \eqref{nogoinequality4d}. A consequence is that {\sl the potential is steep}. In particular, not only de Sitter solutions are excluded this way, but also {\sl quasi-de Sitter}, where the latter refers to an almost flat potential.

These results point towards a characterisation of scalar potentials (through their slope) in a 4d effective theory of string theory in the classical regime, at least for compactifications that obey the no-go theorems assumptions. Such ideas will be crucial to the swampland de Sitter conjectures discussed in Section \ref{sec:swamp}.

\subsubsection{Stability no-go theorems}\label{sec:nogostab}

Perturbative (in)stability of a 4d extremum was introduced and discussed in Section \ref{sec:dimred}. We focus here on de Sitter: a solution is unstable if one mass matrix eigenvalue is negative. This is equivalent to $\eta_V <0$, where $\eta_V$ is defined in \eqref{epseta}. It is useful for cosmology to determine the (in)stability of a de Sitter extremum, together with a quantitative estimate of $\eta_V$ (see Section \ref{sec:intro}). In addition, all known examples of de Sitter solutions, verifying the 10d ansatz of Section \ref{sec:10dansatz}, and found on ${\cal M}$ being a group manifold, have been found unstable (see Section \ref{sec:dSsolstab}). Several works have tried to understand or prove such an apparent systematic instability. This has led to the following proposals for no-go theorems
\begin{enumerate}[label=\textbf{\roman*.}]
  \item \label{nogostab1}{\bf Conjectured stability no-go theorem for classical de Sitter}
  
  Summarizing the status, it has been proposed in \cite{Andriot:2019wrs} (conjecture 2) that {\sl all solutions with a de Sitter spacetime, obtained as a string background in a classical regime, are perturbatively unstable}. In other words, they all admit at least one scalar field direction along which the potential is a maximum, giving $\eta_V<0$.
\end{enumerate}
This is related to an earlier and more precise conjectured no-go theorem, initially proposed in \cite{Danielsson:2012et} for massive type IIA with $p=6$ sources  
\begin{enumerate}[label=\textbf{\roman*.},resume]
  \item \label{nogostab2}{\bf Conjectured $(\rho,\tau,\sigma_I)$ tachyon}
  
   For classical de Sitter solutions verifying the 10d ansatz of Section \ref{sec:10dansatz}, the claim \cite{Danielsson:2012et} is that a systematic tachyon is present, and can be found within the reduced set of fields $(\rho,\tau,\sigma_I)$ (introduced in Section \ref{sec:rhotau}). By now, this conjecture has been tested on de Sitter solutions obtained in various group manifold compactifications \cite{Danielsson:2012et, Junghans:2016uvg, Andriot:2020wpp, Andriot:2021rdy, Andriot:2022yyj}, without any counter-example up to now, as long as one considers a compact ${\cal M}$. Note that restricting further to $(\rho,\tau)$ does not work, $\sigma_I$ has to be included (see e.g.~\cite{Andriot:2020wpp}).
\end{enumerate}
We will discuss these conjectures in more details in Section \ref{sec:dSsolstab}. Trying to prove them has motivated formal work on stability, that we present below. Before entering these technicalities, let us mention another, proven, no-go theorem
\begin{enumerate}[label=\textbf{\roman*.},resume]
  \item \label{nogostab3}{\bf Tachyon in nearly no-scale de Sitter}
  
  Building on \cite{Covi:2008ea, Danielsson:2012et, Kallosh:2014oja, Junghans:2016uvg}, a stability no-go theorem was proven in \cite{Junghans:2016abx} for a de Sitter solution in 4d ${\cal N}=1$ supergravity, in the case where this extremum is close to a no-scale Minkowski solution. A large class of such de Sitter solutions was shown to admit a systematic tachyon, with $\eta_V < - \frac{4}{3}$. The de Sitter tachyon is different from the sgoldstino, but aligns with it in the Minkowski limit. A question is then to identify 10d compactifications that fall in this 4d class. This interpretation of the tachyon was shown to hold for some de Sitter solutions on group manifolds \cite{Junghans:2016uvg}, namely some examples in type IIA with $p=6$ and IIB with $p=5\& 7$. Similarly, a 4d ${\cal N}=1$ gauged supergravity, that should correspond to a compactification of type IIA on group manifolds with $p=6$, was shown in \cite{Dibitetto:2011gm} to admit a family of de Sitter maxima. The latter turned out to verify precisely the same upper bound $\eta_V < - \frac{4}{3}$, as can be seen numerically in \cite[Fig.4]{Dibitetto:2011gm}.
\end{enumerate}  

In the following, we introduce formal tools and methods to prove stability no-go theorems, and eventually give some examples. This is mostly based on \cite{Andriot:2021rdy}, building on \cite{Shiu:2011zt, Andriot:2018ept, Andriot:2019wrs, Andriot:2020wpp}. Other (early) papers discussing (in)stability include \cite{Silverstein:2007ac, Caviezel:2008tf, Flauger:2008ad, Caviezel:2009tu, Danielsson:2011au, Danielsson:2012by}. Similarly to conditions for existence, the stability no-go theorems give constraints on ingredients necessary for stability, or sufficient for instability, and conditions to be obeyed.\\

The stability of a de Sitter critical point of a potential $V$ is captured by the eigenvalues of the mass matrix. This spectrum is independent of the field basis, but it can be convenient to work in a canonical basis when it exists. Indeed, the mass matrix gets written there $\hat{M}^i{}_j= \delta^{ik}\del_{\hat{\varphi}^k} \del_{\hat{\varphi}^j} V$ and it is then a (real) symmetric matrix, which has interesting properties as we will see. In addition, in that basis, it can be identified with the Hessian. As a starting point, a good illustration is given by the $(\rho,\tau)$ theory \eqref{rhotautheory}, for which a canonical field basis can be reached (see \cite[App.A]{Andriot:2021rdy} for explicit transformations). Since $\hat{M}$ is then a $2\times2$ square matrix, eigenvalues can be determined analytically. The lowest one is given by
\begin{gather}
\label{lambdaminus}
\begin{aligned}
\lambda_- & = \frac{1}{2} \left({\rm tr} \hat{M} - \sqrt{({\rm tr} \hat{M})^2 - 4\, {\rm det} \hat{M}} \right)\\
&= \frac{1}{2} \left(\del_{\hat{\rho}}^2 V + \del_{\hat{\tau}}^2 V - \sqrt{(\del_{\hat{\rho}}^2 V - \del_{\hat{\tau}}^2 V)^2 + 4 (\del_{\hat{\rho}}\del_{\hat{\tau}} V)^2  }  \right) \ .
\end{aligned}
\end{gather} 
It is then straightforward to show \cite{Shiu:2011zt, Andriot:2021rdy} that
\beq
\text{{\sl Unstable de Sitter (among $\rho,\tau$)}} \ \Leftrightarrow \ \lambda_- <0 \ \Leftrightarrow\ {\rm tr} \hat{M}<0 \ ,\ \ {\rm or}\ {\rm tr} \hat{M}\geq 0\ \& \ {\rm det} \hat{M} <0 \ .
\eeq
Considering the $(\rho,\tau)$ theory \eqref{rhotautheory} for sources of single dimensionality $p$, we extend here the result obtained in \cite[(3.5)]{Andriot:2021rdy} for $p=5$ to the following relation
\begin{gather}
\label{traceH}
\begin{aligned}
& \frac{2}{M_p^2}\left(\frac{10-3p}{3} \rho\del_{\rho}V + \left(\frac{3+16p-2p^2}{18}\right) \tau \del_{\tau}V + M_p^2 \left( \del_{\hat{\rho}}^2 V + \del_{\hat{\tau}}^2 V \right) \right) \\[3pt]
=\ & \begin{array}{|lc}  - \frac{2}{3}\tau^{-2} \rho^{-1}\, {\cal R}_6 + \frac{14}{3} \tau^{-2} \rho^{-3}\, |H|^2 + \tau^{-4} g_s^2 \frac{1}{3}\left(|F_3|^2 + 10 \rho^{-2}\, |F_5|^2\right) & \quad (p=5) \\[5pt]   - \frac{7}{3}\tau^{-2} \rho^{-1}\, {\cal R}_6 + \frac{13}{2} \tau^{-2} \rho^{-3}\, |H|^2 + \tau^{-4}  g_s^2\left(\frac{8}{3} \rho^{-1}\, |F_4|^2 + 8 \rho^{-3}\, |F_6|^2\right) & \quad (p=6) \end{array} \ .
\end{aligned}
\end{gather} 
Considered at an extremum \eqref{extremumsol}, and using that de Sitter requires ${\cal R}_6 <0$ \eqref{summarydS}, we deduce for $p=5,6$ that $(\del_{\hat{\rho}}^2 V + \del_{\hat{\tau}}^2 V)|_0  = {\rm tr} \hat{M}|_0>0$. In other words,
\beq
\text{{\sl Unstable de Sitter (among $\rho,\tau$) with}}\ p=5,6 \ \Leftrightarrow \ \lambda_- <0 \ \Leftrightarrow\ {\rm det} \hat{M} <0 \ .
\eeq
From there, characterising the instability becomes delicate because the determinant involves powers of terms in the potential. This is more difficult to handle than linear conditions from the trace. Some configurations avoiding such determinant constraints were still identified in \cite{Shiu:2011zt}, while few conditions from such constraints were inferred in \cite[Sec. 3.2]{Andriot:2021rdy}.\\

To go beyond such a determinant constraint, or in order to face situations with more fields, where analytical expressions of eigenvalues are less available, it is useful to rely on some mathematical results. Two of them, mentioned already in Section \ref{sec:dimred}, are particularly relevant. They deal with negative eigenvalues of a real symmetric matrix. The first one is Sylvester's criterion \cite{Shiu:2011zt}, that we formulate here as follows
\begin{gather}
\label{Sylvester}
\begin{aligned}
& \text{{\sl A real symmetric matrix is definite-positive if and only if the determinants of all}}\\
& \text{{\sl its upper-left square submatrices (a.k.a~leading principle minors) are positive.}}
\end{aligned}
\end{gather}  
The second one is a lemma, established in \cite[Sec.3.3]{Andriot:2020wpp}, reformulated as follows
\begin{gather}
\label{Lemma}
\begin{aligned}
& \text{{\sl Let $\hat{M}$ be a real symmetric matric, and $A$ an upper-left square submatrix. If $\mu$ is}}\\
& \text{{\sl the minimal eigenvalue of $\hat{M}$, and $\alpha$ any eigenvalue of $A$, then $\mu \leq \alpha$.}}
\end{aligned}
\end{gather} 
A first important consequence, discussed in Section \ref{sec:dimred}, is that considering more fields can only make a solution more unstable. Therefore, if an instability is found within a finite set of fields (e.g.~$\rho,\tau$), it is sufficient to conclude on an unstable solution in a more complete theory, with a more negative $\eta_V$. This will be crucial when concluding on an unstable de Sitter solution. A second important consequence of the two mathematical results is that if a single diagonal entry of the matrix is negative, then one eigenvalue is negative. In other words, one has
\begin{gather}
\label{Instab1field}
\begin{aligned}
& \text{{\sl If there exists a canonical field $\hat{\varphi}$ such that $\del_{\hat{\varphi}}^2 V|_0 <0$,}}\\
& \text{{\sl then the de Sitter extremum is unstable.}}
\end{aligned}
\end{gather}

Inspired by the criterion \eqref{Instab1field} and the existence no-go theorems, various combinations of derivatives of $V$ and inequalities were considered to prove stability no-go theorems. A first kind \cite{Andriot:2018ept, Andriot:2019wrs} is 
\beq
\frac{a}{M_p^2} V + \frac{1}{M_p} \sum_i b_i\, \del_{\hat{\varphi}^i} V + \sum_i c_i\, \del_{\hat{\varphi}^i}^2 V < 0 \ ,\quad a, c_i \geq 0 \ ,\ \exists\, c_i \neq0 \ ,\label{firstcombi}
\eeq
possibly completed with terms appearing in flux BI rewritings. Considered on a de Sitter extremum, this inequality implies an instability. This illustrates the idea of the combination of derivatives to prove a no-go theorem. In practice however, this did not lead to instability results in \cite{Andriot:2018ept, Andriot:2019wrs}.

Still, considering rather $c_i\leq 0$ led to show some stability necessary conditions $\del_{\hat{\varphi}}^2 V|_0 >0$ in group manifold compactifications \cite[Sec.3.3]{Andriot:2018ept}. Inspired by these results (and correcting some typos), we show here the following general expression for the $(\rho,\tau)$ potential, with sources of single dimensionality $p$:
\begin{gather}
\label{del2rho}
\begin{aligned}
& \frac{2}{M_p^2} \left( \left(p(p-4) + 1\right)\, V + \frac{1}{2} (p^2 - 6p + 8)\, \tau \del_\tau V + 2 (p-4)\, \rho \del_{\rho} V - \rho^2 \del_{\rho}^2 V \right) \\
=\ & - 2(p-2) \tau^{-2} \rho^{-3}\, |H|^2 - \frac{(p-4)^2}{4} \tau^{-3} \rho^{\frac{p-6}{2}}\, g_s \frac{T_{10}}{p+1}\\
& - \frac{1}{2} \tau^{-4} \sum_{q=0}^6 (p+q-6)(p+q-8) \rho^{3-q}\, g_s^2 |F_q|^2 \ ,\\
& \text{De Sitter extremum}\ \Rightarrow\ \del_{\hat{\rho}}^2 V|_0 >0 \ .
\end{aligned}
\end{gather} 
The conclusion is drawn using existence results of Table \ref{tab:nogosummary}, namely that a de Sitter solution with single $p$ has $T_{10}>0$ and $p=4$, $5$ or $6$. One could infer that the ``volume field'' is always stabilised, and so that the tachyon should be looked for among other fields. Let us still recall that this holds provided the other (canonically normalised) field is the 4d dilaton $\hat{\tau}$; things could differ in a basis $(\rho,\phi)$.

Given the above results \eqref{traceH} on $(\del_{\hat{\rho}}^2 V + \del_{\hat{\tau}}^2 V)|_0 >0$, one may wonder about the sign of $\del_{\hat{\tau}}^2 V$. Attempts to prove it positive have failed in \cite[Sec.3.2]{Andriot:2021rdy}. In addition, having as well $\del_{\hat{\tau}}^2 V>0$ would not guarantee stability among $\rho,\tau$: indeed, one may still get ${\rm det} \hat{M} <0$ because of the off-diagonal contributions $\del_{\hat{\rho}}\del_{\hat{\tau}} V$. An illustration of such a situation can be found in \cite[Fig.7]{Andriot:2023isc}, as well as in \cite[Sec. 3.3]{Andriot:2020wpp}. Having off-diagonal contributions to the mass matrix precisely indicate that the mass eigenmode, in particular a tachyonic field direction, is not aligned with $\hat{\rho}$ or $\hat{\tau}$ but is a combination of them.

Along the same lines, focusing on group manifold compactifications with $p=5$ sources (class $s_{55}$ of \cite{Andriot:2022way}, presented in \eqref{s55}), results were obtained in \cite[Sec.3.2]{Andriot:2021rdy} on the field $\sigma_1$. It was shown that $\del_{\hat{\sigma}_1}^2 V|_0>0$ and $(\del_{\hat{\sigma}_1}^2 V + \del_{\hat{\tau}}^2 V)|_0>0$, suggesting to look for different tachyonic directions.\\

A second kind of combinations considered in \cite{Andriot:2021rdy}, inspired by the criterion \eqref{Instab1field}, appeared to be more useful than \eqref{firstcombi} to prove stability no-go theorems. It is given by
\beq
\frac{a}{M_p^2} V + \frac{1}{M_p} \left(\sum_i b_i\, \del_{\hat{\varphi}^i} \right) V + \left( \sum_i c_i\, \del_{\hat{\varphi}^i} \right)^2 V < 0 \ ,\quad a \geq 0 \ ,\ \exists\, c_i \neq0 \ .\label{secondcombi}
\eeq
If we had $c_i=0$ and $a>0$, we would be back to \eqref{combifields}, giving a constraint on $\epsilon_V$. As explained around \eqref{combifields}, a linear combination of derivatives with respect to canonical fields, $\sum_i c_i\, \del_{\hat{\varphi}^i}$ and $\exists\, c_i \neq0$, can be understood as a single canonical field direction $\hat{t}$, given by
\beq
\del_{\hat{t}} = \sum_i \frac{c_i}{\sqrt{\sum_j c_j^2}}\, \del_{\hat{\varphi}^i} \ .
\eeq 
The relation between $\hat{t}$ and the $\hat{\varphi^i}$ is simply an orthonormal change of (canonical) basis, and the coefficients of this transformation are the $c_i/\sqrt{\sum_j c_j^2}$. In the following, this field direction $\hat{t}$ can be understood as a tachyonic direction. Indeed, considering the combination \eqref{secondcombi} at a de Sitter extremum, we obtain
\beq
\frac{a}{\sum_j c_j^2} V|_0 + M_p^2 \, \del_{\hat{t}}^2 V|_0 < 0 \ \ \Rightarrow \ \ \eta_V < - \frac{a}{\sum_j c_j^2} \ ,\ \text{{\sl unstable de Sitter}} \ ,
\eeq
where the implication uses the lemma \eqref{Lemma} and criterion \eqref{Instab1field}. From this result, no-go theorems against stability were obtained in \cite{Andriot:2021rdy}, either with an upper bound on $\eta_V$ for $a\neq0$ in ``Method 2'' or without for $a=0$ in ``Method 3''.\footnote{A similar technique to Method 3 was used in \cite{Junghans:2016uvg} to identify the tachyon direction, in concrete solution examples.} We provide two examples of the former below, and refer to that paper for more constraints.
\begin{enumerate}[label=\textbf{\roman*.},resume]
  \item \label{nogostab4}{\bf Stability no-go theorems for compactifications in the class $s_{55}$}
  
  The class $s_{55}$ \cite{Andriot:2022way} considers a compactification on group manifold with two sets of $O_5$ and one of $D_5$ that are intersecting (and no source of other dimensionality): see \eqref{s55} for more details. Using \eqref{secondcombi}, the following sufficient conditions for instability were obtained, among others, in \cite{Andriot:2021rdy} for $s_{55}$
  \bea
  C5:&&\quad a=1\ ,\ b_{\hat{\rho}}= \frac{2893}{496}\sqrt{\frac{3}{2}} \ ,\ b_{\hat{\tau}}= \frac{4217}{992}\sqrt{2} \ ,\ c_{\hat{\rho}}=\sqrt{\frac{3}{2}} \ ,\ c_{\hat{\tau}}= \sqrt{2} \ ,\nn\\
  &&\ \Rightarrow \quad F_5=0 \ \ \&\ \ 10.75\, {\cal R}_6 + g_s T_{10} \leq 0 \quad \Rightarrow \quad \eta_V|_0 \leq -\frac{2}{7} \ ,\\
  C10:&&\quad a=\frac{2}{3}\ ,\ b_{\hat{\rho}}= \frac{1}{2}\sqrt{\frac{3}{2}} \ ,\ b_{\hat{\tau}}= \frac{3}{2\sqrt{2}} \ ,\ b_{\hat{\sigma}_1}=\frac{\sqrt{3}}{2} \ ,\ b_{\hat{\sigma}_2}=0 \ ,\nn\\
  && \quad  c_{\hat{\rho}}=\frac{1}{2\sqrt{3}} \ ,\ c_{\hat{\tau}}= \frac{1}{2} \ ,\ c_{\hat{\sigma}_1}=\frac{1}{\sqrt{6}} \ ,\ c_{\hat{\sigma}_2}=0 \ ,\nn\\ 
  &&\ \Rightarrow\quad -6 \, R_3 \leq {\cal R}_4 \quad \Rightarrow \quad \eta_V|_0 \leq - \frac{4}{3} \ ,
  \eea
  where $R_3$ is a certain negative internal curvature term. $C5$ was obtained using only $\rho,\tau$, while $C10$ was obtained using $\rho,\tau,\sigma_1,\sigma_2$. These two conditions were obeyed by several solutions found in  \cite{Andriot:2020wpp}.
\end{enumerate}

This concludes the presentation of the methods to get stability no-go theorems, as well as the somewhat scattered results obtained so far. Let us recall that we do not know of any stable de Sitter solution (within ansatz of Section \ref{sec:10dansatz}). But we also do not have a complete proof a systematic (perturbative) instability, in spite of above methods and results. A more detailed account will be given in Section \ref{sec:dSsolstab}. This situation is in contrast with different compactification settings and string theory regimes: allowing for one-loop Casimir energy in certain compactifications, a systematic tachyon was shown in \cite{Parameswaran:2024mrc} to exist for de Sitter solutions, using the above techniques; see also \cite{ValeixoBento:2025qih} for a recent and similar result.

Finally, it should be mentioned that allowing for anti-branes, thus going beyond ansatz of Section \ref{sec:10dansatz}, could remove the possibly systematic tachyon. This has been shown for a concrete solution of massive type IIA in \cite[Sec.4]{Kallosh:2018nrk}, by adding $\bar{D}_6$.\footnote{This particular example may still suffer from the same issue as known massive type IIA de Sitter maxima, namely that they fail to exist at large volume and small string coupling. We will come back to this ``classicality'' issue in Section \ref{sec:nodSQG} and \ref{sec:dSsolclassical}.} This would suggest anti-branes as a possible necessary ingredient for stability, a point to be shown. However, similar attempts with $\bar{D}_5$ in the class $s_{55}$ have failed \cite{Andriot:2020wpp}.

These results on stability will feed the swampland de Sitter conjectures, to which we now turn to.

\subsection{Swampland de Sitter conjectures}\label{sec:swamp}

The swampland program, initiated in \cite{Vafa:2005ui} and reviewed e.g.~in \cite{Palti:2019pca, vanBeest:2021lhn, Grana:2021zvf, Agmon:2022thq}, aims at characterising effective theories of a quantum gravity theory. A prime example would be $d$-dimensional theories derived from string theory, e.g.~via dimensional reduction and compactification, as described in Section \ref{sec:dimred}. These effective theories are said to populate a space of theories called the {\sl landscape}. On the contrary, models which cannot be derived from quantum gravity would populate the {\sl swampland}. Examples of the latter would be (quantum) field theories which cannot be coupled consistently to (quantum) gravity, for instance due to an anomaly, or more generally an effective theory in the IR which would not find quantum gravity as a UV completion. The terms {\sl landscape} and {\sl swampland} also sometimes refer to the set of vacua, or solutions, which can or cannot be obtained from a quantum gravity theory. The swampland program tries to provide (simple) criteria that characterise populations of the landscape or the swampland. The genericity of this approach has the benefit to go beyond case by case studies of models, and provide theoretical priors among a set of phenomenological models (e.g.~the set of inflation models compatible with observations).

In this program, a characteristic of a quantum gravity effective theory is typically formulated at first as a conjecture. Motivations for it, and subsequent tests of it, are then carried-out in two fashions. The first one is top-down: one starts from a quantum gravity theory, typically string theory, and derives or tests the claimed characteristic of the effective theory. The second one is bottom-up: generic quantum gravity arguments that an effective theory should obey, e.g.~involving black holes or cosmology, are used to justify the proposed characteristic. In some cases, the conjecture gets proven, or at least, the scope of the tests is so large that it becomes difficult to doubt the property. In other cases, it gets refined. Consistency and interplay among the various conjectures also play a role. Eventually, if put on solid footing, the conjectured characteristic may serve as an actual constraint on 4d phenomenological models: this is the motivation here for the conjectures to be discussed.

In this context, the following question is natural
\begin{gather}
\begin{aligned}
& \text{{\sl Can a 4d effective theory from quantum gravity admit}}\\
&\text{{\sl a solution with a de Sitter spacetime?}}
\end{aligned}
\end{gather}
If one considers that our reality can be described by string theory (the working hypothesis in this article), then this question is of prime importance for the cosmology of our universe. Considering ubiquitous the string effective theories of the form \eqref{S4d}, one could broaden the previous question towards
\begin{gather}
\begin{aligned}
&\text{{\sl In a 4d string effective theory, what are the properties}}\\
&\text{{\sl of a positive scalar potential?}}
\end{aligned}
\end{gather}
This includes the question of positive extrema, but also asks about the slope of the potential. In 2018, the following swampland de Sitter Conjecture (dSC) was proposed \cite{Obied:2018sgi}, as an answer to these questions. As explained, it takes the form of a simple characteristic that an effective theory in the landscape should obey, that is (with $V>0$)
\beq
{\rm dSC:} \quad \quad \frac{\nabla V}{V} \geq \frac{c}{M_p} \ ,\quad c \sim {\cal O}(1) \ . \label{dSconj}
\eeq
This inequality is meant to hold at every point in field space. One motivation for this proposal was the existence no-go theorems discussed in Section \ref{sec:nogo4d}, and indeed, the swampland de Sitter conjecture \eqref{dSconj} takes the form of the constraint already discussed in \eqref{nogoinequality4dv2}. Note that when sending $M_p \rightarrow \infty$, the constraint becomes trivial: this is consistent with the idea that decoupling gravity makes the statement empty. As discussed around \eqref{nogoinequality4dv2}, this condition does not only forbid a de Sitter solution, but also prevents the potential from being too flat, by giving a lower bound to its slope (or equivalently to $\epsilon_V$). On this note, it is important to recall that the 4d no-go theorems of Section \ref{sec:nogo4d} all give $c\geq \sqrt{\frac{2}{3}} \approx 0.82 \sim {\cal O}(1)$, so the corresponding lower bound is consistent with \eqref{dSconj}; this argument was used in \cite{Obied:2018sgi} to motivate this proposal. While such a slope constraint will eventually play an important role when considering rolling solutions, the first focus was on the proposed absence of de Sitter extremum, when \cite{Obied:2018sgi} came out, hence the name of the conjecture. In the following, we will first discuss in Section \ref{sec:nodSQG} arguments for such an obstruction against de Sitter in quantum gravity, especially prior to \cite{Obied:2018sgi}, together with an update on this question.

The initial proposal \eqref{dSconj} was quickly considered too strong. It was followed by an intense period of debates, tests and refinements of the conjecture, on which we will not report in detail here. We will simply discuss the main evolutions of the conjecture in Section \ref{sec:reftestdSconj}, together with tests and connections to the no-go theorems of Section \ref{sec:nogo}. An overall and up-to-date summary will be given in Section \ref{sec:dSconjsum}.

\subsubsection{No de Sitter in quantum gravity?}\label{sec:nodSQG}

It is interesting to recall the context and arguments that would lead to a complete obstruction against a solution with a de Sitter spacetime from quantum gravity, as proposed in \cite{Obied:2018sgi}, and give an updated view on this question. 

A starting point is string theory and the attempts to construct a de Sitter solution in a (possibly 4d) effective theory. A tentative list of such attempts will be provided and discussed in Section \ref{sec:listattempts}, but we can already mention few aspects. As made clear by the no-go theorems of Section \ref{sec:nogo}, especially the results \eqref{summarydS}, a de Sitter solution requires many ingredients. Indeed, in the framework of these no-go theorems, we have shown that the compact space ${\cal M}$ of extra dimensions has to be curved and not flat, that it should support some non-zero RR fluxes, and finally that it should have negative tension objects such as orientifolds. In a complete description of the solution, these objects would generate localized singularities and backreact on ${\cal M}$. We are therefore far from an empty and Ricci flat ${\cal M}$, as in a Calabi-Yau compactification that allows for Minkowski solutions, and with few ingredients for anti-de Sitter ones. The ansatz for de Sitter discussed so far, presented in Section \ref{sec:10dansatz}, stands in the classical regime of string theory, where in principle, {\sl corrections are under control}. But in practice, this statement is difficult to check, precisely because of the numerous ingredients present in the compactification; we will come back to this in Section \ref{sec:dSsolclassical}. Other constructions go beyond the classical regime by including perturbative or even non-perturbative contributions. There as well, the number of ingredients necessary to get de Sitter is important, and the fact that different regimes are reached make the question of control even more delicate.

{\sl The control on a de Sitter construction is a central concept} in all related discussions. To go from full string theory towards a 4d effective theory, a serie of approximations is made, that allows to neglect many contributions and retain only the relevant physics. This has to do with regimes, energy scales, value of (perturbative) expansion parameters, etc. An example of such a process has been presented in Section \ref{sec:compactif}, starting with the classical string theory regime towards a 4d low energy effective theory. All quantities neglected are then possible corrections to the main contributions considered. The central question is whether we have {\sl control} on these corrections, by which is meant whether we are sure that they can be neglected? This requires the following points: for a given construction of a de Sitter solution,
\begin{enumerate}[label=\textbf{\arabic*.}]
  \item Have all corrections been identified?
  \item When identified, do we have an actual analytical expression or an estimate for them?
  \item Given some analytical knowledge, can we compute at least an order of magnitude for them, if not their exact value?
\end{enumerate}
Only once the last point is reached can one verify corrections are indeed small compared to the main contributions, and can be neglected. One would then talk of {\sl ``good control''} on the construction. Note also the distinction between a {\sl ``parametric control''} and a {\sl ``numerical control''}: in the former case, a free parameter is available and can be adjusted to make a correction as small as desired; a complete computation of the correction is then not necessary. On the contrary, a numerical control means that numerical values happen to make the correction small, without any freedom to further adjust the hierarchy.

The problem with de Sitter solutions in string theory is therefore not the lack of examples of constructions: despite the involved settings due to the numerous ingredients and regimes, many proposals have been made. The problem is rather the control in all such scenarios that give de Sitter. Up to now, {\sl there is no example of a 4d de Sitter solution (extremum) from string theory that qualifies as ``well-controlled''}.\footnote{An illustration of this point of view can be found in the April's fool paper \cite{Bena:2023sks}.} It is fair to point-out the very recent example of a 5d de Sitter solution, obtained on a Ricci-flat manifold with (quantum) Casimir energy contribution (from M-theory), that claims to have an excellent control on any known correction \cite{ValeixoBento:2025yhz}. While it is too early to assess the validity of this example, it illustrates the main point here, that the difficulty is the control in the constructions with respect to corrections.

In 2018, when the initial de Sitter conjecture \eqref{dSconj} was proposed \cite{Obied:2018sgi}, similar points were discussed. While classical de Sitter solutions were plagued with no-go theorems and few other difficulties, the validity of other attempts, involving (non-)perturbative contributions, was criticized due to various possible corrections. This was especially the case of the KKLT scenario \cite{Kachru:2003aw}, various aspects of which had been under heavy scrutiny in the years before. An account on these discussions, and more generally the situation back in early 2018 can be found in \cite{Danielsson:2018ztv}. In that work, an impossibility of getting de Sitter from string theory was suggested. An earlier version of a swampland de Sitter conjecture had also been proposed, more in passing and in a less quantitative fashion than \eqref{dSconj}, in 2017 in \cite{Brennan:2017rbf}. This situation led to the proposal of \cite{Obied:2018sgi}.\\

Behind the numerous ingredients and the lack of control on corrections for de Sitter lies an important point: the {\sl absence of supersymmetry}. A $d$-dimensional de Sitter spacetime admits as an isometry group SO(1,d), which encodes the positive cosmological constant. The latter turns out to be incompatible with a (standard, linear) supersymmetry algebra: see e.g.~\cite[Sec. 4.2]{DallAgata:2021uvl}. In other words, a de Sitter spacetime does not preserve supersymmetry. This stands in contrast to Minkowski or anti-de Sitter. There are two important consequences: first, a de Sitter solution is a priori not protected from corrections that supersymmetry would typically avoid. The only way-out is thus control arguments, allowing to neglect corrections. Second, de Sitter solutions are not as ``simple'' as supersymmetric solutions can be, resulting in this impression of having many ingredients as described above. Indeed, supersymmetry is inherent to string theories and supergravities, and helps to organise the structures appearing there. This is made manifest in various fashion, especially in solutions where supersymmetry relates various quantities. An illustration of such an organisation, and how de Sitter must go beyond it, can be found e.g.~in \cite[(4)]{Andriot:2016ufg}. 

This relates to the general difficulty in finding de Sitter solutions. To find supersymmetric solutions, it is typically sufficient to solve first order (derivative) equations, which are conditions to preserve supersymmetry (see e.g.~\cite{Koerber:2007hd}). For non-supersymmetric ones however, one is back to equations of motion, which can be second order (Einstein equation). It is therefore technically more difficult to find de Sitter solution, due to the lack of supersymmetry, with numerous ingredients. {\sl Difficulty however does not mean impossibility}, but this still goes in line with the idea that incompatibility with supersymmetry is not natural, to say the least, within a supersymmetric quantum gravity theory (the case of most string theories).

The absence of supersymmetry played a further important role in the development of an obstruction against de Sitter. Indeed, another swampland conjecture had been proposed in 2016, stating that any non-supersymmetric anti-de Sitter solution would be unstable \cite{Ooguri:2016pdq}. The instability would be found either perturbatively, violating the BF bound \eqref{BFbound}, or non-perturbatively; a sample of the latter can be found in e.g.~\cite[Tab.2]{Andriot:2022brg}. An instability in an anti-de Sitter solution was sometimes considered as fatal to the solution itself \cite{Ooguri:2016pdq, Brennan:2017rbf}, i.e.~a non-supersymmetric anti-de Sitter solution would then barely exist. The reason is that an instability, generated for instance at the boundary in the form of a non-perturbative bubble, propagating at the speed of light, can reach any point in the bulk within a {\sl finite} proper time, as an in-falling observer. In short, the anti-de Sitter solution is then doomed to disappear in finite time. Therefore, one could wonder more generally about the fate of non-supersymmetric solutions with maximally symmetric spacetimes, in quantum gravity \cite{Brennan:2017rbf}, suggesting support to the swampland de Sitter conjecture \cite{Obied:2018sgi}. As we will see, this point of view will later be refined to the stability question \cite{Agmon:2022thq}, instead of the actual existence of the solution.\\

More general arguments against de Sitter in quantum gravity can be reported. To start with, the formulation of quantum field theory on curved spacetimes, and especially de Sitter, is notoriously challenging. From this, one may argue that building in the same way (i.e.~perturbatively) a complete quantum gravity on a de Sitter spacetime appears naively out of reach. More generally, following an S-matrix approach in a de Sitter spacetime is compromised by its horizon: indeed, one would need to define asymptotic states, but those stand a priori behind the horizon, and are thus causally disconnected. The cosmological event horizon will actually be an important element, to which we come back at the end of Section \ref{sec:reftestdSconj}. 

What about a non-perturbative definition of quantum gravity following a holographic approach? The celebrated AdS/CFT correspondence provides a concrete example of a holographic duality between (quantum) gravity and a gauge theory. There, certain string theories over anti-de Sitter backgrounds are considered defined non-perturbatively, thanks to the holographic duality, by the adequate conformal field theories living on the boundary. From that perspective, one may wonder whether quantum gravity on a de Sitter spacetime could be defined in the same (non-perturbative) fashion, through a putative dS/CFT correspondence \cite{Hull:1998vg, Strominger:2001pn, Balasubramanian:2001nb}. This also turns out to be difficult to realise. To start with, because of the isometry group of de Sitter, the (would-be) dual CFT is Euclidian, typically with non-unitary representations of the conformal group, or complex conformal weights.\footnote{\label{foot:Delta}Let us consider (A)dS${}_d$ with cosmological constant $\Lambda_d$ of sign $s_{\Lambda}$ and radius $l$, where $\frac{1}{l^2} = \frac{2|\Lambda_d|}{(d-1)(d-2)}$. The relation between the mass $m$ of a scalar mode in the bulk and the conformal weight $\Delta$ of the (boundary) CFT is given as follows
\beq
m^2 l^2 = - s_{\Lambda}\ \Delta(\Delta -(d-1)) \ \Leftrightarrow\ \Delta_{\pm} = \frac{d-1}{2} \pm \sqrt{\left(\frac{d-1}{2}\right)^2 - s_{\Lambda}\, m^2 l^2} \ .
\eeq
In AdS, the unitary representations require $\Delta$ to be positive and real. For irrelevant operators ($\Delta>d-1$), one gets $\Delta_+$ with $m^2>0$, while for relevant operators ($\Delta<d-1$), one gets $\Delta_{\pm}$ with $m^2 <0$ together with the BF bound \eqref{BFbound}. In dS, one is only allowed to have relevant operators with $\Delta_{\pm}$ for $0 < m^2 l^2 < \left(\frac{d-1}{2}\right)^2$, or complex $\Delta$ for higher $m^2$; see e.g.~\cite[App.A]{Guo:2025mlb}.} Regarding the boundary where the field theory should live, there are several proposals: a first one is the (Euclidian) future spacelike boundary ${\cal I}_+$, a second one is the cosmological event horizon of de Sitter (a null hypersurface located at finite distance). In either case, a consequence is that one cannot decouple gravity at the boundary, in contrast to the AdS situation.\footnote{Recent related comments have been made in \cite{Bedroya:2025ltj}, on the absence of a brane picture for dS/CFT; see however \cite{Apers:2026lgi} about the AdS counterpart of the argument.} Finally, having the de Sitter horizon as the boundary leads to a non-zero temperature, a finite entropy, with a possibly finite number of corresponding microscopic degrees of freedom, i.e.~a finite Hilbert space (see \cite{Banks:2000fe, Banks:2001yp, Witten:2001kn, Banks:2002wr} and references therein; see also \cite{Susskind:2021omt}): this all differs from the AdS/CFT situation. In particular, having a finite number of states from string theory is especially challenging.

In spite of these differences and difficulties, we should still mention few proposed realisations of a de Sitter holographic duality. First, a free CFT, dual to a higher spin theory in the bulk, has been considered \cite{Anninos:2011ui, Anninos:2017eib}. It uses e.g.~Vasiliev higher spin theory, which is however not easily connected to string theory (see \cite[Sec. 4.3]{Danielsson:2018ztv} for comments). More recently, progress has also been made on the low dimensional side (dS${}_d$ with $d\leq 3$): for dS${}_3$/CFT${}_2$ see e.g.~\cite{Collier:2025lux} and references therein, while proposals have also been made for dS${}_2$/CFT${}_1$, the latter referring to a theory on the timeline. But gravity is very different in dimension $d\leq 3$ (topological nature with no propagating degree of freedom), and it is conceivable that results in low dimensions differ from the expected physics in $d\geq 4$. As we will see, swampland conjectures also sometimes make this distinction of dimensions. Finally, we should mention recent tentative approaches to a de Sitter quantum gravity, using path integral formulations (see e.g.~\cite{Castro:2023dxp, deBoer:2025tmh} and references therein), making use of lessons from AdS/CFT.

In all above considerations, an important question is that of the stability of the de Sitter spacetime. Implicitly, discussions of quantum gravity on de Sitter often assume a fully stable spacetime, i.e.~an eternal de Sitter. However, realising de Sitter as an extremum of a scalar potential \eqref{extremumsol}, we have seen that a de Sitter solution could also be metastable or even unstable; the concrete string theory constructions also preferably hint at such behaviours. The Dine-Seiberg problem \cite{Dine:1985he}, reviewed in Section \ref{sec:reftestdSconj}, also suggests the absence of a fully stable de Sitter in controlled constructions, but at best metastable, if any. In this case, after departing from the de Sitter solution due to a (non-)perturbative instability, the corresponding horizon would be lost in the future, then possibly preventing from a holographic description. Such an absence of fully stable de Sitter solutions, allowing at best a temporary holographic definition, would be in line with the ``No Cosmological Horizon Conjecture'' \cite{Andriot:2023wvg} discussed in Section \ref{sec:reftestdSconj}. This question of de Sitter stability will also be important in the following for the swampland de Sitter conjectures.\\

The above indicates numerous difficulties encountered when trying to obtain a de Sitter spacetime from a quantum gravity theory, especially string theory, together with a certain ``unnaturalness''. However, it is also fair to say that no argument can sustain a complete obstruction against de Sitter, starting with the fact that a complete exploration of all regimes of string theory is for now out of reach. As a consequence, the proposal of \cite{Obied:2018sgi} was quickly considered too strong, and got refined as described in Section \ref{sec:reftestdSconj}. Nevertheless, it led to an intense activity in testing known de Sitter constructions from string theory, as well as proposing new ones. The question of control, discussed above, became central, and as mentioned, {\sl the current status is an absence of any well-controlled example of a de Sitter extremum in 4d}.

One particular class of examples is worth mentioning here, as it illustrates this situation: the de Sitter solutions of 10d type II supergravities respecting the ansatz of Section \ref{sec:10dansatz}, that will be discussed in more detail in Section \ref{sec:dSsolexistence}. Solutions of this kind were already known in 2018. Few technical criticisms against them were then formulated (see e.g.~\cite{Danielsson:2018ztv, Andriot:2018wzk}): the sources are smeared, i.e.~not localized and backreacted (see Section \ref{sec:backreact}), and most of the solutions, found in type IIA, require a non-zero Romans mass $F_0$, whose string theory origin is sometimes debated. Another, phenomenological reason to set aside these solutions was then that they are not minima, and therefore would not provide a true cosmological constant. They are even too unstable for a slow-roll inflation, with $\eta_V < -1$ (see Section \ref{sec:dSsolstab}). Interestingly however, a more important criticism only came after \cite{Obied:2018sgi}: the known type IIA supergravity solutions, meant to be classical string backgrounds, turned out to fail verifying the corresponding constraints, in particular having a large volume together with a small string coupling \eqref{asymptoticclass}, while having quantized fluxes and an adequate (bounded) number of orientifolds. Indeed, this first got systematically investigated for type IIA with $p=6$ in \cite{Roupec:2018mbn, Junghans:2018gdb, Banlaki:2018ayh}. Further discussions and the current knowledge on this delicate question of ``classicality'' will be the topic of Section \ref{sec:dSsolclassical}. This illustrates the efforts that have been made subsequently to \cite{Obied:2018sgi} regarding the control of proposed de Sitter solutions.

\subsubsection{Refinements and tests}\label{sec:reftestdSconj}

Beyond arguments against de Sitter in quantum gravity, as just listed in Section \ref{sec:nodSQG}, the initial swampland de Sitter conjecture \eqref{dSconj} proposed in \cite{Obied:2018sgi} was mostly based on existence no-go theorems presented in Section \ref{sec:nogo4d}, and the values of $c$ obtained there (further arguments from the strong or the null energy conditions were also given). Strictly speaking however, these existence constraints can be circumvented, while the quantum gravity arguments put forward are subject to discussion. As a consequence, the initial proposal \eqref{dSconj} soon got refined towards various proposals \cite{Andriot:2018wzk, Garg:2018reu, Ooguri:2018wrx, Andriot:2018mav, Rudelius:2019cfh, Bedroya:2019snp, Rudelius:2021oaz}. We give in the following the main elements of these refinements, that go along two main directions: including (in)stability in the conjecture, and restricting to field space asymptotics. We refer to other reviews for more arguments and references \cite{Palti:2019pca, vanBeest:2021lhn, Agmon:2022thq, Vafa:2025nst}.\footnote{Discussions of de Sitter in quantum gravity can be found in \cite{Palti:2019pca}, \cite{Agmon:2022thq} emphasises on (in)stability arguments, and connection to other swampland conjectures, namely the distance conjecture, the emergent string conjecture and further results on the species scale (see below on those) can be found in \cite{Palti:2019pca, Vafa:2025nst}.}\\

A first proposed refinement to \eqref{dSconj} was that de Sitter extrema can actually exist, but they must be perturbatively unstable \cite{Andriot:2018wzk, Garg:2018reu}. A first argument supporting this point was the existence of known de Sitter solutions of type II supergravities, that were tachyonic. As explained in Section \ref{sec:nodSQG} however, it was only later realised that these solutions known in 2018 actually did not fulfill the requirements to be classical string backgrounds. Still, arguments in favor of the existence de Sitter maxima were discussed further, for example with \cite{Conlon:2018eyr} that relies on non-perturbative contributions to string theory constructions. 

More dramatically, the connection to scalar potentials in particle physics appeared crucial. It is commonly considered that a unifying, fundamental theory of Nature, should be a quantum gravity theory, and be able in addition to provide the standard model of particle physics as an effective theory. As a consequence, scalar potentials appearing in the standard model should obey characteristics of a positive scalar potential in a quantum gravity effective theory; indeed, the cosmological dark energy provides a small positive energy over which standard model scalar potentials can be considered. Concretely, this implies that the scalar potentials of the Higgs boson \cite{Denef:2018etk}, the QCD axion \cite{Murayama:2018lie} as well as the (neutral) pion \cite{Choi:2018rze} should obey any kind of swampland de Sitter conjecture. However, these potentials exhibit local maxima, which seems in contradiction with \eqref{dSconj} and calls for a refinement. Among various ways out discussed in those works, one would be to have a second, decoupled scalar field $\hat{\varphi}$ that would be rolling in a transverse direction, giving overall a non-zero $\nabla V$. If that field direction transverse to the standard model fields was steep, this could have affected the EW (Higgs potential) and the QCD (pion potential) phase transitions that have occurred in the early times of our universe. Taking $\hat{\varphi}$ as a quintessence field, the small value of today's dark energy rather hints at a small $\del_{\hat{\varphi}} V$. This however makes it very difficult to accommodate \eqref{dSconj} where $c \sim {\cal O}(1)$ \cite{Denef:2018etk}. Related comments and further criticisms against the initial proposal \eqref{dSconj} can be found in \cite{Cicoli:2018kdo}.

As a consequence, the Refined de Sitter conjecture was proposed in \cite{Garg:2018reu, Ooguri:2018wrx}, and is stated as follows: an effective theory in the landscape, of the form \eqref{S4d}, should obey (with $V>0$)
\begin{gather}
\label{dSconjRef}
\begin{aligned}
{\rm RdSC:}\quad\quad & \frac{\nabla V}{V} \geq \frac{c}{M_p} \ \ {\rm {\it or}}\ \ \, \frac{{\rm min} M}{V} \leq -\frac{c'}{M_p^2} \ ,\quad c,c' \sim {\cal O}(1) \\
\Leftrightarrow \ \ & \epsilon_V \geq \frac{c^2}{2} \ \ {\rm {\it or}}\ \ \, \eta_V \leq -c'  \ ,
\end{aligned}
\end{gather}
and we refer to \eqref{epseta} for all notations. In particular, we recall that the minimal eigenvalue of the mass matrix $M$ is the same as that of this matrix in a canonical basis, the latter being identical to the Hessian in such a field basis $\nabla_{\hat{\varphi}^i} \del_{\hat{\varphi}^j} V$. This refinement now extends the initial proposal \eqref{dSconj} to allow for another condition on $M$. Considering in particular a de Sitter extremum, $\nabla V=0$, one is forced to obey the new condition, namely that it is (perturbatively) unstable, in line with above arguments. In addition, the positive constants $c,c'$ of order $1$ forbid ``slow-roll'' values (at least for a single field) of $\epsilon_V$ and $\eta_V$, the instability is thus severe. Finally, other refinements have also been proposed \cite{Andriot:2018mav, Rudelius:2019cfh}, leading as well to de Sitter maxima, while earlier ideas on a systematic de Sitter instability were proposed in \cite{Dvali:2014gua, Dvali:2017eba}.

Let us mention already some tests of \eqref{dSconjRef}. We pointed-out that existence no-go theorems give $c \sim {\cal O}(1)$, so one may wonder about stability no-go theorems. As discussed in Section \ref{sec:nogostab}, those are more difficult to obtain. Few examples mentioned there give $c'=\frac{4}{3}$ \cite{Junghans:2016abx, Andriot:2021rdy}, thus in agreement with $c' \sim {\cal O}(1)$. Other no-go theorems in \cite{Andriot:2021rdy} nevertheless give smaller $c'$, but those values do not appear relevant in solution examples. Indeed, another test is that all known de Sitter solutions of type II supergravities obey $\eta_V \leq -2.4$. We refer to \cite[Sec.3]{Andriot:2018mav} for solutions known in 2018, and to \cite[Sec.5.2]{Andriot:2022bnb}, \cite[3.3]{Andriot:2024cct} for later solutions. Those values are thus in good agreement with \eqref{dSconjRef}. Finally, let us report on exploration of scalar potentials away from critical points, where \eqref{dSconjRef} is also supposed to hold. We can mention \cite{Blaback:2013fca}: as reported in \cite{Andriot:2018mav}, ignoring possible consistency issues in type IIA compactifications, points are found with a small $-\eta_V$ but $\sqrt{2 \epsilon_V} \sim 1$. Similarly, a type IIB potential is followed away from a de Sitter point along the steepest descent in \cite{Andriot:2023isc}: values of $\epsilon_V, \eta_V$ obey \eqref{dSconjRef}. For example, point 1 there gives $\sqrt{2 \epsilon_V} \approx 1.05$, $\eta_V \approx - 0.95$. At least in the classical regime, \eqref{dSconjRef} is therefore well-verified so far.\\

A second direction of refinement has been the restriction to the asymptotics of field space. By this we mean to consider quantities, such as the scalar potential, in an asymptotic field limit $\hat{\varphi} \rightarrow \infty$.\footnote{Strictly speaking, to define the limit properly, one may also need to specify the behaviour of the other (transverse) fields; those are often implicitly set at a given finite value.} In the context of swampland de Sitter conjectures, such a limit was first discussed in \cite{Ooguri:2018wrx}, inspired by the Dine-Seiberg problem \cite{Dine:1985he}. Let us recall the latter here; an extended discussion can be found in \cite{Denef:2008wq}. The idea is to consider a 4d effective theory obtained in a weakly coupled regime, where main contributions to the scalar potential are obtained at tree-level, and corrections are small and appear as a perturbative (power) expansion of the coupling. Such a situation would give control on the 4d theory. Crucially for string theory, the coupling(s) defining the perturbative expansion are fixed by 4d field values. Indeed, the main coupling considered is the string coupling $g_s$, which is small in the dilaton asymptotic $\phi \rightarrow - \infty$; another option is the expansion of $\alpha'$-corrections, which are small for large lengths, such as the large volume asymptotic $\rho \rightarrow \infty$. We discussed such an asymptotic regime around \eqref{asymptoticclass}. In these asymptotic limits, it can be argued that $V \rightarrow 0$ \cite{Dine:1985he, Denef:2008wq}.\footnote{One can verify this explicitly on $V(\rho,\tau)$ given in \eqref{rhotautheory}, representing here the ``tree-level'', while perturbative (power) corrections are negligible in the limit. This verification requires in some terms to extract the $\rho$ dependence from $\tau$.} Considering $V>0$, this situation implies {\sl a runaway in the potential} for these fields directions. Indeed, avoiding a runaway, meaning having a de Sitter extremum, requires more than one term to contribute to the potential, i.e.~more than one power of the coupling. A de Sitter maximum can be obtained with two terms, while a minimum can be obtained with three. We note in addition that due to the runaway, {\sl the de Sitter minimum can at best be metastable, not fully stable}. The Dine-Seiberg ``problem'' is then that {\sl for the different terms to contribute enough to generate a de Sitter extremum, they should be non-negligible}. Since those are, in the above perspective, generated by different orders of corrections, this means that corrections are non-negligible, so we most likely loose control of the description. There is even a risk that the de Sitter solution lives in a strong coupling regime.

As stressed in \cite{Denef:2008wq}, this problem can be relieved (at least for large but finite field values) by allowing for several contributions at ``tree-level'', namely those from the 10d supergravity approximation of string theory, especially the fluxes and the sources. One verifies indeed in $V(\rho,\tau)$ in \eqref{rhotautheoryexp} that each term alone is a runaway given by exponentials, but several terms can in principle compete to get an extremum (see the Maldacena-Nu\~nez no-go theorem (\ref{nogo1}) for related discussion). Still, {\sl in the asymptotic, one term of the positive potential would dominate and give a runaway}. Such a generalisation of the Dine-Seiberg problem to any scalar field was considered in \cite{Ooguri:2018wrx}. It implies that the slope of the potential does not vanish in (any) asymptotic, thus indicating an inequality of the type \eqref{dSconj} to hold in those limits.\\

More explicit asymptotic statements were obtained in \cite{Bedroya:2019snp, Rudelius:2021oaz}. First, the Trans-Planckian Censorship Conjecture (TCC) \cite{Bedroya:2019snp} considers the following physical argument: a mode of super-Planckian energy should not, through redshift in an expanding universe, reach a 4d accessible energy scale; otherwise the 4d effective theory would not be trustable anymore. A more detailed account and some refinements can be found in \cite{Andriot:2022brg, Bedroya:2025ris}. While this argument has been debated,\footnote{For example, if a super-Planckian mode is formed as a fluctuation in a thermal bath, then it would most likely decay as well there before redshifting. Other comments can be found e.g.~in \cite{Burgess:2020nec}.} the TCC had in any case interesting consequences in the form of a swampland de Sitter conjecture. Considering a single canonical field, the TCC implies that a 4d effective theory of the form \eqref{S4d}, in the landscape, with $V>0$ and $\del_{\hat{\varphi}} V \leq 0$, should obey
\beq
{\rm TCC:}\quad\quad 0 \leq  V(\hat{\varphi}) \leq e^{ - \frac{c}{M_p}\, |\hat{\varphi} - \hat{\varphi}_0|}  \ ,\quad\quad \left\langle \frac{|\del_{\hat{\varphi}} V |}{V} \right\rangle_{\hat{\varphi} \rightarrow \infty}  \geq \frac{c}{M_p} \ ,\quad\quad c=\sqrt{\frac{2}{3}} \ .\label{TCC}
\eeq
In other words, a runaway potential is bounded from above by a decreasing exponential of rate $c$. As a consequence, the logarithmic derivative {\sl in the asymptotic} is also bounded, in average, by $c$. This provides an asymptotic version (refinement/restriction) of the swampland de Sitter conjecture \eqref{dSconj}. Also, there is now a proposed value for the number $c \sim {\cal O}(1)$. Restricting to an exponential potential, as is common in asymptotics of string theory (see \eqref{rhotautheoryexp}),
\beq
V=V_0\, e^{-\frac{\lambda}{M_p}\, \hat{\varphi}} \ ,\quad  V_0,\, \lambda >0 \ , \label{Vexp}
\eeq
one deduces a bound on the exponential rate: $\lambda \geq c$. Restricting to such a single field model with exponential potential, further results were obtained in \cite{Bedroya:2019snp}. Such a setting can be treated using dynamical systems (see \cite{Andriot:2024jsh} for a recent comprehensive account). Considering the only cosmological content to be the scalar field kinetic energy together with $V$, one finds an attractive fixed point for $\lambda < \sqrt{6}$, denoted $P_{\varphi}$ in Table \ref{tab:fixedpoints}. Focusing on the physics at this asymptotic solution, the TCC then implies the requirement $\lambda \geq \sqrt{2}$, thus suggesting an even higher bound $c$.

This same value was proposed in a more general setting, considering a multifield theory with an arbitrary positive potential $V$. Requiring invariance of $c$ under dimensional reduction, i.e.~having a single expression depending the dimension $d$, together with tests on examples, the following Strong de Sitter Conjecture (SdSC) was proposed in \cite{Rudelius:2021oaz, Rudelius:2021azq}. A 4d effective theory of the form \eqref{S4d}, in the landscape, with $V>0$, should obey in any field space asymptotic \cite{Bedroya:2019snp, Rudelius:2021oaz}
\beq
{\rm SdSC:}\quad \quad \hat{\varphi} \rightarrow \infty:\quad \frac{\nabla V}{V} \geq \frac{c}{M_p} \ ,\quad c =\sqrt{2} \ . \label{SdSC}
\eeq
This provides a refinement of the initial de Sitter conjecture \eqref{dSconj}, by restricting to the asymptotics as well as fixing the value of $c$. The statement \eqref{SdSC} is weaker than the initial one \eqref{dSconj}, which was valid at any point in field space; the terminology ``Strong'' is rather related to the strong energy condition \cite{Rudelius:2021azq}, as we explain in Section \ref{sec:accnohor}.

Let us finally mention consequences of the TCC regarding stability of a de Sitter solution. As argued in \cite{Bedroya:2019snp}, the TCC forbids fully stable de Sitter extrema. Perturbatively unstable ones are allowed, while metastable ones are also allowed, contrary to the RdSC \eqref{dSconjRef}. An upper on the lifetime $T$ of a metastable de Sitter is however obtained, given by
\beq
{\rm TCC:}\quad\quad T \leq \frac{1}{H} \ln \frac{M_p}{H} \ , \label{TCCstab}
\eeq
with $H$ the Hubble parameter in an FLRW cosmology. We refer to \cite{Bedroya:2019snp} for more comments. We finally point-out that from the TCC, an asymptotic bound on the second derivative (of the runaway) was also obtained in \cite{Andriot:2022brg}: in 4d, it is given by $\langle \del_{\hat{\varphi}}^2 V / V \rangle_{\hat{\varphi}\rightarrow \infty} \geq 2/3$.

Initially, the slope bounds proposed by the TCC \eqref{TCC} and the SdSC \eqref{SdSC} were argued to be compatible, if the TCC bound $\sqrt{2/3}$ was considered for one field while the SdSC one $\sqrt{2}$ was for all fields, thanks to the difference between $\del_{\hat{\varphi}}$ and $\nabla$. From this perspective, these bounds successfully passed at first numerous checks. To start with, {\sl there is up to now no counterexample to the SdSC}. Tests in the asymptotics of Calabi-Yau compactifications of type II or heterotic string can for instance be found in \cite{Cicoli:2021fsd}. Also, if one considers a potential with only positive terms, in a string perturbative limit, it would depend on the 4d dilaton $\tau$ which would necessarily roll-down. By itself, this field would then already saturate the SdSC bound (see e.g.~\cite{Shiu:2023fhb}): this is due to its canonical normalisation \eqref{hattaurho}. This is also valid in $d$ dimensions, for which the SdSC bound becomes $c=2/\sqrt{d-2}$. Turning to the TCC, the existence no-go theorems (those discussed in Section \ref{sec:nogo4d} and more) remarkably satisfy the TCC bound, sometimes with saturation \cite{Andriot:2020lea}: we indeed pointed-out that they give $c\geq \sqrt{\frac{2}{3}}$. The same result is even true when extending to arbitrary dimension $d\geq 3$: there the TCC bound becomes $c= 2/\sqrt{(d-1)(d-2)}$, and the existence no-go theorems get this same lower bound $c$ \cite{Andriot:2022xjh}, with few exceptions in $d=3$. As stressed in Section \ref{sec:nogo4d} (see \eqref{combifields}), these no-go theorems correspond to slope bounds on a particular single field direction, the latter being a combination of the dilaton and internal volumes. Considering a different field sector then led to a counterexample to the TCC slope bound on a single field: in \cite{Calderon-Infante:2022nxb}, considering compactifications of F-theory or type IIB supergravity on Calabi-Yau manifolds with fluxes, a field direction in the complex structure moduli space was found with an absolute value of the logarithmic derivative of $\sqrt{\frac{2}{7}}$. Note however that adding other fields in this example, namely K\"ahler moduli, would still allow to verify the SdSC bound. Finally, multifield examples were even exhibited in \cite{Andriot:2025gyr} where $|\del_{\hat{\varphi}}V|/V \rightarrow 0 $ in some (off-shell) asymptotic field direction $\hat{\varphi}\rightarrow \infty$. This leaves the multifield SdSC bound \eqref{SdSC} as the one now commonly accepted.\\

Another set of arguments have been put forward to characterise the scalar potential in the asymptotics of field space: those rely on the connection to other swampland conjectures, namely the distance conjecture \cite{Ooguri:2006in, Klaewer:2016kiy, Baume:2016psm}, the scalar weak gravity conjecture \cite{Palti:2017elp}, the emergent string conjecture \cite{Lee:2019wij} and the related behaviour of the species scale (a concept first introduced in \cite{Dvali:2007hz, Dvali:2007wp} and revived in the swampland context in many references). Let us briefly recall here few elements on this topic. In short, for a string effective theory, the distance conjecture proposes that in moduli (and maybe even field) space asymptotics, a tower of massive states decays exponentially in the field distance, with an exponential rate $\gamma \sim {\cal O}(1)$: schematically, $m \sim m_0\, e^{-\frac{\gamma}{M_p} \, \hat{\varphi}}$ when $\hat{\varphi} \rightarrow \infty$. The emergent string conjecture identifies these towers as being either a Kaluza--Klein tower, the limit corresponding then to a decompactification, or a tower of string states. More precisely, these are the lightest towers, namely those that decay fast and would interfere (in terms of energy scale) first with the effective theory. The sharpened distance conjecture \cite{Etheredge:2022opl} proposes for the lightest towers a lower bound $\gamma \geq 1/\sqrt{d-2}$ in $d$ dimensions. This lower bound turns out to correspond to the tower of string states; the Kaluza--Klein one being typically higher (see e.g.~\cite{vandeHeisteeg:2023ubh, Monnee:2025ynn}). The species scale $\Lambda_s$, namely the energy scale at which quantum gravity effect start being relevant, typically smaller than $M_p$, could then be related to the mass scale $m$ of these towers (see e.g.~\cite{Castellano:2023stg} and references therein).

Most of these statements, starting with the distance conjecture, are made in asymptotic field limits. A connection to swampland de Sitter conjectures would then also be made in these limits. Such connections in the asymptotic have been proposed, starting with \cite{Ooguri:2018wrx} and the RdSC; see also \cite{Hebecker:2018vxz} about it. Other connections or comparisons between these conjectures have been made (see e.g.~\cite{Lust:2019zwm, Rudelius:2022gbz, Casas:2024oak}), that we do not try to review here. Let us mention \cite{Andriot:2020lea}, where was proposed the idea that the distance conjecture exponential rate $\gamma$ is related to the de Sitter conjecture bound $c$, as $c= 2 \gamma$. While the evidence there was mainly based on the TCC bound and some (not lightest) massive towers, the relation nowadays still holds between the Strong de Sitter Conjecture and the Emergent String Conjecture/Sharpened Distance Conjecture values, with $2 \gamma = c= 2/\sqrt{d-2}$. Intuitively, this can be understood by saying that preserving the validity of the effective theory requires the scalar potential to decrease faster than (the lightest) towers of massive states appearing in asymptotics \cite{Rudelius:2022gbz}. We also mention \cite{vandeHeisteeg:2023uxj}: there, the TCC bound is recovered, using that the validity of the effective theory would require $V \leq M_p^2\, \Lambda_s^2$. This interplay of swampland conjectures, also referred to as a web, eventually strengthens the various statements, and especially now, the SdSC.\\

A question which arises when looking for characterisations of scalar potentials is whether those should be formulated at the level of the effective theory, meaning off-shell (e.g.~at any point in field space or in any asymptotic), or at the level of its solutions, i.e.~on-shell, along physical field space trajectories. For now, the claims are rather made off-shell, even though their ``derivations'' sometimes use equations of motion. One may wonder whether restricting to an on-shell statement, e.g.~to actual cosmological solutions, could provide a stronger claim (e.g.~a higher bound $c$). Note that this question is not relevant for single field situations, where for runaway potentials, there is no difference between off-shell and on-shell trajectories. For a discussion of the multifield cases, and the asymptotics of physical trajectories, we refer to \cite[Sec.4]{Andriot:2025gyr}. For now, we do not have a better answer than the SdSC bound \eqref{SdSC}, that we use both in off- and on-shell considerations.\\

The situation reached after these various refinements will be summarized in Section \ref{sec:dSconjsum}. But we can already mention two main points, related to existence and stability of de Sitter solutions. First, in field space asymptotics, positive potentials tend to be runaways, obeying a slope bound. This has led the community to study rolling solutions (as in inflation or quintessence). Second, a de Sitter extremum is not expected to be fully stable, implying that the de Sitter solution will eventually decay either perturbatively (unstable) or non-perturbatively/quantum mechanically (metastable) to another cosmological solution; one may talk of a ``transient de Sitter'' phase, or ``no eternal de Sitter''. Before investigating further consequences of these points in Section \ref{sec:circumvent}, let us mention another conjecture inspired by them.

Loosing de Sitter in the future also means loosing the corresponding horizon. More precisely, we refer here to the cosmological event horizon, which is the one defined by the (finite) distance traveled by light until the end of time, in a spacetime described by an FLRW metric; see Section \ref{sec:accnohor} and Appendix \ref{ap:cosmo} for this definition. The far future of the universe therefore matters, and the idea that it will not be de Sitter hints at an absence of (event) horizon. Of course, the cosmological solution after the de Sitter decay may still have a horizon, but we have reason to believe it will not be the case. One may also consider other cosmological solutions obtained from string theory,\footnote{Expressed in terms of the FLRW metric \eqref{FLRW}, one verifies that anti-de Sitter and Minkowski spacetimes do not admit a cosmological event horizon. This agrees with the NCHC, since those are thought to be consistent backgrounds of quantum gravity.} such as rolling solutions: a conclusion there is that also they do not possess a cosmological event horizon, as we will discuss in Section \ref{sec:accnohor}. Therefore, the No Cosmological Horizon Conjecture (NCHC) was proposed in \cite{Andriot:2023wvg}, claiming that (for $d\geq 3$)
\beq
\hspace{-0.1in} {\rm NCHC:}\quad\quad \text{{\sl Solutions of quantum gravity do not admit a cosmological event horizon.}} \label{NCHC}
\eeq
This proposal broadens a ``no (fully stable) de Sitter claim'' to cosmological solutions at large (still with FLRW metric), focusing on the central and possibly more fundamental object that is the event horizon. This idea was recently revisited in \cite{Friedrich:2025aec}, in view of an extension into a ``horizon criterion''. We recall from Section \ref{sec:nodSQG} that this horizon may play the role of a holographic boundary in some attempts of a ``dS/CFT'' correspondence. Its absence may then help understanding the difficulties encountered in those efforts (away from low dimensions). Having no event horizon would also benefit to the construction of a (quantum gravity) S-matrix, thanks to causally connected asymptotic states. These comments are in line with \cite{Bedroya:2022tbh, Bedroya:2024zta}, discussing the TCC in a holographic context. This question of the horizon is also related to that of (eternal) cosmological acceleration; using this relation, similar criticism against cosmological event horizon in string theory were already formulated in \cite{Hellerman:2001yi, Fischler:2001yj}. Along these lines, a milder version of an NCHC was actually formulated in passing in \cite{Townsend:2003qv, Russo:2022pgo}, inspired by a discussion reminiscent of the SdSC \eqref{SdSC}. More recently, related comments were made in \cite{Hebecker:2023qke}, connecting an absence of asymptotic acceleration to an absence of de Sitter in higher dimensions. We will discuss the relation between the cosmological event horizon and cosmological acceleration in detail in Section \ref{sec:accnohor}.

\subsection{Summary - where do we stand?}\label{sec:dSconjsum}

Section \ref{sec:constraints} was mainly dedicated to provide constraints on realisations of dark energy in the classical and asymptotic regimes of string theory. Concretely, this meant constraints against de Sitter solutions, and by extension, on rolling solutions, for which fields evolve in a scalar potential away from extrema. To get a 4d de Sitter spacetime, we first introduced in Section \ref{sec:compactif} a specific ansatz for solutions of 10d supergravity. It corresponds as well to a compactification ansatz, towards a 4d effective theory, with all subtleties associated to the latter. Given this ansatz, one can derive existence no-go theorems against de Sitter. A first outcome of those is to indicate ingredients \eqref{summarydS} in the compactification that are necessary to such a solution, that is a negative 6d curvature, certain orientifolds, and specific RR fluxes, to start with. In the 4d effective theory, these no-go theorems can be extended away from extrema: their second outcome is then to provide bounds on the slope of the potential. Further studies in 4d allow as well to get some stability no-go theorems, i.e.~to show under some conditions perturbative instabilities of de Sitter extrema. The no-go theorems of Section \ref{sec:nogo} have inspired more general claims related to de Sitter solutions and on positive scalar potentials in effective theories of quantum gravity. These swampland de Sitter conjectures, discussed in Section \ref{sec:swamp}, are also motivated by the general difficulties in getting well-controlled de Sitter solutions in string theory. The proposed conjectures have been debated, tested and refined. Today, the bottom line would be twofold. First, a de Sitter extremum in a quantum gravity effective theory is not expected to be fully stable: it is (perturbatively) unstable or metastable; quantitative characterisations of these instabilities are given respectively by the RdSC \eqref{dSconjRef}, namely $\eta_V \lesssim -1$, or the TCC \eqref{TCCstab}. Second, a positive potential is expected in field space asymptotics to be a runaway, with a bound on the slope given by the SdSC \eqref{SdSC}: $M_p\, \nabla V/V \geq \sqrt{2}$ in 4d \cite{Bedroya:2019snp, Rudelius:2021oaz}, i.e.~$\epsilon_V \geq 1$. These proposals have led to the broader claim of the NCHC \eqref{NCHC}, stating that solutions from quantum gravity do not admit cosmological event horizons; this newer proposal will discussed in more depth in Section \ref{sec:accnohor}.

It is certainly in the classical or asymptotic regimes of string theory that all above claims can be better trusted. Indeed, results from 10d supergravity compactification fall in this regime, as also emphasized in 4d with the weak coupling and large volume limits \eqref{asymptoticclass}. In addition, the swampland conjectures have mostly been refined towards field space asymptotics, where corrections are a priori under control. As a result, all these constraints will sharpen the options to realise dark energy in such regimes, as we will detail. Of course, most constraints come with assumptions, starting with the regime or the compactification ansatz: this offers ways to circumvent these obstructions, as we will discuss in Section \ref{sec:circumvent}, opening the door to different options to realise dark energy.\\

Let us now mention some cosmological consequences. Starting with classical de Sitter solutions, discussed in more detail in Section \ref{sec:dSsol}, we first recall that they are not expected to be found in the (strict) asymptotic, even though large field values are a priori possible to ensure the regime validity. Whenever found, it seems from above and through all known (tentative) examples that these solutions are all severely unstable with $\eta_V < -1$. This prevents from a standard cosmological constant scenario. But other scenarios could be compatible: a field could have been maintained at the top of a scalar potential by a Hubble friction until the recent universe, when it would start rolling down. Such a rolling solution, away from a maximum with $\eta_V \sim -1$, could still accommodate observational data \cite{Agrawal:2018rcg}; see \cite{Shlivko:2024llw, Bayat:2025xfr, Andriot:2025los} for comparison of such hilltop quintessence models to DESI data. Turning to inflation, it is interesting to point-out that single field models and observations tend to favor concave potentials, reminiscent of the de Sitter instability just discussed. However, the data is well fitted by slow-roll values: $\eta_V \sim - 0.01$ \cite{Planck:2018jri}. Multifield inflation models could give different results (see below), in better agreement with above de Sitter solutions.

Turning to rolling solutions on potential slopes, and focusing on asymptotic runaways, the SdSC bound gives $\epsilon_V \geq 1$, which is fairly steep. In particular, this forbids slow-roll single field inflation. However, multifield inflation models with rapid-turns and non-geodesic trajectories could accommodate this \cite{Brown:2017osf, Garcia-Saenz:2018ifx, Achucarro:2018vey, Bjorkmo:2019aev, Bjorkmo:2019fls, Cicoli:2020noz, Akrami:2020zfz, Aragam:2020uqi} (see however \cite{Freigang:2023ogu}). There is of course no reason for inflation to take place in asymptotics of field space, and therefore to be subject to this constraint, even though from the perspective of the theoretic construction, the level of control there is more satisfying. Regarding quintessence, the bound makes it a priori too steep for a single field and an exponential potential to have accelerated expansion realised. But there are several loopholes to such a claim (in particular coupling to matter), so we postpone this discussion to Section \ref{sec:quint}.

In short, it seems the classical and asymptotic regimes require dark energy today to be realised by specific quintessence models (e.g.~hilltop or runaway exponential), while reproducing inflation would need special multifield models. {\sl This differs from the paradigm of slow-roll single field inflation and today's cosmological constant.}\\

Let us end this section by discussing the following question, of relevance both to the various constraints obtained as well as to the cosmological models: in $d$ dimensions, $3\leq d\leq 9$, {\sl how many scalar fields should be expected in effective theories \eqref{S4d} from string theory? In particular, does one get single or multifield models?} When obtained from a (standard, i.e.~geometric) compactification of 10d string theory, the lower dimensional effective theory possesses at least two scalar fields: one is the dilaton, present from the start, and the second is the volume of the compact manifold ${\cal M}$; those match the universal $(\rho,\tau)$ discussed in Section \ref{sec:rhotau}. In addition, those two qualify as non-compact scalar fields, in the sense that their range is infinite (contrary to axions). Avoiding this situation requires to consider non-geometric constructions of string theory, i.e.~where one looses the geometric description of the extra dimensions (and thus the volume). Even there, one can still get at least two non-compact scalar fields, as e.g.~the dilaton and a complex structure modulus in Landau-Ginzburg models \cite{Greene:1996cy, Becker:2006ks}. More non-geometric constructions, obtained through asymmetric orbifolds, can however give rise to only one non-compact (and neutral) scalar field in the effective theory: see \cite{Baykara:2024vss, Baykara:2024tjr} for examples in 4d with ${\cal N}\leq 1$ supersymmetry. The latter remain very isolated and tailored examples, as far as we know for now; a more detailed account can be found in \cite{Andriot:2025gyr}, together with a discussion of M-theory (without perturbative string limit). Overall, from string theory, we conclude that {\sl the multifield situation seems to be the most common one.}

However, having a multifield effective theory does not mean that all fields appear in the scalar potential (at tree-level): some field directions can be flat. In \cite[Sec.2.1]{Andriot:2025gyr} is discussed an {\sl almost single field} case: that of a compactification on a curved Einstein manifold ${\cal M}$. The specific geometry only provides the volume as scalar field, while the potential is generated only from the curvature term. The fields $(\rho,\tau)$ \eqref{rhotautheory} combine into one field entering the potential (corresponding to the 6d volume in 10d Einstein frame), and the transverse combination is flat. The exponential rate in $d$ dimensions, for $3 \leq d \leq 9$, is given by $4\sqrt{2}/\sqrt{(d-2)(10-d)}$ (corresponding to a standard Kaluza--Klein rate along $10-d$ dimensions): it obeys, as expected, the SdSC bound \eqref{SdSC}. The same holds for another {\sl almost single field} example, with a flux generated potential, discussed in \cite[Sec. 2.2]{Andriot:2025cyi}. In Section \ref{sec:quint}, we will then consider, among others, single field exponential models \eqref{Vexp} for quintessence with $\lambda \geq \sqrt{2}$.

\newpage

\section{Beyond obstructions: attempts at de Sitter and quintessence}\label{sec:circumvent}

In Section \ref{sec:constraints}, we have established some constraints on realisations of dark energy from string theory in the classical or asymptotic regimes. Those take the form of no-go theorems, based on some assumptions (including a solution ansatz), or swampland conjectures, tested and mostly trusted in the asymptotics. We have summarized these obstructions in Section \ref{sec:dSconjsum}, together with their cosmological consequences for dark energy in the recent universe, and for inflation. In this section, we restrict ourselves to the former, namely we aim at describing the recent accelerated expansion of our universe, either with a de Sitter solution or with a quintessence model. Starting from string theory in the above regimes, we then face the constraints just mentioned. There are thus several options.

\begin{enumerate}[label=\textbf{\arabic*.}]
\item The first option is to stick to the framework of Section \ref{sec:constraints} and embrace the established constraints. Looking for a de Sitter solution in the classical regime, within the ansatz introduced, the existence no-go theorems have indicated the ingredients necessary to find some (see \eqref{summarydS} to start with). Such classical de Sitter solutions will then be the topic of Section \ref{sec:listclassgroup} and \ref{sec:dSsol}; let us recall that they are most likely maxima, i.e.~perturbatively unstable, if they exist at all. The prime setting for these solutions will be the compactifications on 6d group manifolds ${\cal M}$, a set that offers several technical advantages, on top of allowing for negatively curved metrics: ${\cal R}_6<0$. As pointed-out already, a complete solution requires to go beyond the ansatz used, by taking into account the backreaction of $D_p$-branes and orientifold $O_p$-planes, a non-trivial matter that will also be discussed there. Sticking to the previous ansatz however means that the backreaction is not expected to change drastically the solution.
    
    Within the framework and constraints of Section \ref{sec:constraints}, another option is quintessence with rolling solutions. In the classical or asymptotic regime, a natural starting point is to consider an exponential potential (as in \eqref{rhotautheoryexp}), with at first a single field for simplicity, namely the potential \eqref{Vexp}; see Section \ref{sec:dSconjsum} for a discussion of string theory realisations. Restricting to the asymptotic then leads, due to the SdSC \eqref{SdSC}, to the lower bound on the exponential rate: $\lambda \geq \sqrt{2}$ in 4d. The corresponding cosmology, and extensions beyond this simple setting, will be discussed in Section \ref{sec:quint}.

\item The second option is to stick to the string theory (classical or asymptotic) regime, but change the solution or compactification ansatz. Some of the previous constraints are a priori not obeyed anymore, because assumptions have changed. But extensions of prior constraints to the new ansatz or setting may still exist; in other words the difficulty indicated by previous constraints may still be present in some way. Nevertheless, examples of tentative de Sitter constructions falling in this category exist, and will be listed in Section \ref{sec:listattempts}. To already illustrate this idea, one could for instance:
      \begin{itemize}[label=-]
        \item Consider from the start a fully backreacted solution, in which taking a smeared limit would make one loose the de Sitter solution (see Section \ref{sec:listbackreact} and \ref{sec:backreact}).
        \item Include more ``classical'' ingredients, such as $\bar{D}_p$-branes or KK--monopoles (see Section \ref{sec:listclassingr}).
        \item Move away from type IIA/B, and start from a different theory, e.g.~heterotic superstrings (see Section \ref{sec:listMth}, \ref{sec:listhet} and \ref{sec:listnonsusy}).
        \item Go to an asymptotic regime which is not classical, in which one may still have some control: for example F-theory with a non-weak string coupling (see Section \ref{sec:listFth}).
        \item Regarding quintessence, include (non-relativistic) matter, present in our recent universe but not considered in the effective theory \eqref{S4d} (see Section \ref{sec:quint}).
      \end{itemize}

\item The third option, apart from changing completely the setting, is to change the string theory regime. This is meant in a broad sense: for example one could allow for non-perturbative or quantum contributions (instantons, quantum condensates, Casimir energy, etc.), as well as perturbative corrections ($\alpha'$- or loop corrections). We will discuss tentative de Sitter constructions along these lines in Section \ref{sec:listattempts}, in particular in Section \ref{sec:listCas} and \ref{sec:listnonpert}.

\end{enumerate}

Given these options to accommodate or circumvent the constraints of Section \ref{sec:constraints}, we organise this section as follows. We first provide for completeness in Section \ref{sec:listattempts} a tentative list of de Sitter constructions, with the examples just mentioned and more, together with brief comments on control issues building on Section \ref{sec:nodSQG}. We reserve a detailed treatment of classical de Sitter solutions to Section \ref{sec:dSsol}. We then introduce quintessence in Section \ref{sec:quint}, and focus, as a simple starting point, on single field exponential quintessence. A first overview of possible cosmological solutions is provided in Section \ref{sec:fixedpoints} thanks to a dynamical system approach. These solutions illustrate in Section \ref{sec:accnohor} a general discussion on asymptotic acceleration, and its relation to cosmological event horizons, discussed around \eqref{NCHC}. We restrict in Section \ref{sec:matter} to (candidate) realistic solutions by including matter, radiation and possibly spatial curvature. We discuss properties of these solutions, and compare them to recent observational data. A possibly observed phantom regime challenges the latter. We then describe in Section \ref{sec:phantom} how allowing for a coupling to matter can accommodate such a regime and observations, while helping with an asymptotic string theory realisation. Properties of the resulting solutions are briefly discussed. We summarize in Section \ref{sec:sumquint} all these attempts to go beyond previous obstructions and realise dark energy from string theory.

\subsection{Attempts at de Sitter: a tentative list}\label{sec:listattempts}

We propose in this section a tentative list of proposed de Sitter constructions from string theory, with few other dark energy proposals. This list certainly includes examples in regimes different than the classical or asymptotic ones. As such, the description of the attempts may be found brief or incomplete: this should be taken as a warning to the reader. In spite of this risk, we believe this tentative list is of value for two reasons. First, such a list is not common in the literature, and could hopefully provide an inspiring overview, while motivating further studies on one or the other attempt. Second, we aim here at pointing-out weaknesses identified in most of the approaches, often related to the discussion of Section \ref{sec:nodSQG} on control issues. As explained there, we indeed believe that there is currently no example of a de Sitter construction which qualifies as well-controlled, at least in 4d. By this, we mean that all corrections to a solution should be identified, estimated or computed, and found small. It is also true that efforts on building a solid de Sitter solution from string theory are on-going, some constructions being currently under investigation, so we hope the following will also be helpful in that respect. 

The list goes from the most classical and geometric compactification towards the most different string regimes and settings. For an overview, we refer to the table of contents.

\subsubsection{Classical solutions on group manifolds}\label{sec:listclassgroup}

De Sitter solutions of type II supergravities, verifying the ansatz of Section \ref{sec:10dansatz}, have been found in \cite{Caviezel:2008tf, Flauger:2008ad, Caviezel:2009tu, Danielsson:2010bc, Danielsson:2011au, Roupec:2018mbn, Andriot:2020wpp, Andriot:2021rdy, Andriot:2022way, Andriot:2024cct} with a compact 6d manifold ${\cal M}$ being a group manifold. The first of those was $SU(2) \times SU(2)$, and most of the other ${\cal M}$ are not obtained from semi-simple Lie groups but rather from solvable ones (giving solvmanifolds), or mixture of both. While these solutions will discussed in detail in Section \ref{sec:dSsol}, let us give here a few features. All solutions require (at least three) intersecting sets of $D_p$ and $O_p$: by this we mean that these extended objects need to wrap different compact dimensions in these parallelizable manifolds. As a consequence, having a description of their backreaction is difficult, as we will discuss in Section \ref{sec:backreact}, and the solutions are obtained as smeared; this could be perceived as a first weakness. A second feature is that they are all found perturbatively unstable, with $\eta_V \leq -2.4$. We mentioned in Section \ref{sec:dSconjsum} that this is not necessarily a disadvantage for describing today's dark energy. Finally, a more delicate issue is their classicality: as mentioned in Section \ref{sec:nodSQG}, most solutions were shown not to allow for a weak string coupling together with a large volume, while having flux quantized and a bounded number of $O_p$. To those conditions, one should also add a check of the lattice conditions which ensure the compactness of ${\cal M}$. These are all necessary conditions for a supergravity solution to qualify as a classical string background, or in short, to be in the classical string regime; it is therefore unclear whether a classical de Sitter solution of this kind actually exist. While we will discuss this issue in Section \ref{sec:dSsolclassical}, let us mention already a possible counter-example found in \cite{Andriot:2024cct}.

\subsubsection{Classical backreacted solutions}\label{sec:listbackreact}

One way to differ from the ansatz of Section \ref{sec:10dansatz}, where sources are smeared, is to have a de Sitter solution with non-trivial warp factor(s) and dilaton, which capture the backreaction of sources, and where taking a smeared limit is incompatible with a de Sitter solution. It is the case of the 10d type IIA supergravity solutions proposed in \cite{Cordova:2018dbb, Cordova:2019cvf}, especially some with localized $O_8$ sources, which led to comments in \cite{Cribiori:2019clo, Kim:2020ysx, Bena:2020qpa, Horer:2024hgy}; for an account, see e.g.~\cite[Sec. 2.2]{Andriot:2021gwv}. The discussion there had to do with the boundary conditions (of e.g.~the warp factor) close to an $O_8$. Standard $D_8/O_8$ in Minkowski spacetime, or wrapped on a torus, would have so-called restrictive boundary conditions, where two terms are fixed in the expansion of the warp factor in the distance to the source. But \cite{Cordova:2018dbb, Cordova:2019cvf} allowed for permissive boundary conditions, where only one term is fixed, leaving some freedom to the rest of the expansion; this is similar to some supersymmetric anti-de Sitter solutions. Deciding on the allowed boundary conditions is delicate, because they are related to the physics close to the source, where one may leave the validity of the supergravity description and start requiring string corrections (see Section \ref{sec:backreact}); in other words, permissive boundary conditions could be due to non-classical contributions close to an $O_8$. Attempts at including string corrections, to describe the non-standard sources of \cite{Cordova:2018dbb, Cordova:2019cvf}, and resulting difficulties, are reported in \cite{Horer:2024hgy}.

Another example of a proposed classical de Sitter solution is reminiscent of the previous situation. In \cite{Burgess:2024jkx} is presented a de Sitter solution of 10d type IIB/F-theory equations, obtained starting in 4d and proceeding with various dimensional uplifts. The resulting 10d solution has warp factors with singularities: those should correspond to some (stringy) sources. But the latter are not easily identified, as e.g.~standard $D_p/O_p$. Once again, a non-standard backreaction sheds doubt on whether one faces a de Sitter solution of string theory and its fundamental objects, or just supergravity as an independent theory.

\subsubsection{Classical solutions with further ingredients}\label{sec:listclassingr}

The ansatz of Section \ref{sec:10dansatz} contained as stringy extended objects $D_p$-branes and orientifold $O_p$-planes only. While staying in a classical regime, one could consider including more. To start with, one could consider anti-$D_p$-branes, $\bar{D}_p$. One should then be sure of the absence of parallel $D_p$, to avoid the subsequent instability; depending on the formalism used when looking for solutions, this is not necessarily easy. As discussed in Section \ref{sec:nogostab}, $\bar{D}_6$ were included in a massive type IIA supergravity example, to remove the de Sitter tachyon \cite[Sec.4]{Kallosh:2018nrk}. It remains unclear whether this de Sitter supergravity solution stands in the classical regime of string theory: as discussed in Section \ref{sec:nodSQG} for other type IIA de Sitter solutions, it could fail to have a large volume and a weak string coupling.

Other extended objects that have been included for de Sitter are Kaluza--Klein monopoles: in \cite{Silverstein:2007ac}, it was argued from 4d that they should help in getting de Sitter solutions. A difficulty would then be to have a 10d realisation, with localized objects, together with possible $D_p/O_p$. In addition, in analogy to RR Bianchi identities sourced by $D_p/O_p$, Kaluza--Klein monopoles are sources for ``geometric fluxes'' and the corresponding Bianchi identity needs to be obeyed. The latter should be understood as a source-corrected version of the Riemann Bianchi identity \eqref{BIRiemann}, ${\cal R}^a{}_{[bcd]}=0$, as shown e.g.~in \cite[(3.5)]{Andriot:2014uda} (see also \eqref{BIRiemannNotconstant}). On a group manifold, this amounts to a (localized) violation of the Jacobi identity due to the source terms. This idea was used in \cite{Blaback:2018hdo} to find 4d (classical) de Sitter solutions (in particular non-unstable ones, with $\eta_V=0$), by relaxing this Jacobi identity. There as well, one may wonder about the 10d realisation: in particular, having a specific number of sources matching the non-vanishing identity would be necessary. A last point would be the necessity of objects of opposite charge, analogously to $O_p$ for $D_p$: those are sometimes requires for tadpole cancelation, namely satisfying the integrated RR Bianchi identity. Such objects, denoted KKO, were introduced at an effective level in\cite{Villadoro:2007yq} (see also \cite{Dibitetto:2019odu}); they could be argued to be the T-dual to the S-dual of an $O_5$, following the chain from $D_5$ to Kaluza--Klein monopoles. Such objects were also used in \cite{Blaback:2019zig} to find (non-classical, with higher curvature terms) de Sitter solutions. Finally, similar questions would be raised when allowing for an $N\!S_5$-brane.

\subsubsection{Classical solutions in $d\neq 4$}\label{sec:listclassdim}

Most of the discussions in this article focus on $d=4$, but it is legitimate to ask about de Sitter solutions in different dimensions. To start with, the ansatz of Section \ref{sec:10dansatz} can be extended to a compactification towards a $d$-dimensional de Sitter spacetime. Existence no-go theorems can as well be extended \cite{VanRiet:2011yc, Andriot:2022xjh}, providing various constraints: de Sitter solutions are then excluded in $8 \leq d \leq 10$, as well as in type IIB in $d=7$. Restricting further to supersymmetric source configurations, they get further excluded in $d=7$, and in type IIB in $d=6$. As mentioned in Section \ref{sec:nogo10d}, it was conjectured \cite{Andriot:2022way} that they do not exist in $d\geq 5$, unless the $d$-dimensional effective theory is non-supersymmetric. We discuss these ideas further in Section \ref{sec:dSsolexistence}. Related observations and arguments were given in \cite{Cribiori:2023ihv, Hebecker:2023qke}. While classical de Sitter solutions may not exist in $d>4$, let us point-out a recently proposed $5$-dimensional de Sitter solution beyond this regime, using Casimir energy \cite{ValeixoBento:2025yhz}, obtained from M-theory, however found in a non-supersymmetric effective theory.

While gravity becomes peculiar in $d\leq 3$ (see Section \ref{sec:nodSQG}), let us indicate possibilities as well as difficulties for classical de Sitter solutions in $d=3$, mentioned in \cite{Farakos:2020idt, Emelin:2021gzx, Farakos:2025shl, Tringas:2025uyg}. The difficulties often have to do with the question of classicality mentioned above: for instance problems are met when trying to maintain the supergravity solution in a classical string regime once fluxes and sources get quantized. Existence no-go theorems were also extended to $d=3$, sometimes with peculiarities specific to this dimension \cite{Andriot:2022xjh}.

\subsubsection{M-theory}\label{sec:listMth}

Leaving the realm of type II supergravities, one may wonder about classical de Sitter solutions from M-theory. Eleven-dimensional supergravity only has in its bosonic sector gravity and a four-form abelian flux, so the Maldacena-Nu\~nez no-go theorem (\ref{nogo1}) can easily be extended there for smooth compact manifolds (see e.g.~\cite[Sec.2.1]{Obied:2018sgi} for a no-go theorem). Circumventing it with singular and negative tension objects, as orientifolds in 10d, is not obvious in M-theory, given one only has there $M$-branes. We mentioned already type IIA supergravity de Sitter solutions with $p=6$ $D_p/O_p$ sources. $O_6$-planes famously uplift in M-theory to (four-dimensional) Atiyah-Hitchin geometry, but the latter is smooth, and the corresponding charge is not localized. Therefore this uplift does not allow circumvent the Maldacena-Nu\~nez no-go theorem in 11d. Another issue is that such solutions require a Romans mass $F_0 \neq 0$ to exist (see no-go (\ref{nogo2})). Massive type IIA supergravity does not admit an uplift to M-theory; indeed, weak string coupling appears to be related to weakly curved spaces \cite{Aharony:2010af}. One way out might be to consider type IIA de Sitter solutions without Romans mass, forcing one to go beyond the pure $p=6$ source case. It turns out supergravity de Sitter solutions of this kind were found, with both an $O_6$ and an $O_4$ (class $m_{46}$) \cite{Andriot:2022way}, the latter setting $F_0=0$. While an $O_4$ could be uplifted to an $M_5$-brane orbifold, uplifting a whole solution remains challenging.

There could be room for de Sitter solutions from M-theory beyond the classical regime, such as \cite{ValeixoBento:2025yhz} in 5d that we will come back to. Obtaining them in 4d however leads to the complication of compactifying on a seven-dimensional manifold, which are not necessarily easy to deal with. A canonical candidate there are $G_2$-manifolds, which preserve some supersymmetry analogously to Calabi-Yau manifolds. But they are mostly studied for supersymmetric solutions, so not de Sitter.

Finally, a fully quantum proposal in M-theory has been made in a series of papers, a few representative being \cite{Chakravarty:2024pec, Bernardo:2021rul, Brahma:2020tak, Dasgupta:2019gcd}, with a recent implementation in heterotic string \cite{Dasgupta:2025ypg}. The idea is that the de Sitter spacetime appears in a 4d slice from a quantum excited state, known as a Glauber-Sudarshan state. The latter is a coherent state, over a supersymmetric Minkowski vacuum. This is claimed to be realised in M-theory, where many (non-classical) contributions to the action are also discussed (perturbative or not, local or not).

\subsubsection{F-theory}\label{sec:listFth}

F-theory, a non-perturbative, possibly strongly coupled, description of type IIB string theory, is typically studied in supersymmetric compactification, therefore not favorable for de Sitter solutions. We recall in addition the no-go theorem (\ref{nogo2}), obtained in smeared approximation, which forbids de Sitter solutions with $p=7$ sources, extended to $p=3\&7$ in \cite{Andriot:2017jhf}. At least for $p=7$, the obstruction is extendable with non-constant, and non-related, warp factor and dilaton, in the same way it was done for $p=8$ in \cite{Cribiori:2019clo}. One must then go to a fully non-supergravity regime of F-theory to get a chance on de Sitter. In this context, an involved proposal was made in \cite{Heckman:2018mxl, Heckman:2019dsj}.

We should also mention \cite{Grimm:2019ixq}, that uses asymptotic Hodge theory to provide classifications of flux potentials in field space asymptotics, especially in the context of F-theory compactified on Calabi-Yau four-folds. This provides constraints against de Sitter in asymptotics, beyond the classical regime. These results are further used e.g.~in \cite{Calderon-Infante:2022nxb} to study asymptotic slopes of scalar potentials, giving constraints on quintessence realisations there (see Section \ref{sec:reftestdSconj}).

\subsubsection{Supersymmetric heterotic string theories}\label{sec:listhet}

As initiated with no-go theorem (\ref{nogo5}), finding de Sitter solutions in supersymmetric heterotic string theories is highly constrained. No-go theorems and constraints were established against de Sitter at all orders in $\alpha'$ and tree-level $g_s$ \cite{Green:2011cn, Gautason:2012tb, Kutasov:2015eba, Basile:2021krk}, and even considering gaugino condensation and other non-perturbative contributions \cite{Quigley:2015jia, Leedom:2022zdm}. Loopholes involving non-perturbative and instantonic effects were however investigated recently in \cite{Leedom:2022zdm, Alvarez-Garcia:2024vnr}, opening to new possibilities. 

One can also consider orbifolds of supersymmetric heterotic theories; those are interesting because they sometimes allow to reproduce particle physics (e.g.~MSSM). In that context, allowing corrections to the 4d superpotential (in particular non-perturbative ones due to gaugino condensation) and fixing by assumption some related coefficients, while truncating to only few fields, several de Sitter maxima are reported in \cite{Parameswaran:2010ec, Gordillo-Ruiz:2025xcg}. Similarly, truncating to one field (i.e.~fixing the others) in a specific orbifold, a de Sitter minimum is reported in \cite{Florakis:2016ani} from the resulting one-loop string potential. While heterotic orbifolds deserve further study, their complete scalar potentials remain difficult to fully determine and analyse.

\subsubsection{Non-supersymmetric 10d string theories}\label{sec:listnonsusy}

Spacetime supersymmetry is known to remove the tachyon from the 10d string spectrum, an important result for consistency of superstring theories. Using orbifolds or orientifolds of the latter, one can however break supersymmetry at the string scale, and in some cases remain with a tachyon-free theory. Three such non-supersymmetric string theories were constructed in 10d: the heterotic O(16)$\times$O(16) \cite{Alvarez-Gaume:1986ghj, Dixon:1986iz}, the type 0'B \cite{Sagnotti:1995ga, Sagnotti:1996qj}, and the USp(32) string theory \cite{Sugimoto:1999tx}. Recent reviews can be found in \cite{Leone:2025mwo, Dudas:2025ubq}; see also \cite{Fraiman:2023cpa, Baykara:2024tjr} for recent accounts. These three tachyon-free theories have a positive leading order scalar potential depending on the dilaton, while supersymmetric ones do not. For the first two theories, this potential appears at one-loop in the string coupling. Such a positive potential suggests options for dark energy, even though it may just be a runaway. Various tentative dark energy constructions have been made along these lines, so far inconclusive.

Let us point-out that an absence of supersymmetry may at first sight appear more promising for de Sitter solutions. But it also challenges consistency of the theories, and control on the corrections; in particular, a runaway potential for the dilaton may go to strong or to zero coupling, neither being satisfying. Also, considering the heterotic example, the numerous constraints obtained in the supersymmetric theory can make one doubtful on the options for the non-supersymmetric version. And indeed, some existence no-go theorems against de Sitter and Minkowski solutions were established for the non-supersymmetric string theories in \cite{Basile:2020mpt}, typically extending those of Section \ref{sec:nogo} to these new settings. This underlines difficulties faced to get dark energy with these theories.

\subsubsection{Casimir energy in M-theory or type II supergravities}\label{sec:listCas}

The Maldacena-Nu\~nez no-go theorem (\ref{nogo1}) does not necessarily require orientifolds to be circumvented: other contributions with an adequate sign could play the same role. It is the case of the quantum-generated Casimir energy, which appears again as a one-loop contribution. The most important contributions to the latter come from the lightest states running in the loop: here, one should consider 4d lightest states, e.g.~Kaluza--Klein states more than string states. A compactification of M-theory on a hyperbolic (negatively curved) compact manifold with such a Casimir energy was discussed in \cite{DeLuca:2021pej}, giving a de Sitter solution. Among difficulties faced in such a construction, one is that strongly curved regions appear necessary. Further difficulties were indicated in \cite{Parameswaran:2024mrc}, that considered 10d supergravity in certain (curved) compactifications with a Casimir energy contribution. Existence no-go theorems against de Sitter were again established, extending for instance those of Section \ref{sec:nogo}. Control and classicality issues were reported for the de Sitter attempts, with for instance the requirement of quantum effects being of the same order as classical ones, unless a large internal anisotropy is present.

Interestingly, some of these difficulties could be circumvented if one considers for the compactification a flat internal manifold ${\cal M}$. This is the case of \cite{ValeixoBento:2025yhz} that chose for ${\cal M}$ (non-supersymmetric) Riemann-flat manifolds. Compactification of M-theory on such a 6d manifold with Casimir energy and flux is claimed to provide a very well-controlled de Sitter solution in 5d (but for now not in 4d). Computations of the one-loop Casimir energy on such manifolds are also reported in \cite{DallAgata:2025jii}; it depends crucially on the Kaluza--Klein spectrum. Further computations can be found in \cite{Aparici:2025kjj}, and in a dynamical setting in \cite{Paul:2026exi}. These recent results thus appear promising, even though they have not yet given a controlled de Sitter in 4d, and stability no-go theorems proving systematic tachyons have been derived \cite{Parameswaran:2024mrc, ValeixoBento:2025qih}.

\subsubsection{Non-perturbative contributions in type IIB: KKLT and LVS}\label{sec:listnonpert}

This approach is by far the one which has driven most activity in trying to construct a de Sitter solution. The focus is on obtaining a de Sitter vacuum, i.e.~minimum, to realise a cosmological constant. As a consequence, all scalar fields (initially moduli) need to be stabilised: this is the guideline for the constructions. The starting point is a Calabi-Yau (CY) compactification of type IIB supergravity with three-form fluxes, $D_3/O_3$ (and possibly $D_7/O_7$) sources \cite{Dasgupta:1999ss, Giddings:2001yu}. In this setting, one considers a specific (ISD) 4d Minkowski non-supersymmetric solution, in which the complex structure moduli of the CY (and the axio-dilaton) are stabilised by the fluxes. To stabilise the remaining, K\"ahler moduli, corrections are considered. In KKLT (from the name of the authors) \cite{Kachru:2003aw}, one considers non-perturbative contributions to the 4d superpotential $W$, generated for instance by gaugino condensation on $D_7$-branes or Euclidian $D_3$-brane instantons. The resulting 4d, ${\cal N}=1$ theory then admits a supersymmetric anti-de Sitter minimum. In the Large Volume Scenario (LVS) \cite{Balasubramanian:2005zx, Conlon:2005ki}, one considers not only the non-perturbative contributions to $W$ but also perturbative $\alpha'$-corrections to the 4d K\"ahler potential $K$ for K\"ahler moduli. Those allow to get an exponentially large volume, while the 4d theory now admits a non-supersymmetric anti-de Sitter minimum. In KKLT, the value of the AdS cosmological constant, fixed by $|W_0|$, needs to be very small, $|W_0|\ll 1$, while it is not the case in LVS. While moduli are then stabilised, the final ingredient is a positive contribution to the potential which would ``uplift'' towards a de Sitter minimum, i.e.~without destabilising the fields, while breaking supersymmetry for KKLT. Proposed ingredients to generate such an uplift are plenty, the prime example being an anti-brane $\bar{D}_3$ at the bottom of the (warp factor) throat. Detailed reviews on these constructions of de Sitter minima can be found e.g.~in \cite{Danielsson:2018ztv, Cicoli:2023opf, Dudas:2025ubq}.

As can be seen right away, these proposals involve many ingredients, and contributions of a priori different regimes. This brings us back to the question of control discussed in Section \ref{sec:nodSQG}. Criticisms have been raised concerning each of the three steps briefly described above, and it is not the place to provide a comprehensive account on those. Let us just mention a few points here. The concerns prior to 2018 (and the swampland de Sitter conjecture) can be found in \cite{Danielsson:2018ztv}; they mostly had to do with the backreaction of $\bar{D}_3$ in the throat, building on the initial work \cite{Bena:2009xk}; see also \cite{Sethi:2017phn} for a different issue, namely quantum corrections to non-supersymmetric classical solutions. Replies to concerns were also formulated in 2018 e.g.~in \cite{Cicoli:2018kdo, Kachru:2018aqn}. An updated view on these matters can be found in \cite[Sec.5.3, 5.4]{VanRiet:2023pnx}. As mentioned already, the swampland de Sitter conjecture led to a variety of other checks and discussions on these proposals; we report a few of those in the following.\\ 

Different criticisms were later formulated, among which the important ``{\sl singular bulk problem}'' \cite{Gao:2020xqh} for KKLT, building in part on \cite{Carta:2019rhx}. It is well-known that orientifold planes are located at special fixed points in the manifold, and lead to a peculiar backreaction in the metric around their singular locus. Indeed, the warp factor $H \sim e^{4A}$ in standard notations becomes negative close to the source, leading the supergravity metric to appear imaginary! This is not an issue in string theory, where the singularity gets resolved; in other words, the small region close to a standard orientifold where the backreaction becomes non-trivial leaves the realm of 10d supergravity and gets a stringy description. As long as this region is small, one may ignore it and use 10d supergravity to describe most of the solution. The {\sl singular bulk problem} occurs when this region is actually not small, therefore preventing from the standard orientifold answer just described, and indicating another type of singularity. Having a large region also seriously questions the validity of the 10d supergravity description; from this perspective, it can be viewed again as a control issue. It is the specifics of the solution that may fix the size of this region to be large, i.e.~comparable to the overall size of ${\cal M}$. In the case of KKLT, these specifics involve in particular the various flux numbers, details of the non-perturbative contribution and of an $\bar{D}_3$-uplift; those are required to eventually get the de Sitter minimum. The problematic region around the $O_3$ (or equivalent object) does turn out to be large, comparable to the size of the CY. To replace 10d supergravity, first steps in a stringy description of the problematic region were made in \cite{Carta:2021lqg}, where it was further argued the 4d effective description could nevertheless remain valid. As we will see, such a problem could be more general, as it appears in different de Sitter constructions (see the discussion in \cite{Junghans:2023lpo}), although not in LVS. 

We have briefly presented above the KKLT construction in three successive steps, but it is clear that for consistency, it should be achieved at once. Important efforts in this program have been made and results have recently been obtained, as e.g.~\cite{Demirtas:2021nlu} for $|W_0|\ll 1$. The recent status can be found in \cite{McAllister:2024lnt} (see \cite{McAllister:2025qwq} for an overview) that makes progress in trying to achieve the complete construction at once, and where admittedly, control issues remain (as defined here in Section \ref{sec:nodSQG}). Among those, one may mention the question of (unknown) corrections to the K\"ahler potential, especially those that are genuinely obtained in an ${\cal N}=1$ 4d theory (instead of inherited from an ${\cal N}=2$ origin). Recent related works include \cite{Kim:2023sfs, Kim:2023eut, Cvetic:2024wsj}.\\

Turning to LVS, an interesting simple and explicit realisation of a de Sitter vacuum within this scenario was proposed in \cite{Crino:2020qwk} (see e.g.~\cite{Cicoli:2012vw, Cicoli:2013mpa, Cicoli:2013cha, Cicoli:2017shd} for earlier realisations, \cite{Gallego:2017dvd} for a different, 4d one, and \cite{Cicoli:2023opf} for a review). This work helped to study possible corrections and the question of control as presented in Section \ref{sec:nodSQG}. Several possibly dangerous or unknown corrections were pointed-out in \cite{Junghans:2022exo, Gao:2022fdi, Junghans:2022kxg}. One important point there was the distinction between parametric and numerical control (see \cite[Tab.1]{Junghans:2022exo}). Some of these doubts were addressed in \cite{ValeixoBento:2023nbv} (see also \cite{Cicoli:2021rub}), with several corrections remaining to be estimated.\\

Using perturbative and non-perturbative contributions in 4d theories, many other proposals for de Sitter solutions have been made along similar lines (see e.g.~\cite{Rummel:2011cd, Kallosh:2014oja} or more recently \cite{Terrisse:2019usq, Bernardo:2020lar}), that cannot be summarized here; see however the review \cite{Cicoli:2023opf}. Typical questions arising are the string theory origin or realisation of the various 4d proposals, usually discussed at the level of ${\cal N}=1$ supergravity, and the control on corrections. Let us end the discussion by mentioning as well proposals for quintessence that involve many such corrections: see e.g.~\cite{Cicoli:2024yqh}, where the resulting involved 4d model is meant to describe the universe history from inflation until today's quintessence.

\subsubsection{Non-geometry and non-geometric fluxes}\label{sec:listnongeo}

Superstring theories (in critical dimension) are 2d SCFT, typically formulated as a sigma-model, describing the worldsheet of the string. The target space, so far considered here as the 10d spacetime, is an interpretation of SCFT elements (field indices, symmetries, etc.). In particular, it does not need to be a differentiable manifold. Of course, the standard 10d spacetime prevails in the classical regime, allowing to match with a supergravity (target space) description and discuss compactifications, etc. Away from this regime however, one can consider more general (target) spaces allowed by stringy symmetries, or even loose completely a geometric picture for ``the extra dimensions'' by using purely a worldsheet description. We briefly discuss both options in the following in view of 4d de Sitter constructions; these approaches are broadly referred to as ``non-geometry'', since their target space cannot be described by 10d differential geometry.\\

A first freedom available are stringy symmetries beyond target space diffeomorphisms or gauge transformations, that can be used to play the same role, namely act as transition functions on the target space fields between patches. A first example is T-duality that exists on backgrounds with $n$ isometry directions (fields are independent of those). The prime example is to consider a circle of radius $R$, which becomes a circle of radius $l_s^2/R$ under a certain T-duality (symmetry) transformation. Using the latter on the target space metric from one patch to the other gives as a resulting target space a cylinder which ``suddenly'' changes size. Such a non-geometry is known as a T-fold: diffeomorphisms are replaced by T-duality transformations, and the resulting target space is admissible for string theory. It is often argued that such geometries loose control on $\alpha'$-corrections, since they mix small and large lengths. This is true for the previous simple example, but it is less obvious for more complicated ones where the transition functions are picked among the full T-duality group O(n,n). Generalisations of this idea exist with other symmetries, giving rise to S-folds or U-folds. We refer here and in the following to \cite{Plauschinn:2018wbo} for a review and references. Using such spaces as the extra dimensions, it appears less clear how to proceed with a dimensional reduction as in Section \ref{sec:dimred}, since the 6d integral does not seem well-defined. Nevertheless, 4d effective theories have been proposed, where the effect of the non-geometry is captured by so-called non-geometric fluxes. Those enter as constants that generate terms in the scalar potential, similarly to the way standard 10d supergravity fluxes do, as e.g.~in $V(\rho,\tau)$ \eqref{rhotautheory}. For example, a T-fold is captured by a 4d $Q$-flux $Q_a{}^{bc}$. Non-geometric fluxes also correspond to specific gaugings in 4d gauged supergravities. One reason for associating these new terms in the potential to the 10d stringy geometries is the following. Some of the latter, non-geometric string backgrounds, can be obtained by T-duality transformations of standard geometric backgrounds. It is the case of a torus with an $H$-flux, which gets transformed upon two T-dualities into a T-fold (different than the above cylinder). Applying the analogous transformation on 4d fields maps the 4d potential $H$-flux term into a $Q$-flux term.

The above presentation gives a stringy origin, as a non-geometry, to 4d theories with non-geometric fluxes. With an important caveat: while it is understood how some non-geometries give rise to some non-geometric fluxes in 4d, the opposite is not true. This raises the first issue: as we will see, one may use scalar potentials with many terms, argued to be non-geometric fluxes (e.g.~from gauged supergravity), but the 10d string theory origin is not necessarily clear or even existing, especially when all fluxes are turned-on together. Progress has been made on this matter with the formalisms of $\beta$-supergravity, double and exceptional field theory, as well as generalized and exceptional geometry. Those provide a higher dimensional, geometric description of lower dimensional theories with non-geometric fluxes. Whether the higher dimensional geometry is a mathematical artefact, or beyond that, an actual stringy (non-) geometry, has been subject to debate. One possibility is that the only 4d theories that get a string theory origin are those that can get transformed back to a purely geometric setting, as in the above $H$-flux / $Q$-flux example: those 4d theories / 10d backgrounds are said to be on geometric (T-duality) orbits (see e.g.~\cite[Sec.4.3]{Andriot:2014uda}). In that case however, no new physics is obtained by considering non-geometric fluxes, beyond the one captured by standard compactifications; in particular no new de Sitter solution. The question is thus whether purely non-geometric orbits exist, with only stringy non-geometry origin: those could give new 4d physics.

Mostly ignoring the question of the string theory realisation, many works have looked for de Sitter solutions using 4d potentials incorporating many non-geometric fluxes. One motivation for using such a framework is that the new potential terms can help stabilising further moduli. It is the case for instance of K\"ahler moduli in the (non-geometrically extended) version of type IIB Calabi-Yau compactification, which were otherwise stabilised using non-perturbative contributions. De Sitter minima were then found in \cite{deCarlos:2009fq, deCarlos:2009qm, Damian:2013dwa}. Further de Sitter extrema were reported, see e.g.~for an early sample \cite{Dibitetto:2010rg, Dibitetto:2011gm, Danielsson:2012by, Blaback:2013ht}. However, {\sl a delicate issue with such solutions are Bianchi identities} (BI). Indeed, those need to be verified in addition to the extremization conditions of the 4d potential, and can be understood as having a 10d string origin. In the geometric compactification discussed in Section \ref{sec:10dansatz}, they were given in \eqref{BI}, as the NSNS one $\d H=0$ together with the Riemann BI, and the RR ones $\d F_q - H\w F_{q-2} = D_p/O_p\ {\rm contributions} $. Extended versions exist in presence of non-geometric fluxes (see e.g.~\cite{Andriot:2014uda}). In \cite[Sec.6]{Plauschinn:2020ram}, several references with de Sitter minima were found problematic in that respect: one or the other type of BI was not respected. In particular, the focus was on the sign of the RHS of the RR BI, which was argued to be fixed for a stringy origin (a matter also known as tadpole cancelation). That property was not respected by solutions of \cite{deCarlos:2009fq, deCarlos:2009qm}. Similar issues were reported e.g.~in \cite{Gao:2018ayp}, where some BI were shown to be missing, i.e.~not taken into account, in the formulation used (possibly corresponding to ignoring non-harmonic forms). Related de Sitter constructions like \cite{Shukla:2022srx} could then be plagued by this problem.\\

As explained above, more formal treatments of the ``extra dimensions'' can be made, using purely the worldsheet description and loosing completely a target space geometric picture, delving even more than before into ``non-geometry''. In spite of this, 4d effective theories can still be obtained. It is for instance the case of asymmetric orbifolds. Other examples include specific Landau-Ginzburg models, whose effective theory is similar to the one obtained with a type IIB Calabi-Yau compactification, however without K\"ahler moduli \cite{Becker:2006ks, Becker:2007dn}. The latter typically describing the volume of the 6d compact manifold, having none makes the setting intrinsically non-geometric. In such a framework, a de Sitter maximum was found in \cite{Ishiguro:2021csu}; comments on the latter, as well as on de Sitter minima that do not satisfy the BI (tadpole cancellation) were made in \cite{Bardzell:2022jfh}. In a similar setting, promising de Sitter maxima were recently argued to exist in \cite{Chen:2025rkb}. Finally, let us point-out discussions in \cite{Cremonini:2023suw} on the scalar potential in such models, studied in the asymptotics of field space, in view of a quintessence realisation of dark energy.
 
\subsubsection{Supercritical string}\label{sec:listcrit}

It is well-known that $\beta$-functionals of (super)string theories, once expanded in $\alpha'$, give the (super)gravity equations of motion at order $(\alpha')^0$; the higher orders give the $\alpha'$-corrections. In other words, getting a conformal field theory amounts to have a string background that verifies these equations. The dilaton $\beta$-functional however possesses a leading term in $\frac{D-D_{{\rm crit.}}}{\alpha'}$, where $D$ is the target spacetime dimension and $D_{{\rm crit.}}$ its critical value, namely $D_{{\rm crit.}}=26$ or $10$ for bosonic or super-string. Setting $D=D_{{\rm crit.}}$ and considering the 10d supergravity dilaton equation at $(\alpha')^0$ is therefore a first option to ensure conformal invariance. Another option is to have both non-zero but compensating each other to make the $\beta$-functional vanish. This implies having a non-critical string, where $D \neq D_{{\rm crit.}}$. If one sticks to an $\alpha'$-expansion (for higher order terms) and low energy approximation, then the (classical) effective theory is like 10d (or bosonic) supergravity but in $D$ dimensions, corrected by a cosmological constant term in $\frac{D-D_{{\rm crit.}}}{\alpha'}$. Indeed, the corresponding equations of motion then match the leading $\beta$-functionals. More precisely, the type II NSNS action \eqref{SNSNS} in string frame gets modified as ${\cal R}_{10} \rightarrow {\cal R}_{D} -\frac{D-10}{\alpha'}$ (or $-\frac{2}{3}\frac{D-26}{\alpha'}$ for bosonic string) \cite[Sec. 14.1]{Blumenhagen:2013fgp}. Since the new term enters as a cosmological constant term, it could contribute to give a de Sitter solution, provided it is positive. This is the case for supercritical string, where $D > D_{{\rm crit.}}$. Such de Sitter solutions have been reported in \cite{Silverstein:2001xn, Maloney:2002rr, Dodelson:2013iba, Harribey:2018xvs}, and recently discussed in \cite{Junghans:2023lpo} to which we refer for more details and references.

Immediate questions can be raised about such constructions. Let us first recall that textbook string quantization usually leads one to pick $D=D_{{\rm crit.}}$: this is because the construction is done on a $D$-dimensional Minkowski spacetime, with a constant dilaton and no further content. With such a background, all terms in the $\alpha'$-expansion of the $\beta$-functionals vanish, except $\frac{D-D_{{\rm crit.}}}{\alpha'}$: conformal symmetry forces one to have a critical dimension. Beyond Minkowski however (e.g.~on a curved background with fluxes as typically considered here), one could in principle have non-vanishing terms which compensate the leading one and allow for another dimension $D$. One however faces another issue: to preserve the validity of the $\alpha'$-expansion, hierarchy among the terms is needed. Typically, one requires curvatures and other supergravity fields to be small, as mass scales, compared to $1/\alpha'= l_s^{-2}$; equivalently, typical lengths are large in string units. In \cite[Sec.14.1]{Blumenhagen:2013fgp}, this point is referred to as maintaining a large volume limit. Preserving the $\alpha'$-expansion then singles out the leading term $\frac{D-D_{{\rm crit.}}}{\alpha'}$ and imposes $D=D_{{\rm crit.}}$. Such an expansion is in principle not necessary to string theory, and some CFTs, typically strongly coupled, can be defined without it \cite{Blumenhagen:2013fgp}, in particular in other $D$. But we recall that Einstein equation (and not another gravity) is recovered through this expansion as a low energy limit.

A similar point was raised and discussed in \cite{Junghans:2023lpo}: maintaining an $\alpha'$-expansion makes it a priori difficult to satisfy the equations of motion of the new supercritical effective theory, where different $\alpha'$ orders are present. A way-out is however mentioned in \cite{Junghans:2023lpo} when $D$ is very large.\footnote{Given that $D-D_{{\rm crit.}}$ appears in many different places when constructing and quantizing string theories, one may wonder whether any $D \neq D_{{\rm crit.}}$ allows for a consistent theory. Generalizing type II superstrings to a non-critical version as just described, \cite{Dodelson:2013iba} reports on the condition $D= 10 \ ({\rm mod}\, 16)$, stronger than another one sometimes encountered, $D= 2 \ ({\rm mod}\, 8)$.} Alternatively, in \cite{Harribey:2018xvs}, one-loop contributions in bosonic string, added to the above ``classical'' action, seem to help with respect to such hierarchy questions; note though that the bosonic string remains tachyonic.

Ignoring loop contributions, and sticking to the above ``classical'' (supercritical) effective string theory in spite of the questions raised above, the Maldacena-Nu\~nez no-go theorem (\ref{nogo1}) got straightforwardly extended in \cite{Junghans:2023lpo}, leading the author to consider orientifolds. Interestingly, an issue analogous to the singular bulk problem discussed above for KKLT was then shown to appear in the present framework. Indeed, it was argued in \cite{Junghans:2023lpo} that in many cases, the region around orientifolds, where a naive backreaction is problematic and requires a stringy description, is large, meaning comparable to the overall size of the compact space. In other words, in a de Sitter solution of such a supercritical classical string with orientifolds, one looses the classical description in a large part of space, making the solution a priori inconsistent.

\subsubsection{Dark bubble}

We end our list with a construction which leaves the realm of compactification. Starting with \cite{Banerjee:2018qey, Banerjee:2019fzz}, it is proposed that our universe, a 4d de Sitter spacetime (with matter and radiation), is realised as a bubble which expands in a 5d anti-de Sitter bulk. More precisely, it can be viewed as a Coleman-de Luccia bubble that separates two AdS${}_5$ (outside and inside the bubble), and nucleates as a non-perturbative instability, mediating the decay of e.g.~a non-supersymmetric AdS${}_5$. This proposal is reminiscent of braneworld models as Randall-Sundrum \cite{Randall:1999vf}, where we live on a brane, transverse to a non-compact dimension, where only gravity can access; differences are however pointed-out in \cite{Banerjee:2022ree}. As in any braneworld scenario, one needs to face the question of having a massless 4d graviton (or even gravitational wave), which requires a normalisable warp factor in the non-compact transverse space (see e.g.~\cite[(2.5)]{Andriot:2019hay}). For the dark bubble, this is discussed in \cite{Banerjee:2020wov, Banerjee:2023uto}. Further realistic ingredients on the bubble, as well as string theory realisations, are discussed in subsequent works.

\subsection{(Exponential) Quintessence}\label{sec:quint}

In Section \ref{sec:listattempts}, we have listed attempts at constructing de Sitter solutions from string theory, circumventing the constraints pointed out in Section \ref{sec:constraints}. In most cases, weaknesses in the proposed de Sitter construction have been identified, often related to control issues discussed in Section \ref{sec:nodSQG}. In short, obtaining a 4d well-controlled de Sitter solution from string theory is difficult and not yet achieved. This suggests a different realisation of dark energy in the recent universe: we focus here on {\sl quintessence}, where fields are rolling along a positive scalar potential. This option was also advocated in Section \ref{sec:dSconjsum}. As pointed-out there, it could still include hilltop quintessence, namely a de Sitter maximum and departure from it.

Let us mention that from string theory, quintessence models have certain advantages over de Sitter solutions; this does not mean they are free of disadvantages, that we also discuss below. First, considering the landscape of string effective theories and their solutions, it is statistically much easier to find a point on the slope of a scalar potential, i.e.~considering rolling solutions, than finding an extremum of the potential. This of course does not guarantee the observational validity of the former, but the {\sl ubiquity of rolling solutions} should be emphasized. Second, the Maldacena-Nu\~nez no-go theorem (\ref{nogo1}) can be circumvented by rolling solutions while still having (transient or eternal) accelerated expansion \cite{Townsend:2003fx, Ohta:2003pu, Emparan:2003gg}. It is important to note that those rolling solutions can be obtained {\sl without any orientifold} \cite{Andriot:2023wvg}, thus avoiding the difficult treatment of the backreaction of these objects in compactifications (discussed in Section \ref{sec:listbackreact} and \ref{sec:backreact}). We also recall that asking for backreacted orientifolds may lead to the singular bulk problem, mentioned in Section \ref{sec:listnonpert} and \ref{sec:listcrit}. Avoiding such issues is therefore an important benefit of the cosmological rolling solutions. Last but not least, rolling solutions in stringy exponential models allow for {\sl scale separation} in the asymptotics, as emphasized in \cite{Andriot:2025cyi}. This means there is an energy gap between the 4d scale, given by the cosmological constant or the Hubble parameter, and the 6d internal or Kaluza--Klein scale. Such a gap is necessary to ensure the validity of the 4d effective theory, as discussed in Section \ref{sec:dimred}; see \cite{Coudarchet:2023mfs} for a review on scale separation. Such a property is not easily achieved by string theory solutions with a cosmological constant, so this is again an advantage of quintessence solutions.

Beyond the previous theoretical and stringy motivations to consider quintessence, let us recall that recent cosmological observations also suggest it. Indeed, the surveys DES \cite{DES:2024jxu} and DESI \cite{DESI:2024mwx, DESI:2025zgx} have reported since 2024 the possibility that their data gets well fitted by a dynamical dark energy, instead of a cosmological constant. In more detail, a key outcome of these observations is the measurement of the dark energy equation of state parameter $w_{{\rm DE}}$. A cosmological constant would give the constant value $w_{{\rm DE}}=-1$, while a varying $w_{{\rm DE}}$ is consistent with a dynamical dark energy. Both options for $w_{{\rm DE}}$ currently fit well the data, but they now exclude each other, as they stand apart at about 3$\sigma$. These observations are achieved at unprecedented precision, which shall further increase as surveys are still on-going. Considering the dynamical dark energy option, one way to realise it is through quintessence models, hence the motivation to study those in detail in this section.\\

Quintessence, or the idea that dark energy in the recent universe is realised by rolling scalar field(s), was first proposed soon after inflation \cite{Ratra:1987rm, Peebles:1987ek, Wetterich:1994bg, Caldwell:1997ii}. A renewed interest can be noticed in the literature in the early 2000's, probably due to the observation of dark energy \cite{SupernovaSearchTeam:1998fmf, SupernovaCosmologyProject:1998vns}. The recent observations \cite{DES:2024jxu, DESI:2024mwx, DESI:2025zgx} have led again to a burst of activity on this topic. We give a brief review of the formalism for (classical and background) quintessence in Appendix \ref{ap:cosmo}. As a starting point, the 4d action to be used is \eqref{Scosmo}, that we repeat here for convenience
\beq
{\cal S}= \int \d^4 x \sqrt{|g_4|} \left(\frac{M_p^2}{2} {\cal R}_4 - \frac{1}{2} g_{ij} \del_{\mu}\varphi^i \del^{\mu}\varphi^j -V (\varphi) - {\cal L}_m - {\cal L}_r \right) \ . \label{S4dquint}
\eeq
This action is the same as the one considered so far for 4d string effective theories, \eqref{S4d}, together with matter and radiation contributions. As an ansatz for the resulting cosmological solutions, we take the 4d metric to be FLRW \eqref{FLRW}, and the scalar fields only depend on time, $\varphi^i(t)$ and $\del_{t} \varphi^i \equiv \dot{\varphi}^i$. An output of this formalism is the equation of state parameter for the scalars
\beq
w_{\varphi} = \frac{\frac{1}{2} g_{ij}\, \dot{\varphi}^i \dot{\varphi}^j - V(\varphi)}{\frac{1}{2} g_{ij}\, \dot{\varphi}^i \dot{\varphi}^j + V(\varphi)} \ , \label{wphi}
\eeq
whose value evolves along the solution describing the cosmological history. In quintessence models, the scalar contribution plays, in first approximation, the role of dark energy, giving $w_{{\rm DE}}=w_{\varphi}$; more sophisticated options will be discussed in Section \ref{sec:phantom}. With this dynamical dark energy realisation, we then get an evolving $w_{{\rm DE}}$, that may or may not match the latest observational data.

The first models considered are known as freezing models (see reviews \cite{Caldwell:2000wt, Tsujikawa:2013fta}). These models were typically single field, with scalar potentials of the form $V(\varphi) \sim 1/ \varphi^p$ with $p>0$, or $\sim e^{M_p/\varphi}-1$, for a canonically normalised field $\varphi$; those are uncommon from a string theory perspective (see however \cite{Cicoli:2012tz}). They should be distinguished from thawing models (see \cite{Andriot:2024sif} for a recent account) that include exponential potentials ($V(\varphi) \sim e^{-\lambda\, \varphi/ M_p}$ as in \eqref{Vexp}, with not too large $\lambda$), axionic ($\sim 1 + \cos{\varphi/f}$) or even hilltop ones ($\sim 1 - \kappa^2 \varphi^2$). The freezing models typically admit cosmological solutions where the scalar field slows down. Therefore $w_{\varphi}$ goes from $0$ (matter-like), or a slightly negative value, down towards $-1$ (cosmological constant-like) in the recent universe: see e.g.~\cite[Fig.1,2]{Tsujikawa:2013fta}. On the contrary, solutions from thawing models have a field previously frozen by Hubble friction which starts rolling. Then, $w_{\varphi}$ goes from $-1$ up to a higher value, e.g.~$-0.7$: see e.g.~\cite[Fig.3]{Tsujikawa:2013fta}. Freezing models were initially appealing, as they mimic the recent evolution of the universe from a matter dominated phase to a dark energy dominated one, possibly explained by a cosmological constant. However, the recent observations \cite{DES:2024jxu, DESI:2024mwx, DESI:2025zgx}, as well as older ones by the Planck collaboration \cite{Planck:2018vyg}, suggest that a dynamical dark energy should have a growing $w_{{\rm DE}}$ in the recent universe. This favors thawing models. This can be seen e.g.~via a standard parametrisation of a varying $w_{{\rm DE}}$, known as CPL parametrisation \cite{Chevallier:2000qy, Linder:2002et}. It is given by a two parameters expansion in terms of the scale factor $a$ away from today's value $a(t_0)=1$
\beq
w_{{\rm DE}} = w_0 + w_a\, (1-a) \ . \label{CPL}
\eeq
The observations consistently report a value today $w_0 > -1$, and a (large) positive slope $-w_a >0$. Such a growing $w_{{\rm DE}}$ is in agreement with the behaviour described by solutions of thawing models.\\

With this motivation in mind, we focus in this section on the simplest thawing quintessence model, namely the single field exponential model \eqref{Vexp} that we recall here
\beq
V(\varphi) = V_0\, e^{- \frac{\lambda}{M_p}\, \varphi} \ ,\quad V_0,\, \lambda >0 \ .\label{Vexp2}
\eeq
For simplicity in this section, we denote by $\varphi$ the single, canonically normalised field. Another motivation for considering such a model is its string theory realisation. As discussed in Section \ref{sec:dSconjsum}, it can be derived from string theory with ``almost single field'' models. This is for instance the case of models obtained from type IIA/B compactifications, recently discussed in \cite{Marconnet:2022fmx, Andriot:2023wvg, Andriot:2025cyi}. Early references can also be found in the review \cite{Townsend:2003qv}, while an M-theory realisation of this model was obtained in \cite{Andersson:2006du}. More generally, an exponential potential is expected in the asymptotic regime of string theory, as in $V(\hat{\rho},\hat{\tau})$ in \eqref{rhotautheoryexp}. In addition, in such a regime, corrections are expected to be under control, which makes it a promising setting. But we should also recall that in field space asymptotics, the SdSC bound \eqref{SdSC} applies, and imposes $\lambda \geq \sqrt{2}$, i.e.~a steep potential; we will keep that constraint in mind. This simple and string-motivated quintessence model provides a first illustration of the options and the challenges faced to reproduce dark energy and accelerated expansion.

Many other quintessence models have been studied. In particular, extensions of the previous simple model have been discussed in the string theory literature; we mention here few recent references. First, one may consider several fields and several exponentials, as done in \cite{Collinucci:2004iw, Shiu:2023nph, Shiu:2023fhb, Marconnet:2025vhj}. Second, while the above $\varphi$ is typically a saxion (as in \eqref{rhotautheoryexp}), another multifield extension is to consider as well axion fields. Their (stringy) kinetic terms include a non-normalisable coupling to saxions \cite{Cicoli:2020noz, Russo:2022pgo, Brinkmann:2022oxy, Revello:2023hro, Seo:2024qzf, Shiu:2024sbe, Licciardello:2025fhx, Grimm:2025cpq, Gallego:2026xqq}, therefore a curved field space. Third, one may also change the scalar potential, towards e.g.~a hilltop \cite{Cicoli:2021skd, Gordillo-Ruiz:2025xcg} or an axionic one \cite{Anchordoqui:2025fgz}. Multifield string-inspired models involving several of the previous aspects were recently discussed in \cite{Cicoli:2024yqh, Anchordoqui:2025epz}. Finally, another extension is to consider a classical coupling to matter; we will discuss this in detail in Section \ref{sec:phantom}. All these extensions offer new and interesting physics, worth being explored. The results presented here on the single field exponential model shall provide a first handle for such investigation.\\

Before delving into the study of this model for dark energy, let us finally mention that quintessence suffers as well from general criticisms or challenges, summarized in \cite[Sec. 6.2.1]{Cicoli:2023opf}. We discuss here below two related points: the belief that the quintessence field should be {\sl ultralight}, and the questions raised by its possible {\sl couplings}.
\begin{itemize}
\item The quintessence field is generally believed to be {\sl ultralight}. One justification comes as follows. The cosmological evolution can be considered as slow, compared e.g.~to an Earth-based particle physics experiment. As a consequence, a (canonical) rolling scalar field $\varphi$ driving quintessence can in good approximation be taken as a constant today, plus a small (quantum) fluctuation: $\varphi_0 + \delta \varphi$. This leads to the following expansion of $V(\varphi)$, where we denote $V(\varphi_0)\equiv V_0$ and $\del_{\varphi} V \equiv V'$ 
\beq
V(\varphi) \simeq V_0 + \delta \varphi \, V_0' + \frac{1}{2} \delta \varphi^2 \, V_0'' = V_0 - \frac{1}{2} \frac{(V_0')^2}{V_0''} \ +  \frac{1}{2} \left(\delta \varphi + \frac{V_0'}{V_0''} \right)^2\, V_0'' \ .
\eeq
From this one reads a mass term for the (shifted) field fluctuation, with an effective mass $m$, such that $m^2 = V_0''$. Finally, one would say that $|V_0''| \sim V_0 / M_p^2 \sim \Lambda \sim H_0^2$, up to order $1$ factors, where $\Lambda$ is today's would-be cosmological constant, and $H_0$ today's Hubble parameter (see Section \ref{sec:topicsnotcovered}). One concludes on $m \sim H_0 \sim 10^{-33} \, {\rm eV}$, i.e.~an ultralight field.

Let us indicate {\sl three loopholes} to this conclusion. First, the shift in $\delta \varphi $, or equivalently the fact there is a slope in the potential, is often ignored, based on the idea that the potential should be fairly flat for quintessence. But the latter is not necessarily true as we will see in this section. In that case the shift does not have to be small, then questioning in particular the validity of the expansion. Second, it is a priori unclear how to interpret the effective mass in a tachyonic case, where $V_0''<0$. Actually, these first two loopholes appear when going from a standard intuition based on a minimum of the potential (as for a cosmological constant, or where to define properly a quantum field theory), to a (concave) runaway potential, which could be relevant to quintessence. Third, the assumption that $|V_0''| \sim V_0 / M_p^2$ is correct for a cosine or an exponential potential, both with an order $1$ rate, but it is not always true, as e.g.~for a linear runaway potential.

Ignoring those loopholes, and developing a quantum field theory for this scalar fluctuation, valid e.g.~for a particle physics experiment today, having an ultralight scalar field poses several challenges. One is the protection of this mass against loop corrections. This relates to the question of the coupling to other fields, discussed below. But a universal coupling, possibly serving as a mediator, is always the gravitational one.

\item Having a new scalar field immediately raises the question of its {\sl coupling} to any other constituent in the universe. If obtained through ``sequestering'', or protected by some symmetry, it may avoid coupling to anything else but gravitation. As mentioned, the latter may still provide, by quantum (loop) corrections, indirect coupling to other constituent, even though suppressed by the smallness of the gravitational coupling constant. If however there is a direct coupling to part of (dark) matter or radiation, many questions can be asked. A typical concern is that of a fifth force: a light scalar could lead, through a Yukawa coupling and potential, to a long range force, unobserved so far. Protection mechanisms such as having a symmetry, an axion-like particle and coupling, or screening, can then be thought of. Coupling only to dark matter could also be less constraining, although it could lead to an undesired dark matter decay. Another concern appears from a more classical perspective: having a varying scalar field (as in quintessence), coupled to other constituents, could lead to time-varying fundamental constants \cite{Uzan:2010pm, Uzan:2024ded}, as well as to violations of the equivalence principle \cite{Hees:2018fpg, MICROSCOPE:2022doy, Elder:2025tue}, both subject to many constraints. We refer to \cite[Sec.6.2.1]{Cicoli:2023opf} for a more detailed discussion on these questions. 
    
    In Section \ref{sec:phantom}, we will consider a (classical) coupling to matter. This is motivated by an explanation for a cosmological phantom regime, as detailed there. Doing so, we will for now set aside all possible issues or constraints just raised with such a coupling. Let us note that several of these questions are of quantum nature, which to some extent, goes beyond a pure classical regime as considered there. Also, a quantum versus classical nature of $\varphi$ relates to the loopholes mentioned above regarding its believed ultralight mass. Finally, many of the questions raised are very model dependent: which field, which coupling, how strong, how light, how steep? While we do not deny the relevance of the indicated challenges, it appears premature to really address them without a complete model of quintessence together with detailed particle physics; the latter part is often missing, at least in this article.\\
\end{itemize}

We organise our study of single field exponential quintessence as follows. We first present in Section \ref{sec:fixedpoints} a dynamical system approach to this model. This provides specific cosmological solutions that correspond to fixed points. A realistic cosmological solution will successively pass close to these various fixed points, which then serve as approximations along the cosmological history. We then study in Section \ref{sec:accnohor} on general grounds whether cosmological solutions can have asymptotic acceleration, and how this is related to the question of the cosmological event horizon. The exponential quintessence solutions will serve as illustration. We turn in Section \ref{sec:matter} to tentative realistic solutions, including matter and radiation, and discuss their properties. We also offer a first comparison to observations. Facing the challenge of a possible phantom regime, we discuss in Section \ref{sec:phantom} how a coupling to matter addresses it, with benefits for a string theory realisation. A summary of the situation will be provided in Section \ref{sec:sumquint}.

\subsubsection{Cosmological solutions and fixed points}\label{sec:fixedpoints}

We consider a model of exponential quintessence, with the potential \eqref{Vexp2} and canonical field $\varphi$. The 4d action \eqref{S4dquint} boils down here to
\beq
\int \d^4 x \sqrt{|g_4|} \left( \frac{M_p^2}{2} {\cal R}_4 - \frac{1}{2} \del_{\mu} \varphi \del^{\mu} \varphi - V_0\, e^{-\frac{\lambda}{M_p} \varphi} - {\cal L}_m - {\cal L}_r \right) \ ,
\eeq
with $V_0,\, \lambda>0$. This model can be studied using methods of dynamical systems, as we recall in this section. This will provide powerful tools to study the general cosmological solutions of such a quintessence model, as we will do in the next sections. While various solutions of this model appeared in the literature, a systematic treatment with dynamical system can be found e.g.~in \cite{Halliwell:1986ja, Copeland:1997et, Ferreira:1997hj, vandenHoogen:1999qq, Gosenca:2015qha, SavasArapoglu:2017pyh}. A review was given in \cite{Bahamonde:2017ize}, while a recent and comprehensive account can be found in \cite{Andriot:2024jsh}. We follow the latter here, including all constituents of the universe listed in Table \ref{tab:rhow}: radiation, matter, curvature, scalar. We work in 4d, while a $d$-dimensional treatment can be found e.g.~in \cite[App.A]{Andriot:2024jsh}.\\

The 4d dynamical system variables are given by
\beq
\hspace{-0.1in} x = s_{\dot{\varphi}}\sqrt{\Omega_{{\rm kin}}}= s_{\dot{\varphi}} \sqrt{\frac{\Omega_{\varphi}(1+w_{\varphi})}{2}} \ ,\ y = \sqrt{\Omega_V} = \sqrt{\frac{\Omega_{\varphi}(1-w_{\varphi})}{2}}  \ ,\ z= \sqrt{\Omega_k} \ ,\ u = \sqrt{\Omega_r} \ ,
\eeq 
where we recall that $\Omega_n = \rho_n / (3 M_p^2 H^2)$, and refer to Appendix \ref{ap:cosmo} for definitions. We denote by $s_{\dot{\varphi}}$ the sign of $\dot{\varphi}$; only the variable $x$ can be negative. We restrict for simplicity to a flat or open universe, $k=0$ or $-1$, giving $\Omega_k \geq0$; a treatment of a closed universe can be found e.g.~in \cite{Andriot:2023wvg}. Using the second Friedmann equation $F_2=0$ and the (single) scalar equation of motion $E^1=0$ \eqref{F1F2Ei}, one obtains the following system of first order differential equations
\begin{gather}
\label{dynsys}
\begin{aligned}
x' = & x \left(-3 + 3 x^2 + z^2 + 2 u^2 + \frac{3}{2} \Omega_m \right) + \sqrt{\frac{3}{2}}\lambda\, y^2 \\
y' = & y \left(-\sqrt{\frac{3}{2}}\lambda\, x + 3 x^2 + z^2 + 2 u^2 + \frac{3}{2} \Omega_m \right) \\
z' = & z \left(-1 + 3 x^2 + z^2 + 2 u^2 + \frac{3}{2} \Omega_m \right) \\
u' = & u \left(-2 + 3 x^2 + z^2 + 2 u^2 + \frac{3}{2} \Omega_m \right)
\end{aligned}
\end{gather}
Here, $x'\equiv  \frac{\del x}{\del N}$ with the number of e-folds $N = \ln a$, where $N=0$ today. The first Friedmann equation $F_1=0$ is used to define $\Omega_m$
\beq
\Omega_m = 1 - x^2 - y^2 -z^2 - u^2 \ . \label{OmF1}
\eeq
Because $\del_{\varphi}V /V = -\frac{\lambda}{M_p}$ is constant for the exponential potential, this quantity does not enter as an extra variable in the system. This is not necessarily true in studies of other potentials, where extra variables are introduced, to capture in particular the second derivative. Here, combining \eqref{dynsys} and \eqref{OmF1}, we obtain a closed first order differential system on the above variables. This allows to use a dynamical system approach. 

The first consequence is the determination of fixed points. In the (phase) space spanned by the coordinates $\{x,y,z,u\}$, these are the points for which $\{x',y',z',u'\}=\{0,0,0,0 \}$. Physically, these correspond to specific cosmological solutions, that exist provided some conditions are obeyed. Typically, they correspond to having only one or two constituents in the universe, so they are not realistic. Realistic cosmological solutions nevertheless pass close to, or even asymptote to, those fixed points at different moments of the universe: see \cite[Fig.7,8]{Andriot:2024jsh} for an illustration. In other words, the fixed points serve as approximate solutions at certain times in the cosmological history. For example, the fixed point $P_r$ corresponds to a universe filled with radiation, and approximates a realistic cosmology during an epoch of radiation domination. We detail the fixed point solutions in Table \ref{tab:fixedpoints}. We recall that we consider solutions with $H>0$ (expanding) and that we took $\lambda>0$.

\begin{table}[ht!]
\begin{center}
\begin{tabular}{|c||c|c|c|c|c|}
\hline
&&&&&\\[-8pt]
Fixed point & \multirow{2}{*}{$(\Omega_{{\rm kin}},\Omega_V,\Omega_k,\Omega_m,\Omega_r)$} & \multirow{2}{*}{$a(t)$} & \multirow{2}{*}{$\varphi(t)$} & Existence & \multirow{2}{*}{$w_{{\rm eff}}$} \\
solution & & & & conditions & \\[4pt]
\hline
&&&&&\\[-8pt]
\multirow{2}{*}{$P_{{\rm kin}}^+$} & \multirow{2}{*}{$\left(1, 0, 0, 0, 0\right)$} & \multirow{2}{*}{$a_0\, t^{\frac{1}{3}}$} & \multirow{2}{*}{$\varphi_0 + \sqrt{\frac{2}{3}}\ln t$} & & \multirow{2}{*}{$ 1$} \\
 & & & & & \\[4pt]
\hline
&&&&&\\[-8pt]
\multirow{2}{*}{$P_{{\rm kin}}^-$} & \multirow{2}{*}{$\left(1, 0, 0, 0, 0\right)$} & \multirow{2}{*}{$a_0\, t^{\frac{1}{3}}$} & \multirow{2}{*}{$\varphi_0 - \sqrt{\frac{2}{3}}\ln t$} & & \multirow{2}{*}{$1$} \\
 & & & & & \\[4pt]
\hline
&&&&&\\[-8pt]
\multirow{2}{*}{$P_r$} & \multirow{2}{*}{$\left(0 , 0 , 0 , 0 , 1 \right)$} & \multirow{2}{*}{$a_0 \, t^{\frac{1}{2}}$} & \multirow{2}{*}{$\varphi_0 $} &  & \multirow{2}{*}{$ \frac{1}{3}$} \\
 & & & & & \\[4pt]
\hline
&&&&&\\[-8pt]
\multirow{2}{*}{$P_{r\varphi}$} & \multirow{2}{*}{$\left(\frac{8}{3\lambda^2} , \frac{4}{3 \lambda^2} , 0 , 0 , 1 - \frac{4}{\lambda^2} \right)$} & \multirow{2}{*}{$a_0 \, t^{\frac{1}{2}}$} & \multirow{2}{*}{$\frac{1}{\lambda}\ln (\lambda^2 V_0\, t^2) $} & \multirow{2}{*}{$\lambda > 2$} & \multirow{2}{*}{$ \frac{1}{3}$} \\
 & & & & & \\[4pt]
\hline
&&&&&\\[-8pt]
\multirow{2}{*}{$P_{m}$} & \multirow{2}{*}{$\left(0 , 0 , 0 , 1 , 0 \right)$} & \multirow{2}{*}{$a_0 \, t^{\frac{2}{3}}$} & \multirow{2}{*}{$\varphi_0 $} &  & \multirow{2}{*}{$ 0 $} \\
 & & & & & \\[4pt]
\hline
&&&&&\\[-8pt]
\multirow{2}{*}{$P_{m\varphi}$} & \multirow{2}{*}{$\left( \frac{3}{2 \lambda^2} , \frac{3}{2 \lambda^2} , 0 , 1-\frac{3}{\lambda^2} , 0 \right)$} & \multirow{2}{*}{$a_0 \, t^{\frac{2}{3}}$} & \multirow{2}{*}{$\frac{1}{\lambda}\ln \frac{\lambda^2 V_0\, t^2}{2} $} & \multirow{2}{*}{$\lambda > \sqrt{3} $} & \multirow{2}{*}{$ 0$} \\
 & & & & & \\[4pt]
\hline
&&&&&\\[-8pt]
\multirow{2}{*}{$P_{k}$} & \multirow{2}{*}{$\left( 0 , 0 , 1 , 0 , 0 \right)$} & \multirow{2}{*}{$ t$} & \multirow{2}{*}{$\varphi_0 $} & \multirow{2}{*}{$ k=-1 $} & \multirow{2}{*}{$ -\frac{1}{3}$} \\
 & & & & & \\[4pt]
\hline
&&&&&\\[-8pt]
\multirow{2}{*}{$P_{k\varphi}$} & \multirow{2}{*}{$\left(\frac{2}{3 \lambda^2}, \frac{4}{3 \lambda^2}, 1-\frac{2}{\lambda^2}, 0,0 \right)$} & \multirow{2}{*}{$\frac{\lambda}{\sqrt{\lambda^2 - 2}}\, t$} & \multirow{2}{*}{$\frac{1}{\lambda}\ln \frac{\lambda^2 V_0\, t^2}{4}$} & \multirow{2}{*}{$\lambda > \sqrt{2}$, $k=-1$} & \multirow{2}{*}{$-\frac{1}{3}$} \\
 & & & & & \\[4pt]
\hline
&&&&&\\[-8pt]
\multirow{2}{*}{$P_{\varphi}$} & \multirow{2}{*}{$\left(\frac{\lambda^2}{6} , \frac{6-\lambda^2}{6} , 0 , 0 , 0 \right)$} & \multirow{2}{*}{$a_0 \, t^{\frac{2}{\lambda^2}}$} & \multirow{2}{*}{$\frac{1}{\lambda}\ln \frac{\lambda^4 V_0\, t^2}{2(6-\lambda^2)} $} & \multirow{2}{*}{$\lambda < \sqrt{6} $} & \multirow{2}{*}{$ \frac{\lambda^2}{3} - 1 $} \\
 & & & & & \\[4pt]
\hline
\end{tabular}
\end{center}\caption{Fixed point solutions for single field exponential quintessence. The constants $a_0, \varphi_0$ are arbitrary. The existence conditions may also require $V_0, k, \Omega_{m}$ or $\Omega_r$ to be zero: this can be read in the solution information. To express the solutions as $a(t), \varphi(t)$, the time coordinate has been shifted such that $a(t=0)=0$. The sign of $\dot{\varphi}$, $s_{\dot{\varphi}}$, can be read from the solution: it is negative only for $P_{{\rm kin}}^-$.}\label{tab:fixedpoints}
\end{table}

In more detail, the fixed points $P_n$, for $n={\rm kin}, r, m, k, \varphi$, corresponds to a universe filled with the constituent $n$.\footnote{\label{foot:Pn}The general solution for $a(t)$ to equations $F_1=F_2=0$ \eqref{F1F2Ei} for a single constituent $n$, meaning $\Omega_n=1$, with constant $w_n\neq-1$, is given by $a^{3(1+w_n)} = t^2\, 3(1+w_n)^2 \rho_{n0} /(4 M_p^2)$, up to a constant shift in $t$, and $\rho_n = \rho_{n0}\, a^{-3(1+w_n)}$ with a free constant $\rho_{n0}$. The fixed points $P_n$ here are examples.} The fixed points $P_{n\varphi}$ are called ``$n$''-scaling or ``$n$''-tracker solutions, as they mimic the expansion evolution in presence of $n$, as can be seen through $w_{{\rm eff}}$, while having as well an evolving scalar field: the total energy density scales as the one of $n$. The solutions $P_{n\varphi}$ interpolate between $P_{\varphi}$ and $P_n$ at small (lower bound) and large (infinite) $\lambda$ values. The solution $P_{\varphi}$ can also be understood as interpolating between a pure cosmological constant and a pure kination solution at small (vanishing) and large (upper bound) $\lambda$ values. Note that the former solution cannot appear here, since $\lambda>0$. Finally, recalling the acceleration condition \eqref{acccond} $w_{{\rm eff}} < -\frac{1}{3}$, we see that only $P_{\varphi}$ can be accelerating for $\lambda < \sqrt{2}$. $P_{\varphi}$ at $\lambda = \sqrt{2}$, as well as $P_k$ and $P_{k\varphi}$, are solutions at the no-acceleration boundary, namely $\ddot{a}=0$.\\

A second outcome of the dynamical system approach is the determination of the stability of the fixed points. This will have important physical consequences, such as the determination of late time attractors. To assess the stability, one first considers $\{ x',y',z',u'\}$ as four multivariate functions of $\{x,y,z,u\}$, once $\Omega_m$ is replaced with \eqref{OmF1}. One then considers the following Jacobian, evaluated at each fixed point
\beq
\left(\begin{array}{cccc}
\del_x x' & \del_y x' & \del_z x' & \del_u x' \\
\del_x y' & \del_y y' & \del_z y' & \del_u y' \\
\del_x z' & \del_y z' & \del_z z' & \del_u z' \\
\del_x u' & \del_y u' & \del_z u' & \del_u u' 
\end{array}\right)  \label{Jacobian}
\eeq
The eigenvalues of the resulting matrix give the stability of the fixed point. One negative eigenvalue indicate a stable, i.e.~attractive direction (in phase space) while a positive one indicates an unstable, i.e.~repelling direction. It is also possible to get a pair of complex conjugate eigenvalues, indicating a spiral. Its (un)stable nature is then determined by the sign of their real part. 

Crucially, a fully unstable point must serve as a universal starting point to any cosmological solution, while a fully stable one is the universal final point, or late time attractor: any solution asymptotes to both. A fixed point with a set of eigenvalues of mixed signs corresponds to a saddle, close to which a cosmological solution can only pass: at best, it only serves as a temporary approximation of the solution. 

We give those results in the following, starting with the above setting: we consider that all four constituents ($r,m, k, \varphi$) can be present, meaning $\Omega_{r}, \Omega_{m}, \Omega_{k}, \Omega_{\varphi}$ can be non-zero, such that all fixed points in Table \ref{tab:fixedpoints} are possible. In that case, depending on the initial condition on $s_{\dot{\varphi}}$ as well as on $\lambda$, the fully unstable (thus starting) point is $P_{{\rm kin}}^+$ or $P_{{\rm kin}}^-$, and the fully stable (final) one is $P_{\varphi}$ or $P_{k\varphi}$ \cite[Sec.2.4]{Andriot:2024jsh}. In between, the cosmological solutions may or may not pass close to the other fixed points, which are saddles. Analytical solutions close to each fixed point can be found in \cite[App. C]{Andriot:2024jsh} (see also \cite{Andriot:2023wvg} when restricted to $k,\varphi$). We summarize the general cosmological solutions as follows, indicating the starting and final fixed point solutions when allowing for all four constituents 
\beq
(r,m,)\, k, \varphi:\qquad \left|\begin{array}{ll}\quad\! 0< \lambda \leq \sqrt{2}:\quad &  P_{{\rm kin}}^{\pm} \ \longrightarrow P_{\varphi} \\[4pt]
\sqrt{2}< \lambda \leq \sqrt{6}:\quad & P_{{\rm kin}}^{\pm} \ \longrightarrow P_{k\varphi} \\[4pt]
\sqrt{6}< \lambda :\quad & P_{{\rm kin}}^- \ \longrightarrow P_{k\varphi}
\end{array}\right. \label{rmkp}
\eeq
The parentheses $(r,m,)$ indicate that the result does not change if one removes either/or $r,m$, by which is meant setting $\Omega_r=0$, $\Omega_m=0$. Removing one constituent can be done as long as it provides a consistent subsystem (also known as an invariant submanifold), which is the case here; we refer to \cite[Sec.2.4]{Andriot:2024jsh} for a more detailed discussion. In that case, one eigenvalue is also removed, which can in principle change the stability. However, here, since the initial and final points in \eqref{rmkp} do not involve $r,m$, it makes sense that removing the latter has no impact.

What should have impact is to remove curvature: $\Omega_k=0$. Keeping matter, and optionally radiation, we get the following results \cite[Tab.1]{SavasArapoglu:2017pyh}
\beq
(r,)\, m, \varphi:\qquad \left|\begin{array}{ll}\quad\! 0< \lambda \leq \sqrt{3}:\quad &  P_{{\rm kin}}^{\pm} \ \longrightarrow P_{\varphi} \\[4pt]
\sqrt{3}< \lambda \leq \sqrt{6}:\quad & P_{{\rm kin}}^{\pm} \ \longrightarrow P_{m\varphi} \\[4pt]
\sqrt{6}< \lambda :\quad & P_{{\rm kin}}^- \ \longrightarrow P_{m\varphi}
\end{array}\right. \label{rmp}
\eeq
We then also remove matter and get with radiation \cite[Tab.5]{Bahamonde:2017ize}
\beq
r, \varphi:\qquad \left|\begin{array}{ll}\quad\! 0< \lambda \leq 2:\quad &  P_{{\rm kin}}^{\pm} \ \longrightarrow P_{\varphi} \\[4pt]
\quad\! 2< \lambda \leq \sqrt{6}:\quad & P_{{\rm kin}}^{\pm} \ \longrightarrow P_{r\varphi} \\[4pt]
\sqrt{6}< \lambda :\quad & P_{{\rm kin}}^- \ \longrightarrow P_{r\varphi}
\end{array}\right. \label{rp}
\eeq
The last option is to have $\varphi$ as the only constituent. We then verify\footnote{One way to treat this case is to consider only the equation $x'=3(x^2 -1) (x - \lambda/\sqrt{6})$, where \eqref{OmF1}, namely $x^2+y^2=1$, has been used. Indeed, the equation $y'=\dots$ can be obtained from the latter. Then, $x'(x)$ gives the expected fixed points $P_{{\rm kin}}^{\pm}$ and $P_{\varphi}$, while one reads their stability from the $1\times 1$ Jacobian $\del_x x'$.} that cosmological solutions are as follows
\beq
\varphi:\qquad \left|\begin{array}{ll}\quad\! 0< \lambda < \sqrt{6}:\quad &  P_{{\rm kin}}^{\pm} \ \longrightarrow P_{\varphi} \\[4pt]
\sqrt{6}\leq \lambda :\quad & P_{{\rm kin}}^{-} \ \longrightarrow P_{{\rm kin}}^{+}
\end{array}\right. \label{p}
\eeq
This concludes the stability study of the fixed points, which as explained, has important physical consequences on the general cosmological solutions. We will use these results in the next sections to discuss properties of exponential quintessence cosmological solutions.

\subsubsection{Asymptotic acceleration and no event horizon}\label{sec:accnohor}

In this section, we discuss the notion of asymptotic acceleration in cosmological solutions, and the relation to the cosmological event horizon, defined around \eqref{horizon}. This will allow us to investigate further the {\sl No Cosmological (event) Horizon Conjecture} (NCHC), stated in \eqref{NCHC}. Our discussion builds on solutions of exponential quintessence, but some results are established at a more general level.\\

We first focus on single field exponential quintessence, motivated at the beginning of Section \ref{sec:quint}. We are interested by a string theory realisation in the asymptotics of field space. In that case, the SdSC bound \eqref{SdSC} gives $\lambda \geq \sqrt{2}$. A cosmological solution at large field $\varphi \rightarrow \infty$ is not necessarily the same as a solution in future asymptotics, meaning at late time $t \rightarrow \infty$. Nevertheless, we consider in this section both asymptotics together: this is justified for a single field runaway potential as here, where the field eventually rolls-down to a large value. In that case, the cosmological solutions ``in the asymptotics'' are well-approximated by the late time fixed point attractors discussed in Section \ref{sec:fixedpoints}. Those are $P_{\varphi}, P_{k\varphi}, P_{m\varphi}, P_{r\varphi}, P_{{\rm kin}}^+$, depending on the precise setting.

To have a chance of being realistic, a cosmological solution should provide a universe today in accelerated expansion.\footnote{This point is sometimes disputed, as in \cite{Shlivko:2024llw}, arguing that observations of acceleration are made in the recent past, and not strictly today. For simplicity, we discard this point of view in this article.} Considering exponential quintessence solutions in the asymptotics, the question is then whether they are accelerating at late time, namely whether they are {\sl in asymptotic acceleration}, and if this can be realised in string theory. Looking at $a(t)$ or $w_{{\rm eff}}$ in Table \ref{tab:fixedpoints}, we see that $P_{m\varphi}, P_{r\varphi}, P_{{\rm kin}}^+$ are decelerating, and $P_{k\varphi}$ is at the boundary of acceleration. At first sight, this leaves only $P_{\varphi}$, for which one has the following condition
\beq
\text{Acceleration with }P_{\varphi}\quad \Leftrightarrow \quad \lambda < \sqrt{2} \ . \label{accPp}
\eeq
This upper bound gets generalized in $d$ dimensions to $2/\sqrt{d-2}$ \cite{Rudelius:2022gbz, Andriot:2022xjh}. This is precisely the same value as in the SdSC, as noticed in \cite{Obied:2018sgi, Rudelius:2021oaz}, where the swampland bound is however a lower bound on $\lambda$. One would then conclude on {\sl no asymptotic acceleration from string theory}, due to a too steep (asymptotic) stringy potential, in this example of exponential quintessence but also maybe more generally.\footnote{The strong energy condition (SEC) also corresponds, for a $\rho >0$, to $w \geq -1/3$, i.e.~a no-acceleration lower bound. Combined with the $P_{\varphi}$ example \eqref{accPp}, the SEC then inspired the terminology ``Strong'' of the SdSC \cite{Rudelius:2021azq}.} Having no asymptotic acceleration could be viewed as agreeing with an absence of a fully stable de Sitter solution (see Section \ref{sec:dSconjsum}), while a relation was established in \cite{Hebecker:2023qke} to the absence of de Sitter solution in higher dimensions. We give a different view on this in the following, relating to the absence of a cosmological event horizon.

Loopholes to the above conclusion can however be thought of, such as considering different potentials, or different multifield dynamics, even though those should still be found with a string theory realisation while invalidating the statement. For example, the works \cite{Shiu:2023fhb, Shiu:2023nph} (see also \cite{VanRiet:2023cca, Seo:2024fki}) performed a dynamical system analysis of a multifield (canonical) multi-exponential potential, sticking to only positive terms, without including any other constituent $r,m,k$. In that framework, they reached the same conclusion: acceleration is obtained in the late time solution (analogue of $P_{\varphi}$) if and only if the multifield analogue of $\lambda$ is smaller than the SdSC bound, therefore preventing from asymptotic acceleration in such stringy settings.\footnote{More precisely, $\epsilon$, defined in Appendix \ref{ap:cosmo} and governing acceleration \eqref{acccond}, is proven equal in the late time solution to $\epsilon_V$ \eqref{epseta}, that relates to the SdSC bound \eqref{SdSC}.} One important loophole was however pointed-out in \cite{Cicoli:2020noz, Brinkmann:2022oxy, Payeur:2024kyy}, thanks to a multifield aspect: considering a curved field space metric can in principle avoid the obstruction to asymptotic acceleration. This is possibly realised with saxions entering the axionic metric and preventing the latter from being canonically normalised.

A different loophole was recently discussed in \cite{Marconnet:2022fmx, Andriot:2023wvg}, allowing to stay in the realm of single field exponential quintessence. It consists in having spatial curvature, namely an open universe ($k=-1$),\footnote{Some stringy arguments favor $k=-1$ in a late universe \cite{Freivogel:2005vv}: this is due to tunneling between vacua in the landscape which should end in one with $k=-1$. Counter-examples to the latter were however discussed in \cite{Buniy:2006ed, Horn:2017kmv, Cespedes:2020xpn}.} thereby having access to the fixed point $P_{k\varphi}$. Indeed, most studies ignored spatial curvature, often because observations put tight upper bounds on $|\Omega_k|$. When setting $k=0$, one is led to cosmological solutions described in \eqref{rmp}, \eqref{rp} or \eqref{p}. A strict SdSC bound $\lambda> \sqrt{2}$ then imposes decelerating late time solutions as mentioned above; in addition, getting an asymptotic string theory potential which saturates the bound, $\lambda= \sqrt{2}$, is difficult, due to the 4d dilaton $\tau$ with other fields (see Section \ref{sec:reftestdSconj}). On the contrary, with $k=-1$, one accesses solutions described by \eqref{rmkp}: for $\lambda> \sqrt{2}$, those are given at late time by $P_{k\varphi}$. The solution $P_{k\varphi}$ is at the boundary of acceleration, namely $\ddot{a}=0$, so in the strict limit, there is no acceleration either:
\beq
k=-1,\ \lambda > \sqrt{2}: \quad \text{Late time dynamics close to }P_{k\varphi}\text{ for which }\ddot{a}=0 \ .
\eeq
But \cite{Marconnet:2022fmx, Andriot:2023wvg} exhibited stringy solutions which approach the fixed point while always accelerating: one talks of {\sl eternal acceleration}; see \cite{Chen:2003dca, Andersson:2006du} for early M-theory realisations. These (string theory realised) solutions also qualify as being in {\sl asymptotic acceleration}, in the sense that they have $\ddot{a}>0$ for an infinite amount of time as they asymptote to the limit solution $P_{k\varphi}$. {\sl Asymptotic acceleration is then possible from string theory.}

This discussion on asymptotic acceleration does not involve matter nor radiation. Indeed, possibly asymptotically accelerating solutions are those close to $P_{\varphi}$ or $P_{k\varphi}$, therefore having little or no matter and radiation contribution; rather, when the latter would play a role asymptotically, namely with $P_{m\varphi}, P_{r\varphi}$, then the solution is decelerating. This is why the present discussion on acceleration in the asymptotics can be considered as ``without matter''; in fact, most related works do not include any matter or radiation. The drawback is of course that the resulting cosmological solutions, even if found (asymptotically) accelerating, cannot be realistic for today's universe, since I am here typing these words, and the rest of the 30$\%$ of matter is also present. We will come back to this in Section \ref{sec:matter}.\\

The question of asymptotic acceleration nevertheless plays a role when discussing the {\sl cosmological event horizon}. The latter is defined in Appendix \ref{ap:cosmo}: a cosmological solution, with FLRW metric, admits an event horizon if the distance $d_e$ defined in \eqref{horizon} is finite. This is the distance traveled by light emitted at $t_0$ until the end of time, considered as $t=\infty$: if finite, then the universe region is causally disconnected from the rest, hence the horizon. If infinite, then there is no event horizon. Asymptotic acceleration appears related because it deals with $\ddot{a}(t)$ for $t \rightarrow \infty$, while $d_e$ depends on the behaviour of $a(t)$ in this limit (see \cite{Hellerman:2001yi, Fischler:2001yj, Boya:2002mv, Townsend:2003qv} for early works on such relations). As a concrete example below \eqref{horizon}, a pure de Sitter spacetime is eternally accelerating and admits such a horizon. We will make the connection between asymptotic acceleration and event horizon more precise in the following.

We questioned above the possibility of having or not an asymptotic acceleration from string theory, with eventually both options being possible. But the possibility of having a cosmological event horizon from quantum gravity has been conjectured to be impossible. We gave an account on this {\sl No Cosmological (event) Horizon Conjecture} (NCHC) \cite{Andriot:2023wvg} around \eqref{NCHC}, mentioning several related works. Motivations for this conjecture were discussed in Section \ref{sec:nodSQG} and \ref{sec:reftestdSconj}: a necessary absence of an event horizon could be central for quantum gravity, by e.g.~allowing to define asymptotic states needed for an S-matrix, and forbidding a fully stable (i.e.~eternal) de Sitter spacetime. Let us now enter the detailed relation to acceleration to further study and justify this conjecture.

To start with, \cite{Boya:2002mv} considered a flat FLRW cosmology, with a single constituent having a {\sl constant} equation of state parameter $w$. In that case, one can show that the solution obeys $a(t) \sim t^{\frac{2}{3(1+w)}}$ for $w\neq -1$ (see Footnote \ref{foot:Pn}), from which one infers results on acceleration and event horizon. Having a constant $w$ is a simplifying assumption: even though the specific fixed points of Table \ref{tab:fixedpoints} have constant $w_{{\rm eff}}$, a general cosmological solution from quintessence has a varying $w_{\varphi}$. Nevertheless, within this simple model, it is shown that decelerating and not accelerating universes ($w\geq -1/3$), in expansion, have no event horizon, while accelerating ones ($w<-1/3$) do. This is the right tendency, but is not fully correct in general; we now revisit this question in detail, without any assumption on any $w$. To that end, we focus on four cases in the following; a summary of our results can be found afterwards.

\begin{itemize}
  \item Consider at first a cosmological solution in expansion, that asymptotes at large $t$ to the linear solution $a(t) =a_l\, t$. This means that after a finite time $t_1$, the correction to this behaviour is subdominant to the asymptotic one, namely for the cosmological solution at $t> t_1$, $a(t)= a_l\, t + o(a_l\, t) = a_l\, t ( 1+ o(1) ) < 2\, a_l\, t $. In that case, one gets
\beq
d_e = a(t_0) \int_{t_0}^{\infty} \frac{\d t}{a(t)} > a(t_0) \int_{t_0}^{t_1} \frac{\d t}{a(t)} + a(t_0) \int_{t_1}^{\infty} \frac{\d t}{2\, a_l\, t} = {\rm finite} + \infty \ .
\eeq
The first term is finite thanks to the expanding assumption, which guarantees a simple monotonic function with $\dot{a}>0$. The second term is infinite due to the asymptotic behaviour of $a(t)$. With $d_e=\infty$, one concludes on the absence of an event horizon. Importantly, from the final solution, one gets no acceleration in the limit: $\ddot{a} \rightarrow 0$. However, one may still have asymptotic acceleration in the sense described above, due to the time dependent corrections: the cosmological solution can be eternally accelerating if these corrections at any finite time are positive, $\ddot{a}(t) >0$. Therefore we see that {\sl asymptotic (or eternal) acceleration is compatible with an absence of event horizon, in the case $\ddot{a} \rightarrow 0$}.

The above proof, proposed in \cite[Sec.3.3.1]{Andriot:2023wvg}, was used there to show that stringy solutions in single field exponential quintessence, asymptoting to $P_{k\varphi}$, would have no cosmological event horizon. The same can be done for solutions with $k=0$ asymptoting to $P_{\varphi}$ with value $\lambda= \sqrt{2}$, in case they can be found from string theory. Indeed, the final solution $P_{\varphi}$ in Table \ref{tab:fixedpoints} is then given by $a(t) =a_l\, t$, for which the above reasoning applies. Let us note that this solution $P_{\varphi}$ with $k=0$ and $\lambda= \sqrt{2}$ was discussed in \cite{Townsend:2003qv}, in relation to a milder form of the NCHC. Interestingly, this value of $\lambda$ was there suggested to be a possible lower bound value, because it is the last one avoiding a cosmological event horizon (as we discuss next): this is reminiscent of the SdSC \eqref{SdSC}, with different motivations. The connections of these ideas were recently revisited in \cite[Sec.3.3]{Russo:2022pgo}. So what about stringy solutions asymptoting to $P_{\varphi}$ with $\lambda > \sqrt{2}$ ? Those are {\sl expanding but decelerating: therefore, they do not provide an event horizon either}, as we prove in full generality in the following.

  \item Consider a decelerating but expanding cosmological solution: $\ddot{a} (t)<0,\, \dot{a}(t)>0$. The former implies that $\dot{a}$ decreases. In the limit $t \rightarrow \infty$, because of its monotonous behaviour and lower bound, one must have $\dot{a}(t) \rightarrow {\rm constant} \equiv b_1$, the latter being possibly $0$. This implies that $a(t) \rightarrow b_1\, t + b_2$. If $b_1 \neq 0$, then the solution asymptotes to the linear solution discussed above, and one concludes on $d_e=\infty$. If $b_1=0$, then it asymptotes to the constant $b_2$, possibly $0$. In that case, there exists a time $t_1$ after which $a(t) < t$. We then conclude as above that $d_e=\infty$. In either case, one deduces the absence of an event horizon. Let us also note that $\dot{a}(t) \rightarrow b_1$ implies $\ddot{a}(t) \rightarrow 0$ in full generality. For illustration, we can look at $P_{\varphi}$ with $\lambda > \sqrt{2}$, or at $P_{m\varphi}, P_{r\varphi}, P_{{\rm kin}}^+$: they are all decelerating and expanding. They are of the form $a(t) \sim t^{1-\varepsilon}$, with $1>\varepsilon>0$, from which it is straightforward to verify the results just obtained. This is all consistent with the NCHC.

  \item We now prove that {\sl a universe strictly accelerating in the limit admits an event horizon} (as e.g.~de Sitter). By this we mean that in the limit $t\rightarrow \infty $, $\ddot{a}(t) >  4\, b_3>0$, for a constant $b_3$; note this may even include e.g.~an oscillating behaviour. This gives in the limit $a(t) > 2 b_3\, t^2 + b_4\, t + b_5$ with constants $b_4, b_5$. For sufficiently large $t$, one has $b_3\, t^2 + b_4\, t + b_5 > 0$ because the first term, positive, is dominating. We define the finite time $t_1$ as the one after which the two previous equalities both hold. We deduce that for $t>t_1$, $a(t) > b_3\, t^2$. It is then straightforward to conclude on the presence of an event horizon
\beq
d_e = a(t_0) \int_{t_0}^{\infty} \frac{\d t}{a(t)} < a(t_0) \int_{t_0}^{t_1} \frac{\d t}{a(t)} + a(t_0) \int_{t_1}^{\infty} \frac{\d t}{b_3\, t^2} = {\rm finite} \ .
\eeq

  \item Finally, let us come back to cases where $\ddot{a}(t) \rightarrow 0$. So far, we have only seen (stringy) examples of those giving no event horizon, because they would asymptote to the linear $a(t)$. We show below two other examples, dominant over the latter, with different results on the horizon:
\beq
\hspace{-0.3in} a(t)\gg t,\ \ddot{a}\rightarrow 0: \left|\begin{array}{ll} a(t) = t\, \ln t,\ \ddot{a}=\frac{1}{t}:&\ d_e \sim \ln \ln t \rightarrow \infty \\[4pt]
a(t) = t^{1+\varepsilon},\ 0<\varepsilon<1,\ \ddot{a}=\frac{\varepsilon (1+\varepsilon)}{t^{1-\varepsilon}}:&\ d_e \sim {\rm finite} + \frac{1}{t^{\varepsilon}} \rightarrow {\rm finite}
\end{array}\right. \label{strangeexamples}
\eeq
This shows that having asymptotic acceleration with $\ddot{a}(t) \rightarrow 0$ does not guarantee an absence of cosmological event horizon. It remains to be seen whether the last example is realised in string theory. It is not the case if the NCHC holds true.
\end{itemize}

We summarize our general results on the relation between (asymptotic) acceleration and event horizon as follows
\begin{itemize}
  \item $\ddot{a}(t) < 0$: decelerating (at any finite, large time) and expanding cosmological solutions have no event horizon. In addition, these solutions obey $\ddot{a}(t) \rightarrow 0$. We know several string solutions of this kind (e.g.~asymptoting to $P_{\varphi}$ with $\lambda> \sqrt{2}$, or to $P_{{\rm kin}}^+$).
  \item $\ddot{a}(t) > {\rm constant}>0$: accelerating strictly (at least in the large time limit) ensures the cosmological solution to have an event horizon. We do not know of such string solutions (e.g.~no fully stable, a.k.a.~eternal, de Sitter solutions, or some asymptoting to $P_{\varphi}$ with a given $\lambda \leq 1$).
  \item $\ddot{a}(t) \rightarrow 0$: expanding cosmological solutions which asymptote to the linear solution $a(t) =a_l\, t$ have no event horizon. We know several string solutions of this kind (e.g.~asymptoting to $P_{k\varphi}$), some being in asymptotic (i.e.~eternal) acceleration.
  \item $\ddot{a}(t) \rightarrow 0$: there exists expanding cosmological solutions dominant over the linear solution, eternally accelerating while $\ddot{a}(t) \rightarrow 0$, which have or do not have event horizons. We do not know of such string solutions (e.g.~solutions asymptoting to $P_{\varphi}$ with $1< \lambda < \sqrt{2}$).
\end{itemize}
These results are consistent with the absence of cosmological event horizon from quantum gravity, as formulated in the NCHC \eqref{NCHC}.

To conclude, we can also invert the results. Considering expanding cosmological solutions, i.e.~$\dot{a}(t)>0$ at any finite time, with an acceleration $\ddot{a}(t)$ that admits a limit at $t\rightarrow \infty$, then {\sl the NCHC implies that this limit must be vanishing: $\ddot{a} \rightarrow 0$}. Approaching the limit, the solution is allowed to be (eternally) decelerating. It is also allowed to asymptote to the linear solution $a(t) \rightarrow a_l\, t$ in any fashion, in particular in asymptotic acceleration. Different behaviours would require further investigation, but we then do not know of stringy examples.

\subsubsection{Including matter: getting realistic}\label{sec:matter}

In section \ref{sec:fixedpoints}, we have presented a general study of cosmological solutions of single field exponential quintessence. A dynamical system approach has allowed us to determine fixed points and their stability. We could then identify the universal starting and final points of every solution, depending on the precise setting: see \eqref{rmkp}, \eqref{rmp}, \eqref{rp} and \eqref{p}. Among all these possible solutions, we are now interested in finding and characterising those that provide a realistic cosmology: this is what we tackle in this section. Such solutions will then be compared to observational data.\\

As indicated by the fiducial values of $\Omega_{n0}$ in \eqref{Onfiducial}, the universe today contains about $30\%$ of (non-relativistic) matter and $70\%$ of dark energy. Realistic solutions would therefore be among those described by \eqref{rmkp} and \eqref{rmp}. In addition, standard universe history requires a past radiation domination phase, followed by a matter domination phase. For a $100\%$ domination, those correspond to the fixed points $P_r$ and $P_m$ of Table \ref{tab:fixedpoints}. A realistic cosmological solution must therefore pass close to those two saddle points. We summarize the candidate realistic solutions as follows
\beq
\xymatrix{ P_{{\rm kin}}^+\ {\rm or}\ P_{{\rm kin}}^- \ \ar[rr] & & \ P_r \ \ar[rr] & & \ P_m \ \ar[rr]^{\hspace{-0.4in}\text{(today)}} & & \ P_{\varphi},\ P_{k\varphi}\ {\rm or} \ P_{m\varphi} } \ , \label{realsol}
\eeq
indicating where today's universe should stand. Further intermediate steps in the solution are possible. The choice among the starting and final points depends on $k,\, \lambda,$ and the initial $s_{\dot{\varphi}}$; as we will see, a final $P_{m\varphi}$ is very unlikely.

Another requirement for a realistic solution is that today's universe is accelerating. In Section \ref{sec:accnohor}, we have discussed the possibility of having asymptotic acceleration, meaning acceleration at late time, namely when the solution is close to the final fixed point. As pointed-out, an asymptotically accelerating solution then has little matter content, so this is not fitting today's universe. The acceleration today could then be transient, meaning having started recently (at the end of matter domination, i.e.~away from $P_m$) and ending within a finite time in the future, or it can be (semi-)eternal and go on asymptotically. The numerical study of all candidate realistic solutions performed in \cite{Andriot:2024jsh} allowed to characterise today's acceleration, depending on $\lambda$ and $k$. As we will explain, for $\lambda> \sqrt{2}$, it was found that {\sl realistic solutions only have a transient acceleration today}.

Before discussing in more detail the cosmological solutions, we make a side comment. Contrary to the asymptotic discussion of Section \ref{sec:accnohor}, let us distinguish here a large field from a large time. We consider, from the string theory perspective, to be in field space asymptotics, therefore have a sufficiently large field to approximate the potential by a single exponential term, i.e.~neglecting other terms and corrections, and apply in addition the SdSC bound \eqref{SdSC} such that $\lambda \geq \sqrt{2}$. With such a large field approximation, we fall in the single field exponential model, in which we would like to discuss general cosmological solutions, in particular candidate realistic ones of the kind \eqref{realsol}. Such solutions are however thought to describe the universe history at many different epochs, at least between radiation domination and today. In that context, a large time refers to the solution close to the final fixed point only, hence the distinction here with a large field. Finally, we mentioned already that the saturation $\lambda= \sqrt{2}$ is not easily achieved in string theory. So we restrict for simplicity in the following to $\lambda > \sqrt{2}$, compatible with a (field asymptotic) string theory realisation. In this framework, we now turn to the detailed study of candidate realistic solutions of exponential quintessence.\\

We start with $k=0$. In that case, we fall in the solutions \eqref{rmp}, for which the final point is $P_{\varphi}$ or $P_{m\varphi}$. With $\lambda> \sqrt{2}$, both are eternally decelerating. So {\sl the acceleration today of such a realistic cosmological solution would at best be transient}. It turns out that such quintessence solutions actually exist, even though this is in principle not guaranteed. Indeed, one may suspect a realistic solution with successive radiation and matter domination phases, followed by dark energy domination today with acceleration, to exist in a setting close to $\Lambda$CDM model. The latter can be mimicked by a small $\lambda$ value (the limit $\lambda \rightarrow 0$ being nevertheless discontinuous): the potential then varies very slowly and mimics a cosmological constant. It is unclear whether the same qualitative physics takes place when raising $\lambda$ and making the potential steeper. An upper bound on $\lambda$ was actually determined in \cite{Andriot:2024jsh}: requiring a solution with radiation and matter domination in the past, and acceleration today ($-w_{{\rm eff}} -1/3 >0$ as in \eqref{acccond}) can only be obtained for
\beq
\lambda \lesssim \sqrt{3} + 0.04 \approx 1.77 \ ,\label{lambdaupper}
\eeq
using the fiducial values \eqref{Onfiducial}. We will refer to this bound as $\lambda \lesssim \sqrt{3}$. Beyond this value, the potential is too steep: $V$ becomes quickly small while the field picks up speed, in such a way that $|w_{\varphi0}|$ is too small to provide acceleration. Due to this bound, the final point for $k=0$ is mostly $P_{\varphi}$, as $P_{m\varphi}$ only appears for $\lambda> \sqrt{3}$. This bound leaves room for candidate realistic solutions with $\lambda > \sqrt{2}$.

\begin{figure}[ht!]
\begin{center}
\begin{subfigure}[H]{0.48\textwidth}
\includegraphics[width=\textwidth]{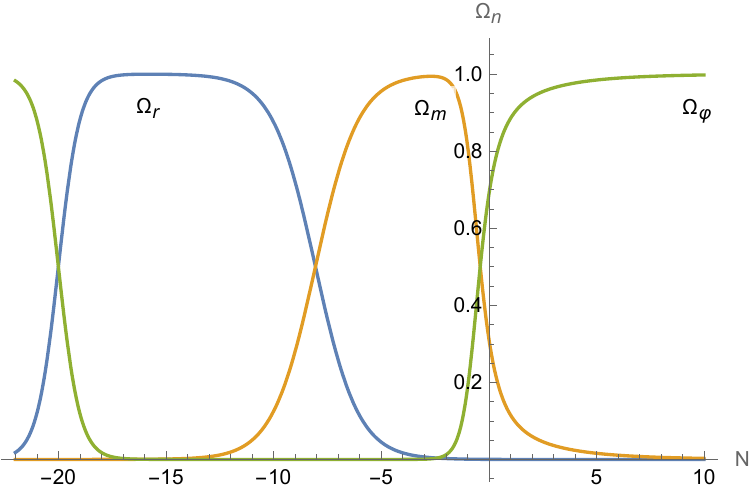}\caption{$\Omega_n(N)$}\label{fig:solO}
\end{subfigure}\quad
\begin{subfigure}[H]{0.48\textwidth}
\includegraphics[width=\textwidth]{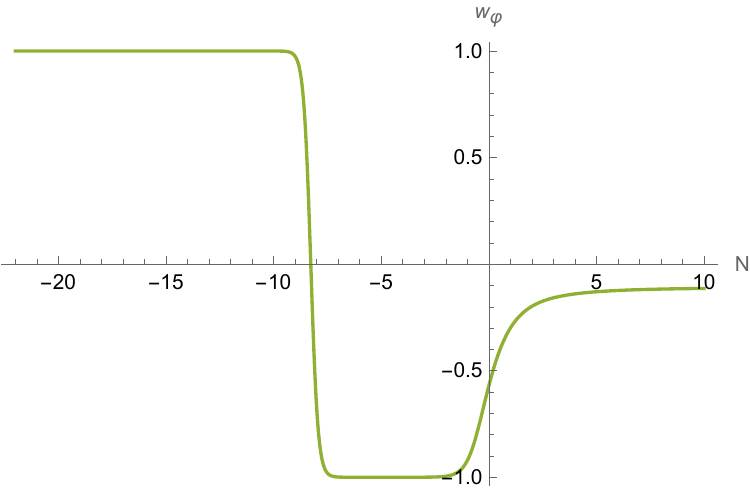}\caption{$w_{\varphi}(N)$}\label{fig:solwp}
\end{subfigure}\\
\begin{subfigure}[H]{0.48\textwidth}
\includegraphics[width=\textwidth]{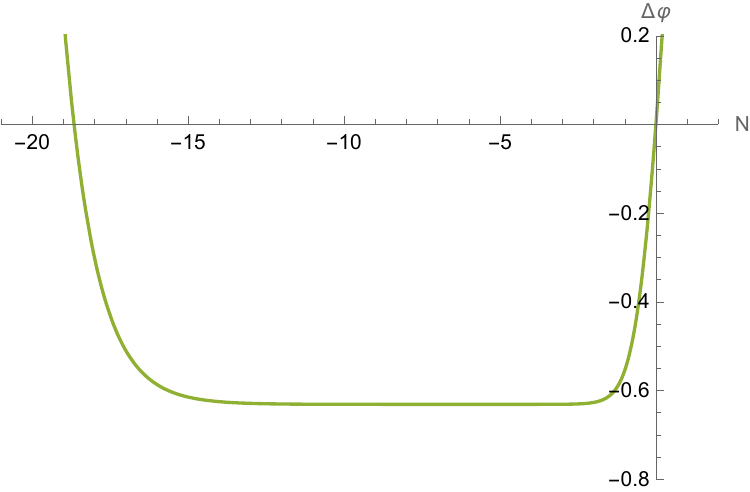}\caption{$\varphi(N)-\varphi_0$}\label{fig:solp}
\end{subfigure}\quad
\begin{subfigure}[H]{0.48\textwidth}
\includegraphics[width=\textwidth]{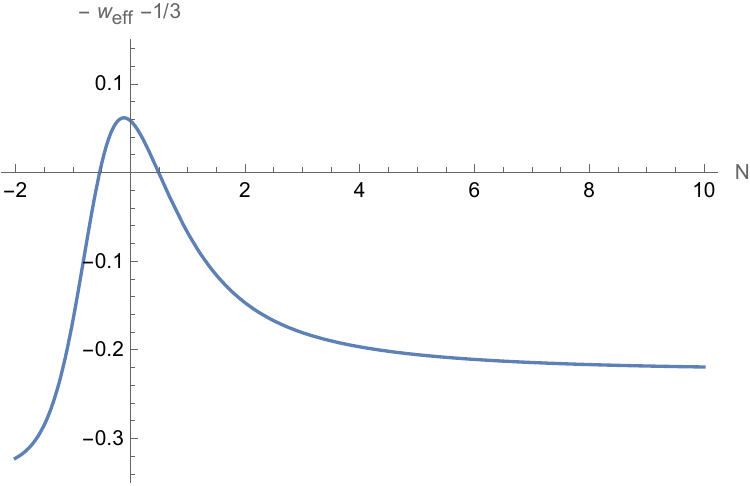}\caption{$-w_{{\rm eff}}-\frac{1}{3} $}\label{fig:solacc}
\end{subfigure}
\caption{Candidate realistic cosmological solution of single field exponential quintessence, with $\lambda=\sqrt{\frac{8}{3}}$, $k=0$, $\Omega_{n0}$ in \eqref{Onfiducial} and $w_{\varphi0}=-0.57196843265$. We observe the successive radiation, matter and dark energy domination phases in Figure \ref{fig:solO}, in terms of e-folds with $N=0$ today. The evolution of the field $\varphi$ (shifted by today's value) and of its equation of state parameter $w_{\varphi}$ along the solution history are given in Figure \ref{fig:solp} and \ref{fig:solwp}. The evolution of the quantity $-w_{{\rm eff}} -1/3$ is displayed in Figure \ref{fig:solacc}, indicating acceleration when it is positive. This emphasizes the transient acceleration phase today. The asymptotic values can be verified to match those of $P_{\varphi}$: for example, $-w_{{\rm eff}} -1/3 \rightarrow -2/9$.}\label{fig:solutionquint}
\end{center}
\end{figure}

Let us display an example of such a solution in Figure \ref{fig:solutionquint}, for $\lambda=\sqrt{\frac{8}{3}}$.\footnote{The value $\lambda=\sqrt{8/3}$ provides a good illustration for $\sqrt{2} < \lambda < \sqrt{3}$ as needed here. Let us mention in passing that this value is specific for several reasons. First, it has a string/M-theory origin (see e.g~\cite{Andersson:2006du, Marconnet:2022fmx, Andriot:2023wvg}) as coming from a compactification on a negatively curved internal manifold, the exponential being generated from the curvature term only (see the end of Section \ref{sec:dSconjsum}). Such an origin however does not allow for a parametric scale separation at late time \cite{Andriot:2025cyi}. Second, it is also a specific value from a dynamical system perspective. Indeed, for $k=-1$ and $\lambda> \sqrt{2}$, the fixed point $P_{k\varphi}$ is stable, but the details of this stability change precisely at this value. For $\lambda \leq \sqrt{8/3}$, the point is an attractive node, but for $\lambda > \sqrt{8/3}$, it is an attractive spiral. Details can be found e.g.~in \cite[Sec.3]{Andriot:2023wvg} or \cite[Sec. 2.4]{Andriot:2024jsh}; we do not comment further here on these different stabilities.} The successive domination phases can be seen in Figure \ref{fig:solO}, while the transient acceleration today can be seen in Figure \ref{fig:solacc}. The latter happens between the e-folds $-0.531 \leq N\leq 0.495$. The beginning of acceleration is slightly earlier than that of flat $\Lambda$CDM, given by $N\approx -0.490$ \cite[(3.16)]{Andriot:2024jsh}; this seems systematic in thawing quintessence models, due to an earlier rise of dark energy \cite[(3.15)]{Andriot:2024sif}. We will comment later on the evolution of $w_{\varphi}$ and $\varphi$ along the solution, displayed in Figure \ref{fig:solwp} and \ref{fig:solp}. One may note already the evolution of $w_{\varphi}$ discussed at the beginning of Section \ref{sec:quint} for thawing models: in the recent universe, it grows from $-1$  to a higher value. Finally, let us also indicate an initial kination phase, as can be seen in Figure \ref{fig:solO} with a dominant $\Omega_{\varphi}$: it corresponds to the initial $P_{{\rm kin}}^{-}$. We know that the cosmological solution discussed here should only be realistic from radiation domination onwards: the early universe, with possible inflation and reheating phases, require ingredients beyond the present framework. Therefore, we do not need to consider the initial kination phase here, even though it may still play a role in a realistic cosmology (see e.g.~\cite{Conlon:2022pnx, Apers:2024ffe}).

Technically, a solution is obtained by solving equations \eqref{F1F2Ei} while fixing its initial conditions, here $a(t_0)=1$, $s_{\dot{\varphi}0}=1$, and $\varphi(t_0), \dot{\varphi}(t_0)$; details on the resolution can be found in \cite[Sec.2.2, 3.2]{Andriot:2024sif}. The parameters in the equations must also be fixed: this is done by taking the fiducial values of the $\Omega_{n0}$ \eqref{Onfiducial} and fixing $\lambda, k$. Note that in the exponential potential, the parameter $V_0$ can be absorbed by a redefinition of $\varphi$, so it can be set to $1$ without loss of generality. The two initial conditions $\varphi(t_0), \dot{\varphi}(t_0)$ remain to be fixed: those can be traded for $\Omega_{\varphi0}$ and $w_{\varphi0}$. Eventually, the solution is essentially determined by $w_{\varphi0}$: by fine-tuning this value, we find a cosmological solution that has radiation domination in the past, systematically followed by matter domination. The tuning adjusts the duration of radiation domination, as well as the initial sign $s_{\dot{\varphi}}$. As expected, a tuned $w_{\varphi0}$ should be close to $-1$ for small $\lambda$, and increases for higher $\lambda$ values: see \cite[(3.9)]{Andriot:2024jsh}. An upper bound is given by the requirement of acceleration today, namely
\beq
w_{\varphi0} \leq -\frac{1}{3} \frac{1}{\Omega_{\varphi0}} \approx - 0.4866 \ ,
\eeq
which corresponds to the upper bound \eqref{lambdaupper}: $\lambda \lesssim \sqrt{3}$.\\

We turn to the case $k=-1$, with $\lambda > \sqrt{2}$. Then, one can see from \eqref{rmkp} that the final point is $P_{k\varphi}$. The recent and future universe described by a realistic solution \eqref{realsol} may then change due to this change of attractor, but the past remains similar. There is a simple reason for that: due to the scaling in $a$, matter and radiation are necessarily dominant over curvature in the past. In particular, since $\Omega_m$ in matter domination gets close to $100\%$, $\Omega_k$ quickly gets negligible in the past, and can only get smaller backwards in time. Therefore the whole past of solution \eqref{realsol}, until the recent universe (shortly after $P_m$) is unaltered with respect to the case $k=0$. Considering today's universe, a generous upper bound on curvature contribution from observations would be $\Omega_{k0} < 0.1$ \cite{Andriot:2024jsh}. Therefore even in the recent universe, $\Omega_k$ remains small and we do not expect qualitative differences in the cosmological solution with respect to $k=0$, even though small quantitative differences can be found as we detail below. The future of the solution gets changed, but this is less interesting in view of observations and time scales. We refer to \cite[Sec.4]{Andriot:2024jsh} for a detailed discussion.

\begin{figure}[ht!]
\begin{center}
\begin{subfigure}[H]{0.48\textwidth}
\includegraphics[width=\textwidth]{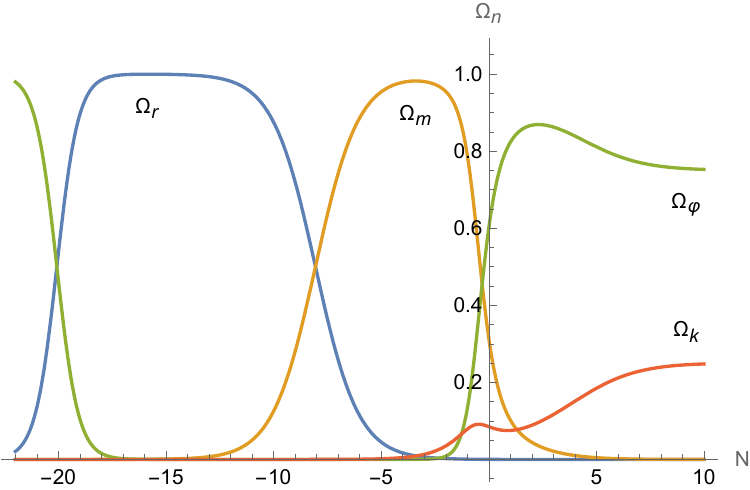}\caption{$\Omega_n(N)$}\label{fig:solkO}
\end{subfigure}\quad
\begin{subfigure}[H]{0.48\textwidth}
\includegraphics[width=\textwidth]{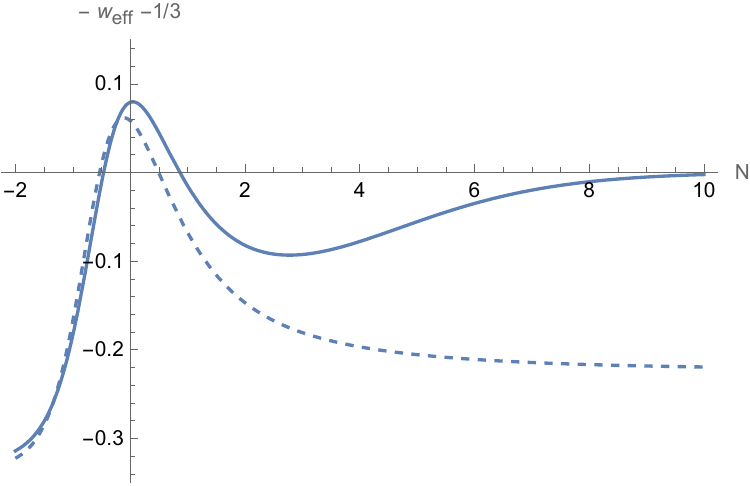}\caption{$-w_{{\rm eff}}-\frac{1}{3} $}\label{fig:solkacc}
\end{subfigure}
\caption{Cosmological solution of single field exponential quintessence, with $\lambda=\sqrt{\frac{8}{3}}$, $k=-1$, $\Omega_{k0}=0.0850$, $\Omega_{\varphi0}=0.6000$, $\Omega_{m0}=0.3149$, and $w_{\varphi0}=-0.63998750867$. It should be compared to the case without curvature, $k=0$, of Figure \ref{fig:solutionquint}: we see here in Figure \ref{fig:solkO} the modest contribution of $\Omega_k$ in the recent universe, and the modified future, that corresponds to $P_{k\varphi}$. The quantity $-w_{{\rm eff}} -1/3$, given as the plain curve in Figure \ref{fig:solkacc}, indicating (transient) acceleration, can be compared to the case $k=0$, i.e.~the curve of Figure \ref{fig:solacc} represented here as dashed. The future of the two curves differ due to the change of attractor.}\label{fig:solutionquintk}
\end{center}
\end{figure}

Observations focus on measuring $\Omega_m$, and always get similar values (at percent level) to the fiducial one $\Omega_{m0}=0.3149$. It is therefore natural, when including $\Omega_{k0}\neq 0$, to diminish accordingly $\Omega_{\varphi0}$. Doing so in \cite[Sec.4.2]{Andriot:2024jsh} with $\Omega_{k0}=0.085$, we obtain as announced small quantitative changes with respect to $k=0$. The upper bound $\lambda \lesssim \sqrt{3}$ \eqref{lambdaupper} gets raised to $\lambda \leq 1.87$. The $w_{\varphi0}$ for a given $\lambda$ get more negative, which should help fitting to data higher $\lambda$ values. For example, $\sqrt{8/3}$ considered in Figure \ref{fig:solutionquint} gets now $w_{\varphi0} \approx -0.64$, instead of $-0.57$ for $k=0$. The corresponding solution with $k=-1$ is displayed in Figure \ref{fig:solutionquintk}. Finally, acceleration today remains transient, with slightly changed duration, as illustrated in Figure \ref{fig:solkacc}: one gets there acceleration for $-0.462 \leq N \leq 0.866$.

The behaviour of acceleration deserves further discussion. Indeed, we mentioned in Section \ref{sec:accnohor} solutions asymptoting to $P_{k\varphi}$ while accelerating: could those be the realistic ones \eqref{realsol}? As pointed-out in \cite[Sec.5]{Andriot:2023wvg}, these solutions in asymptotic acceleration, found without matter nor radiation, have a non-negligible $\Omega_k$ (e.g.~$>0.5$) for a long duration before approaching $P_{k\varphi}$. As explained above, this is impossible in presence of matter and radiation, due to the scaling in $a$ which makes $\Omega_k$ negligible in the past. In addition, the value today is certainly not so large. One may still wonder about the future: we however observe numerically that the candidate realistic solutions approach $P_{k\varphi}$ while decelerating, as illustrated in Figure \ref{fig:solkacc}. This is made obvious by the phase space representations of solutions in \cite{Andriot:2024jsh}, where these two kinds of solutions are clearly distinguished. Therefore, the solutions in asymptotic acceleration of \cite{Andriot:2023wvg} with $k=-1$ are not the realistic ones \eqref{realsol}: the latter have only transient acceleration today. This emphasizes the fact that including matter changes the physics: it plays here a crucial role in realising dark energy today as quintessence, and one may wonder whether the same could hold regarding de Sitter.\\

\begin{figure}[ht!]
\begin{center}
\begin{subfigure}[H]{0.3\textwidth}
\includegraphics[width=\textwidth]{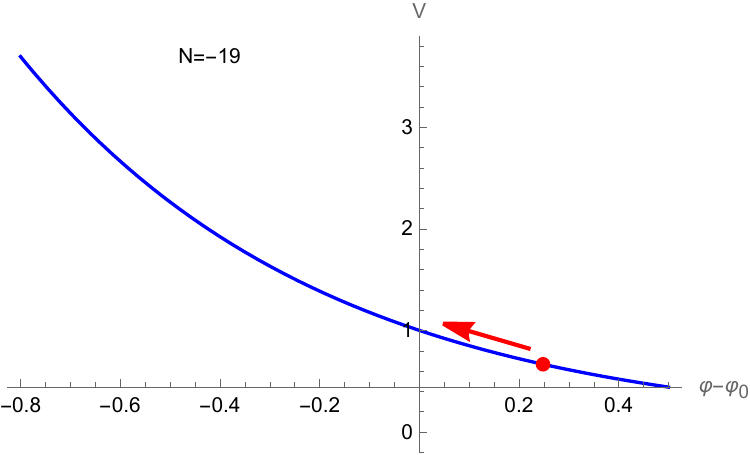}\caption{\footnotesize{$N=-19$}}\label{fig:thaw1}
\end{subfigure}\quad
\begin{subfigure}[H]{0.3\textwidth}
\includegraphics[width=\textwidth]{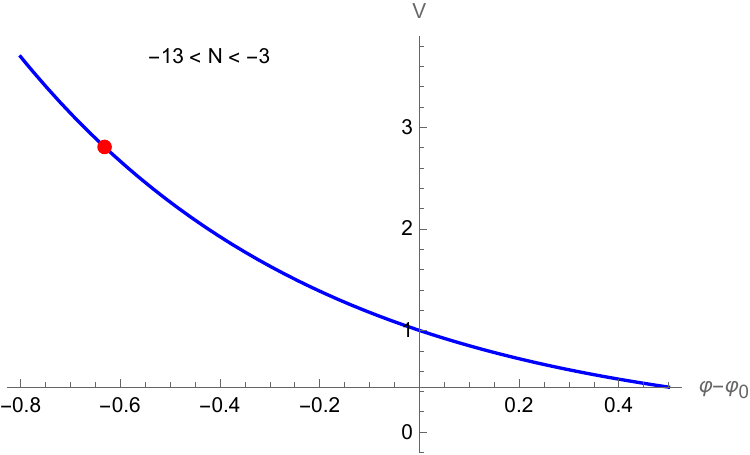}\caption{\footnotesize{$-13 < N < -3$}}\label{fig:thaw2}
\end{subfigure}\quad
\begin{subfigure}[H]{0.3\textwidth}
\includegraphics[width=\textwidth]{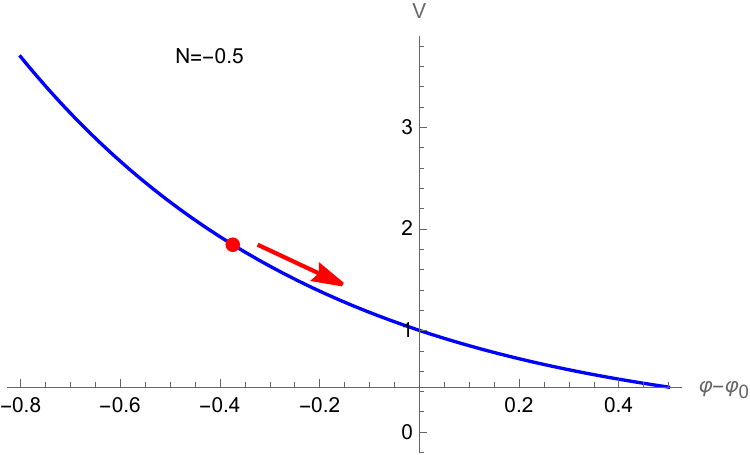}\caption{\footnotesize{$N=-0.5$}}\label{fig:thaw3}
\end{subfigure}
\caption{Universal evolution of the scalar field (red dot), here at different e-folds, in a candidate realistic solution of a thawing quintessence model. The example here is the solution of the model of Figure \ref{fig:solutionquint} with exponential potential $V(\varphi)$. The field is first rolling in a kination regime due to some initial speed (here starting with $P_{{\rm kin}}^-$), it then gets slowed-down and frozen at a point on the slope (here $\varphi-\varphi_0 \approx -0.63$) due to the high Hubble friction during the radiation-matter domination phase. When the friction drops at the end of matter domination, the field ``thaws'' and rolls-down the potential while dark energy is rising.}\label{fig:thaw}
\end{center}
\end{figure}

{\sl The behaviour of the (candidate) realistic solution displayed in Figure \ref{fig:solutionquint} is universal to thawing models} discussed at the beginning of Section \ref{sec:quint}, meaning even beyond an exponential potential \cite{Andriot:2024sif}. We show this universal evolution in Figure \ref{fig:thaw}.

Essentially, the scalar potential contribution is negligible for most of the universe history, until the recent universe: indeed, one has an early kination phase, where $V$ is subdominant compared to $\rho_{{\rm kin}}$, followed by radiation and matter dominations where $\Omega_{r}+\Omega_m$ dominates against $\Omega_V$. In these first phases, the scalar field has some initial speed (kination) but gets slowed down by a high Hubble friction, until it gets frozen by the latter during radiation and matter domination. This leads to the $w_{\varphi} \approx -1$ phase. This field evolution is visible in Figure \ref{fig:solp}. Being independent of $V$ (as long as it is negligible) and thus of the detailed model, the solution is therefore universal. The corresponding functions $a(t)$ (not $t(a)$) and $\varphi(t)$ were worked-out analytically in \cite{Andriot:2024sif}, allowing in particular to evaluate the field displacement during the radiation-matter phase: we obtained analytically
\beq
\text{Radiation - matter:}\qquad \Delta \varphi < 0.04 \, M_p \ ,
\eeq
which effectively corresponds to a frozen field. This result is valid as long as $|\del_{\varphi} V|/V \lesssim 100$ during this phase. Let us note that such a freezing of the field helps with respect to the variation of fundamental constants, in case $\varphi$ is involved in these; one may still ask this question for the later evolution, that we now turn to.

The final part of the solution, in the recent universe, is more model dependent: the Hubble friction has dropped enough by the end of matter domination to allow the field to ``thaw'', and start rolling-down the potential. Doing so, it gains kinetic energy and $w_{\varphi}$ raises. The qualitative physics remains universal. A quantitative upper bound on the (multi)field displacement in this recent phase could still be worked-out. Indeed, $\Delta \varphi$ is related to an integral of $\Omega_{{\rm kin}}$. The latter grows from $0$ to $\Omega_{{\rm kin}0}$, the value of which is bounded from above to guarantee acceleration today. One deduces an upper bound for $\Omega_{{\rm kin}}$. Assuming an average growth for $\Omega_{{\rm kin}}$, it can be shown \cite{Andriot:2024sif} that the field displacement of thawing models in the recent universe obeys
\beq
\text{Matter - dark energy:}\qquad \Delta \varphi < 1 \, M_p \ ,
\eeq
i.e.~the traveled field distance is sub-Planckian. This is verified in the example of Figure \ref{fig:solp}, or in the axion-like model of DESI latest observational results \cite{DESI:2025fii} for which it is reported $\Delta \varphi \sim 0.2-0.4 \, M_p$. This result is important with respect to possible quantum gravity corrections: corrective terms suppressed by $M_p$, or massive modes of the refined distance conjecture \cite{Ooguri:2006in, Klaewer:2016kiy, Baume:2016psm}, could otherwise become relevant for the physics and invalidate the 4d effective theory used.\\

Having presented candidate realistic solutions of exponential quintessence and their properties, we are now left to {\sl compare those to observations}. A first way to proceed is to specify the quintessence model, especially the $\lambda$ value, and compare the resulting cosmological solutions directly to a set of observational data, throughout the whole history of the universe. As argued above, models with lowest $\lambda$ value would typically provide a good fit to the data. The comparison thus provides an upper bound to $\lambda$. Without spatial curvature, \cite{Agrawal:2018own, Akrami:2018ylq, Raveri:2018ddi} obtained the upper bound $\lambda < 0.5-0.6$ at $95 \%$ confidence level, while more recent data sets led \cite{Schoneberg:2023lun} to find slightly higher bounds. Allowing for $k\neq0$ and using more recent data sets, \cite{Bhattacharya:2024hep, Alestas:2024gxe, Akrami:2025zlb} found central values $\lambda \approx 0.4-0.8$ at $68 \%$ confidence level. We see that having the real, observed acceleration is more constraining than only requiring some acceleration, the latter leading above to $\lambda \lesssim \sqrt{3}$. The conclusion from these general studies is that single field exponential quintessence models, with an origin from asymptotic string theory ($\lambda \geq \sqrt{2}$), are fairly {\sl disfavored by observations}. The corresponding candidate realistic solutions described above, with transient acceleration, are ruled-out.

\begin{figure}[ht!]
\begin{center}
\begin{subfigure}[H]{0.48\textwidth}
\includegraphics[width=\textwidth]{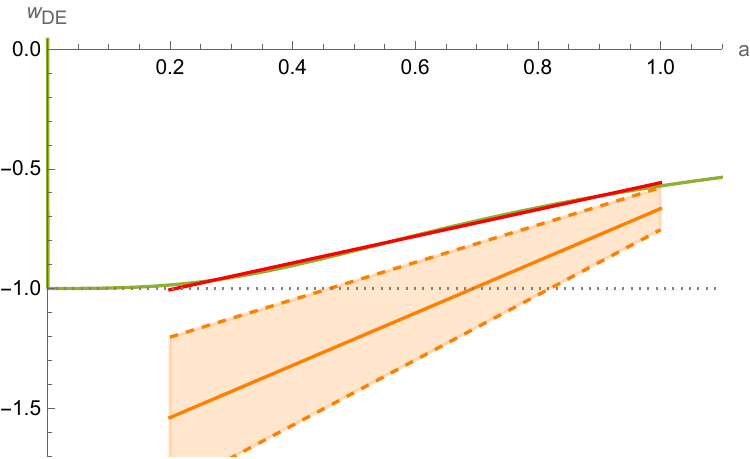}\caption{$w_{{\rm DE}}(a)$}\label{fig:solwa}
\end{subfigure}\quad
\begin{subfigure}[H]{0.48\textwidth}
\includegraphics[width=\textwidth]{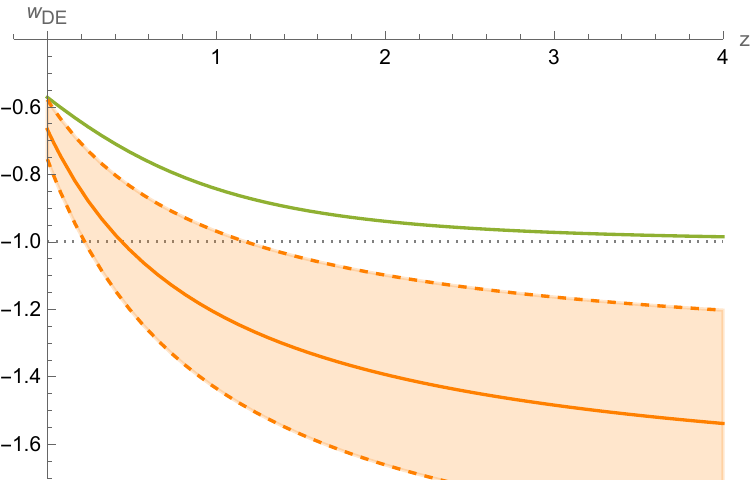}\caption{$w_{{\rm DE}}(z)$}\label{fig:solwz}
\end{subfigure}
\caption{Evolution of the equation of state parameter of dark energy in terms of the scale factor $a$ in Figure \ref{fig:solwa} and the redshift $z$ in Figure \ref{fig:solwz}, where $1+z=1/a$. Today corresponds to $a=1$ and $z=0$. The green curve, above $-1$, corresponds to $w_{\varphi}$ for the candidate realistic solution of exponential quintessence with $\lambda=\sqrt{8/3}$, $k=0$, displayed in Figure \ref{fig:solutionquint}. Its linear fit over the period $0\leq z \leq 4$ corresponds to the red line, with parameters given by $w_0\approx w_a \approx -0.558$. The orange lines correspond to $w_0,\,w_a$ obtained from the observational data sets DESI+CMB+Union3 \cite{DESI:2025zgx}, namely $w_0= -0.667^{+0.088}_{-0.088},\, w_a = -1.09^{+0.31}_{-0.27}$; the plain line is the central value, the dashed ones correspond to the errors. Those largely go below $-1$, corresponding to what is known as the phantom regime.}\label{fig:solutionw}
\end{center}
\end{figure}

We argued above that the evolution of $\Omega_n$ appears adequate, but we have not commented on that of $w_{\varphi}$, displayed e.g.~in Figure \ref{fig:solwp}, beyond its correct qualitative behaviour in the recent universe. In particular, we recall that the value $w_{\varphi0}$ was not free but tuned in order to have an appropriate radiation domination phase. For a candidate realistic solution, comparing quantitatively $w_{\varphi}$, considered as $w_{{\rm DE}}$, to the observations is therefore another way to proceed. To that end, a standard approach is to use the CPL parametrisation \eqref{CPL}: the two parameters $w_0, \, w_a$ are provided directly from the various observational data sets. For instance, one can read from the sets DESI+CMB+Union3 \cite{DESI:2025zgx} the values $w_0= -0.667^{+0.088}_{-0.088},\, w_a = -1.09^{+0.31}_{-0.27}$. They can be compared to $w_0, \, w_a$ obtained by a linear fit of the curve $w_{\varphi}(a)$ for a certain solution in a given model: we provide an example of this here in Figure \ref{fig:solutionw}.\footnote{There are actually several ways to perform such a fit: either by finding the best (least-squares) linear fit to $w_{\varphi}(a)$ for a whole period of time (e.g.~for $0\leq z \leq 4$ corresponding to the observation period of DESI), as we do in Figure \ref{fig:solwa}, or by considering the CPL parametrisation as a Taylor expansion in $a$ close to today, thus using as linear fit today's tangent to the curve $w_{\varphi}(a)$, or finally by fitting another data, e.g.~$H(z)$, and deduce the corresponding $w_0,\,w_a$ (see e.g.~\cite{Shlivko:2024llw}, but also \cite{Bayat:2025xfr}).} With this method, values for the pairs $w_0,\,w_a$ were obtained for the above candidate realistic solutions, for a sample of $\lambda$ values, without or with spatial curvature in e.g.~\cite[Tab.8,9]{Andriot:2024jsh}. The result is that the pairs $w_0,\,w_a$ obtained this way for exponential quintessence are overall difficult to match to the values from recent observations: the latter typically have a too large slope parameter $-w_a$. This is clear in the example of Figure \ref{fig:solutionw}, which gets $w_a \approx -0.558$, compared to the value $w_a = -1.09^{+0.31}_{-0.27}$ from observations. Other examples may provide a slightly better match in the recent universe (small $(1-a)$ and $z$): a lower $\lambda$ gives a lower $w_{\varphi0}$ and $w_0$, which would allow the green curve to enter the orange zone in Figure \ref{fig:solutionw}. Doing so however, the slope becomes flatter, i.e.~$-w_a$ becomes even smaller, and further from the value of observations.

The difficulty in matching these values, especially the slope $-w_a$, is a direct consequence of the fact that the $w_0,\, w_a$ from observations impose a large phantom regime, that is $w_{{\rm DE}}<-1$. Indeed, having a large positive slope $-w_a$ quickly forces $w_{{\rm DE}}$ to enter the phantom regime in the past; the DESI data gives this ``phantom crossing'' at $z_c\approx 0.5$, therefore very early compared to the spanned redshifts. It can easily be shown that having a phantom crossing is equivalent to having $w_0+w_a < -1$, provided the CPL parametrisation is valid and $w_{{\rm DE}}(a)$ is growing (see e.g.~\cite[(4.69)]{Andriot:2024sif}). This inequality always holds for observational values, indicating a systematic phantom regime. We also note that such values and regime were already reported in Planck results \cite[Tab.6]{Planck:2018vyg}, and more recently by ACT \cite[(62)]{AtacamaCosmologyTelescope:2025nti}. Early reviews on phantom regime observations and theoretical options can be found in \cite{Ludwick:2017tox, Escamilla:2023oce}. Having a phantom regime appears theoretically problematic for various reasons we will discuss in Section \ref{sec:phantom}, starting with the fact that $w_{\varphi} \geq -1$ in the standard quintessence models discussed so far. Therefore, even though we picked a simple quintessence model, changing the number of fields or the potential, with possibilities of richer physics, will not help in getting a better match to observations with respect to the phantom regime. The fit of the most recent universe, for which $w_{{\rm DE}}\geq -1$, could however be improved, at the cost of ignoring the earlier data.

This situation has led to various discussions on the observational data and its treatment. Beyond possible questions on systematics or error estimations, let us mention here the question of the parametrisation. The simplicity of the CPL parametrisation \eqref{CPL} may appear not adequate to capture the dynamics of dark energy. For example, viewed as a Taylor expansion close to today, one could include higher terms, whose coefficients correspond to extra parameters. To answer such criticism, the DESI collaboration has studied the benefit of using more parameters, concluding in \cite[Fig.6]{DESI:2025fii} that the fit of the data is not significantly improved beyond two parameters. We refer to e.g.~\cite{Nesseris:2025lke, Malekjani:2025alf} for more discussions on this point. Another possible criticism would be with respect to the linear shape. In \cite[Tab.II]{DESI:2025fii} are listed four other two-parameter models with quadratic, logarithmic, exponential, or fractional dependence on $a$. These parametrisations provide fits of similar quality, and all give a phantom crossing around $z_c\lesssim 0.5$. Finally, different reconstructions of $w_{{\rm DE}}$ are proposed in \cite{DESI:2025fii}, mostly giving a phantom regime. Therefore, even though the CPL parametrisation is simplistic, it could be sufficient to capture the relevant physics of a dynamical dark energy. In particular, it indicates a large phantom regime, that cannot be explained by the simple quintessence models and solutions considered in this section: we turn to this matter in the following.

\subsubsection{Coupling to matter: tackling the phantom regime}\label{sec:phantom}

In Section \ref{sec:matter}, we have discussed candidate realistic cosmological solutions of single field exponential quintessence, and more generally of thawing quintessence models. These solutions admit successive radiation, matter and dark energy domination phases, and allow for acceleration today, as required for any realistic solution. Asking for a string theory origin of the model in field space asymptotics (large field), we have focused on solutions obtained with $\lambda \geq \sqrt{2}$ as required by the SdSC bound \eqref{SdSC}. The comparison to the latest observational data however ruled-out these solutions and models on several levels: first, lower $\lambda$ values are preferred, and second, a large phantom regime ($w_{{\rm DE}} < -1$) is arguably observed, as detailed at the end of Section \ref{sec:matter}. The latter is problematic more generally for any standard quintessence model, with kinetic energy $\rho_{{\rm kin}} \geq 0$ and potential $V>0$, since those only allow for $w_{\varphi} \geq -1$ (see the definition \eqref{wphi}). In this section, we focus on this problem of having a phantom regime, and present a simple resolution to this puzzle: allowing for a coupling to matter. Interestingly, this way-out can favor larger $\lambda$ value, of interest to an asymptotic string theory realisation \cite{Andriot:2025los}: this is in contrast with the previous models without any such coupling. Let us recall that coupling to (dark) matter is subject to many possible observational constraints, mentioned at the beginning of Section \ref{sec:quint}. We are here agnostic about those, focusing for now on the consequence of such a coupling with respect to a possibly observed phantom regime. Furthermore, we discuss properties of the corresponding cosmological solutions.\\

Before presenting this idea in more detail, let us recall few issues with a phantom regime. We observe today that $\Omega_{{\rm DE} 0}>0$, i.e.~$\rho_{{\rm DE} 0}>0$, and $w_{{\rm DE}0} > -1$ (in a dynamical dark energy scenario as considered here). Identifying dark energy with scalar field contributions of a standard quintessence model, the continuity equation $\dot{\rho}_{\varphi} = -3 H (1+w_{\varphi}) \rho_{\varphi}$ \eqref{continuityeq} then indicates that $\rho_{\varphi}$ was larger in the past, in particular $\rho_{\varphi}>0$. In that context, a phantom regime in the past has two unpleasant consequences. First, $w_{\varphi} < -1$ with $\rho_{\varphi}>0$ violates the null energy condition (NEC): indeed, for a perfect fluid with $\rho>0$, the latter is simply $w \geq -1$. Second, $w_{\varphi} < -1$ with $\rho_{\varphi}>0$ requires $\rho_{{\rm kin}}<0$ \cite{Caldwell:1999ew, Feng:2004ad}. Both points signal an unhealthy theory. Avoiding this necessarily requires to change the model. While various, often complicated, options exist in the literature, we focus in the following on a simple change: having a coupling to matter. Coupling quintessence to matter was first discussed in \cite{Amendola:1999qq, Billyard:2000bh}, while the fact that it leads to an effective phantom regime, as shown below, was first pointed-out in \cite{Huey:2004qv, Das:2005yj}. Recent investigations of such models include \cite{Smith:2024ibv, Khoury:2025txd, Chakraborty:2025syu}.

The standard quintessence formalism presented in Appendix \ref{ap:cosmo} and used so far assumed a minimal coupling of the (canonical) quintessence scalar field $\varphi$ to gravity, and no coupling to matter and radiation: this assumption was necessary to the derivation of the equations \eqref{F1F2Ei}. Let us now include a coupling function $A(\varphi)>0$ to the matter Lagrangian ${\cal L}_m = A(\varphi)\, \bar{{\cal L}}_m$, where $\bar{{\cal L}}_m$ does not depend $\varphi$ but depends as before on the spatial metric: $\bar{{\cal L}}_m \sim 1/\sqrt{{\rm det} g_{ij}}$. Then, following the definitions \eqref{rhop} and using FLRW metric, we obtain
\beq
\rho_m = A(\varphi)\, \bar{\rho}_m \ ,\ \bar{\rho}_m = \bar{\rho}_{m0}\, a^{-3} \ ,\quad p_m = \bar{p}_m = 0 \quad \Rightarrow \quad w_m = \bar{w}_m = 0 \ .
\eeq
The key point is that the true matter energy density $\rho_m $ now evolves in time not only due to the standard $a^{-3}$ dependence, but also to an evolving field $\varphi(t)$. However, observations (as in DESI publications, e.g.~\cite{DESI:2025fii}) usually identify matter as the $\Omega_n$ which evolves as $a^{-3}$, i.e.~as $\bar{\rho}_m$; any other contribution in the recent universe (considering for simplicity $\Omega_r=\Omega_k=0$) is due to dark energy. This provides an {\sl effective definition of dark energy} to be compared to observations, denoted ${\rm DE}$, that can be summarized as follows: the true $\rho_m$ and $\rho_{\varphi}$ contributions get traded for the effective and observed $\rho_{{\rm DE}}$ and $\bar{\rho}_m$, as
\beq
\rho_m + \rho_{\varphi} \equiv \bar{\rho}_m + \rho_{{\rm DE}} \quad \Rightarrow\quad \rho_{{\rm DE}} = \rho_{\varphi} + (A(\varphi) - 1)\, \bar{\rho}_m \ .
\eeq
The latter captures the extra evolution of matter due to the coupling. The two Friedmann equations $F_1 = 0$, $F_2 =0$ are Einstein equations: they are unchanged since the metric dependence is the same. They involve the sums of $\rho_n$ and of $w_n\, \rho_n$. The first sum can thus be re-expressed as $\bar{\rho}_m + \rho_{{\rm DE}}$. Since $w_m = \bar{w}_m = 0$, the second sum boils down to the dark energy contribution, and allows to define the effective equation of state parameter $w_{{\rm DE}}$ as
\beq
w_{\varphi}\, \rho_{\varphi} \equiv w_{{\rm DE}} \, \rho_{{\rm DE}} \quad \Rightarrow \quad w_{{\rm DE}} = \frac{w_{\varphi}}{1 + (A(\varphi) - 1) \,\frac{\bar{\rho}_m}{\rho_{\varphi}} } \ .
\eeq
This parameter stands a chance to provide an effective phantom regime: $w_{{\rm DE}} < -1$. A necessary condition is that the denominator is smaller than $1$, thus that $A(\varphi)< 1$. In practice, this will be sufficient and {\sl an effective phantom regime can be reached}. Let us emphasize that this simple mechanism is obtained in a healthy model without any violation of the NEC.\\

We pause here to indicate few subtleties with this idea. There is first an ambiguity in choosing $\bar{\rho}_m$: it varies as $a^{-3}$, but there is freedom in its normalisation. The latter is directly related to the normalisation of $A(\varphi)$. One requirement is to match an observed value of $\rho_n$ or $\Omega_n$ at a given time. Here, to avoid any confusion, we require a matching today, in the form $\rho_{m0} = \bar{\rho}_{m0}$, $\rho_{\varphi 0} = \rho_{{\rm DE}0}$. This means that $A(\varphi_0)=1$ and $w_{{\rm DE}0} = w_{\varphi0}$. In that case, $A(\varphi)< 1$ means that {\sl $A(\varphi)$ has increased in the recent universe}. However, a different time choice for the equality between the true and the observed matter densities could lead to $A(\varphi)=1$ at some point in the past; in that case, a recent phantom regime would rather require $A(\varphi)$ to decrease in the recent universe, as considered in \cite{Agrawal:2019dlm, Bedroya:2025fwh}.

Another subtlety is that this mechanism relies on observations interpreting matter as evolving as $a^{-3}$, or equivalently, having $w_{{\rm matter}}=0$ with a standard continuity equation. However, we will see that the continuity equations of the true densities $\rho_m, \rho_{\varphi}$ are modified. These modifications could be reabsorbed in a different definition of effective equation of state parameters, while maintaining a standard form of the continuity equation. In short, an alternative approach is to identify the observed matter through $\rho_m$, together with either a deformed continuity equation, or a standard one but with an effective $w_{{\rm matter}} \neq 0$: see e.g.~\cite{Silva:2025hxw, Linder:2025zxb}. Here we stick to the above matter identification with $\bar{\rho}_m$, allowing to compare to DESI published results, but a different treatment of the observational data is thus possible.

Finally, the coupling to matter considered here in Einstein frame can be turned into a non-minimal coupling to gravity in Jordan frame, by a conformal transformation of the metric. The same phenomenon of an effective phantom regime was noticed in the latter context \cite{Carvalho:2004ty, Martin:2005bp}, while the DESI data may also be interpreted through models with non-minimal gravity coupling (see e.g.~\cite{Wolf:2025jed}). We refer to \cite[Sec.2.2]{Andriot:2025los} for more details, where we allow in addition for a radiation coupling, motivated by string theory. For the latter, the same formalism as with the matter coupling goes through. But radiation should not have much impact on the recent universe physics in any case. Finally, note that a specific coupling to only part of the matter Lagrangian may not be expressed as an overall function $A(\varphi)$, in which case there is not necessarily a translation into a non-minimal gravity coupling.\\

We have not yet mentioned the scalar field equation $E^1=0$ \eqref{F1F2Ei}. It gets modified by the new field dependence. The latter leads to what can be perceived as a new effective potential, namely $V_{{\rm eff}}(\varphi) = V(\varphi) +  A(\varphi)\, \bar{\rho}_m $. Indeed, the new field equation is given by
\beq
\ddot{\varphi} + 3H \dot{\varphi} + \del_{\varphi} V + \bar{\rho}_m\, \del_{\varphi} A = 0 \ . \label{Eicoupling}
\eeq
From that equation, it is straightforward to get the two following continuity equations
\bea
\dot{\rho}_{\varphi} &=& -3 H (1+ w_{\varphi}) \rho_{\varphi} - \dot{A}\, \bar{\rho}_m \ ,\label{contrhop}\\
\dot{\rho}_{{\rm DE}} &=& -3 H (1+ w_{{\rm DE}}) \rho_{{\rm DE}} \ .\label{contrhoDE}
\eea
As announced, the one for $\rho_{\varphi} $ is deformed, but the one for the effective dark energy $\rho_{{\rm DE}}$ is the standard one. This is consistent with the use of \eqref{contrhoDE} in DESI publications, and the interpretation of the observational data.

The new field equation \eqref{Eicoupling} has important consequences. In our conventions, we require $A(\varphi)$ to be increasing in the recent universe, to generate an effective phantom regime. This means that $\del_{\varphi} A >0$, implying that {\sl a runaway potential with $\del_{\varphi} V <0$ can be steeper than in the absence of coupling}. As shown in \cite{Andriot:2025los}, an exponential potential with $\lambda=\sqrt{2}$ can be found consistent with DESI data. This is promising for asymptotic stringy realisations. This is also a striking difference with the absence of matter coupling, for which low $\lambda$ values were preferred: here {\sl the steep potentials are favored by observations.}\\

\begin{figure}[ht!]
\begin{center}
\begin{subfigure}[H]{0.48\textwidth}
\includegraphics[width=\textwidth]{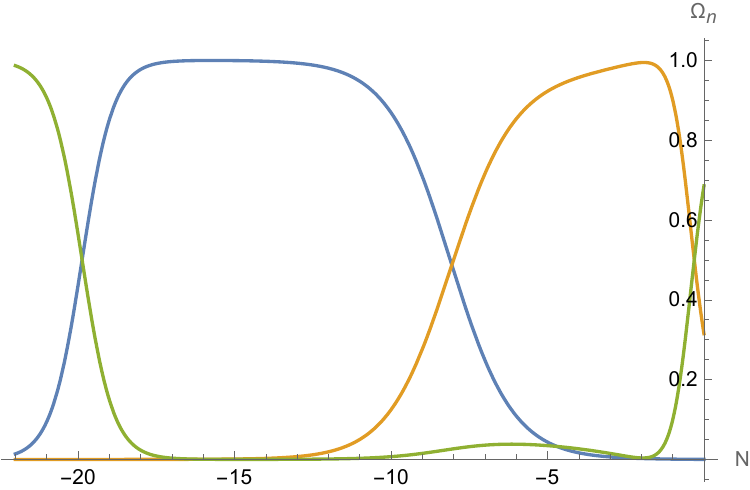}\caption{Exponential $\Omega_n(N)$}\label{fig:couplingExpO}
\end{subfigure}\quad
\begin{subfigure}[H]{0.48\textwidth}
\includegraphics[width=\textwidth]{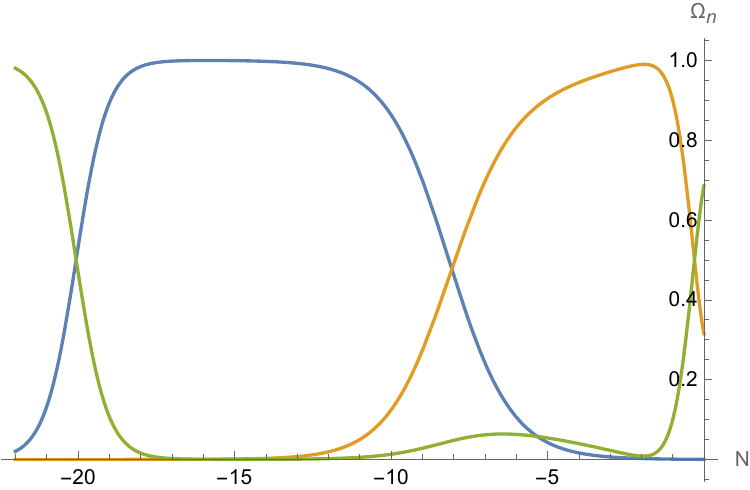}\caption{Hilltop $\Omega_n(N)$}\label{fig:couplingHillO}
\end{subfigure}
\\
\begin{subfigure}[H]{0.48\textwidth}
\includegraphics[width=\textwidth]{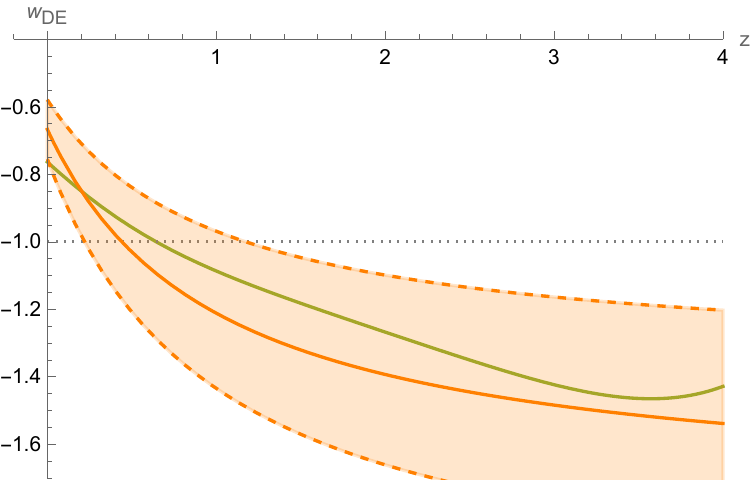}\caption{Exponential $w_{{\rm DE}}(z)$}\label{fig:couplingExpwz}
\end{subfigure}\quad
\begin{subfigure}[H]{0.48\textwidth}
\includegraphics[width=\textwidth]{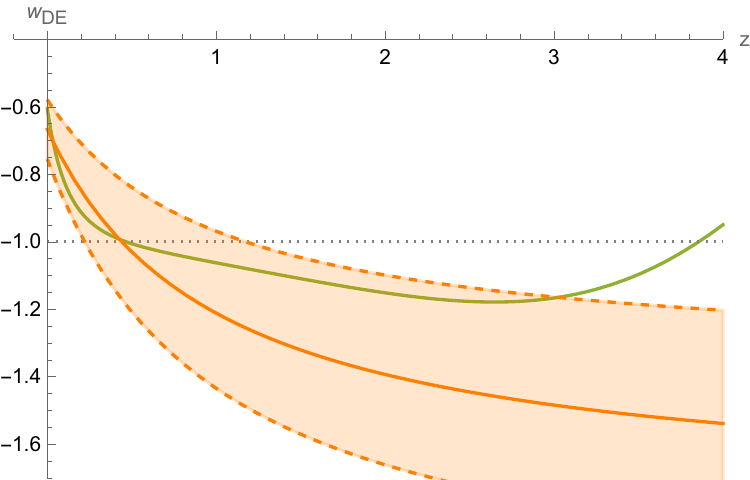}\caption{Hilltop $w_{{\rm DE}}(z)$}\label{fig:couplingHillwz}
\end{subfigure}
\caption{Cosmological solutions for models with a coupling $A(\varphi)$ to matter, and $\Omega_{n0}$ given by the fiducial values \eqref{Onfiducial}. Figure \ref{fig:couplingExpO} and \ref{fig:couplingExpwz} depict the solution of the exponential model $V(\varphi)= V_0\, e^{-\sqrt{2}\, \varphi}$, $A(\varphi)= 1+ 1/8\, (1-e^{-\sqrt{2/3}\, \varphi})$, with $w_{\varphi0} =-0.76365999931$ and $\varphi_0=0$. Figure \ref{fig:couplingHillO} and \ref{fig:couplingHillwz} depict the solution of the hilltop model $V(\varphi)/H_0^2= 5/2\, (1- 9/2\, \varphi^2)$, $A(\varphi)= 1+ 1/10\, (\varphi-\varphi_0)$, with $w_{\varphi0} =-0.60618663271$. Figure \ref{fig:couplingExpO} and \ref{fig:couplingHillO} give $\Omega_r$ (blue), $\bar{\Omega}_m$ (orange), $\Omega_{{\rm DE}}$ (green) in terms of e-folds. Figure \ref{fig:couplingExpwz} and \ref{fig:couplingHillwz} give $w_{{\rm DE}}(z)$ (green) compared to the curve (orange) obtained from the observational data DESI+CMB+Union3 \cite{DESI:2025zgx} with CPL parametrisation. This can be compared to Figure \ref{fig:solutionw}; we refer to the latter and the main text for more details. The solutions are borrowed from \cite{Andriot:2025los}.}\label{fig:coupling}
\end{center}
\end{figure}

We illustrate this mechanism with two models and their candidate realistic solutions from \cite{Andriot:2025los}, depicted in Figure \ref{fig:coupling}. The first one in Figure \ref{fig:couplingExpO} and \ref{fig:couplingExpwz} has exponential potential and coupling, with $\lambda=\sqrt{2}$. The second one in Figure \ref{fig:couplingHillO} and \ref{fig:couplingHillwz} has a hilltop potential and a linear coupling. In both cases, we see that $w_{{\rm DE}}(z)$ does enter the phantom regime, providing in addition a very good match with the observational data, at least for $z<3$. We also see that the solutions go through the expected successive domination phases, and are accelerating today. Finally, it is worth noticing the little maximum in $\Omega_{{\rm DE}}$ slightly after radiation - matter equality: this is reminiscent of an Early Dark Energy model. This feature was argued to be systematic \cite{Andriot:2025los}, and could thus help with respect to the Hubble tension (see also \cite{Teixeira:2024qmw, Smith:2025uaq}).

\begin{figure}[ht!]
\begin{center}
\begin{subfigure}[H]{0.3\textwidth}
\includegraphics[width=\textwidth]{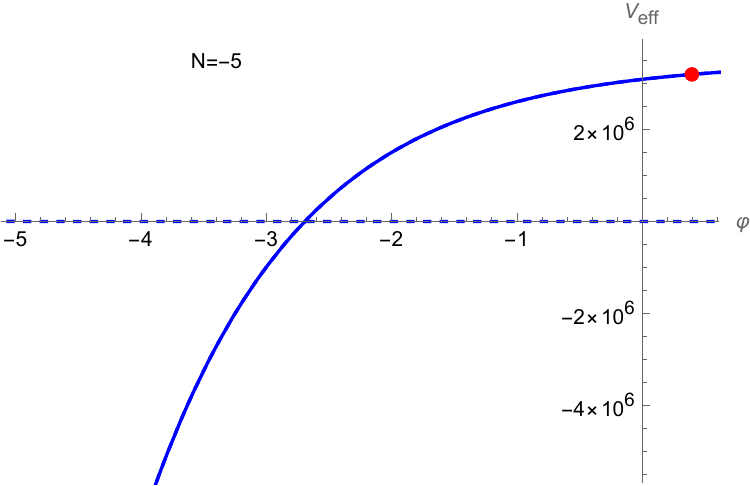}\caption{\footnotesize{$V_{{\rm eff}}$ at $N=-5$}}\label{fig:Veff1}
\end{subfigure}\quad
\begin{subfigure}[H]{0.3\textwidth}
\includegraphics[width=\textwidth]{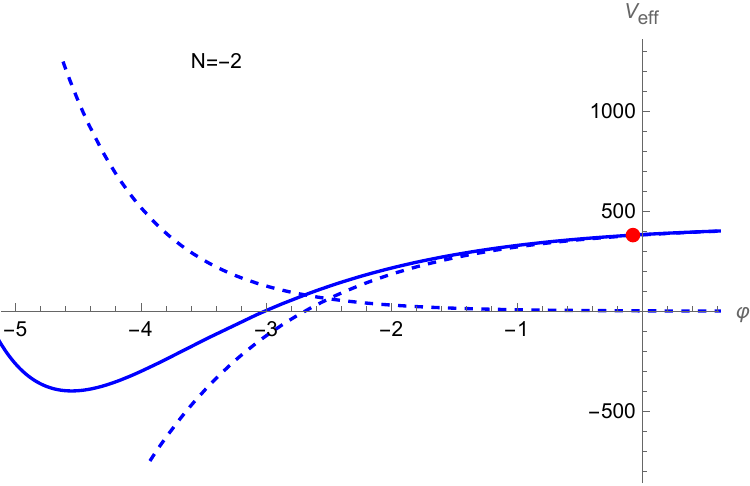}\caption{\footnotesize{$V_{{\rm eff}}$ at $N=-2$}}\label{fig:Veff2}
\end{subfigure}\quad
\begin{subfigure}[H]{0.3\textwidth}
\includegraphics[width=\textwidth]{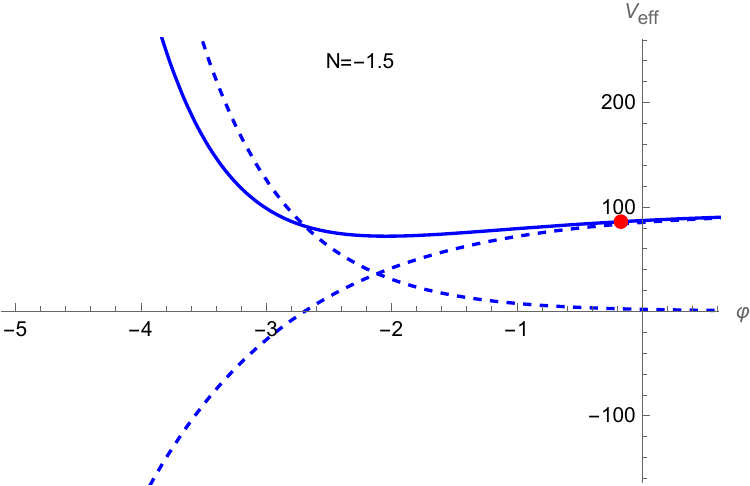}\caption{\footnotesize{$V_{{\rm eff}}$ at $N=-1.5$}}\label{fig:Veff3}
\end{subfigure}
\\
\begin{subfigure}[H]{0.3\textwidth}
\includegraphics[width=\textwidth]{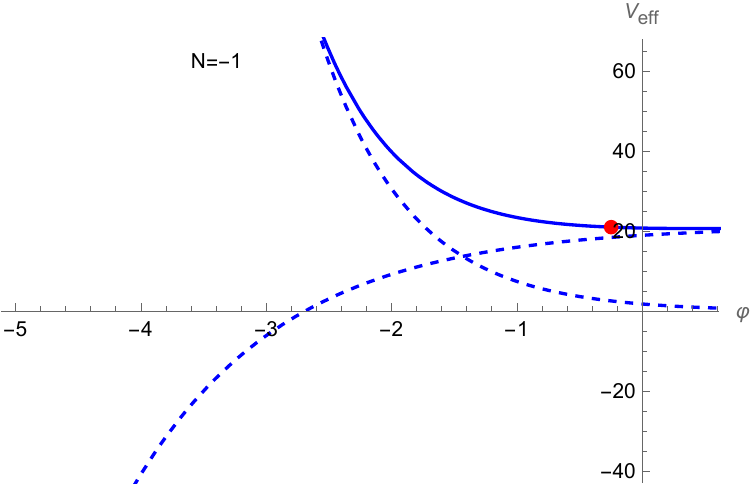}\caption{\footnotesize{$V_{{\rm eff}}$ at $N=-1$}}\label{fig:Veff4}
\end{subfigure}\quad
\begin{subfigure}[H]{0.3\textwidth}
\includegraphics[width=\textwidth]{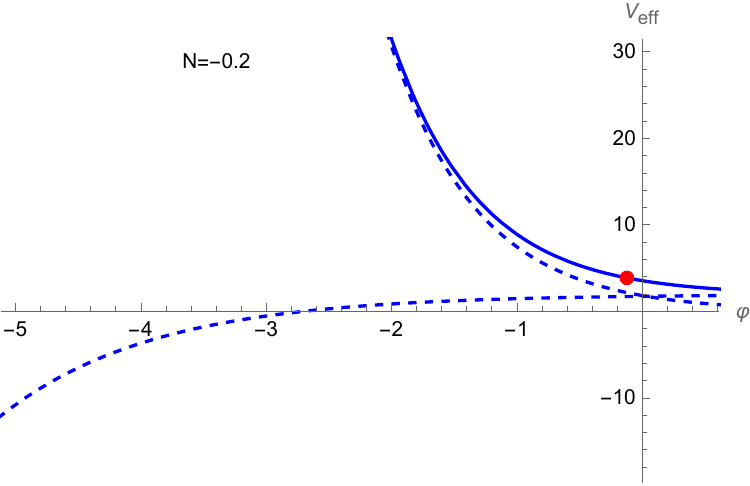}\caption{\footnotesize{$V_{{\rm eff}}$ at $N=-0.2$}}\label{fig:Veff5}
\end{subfigure}\quad
\begin{subfigure}[H]{0.3\textwidth}
\includegraphics[width=\textwidth]{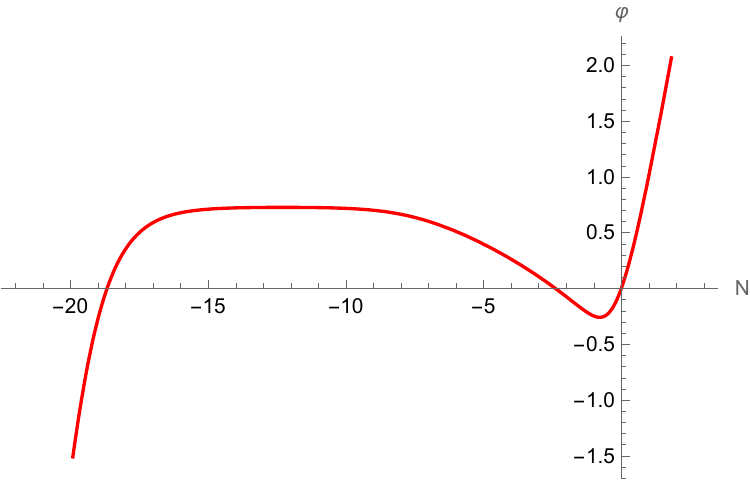}\caption{\footnotesize{$\varphi(N)$}}\label{fig:phiN}
\end{subfigure}
\caption{Evolution of the scalar field in the varying effective potential $V_{{\rm eff}}(\varphi)$ for the solution of Figure \ref{fig:couplingExpO} and \ref{fig:couplingExpwz}, with exponential potential and coupling. The profile of the field in terms of e-folds, $\varphi(N)$, is depicted in Figure \ref{fig:phiN}. $V_{{\rm eff}}(\varphi)$ is depicted at different e-folds in the other figures by a plain curve, its two contributions appearing as dashed curves. The field in this potential then appears as a red dot, allowing to visualize its evolution. This can be compared to the solution of the model without coupling, in Figure \ref{fig:thaw}.}\label{fig:Veff}
\end{center}
\end{figure}

It would be interesting to study in more detail the properties of these solutions, as done in Section \ref{sec:matter} for candidate realistic solutions of exponential quintessence. A first question would be the duration of acceleration. A second, related one would be the behaviour of the scalar field. The latter is dictated by the effective potential $V_{{\rm eff}}(\varphi) = V(\varphi) +  A(\varphi)\, \bar{\rho}_m $ which changes with time, as the second term decreases with $a^{-3}$. Without more investigation for now, we depict the evolution of the field in the potential in Figure \ref{fig:Veff}.

Another string-motivated work, testing an exponential potential and coupling with in-depth comparison to observational data, appeared in \cite{Bedroya:2025fwh}. There, the exponential rates, constrained by data, were different than those of Figure \ref{fig:coupling}, with in particular $\lambda < \sqrt{2}$, suggesting a string model away from asymptotics. It should also be noted that dark matter was there entirely coupled via an exponential function; considering only a partial coupling may give more freedom in the exponential rate values, as here. In case the coupling is only due to one or few terms in the overall particle physics Lagrangian, then classically, the coupling function should contain a constant part, corresponding to the non-coupled Lagrangian. It is the case of the examples of Figure \ref{fig:coupling}. For instance, the coupling could be only through a function $\alpha$, with $\alpha(\varphi_0)=1$, to a density $\bar{\rho}_c$: $\rho_{m}= \bar{\rho}_{nc} + \alpha(\varphi)\, \bar{\rho}_c$, $\bar{\rho}_{m}= \bar{\rho}_{nc} + \bar{\rho}_c$, then $A(\varphi) = 1+ (\alpha(\varphi)-1)\, \bar{\rho}_{c0}/\bar{\rho}_{m0}$, highlighting the constant in $A$ and its interpretation.  

We certainly face here one limitation in this approach: the lack of a detailed particle physics model, which would give all couplings. In principle, string theory can provide such models \cite{Marchesano:2024gul} (see recent progress e.g.~in \cite{Constantin:2026bky}), where scalar fields couple to matter while possibly having scalar potentials.\footnote{Phenomenological works connecting to observations sometimes use the model \cite{Damour:1994zq}, and related follow-ups \cite{Damour:2002mi, Damour:2002nv}. This model describes the couplings of the dilaton to 4d matter (standard model constituents), as they would likely be obtained from string theory. While these proposed couplings are possibly correct, the model is certainly outdated on several levels. It lacks of a clear string theory derivation, in the form e.g.~of a compactification with ingredients generating particle physics. Such an origin would most likely lead to multifield couplings to matter, including, to start with, the volume of the compact space. In addition, the development of flux compactifications in the 2000's (see Section \ref{sec:history} and \ref{sec:dimred}) has led to generally considering string-derived scalar potentials in 4d models, in particular one for the dilaton (see e.g.~\eqref{rhotautheory}); this is also absent from the model \cite{Damour:1994zq}. It would be very useful to provide another, up-to-date model, or at least some universal patterns in the couplings to matter, as intended by \cite{Damour:1994zq}.} An explicit and complete realisation of the latter would allow to make progress here. In the meantime, preliminary studies as above, and more systematic approaches via dynamical systems (as in Section \ref{sec:fixedpoints}), can give a first taste of realistic options.

Dynamical system tools have been applied to this framework with coupling, starting with \cite{Amendola:1999qq, Amendola:1999er}, and more recently in e.g.~\cite{Postolak:2025qmv, Shiu:2025ycw}, discussing fixed points, their stability and their properties, such as acceleration. In \cite{Shiu:2025ycw}, a late time attractor is identified, which can be in strict acceleration. Such an accelerating attractor however did not find a string theory realisation.\footnote{From string theory, one obtains in 4d Einstein frame a coupling to matter and radiation, depending on the volume field in the form of a growing exponential  \cite[(3.10)]{Andriot:2025los}. We can apply this stringy setting to the attractor of \cite{Shiu:2025ycw}, respectively with $\beta= \sqrt{3/8}, w=0$ and $\beta=\sqrt{3/2}, w=1/3$, while the potential generated by the 6d curvature gives $\gamma=\sqrt{8/3}$. We obtain respectively $\epsilon=12/11$ and $\epsilon=8/7$, i.e.~asymptotic deceleration. This agrees with the NCHC.} Moreover, a bottom-up quantum gravity argument is given in \cite{Shiu:2025ycw} against acceleration, based on the idea that the growing dark matter mass could eventually form a black hole. Asking for such a large black hole to be smaller than a Hubble size requires the attractor to be decelerating. All these results are interesting in view of the discussion of Section \ref{sec:accnohor} on asymptotic acceleration and no event horizon.\\

\begin{figure}[ht!]
\begin{center}
\begin{subfigure}[H]{0.48\textwidth}
\includegraphics[width=\textwidth]{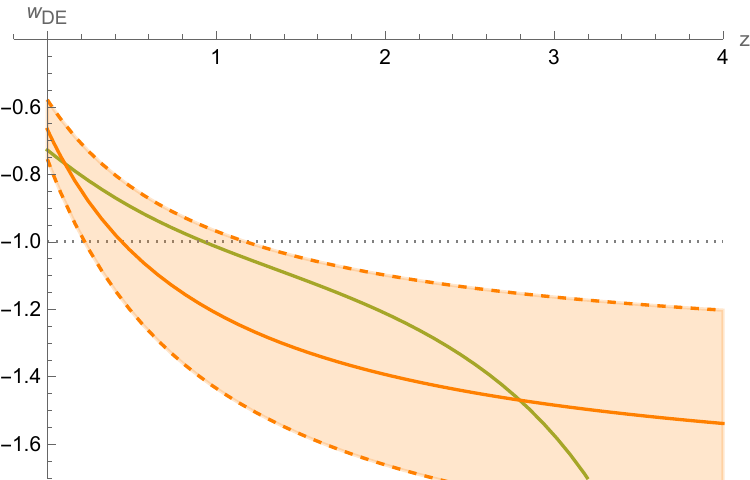}\caption{$w_{{\rm DE}}(z)$}\label{fig:couplingExppolewz}
\end{subfigure}\quad
\begin{subfigure}[H]{0.48\textwidth}
\includegraphics[width=\textwidth]{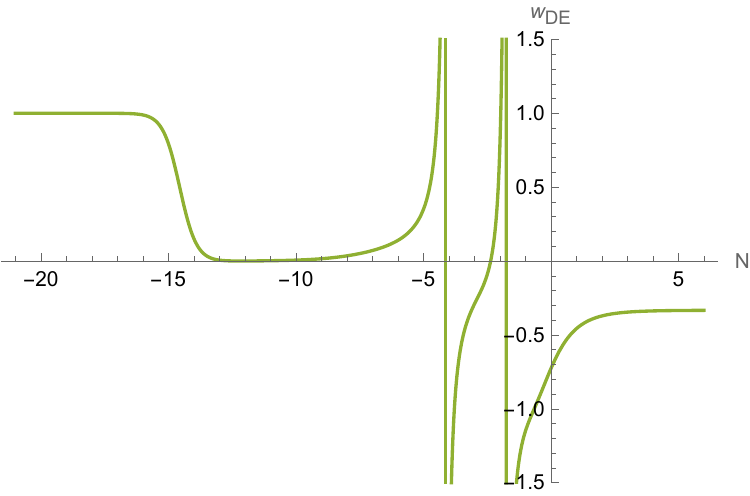}\caption{$w_{{\rm DE}}(N)$}\label{fig:couplingExppolewN}
\end{subfigure}
\caption{Cosmological solution in a quintessence model similar to that of Figure \ref{fig:couplingExpO} and \ref{fig:couplingExpwz}: the potential and coupling are exponential with $V(\varphi)= V_0\, e^{-\sqrt{2}\, \varphi}$, $A(\varphi)= 1+ 1/16\, (1-e^{-\sqrt{2/3}\, \varphi})$, and the solution is obtained with $\Omega_{n0}$ given in \eqref{Onfiducial}, $w_{\varphi0} =-0.72821866117$ and $\varphi_0=0$. The value of $w_{\varphi0}$ allows to get radiation domination in the past as in Figure \ref{fig:couplingExpO}. The only difference is therefore the numerical coefficient in the coupling: it goes from $1/8$ (there) to $1/16$ (here). Figure \ref{fig:couplingExppolewz} and \ref{fig:couplingExppolewN} give $w_{{\rm DE}}$ (green) in terms of redshift $z$ or e-fold $N$. The orange curve is obtained from the observational data DESI+CMB+Union3 \cite{DESI:2025zgx} with CPL parametrisation. The curve for $w_{{\rm DE}}$ has two divergences due to poles, the most recent one being at $z \approx 4.801$ or $N\approx -1.758$.}\label{fig:couplingpole}
\end{center}
\end{figure}

Given the apparent variety of options for a valid model (choice of potential, couplings, field space metric, etc.), an interesting point is that the cosmological observations could provide helpful information, already regarding the background cosmology discussed here, i.e.~not even mentioning perturbations which could bring more. Indeed, the phantom regime, if confirmed, already suggests a coupling to matter, which is constrained by the data. But there is more: as argued in \cite[Sec.4]{Andriot:2025los}, the effective $w_{{\rm DE}}(z)$ could, depending on the model, exhibit two poles (or divergences), due to the vanishing $\rho_{{\rm DE}}=0$ in the past. We illustrate these poles in Figure \ref{fig:couplingpole}: there the most recent pole is at $z \approx 4.801$. These divergences are not physical since they appear for the effective dark energy. In between the two poles, one would observe $\Omega_{{\rm DE}}<0$ and $\bar{\Omega}_m > 1$, which would also be striking. Surprisingly, the appearance of these poles is very sensitive to the model: indeed, one can compare Figure \ref{fig:couplingExpwz} where $w_{{\rm DE}}(z)$ admits a minimum at $z\approx 3.574$ to Figure \ref{fig:couplingExppolewz} where the pole appears at $z \approx 4.801$. The only difference between these two models is a change of coefficient in the coupling function from $1/8$ to $1/16$. In turn, observing either a pole or a minimum in a reconstructed $w_{{\rm DE}}(z)$ for recent redshifts could bring detailed constraints on the model. Interestingly, observing only a {\sl concave or convex behaviour} of this curve around $z\approx 3$ could already be sufficient to conclude. Coming observational data in the near future will therefore be of prime importance.

\subsection{Summary - what now?}\label{sec:sumquint}

Section \ref{sec:circumvent} aimed at presenting concrete attempts to realise dark energy from string theory, going beyond the obstructions summarized in Section \ref{sec:dSconjsum}. Three options were initially proposed: stick to the classical or asymptotic regime of string theory, and to the compactification ansatz introduced in Section \ref{sec:constraints}, or stick to the regime but go beyond that ansatz, or finally consider a different regime and framework allowing for different contributions. In Section \ref{sec:listattempts}, we have provided a list of tentative de Sitter solutions, with examples of all three options. These attempts range from de Sitter solutions of type II supergravities similar to those described in Section \ref{sec:constraints}, to proposed constructions (and constraints) in different string theories, to finally the inclusion of various perturbative and non-perturbative contributions, and broader frameworks beyond standard 6d compactifications. In most cases, weaknesses in the proposed de Sitter construction are identified, related to control issues explained in Section \ref{sec:nodSQG}. Therefore, there is up to now no example of a fully well-controlled 4d de Sitter solution from string theory. This situation should not prevent from further investigations on de Sitter constructions; new ideas and progress are still made on these questions. However, it also justifies the exploration of alternatives, namely dynamical dark energy in the form of quintessence. 

Nowadays, this purely theoretical motivation gets in addition supported by the recent cosmological observations which appear compatible with a dynamical dark energy. We introduced quintessence in Section \ref{sec:quint}, together with some background material in Appendix \ref{ap:cosmo}. We mentioned some advantages over de Sitter solutions from string theory (ubiquity, no orientifold, scale separation), but also some criticisms and challenges for quintessence. As a simple starting point, motivated by the asymptotic regime of string theory, we focused on single field exponential quintessence.

We provided in Section \ref{sec:fixedpoints} an overview of cosmological solutions in such a model. This was achieved thanks to a dynamical system approach that determined fixed points and their stability, providing various approximations of the solutions at different times in their history. Of particular interest are the late time attractor solutions. Those are then used to illustrate in Section \ref{sec:accnohor} a general discussion on asymptotic (meaning late time) acceleration in solutions from string theory. Asymptotic acceleration turns out to be realised in some stringy solutions, in the form of an eternal acceleration while $\ddot{a} \rightarrow 0$. We discuss in great detail the relation to cosmological event horizons, which are conjectured in the NCHC \eqref{NCHC} not to exist in solutions from quantum gravity. We show that the NCHC implies that in an expanding cosmological solution, for which the acceleration admits a finite limit, then this limit must be $\ddot{a} \rightarrow 0$. This limit can be reached in various fashions that we discuss. 

We then restrict in Section \ref{sec:matter} to solutions of single field exponential quintessence which stand a chance of being realistic: those go through the successive radiation, matter and dark energy domination phases, together with acceleration today. We note that these solutions only exist for a ``low'' exponential rate $\lambda$, namely $\lambda \lesssim \sqrt{3}$. Having identified such candidate realistic solutions, with or without spatial curvature, we discuss some of their properties: they fall in the general class of thawing quintessence models, the physics of which is universal. In particular, the field is frozen during radiation-matter domination due to a high Hubble friction (with $\Delta \varphi \leq 0.04\, M_p$), and thaws in the recent universe. For this late rolling, one can show that the field displacement remains sub-Planckian: $\Delta \varphi \leq 1\, M_p$. Another property of the solutions, for $\lambda > \sqrt{2}$, is that their acceleration (today) is transient. 

Comparing to the latest observations however discards these solutions realised in string theory asymptotics, because lower $\lambda$ are favored, i.e.~flatter potentials. Nevertheless, it remains impossible to fit the large phantom regime ($w_{{\rm DE}}< -1$) that appears as well in these observations. This is actually true for any standard quintessence model, for which $w_{\varphi} \geq -1$. We eventually focus on this question in Section \ref{sec:phantom}, where we show that allowing for a coupling to matter can generate an effective phantom regime within a healthy model. We discuss this idea in detail, together with some subtleties. We further argue that a phantom regime requires, in our conventions at least, an increasing coupling function, which in turn allows for a steep scalar potential, of interest for above stringy realisations. We provide candidate realistic solutions in that context, especially one with $\lambda =\sqrt{2}$, which provide a good fit of the data, including the phantom regime. We finally emphasize the possible appearance of (non-physical) divergences in $w_{{\rm DE}}$. Whether or not those are observed could be settled soon, and this would provide important constraints on the model.\\

So what now? Regarding de Sitter, further efforts may be pursued, and identifying the control issues should help in tackling them, and in establishing existence of a well-controlled example. To some extent, this program is pursued on classical de Sitter solutions in Section \ref{sec:dSsol}. Even if existence were to be established to a satisfying level, it is worth remembering that, as argued in Section \ref{sec:dSconjsum}, a fully stable de Sitter is not expected to exist, in line with the NCHC. Systematic destabilising effects of one kind or another may actually, in disguise, be related to the difficulties in proving existence at a good control level. Other frequent difficulties, like the singular bulk problem due to orientifold backreaction, mentioned in Section \ref{sec:listnonpert}, \ref{sec:listcrit} and later \ref{sec:backreact}, may also be reflecting some general difficulty; from this perspective, constructions without orientifold could be more promising. We also mentioned in passing that matter is typically ignored in de Sitter searches, but it could actually play a role. Indeed, we saw with quintessence that including matter could provide very different solutions (see Section \ref{sec:matter}), and adding a coupling to it could even generate an (evolving) effective potential (see Figure \ref{fig:Veff}) resulting in different physics; it could be worth keeping these ideas in mind for de Sitter. Finally, the absence of cosmological event horizon as proposed by the NCHC could stand as a more fundamental principle, that deserves further investigation.

Regarding quintessence, the study of single field exponential models is only the tip of the iceberg, while it already shows a variety of different physics (eternal, transient or no acceleration, including or not the realistic proportions of matter and radiation, fitting or not the observational data, allowing or not for a coupling to matter, with various consequences, etc.). As argued at the end of Section \ref{sec:dSconjsum}, effective string theories tend to be multifield, and non-canonical (curved) field space metrics are also common. Adding to this freedom various possible scalar potentials and couplings to matter, there is clearly room to explore stringy quintessence models; some related works were mentioned in Section \ref{sec:quint}. The question of the regime and control on approximations remains relevant for such models as for de Sitter constructions. The asymptotic regime may thus be appealing from that perspective, allowing for some simplification and universality in the models. The constraints on a coupling to visible or dark matter, or further couplings of the quintessence field, mentioned at the beginning of Section \ref{sec:quint}, should also be investigated and addressed within well-motivated particle physics models. On that front, the coming observations will be crucial: whether dynamical dark energy, with phantom regime, remains and becomes strongly preferred over a cosmological constant, whether divergences or minima are established in $w_{{\rm DE}}$, would provide invaluable observational constraints on the adequate models. The theory side should be ready for those.

\newpage

\section{Classical de Sitter solutions}\label{sec:dSsol}

``Classical de Sitter solutions'' refer to string backgrounds that include a $d$-dimensional de Sitter spacetime, in a classical regime of string theory. Commonly, the latter refers to the double perturbative expansion: first in string coupling $g_s$, allowing to neglect string loop (quantum) contributions (with the approximation $g_s \ll1$), and second in $\alpha'$, corresponding to low energies or large length scales compared to the string scale $l_s$. A further common framework is that of compactification, where the extra dimensions form a compact (internal) manifold ${\cal M}$. Neglecting $\alpha'$-corrections then requires a large internal volume of ${\cal M}$. We already mentioned the asymptotic limits of weak coupling and large volume in \eqref{asymptoticclass}. It is important to note that classical does not mean asymptotic, in the same way that large does not mean infinite. As discussed in Section \ref{sec:reftestdSconj}, building on the Dine-Seiberg problem, a scalar potential in a $d$-dimensional effective theory from string theory is expected to be a runaway in such asymptotics, therefore not allowing for de Sitter extrema. This expectation is by now more precisely formulated in terms of the SdSC \eqref{SdSC}. Therefore, a classical de Sitter solution would rather be found at large volume and $g_s^{-1}$, but not at infinite values. Note that phenomenologically, this is sufficient, since the size of extra dimensions is observationally bounded from above.

The classical regime allows one to approximate string theory by a 10d supergravity, together with some stringy objects such as $D_p$-branes, orientifold $O_p$-planes, and possibly more. This approximation comes with few more requirements, that we will come back to, to ensure the stringy origin. Given the no-go theorems against de Sitter obtained in heterotic strings (see Section \ref{sec:listhet}), one typically uses type IIA and IIB 10d supergravities, and this is what we do in the following. We will restrict ourselves to type IIA/B with only $D_p/O_p$: we refer to Section \ref{sec:listclassingr} for a discussion of options and difficulties when allowing for further ingredients. In the context of a compactification on a $(10-d)$-dimensional compact ${\cal M}$, we recall from the Maldacena-Nu\~nez no-go theorem \ref{nogo1} that $O_p$ are then mandatory to get a dS${}_d$ solution.

To summarize, this section is about solutions of type IIA/B supergravities with $D_p/O_p$, of the form dS${}_d\, \times\, {\cal M}_{10-d}$, where the product can include a warp factor. Let us stress that this does not allow for perturbative or non-perturbative contributions in a $d$-dimensional effective theory. Indeed, those would be due to non-classical string effects; {\sl ``classical'' refers here to the string regime, and not to the effective $d$-dimensional physics}. As we will explain, supergravity solutions of the form just described have been found, but crucially, we recall that this is not sufficient to guarantee their origin in a classical regime of string theory. In addition to a small string coupling and a large volume (or more precisely large lengths implying negligible $\alpha'$-corrections), few more requirements have to be met to ensure ``classicality'': fluxes, sources, and lattice (or compacity) quantizations, as we will detail in Section \ref{sec:dSsolclassical}. There exists up to now no known solution verifying all these conditions at once, part of the reason being that it is difficult to check those in practice. We should nevertheless mention one possible counter-example, solution $s_{55}^+ 29$, that we will come back to. But in short, {\sl up to now, there is no known classical de Sitter solution}. Therefore, this section is about the methods to find and characterise them, rather than actual examples. The tools introduced might even serve to prove the complete absence of such solutions, a result not achieved either so far.\\

Let us further specify the solution ansatz. First, we will mostly focus on $d=4$: we refer to Section \ref{sec:listclassdim} for a discussion of different dimensions. We mentioned there difficulties in finding solutions in $d\geq 5$, and few examples in $d=3$. Second, one needs to specify the compact manifold ${\cal M}$. A main avenue in the search for classical de Sitter solutions has been to consider group manifolds for ${\cal M}$, and we mainly focus on such compactifications here in Section \ref{sec:dSsolgroupman}. Characteristics of such supergravity solutions have been summarized in Section \ref{sec:listclassgroup}, and we will provide more details in the following. The specificity of these manifolds, namely having constant coefficients of the spin connection, suggests an ansatz where most supergravity variables are constant. In turn, this leads to consider smeared sources: we will therefore stick in Section \ref{sec:dSsolgroupman} to the solution and compactification ansatz of Section \ref{sec:10dansatz}, where sources are smeared. Refining such solutions to account for the backreaction of sources will be discussed in Section \ref{sec:backreact}. Other choices for ${\cal M}$ and further options for the backreaction are possible when looking for a classical solution, as discussed in Section \ref{sec:listbackreact}.

The ansatz of Section \ref{sec:10dansatz} has led to no-go theorems discussed in Section \ref{sec:nogo}. The no-go \ref{nogo3} implies, at least for sources $D_p/O_p$ of single dimensionality $p\geq 4$, the necessity of having the internal curvature ${\cal R}_6 <0$. Group manifolds are a simple departure from commonly considered Ricci-flat manifolds (flat torus, Calabi-Yau) that can be negatively curved, while remaining technically easy to handle: this is one motivation to consider them here. The other requirements indicated by the no-go theorems will help us consider the right configurations of ingredients to find solutions.\\

This section is organized as follows. Starting with classical de Sitter solutions on group manifolds in Section \ref{sec:dSsolgroupman}, we first give a brief review on such manifolds in Section \ref{sec:groupman} and Appendix \ref{ap:groupman}. We discuss the adequate solution ansatz and the corresponding dimensional reduction in Section \ref{sec:groupmandimred}, relating to gauged supergravities and 4d effective theories. We then discuss the existence of de Sitter solutions with such an ansatz in Section \ref{sec:dSsolexistence}: we present a classification of possible solutions, a sample of known examples, and expand on conjectured existence no-go theorems. We turn to the question of perturbative stability of these solutions in Section \ref{sec:dSsolstab}: as already mentioned, all known examples are found unstable. We discuss attempts to prove stability no-go theorems in this context. We finally focus on the question of classicality of these supergravity solutions in Section \ref{sec:dSsolclassical}, defining it properly and discussing the possibility of a parametric control via a scaling parameter. Then, we go beyond this ansatz and discuss in Section \ref{sec:backreact} the backreaction of localized sources, through a non-constant warp factor. We summarize the whole situation in Section \ref{sec:sumclassdS}, highlighting the main challenges in the search for classical de Sitter solutions.

\subsection{Solutions on group manifolds}\label{sec:dSsolgroupman}

In this section, we discuss $d=4$ de Sitter solutions of type II supergravities on 6d group manifolds, as motivated and summarized above. A review from 2019 on this topic, together with open problems, can be found in \cite{Andriot:2019wrs}. However, several new results have been obtained since then, which justifies the present update.

Let us first recall that group manifolds have played an important role in string compactifications. They allow to consider tractable dimension reductions over typically {\sl curved} compact manifolds.\footnote{Few Ricci-flat solvmanifolds also exist and were discussed e.g.~in \cite{Andriot:2015sia}. There, one was even shown to be a Calabi-Yau. A 7-dimensional extension as a solvmanifold with special $G_2$-structure was proposed, and recently revisited in \cite{VanHemelryck:2025qok}.} This curvature can be phrased in terms of ``geometric fluxes'', corresponding to non-vanishing spin connection coefficients, or structure constants as we will see below. In a Calabi-Yau language, these new types of fluxes were found useful since they allow to stabilise some K\"ahler moduli. One of the first and simplest example of a group manifold is the 3-dimensional nilmanifold, based on the nilpotent Heisenberg algebra. It appears when T-dualising a 3-torus carrying a constant $H$-flux \cite{Kachru:2002sk}. Since then, many more solutions on such ``twisted tori'' were obtained, with a 4d external spacetime being Minkowski \cite{Grana:2006kf}, as well as anti-de Sitter and de Sitter, on a broader set of group manifolds: an account is given in \cite{Andriot:2022way}. Among those, some solutions found on solvmanifolds (such as $s_{55}^+29$ to be discussed below, with a non-zero pair $f^1{}_{64}, f^6{}_{14}$) are rather T-dual to solutions with non-geometric fluxes, discussed in Section \ref{sec:listnongeo} (e.g.~$H_{146}, Q_4{}^{16}$) \cite{Hassler:2014sba, Andriot:2015sia} when using standard T-duality rules \cite{Shelton:2005cf}. Such geometric solutions on solvmanifolds appear as more isolated, since they do not lie on the same T-duality orbit as previous ones on nilmanifolds; interestingly, they may then exhibit new physics.

We can also mention the bosonic closed string solution obtained on the 3-dimensional nilmanifold, and its canonical treatment, in \cite{Andriot:2012vb}. In addition, the explicit Laplacian spectrum for this manifold was determined in \cite{Andriot:2016rdd, Andriot:2018tmb}, with applications in finding Kaluza--Klein towers \cite{Cribiori:2021djm}.

We now provide a brief review on those manifolds, before entering the details of (de Sitter) supergravity solutions in such compactifications, and discussing their properties.

\subsubsection{Group manifolds: a brief review}\label{sec:groupman}

In this section, we give an account on group manifolds. We start by recalling the vielbein formalism, motivating geometrically the special role of group manifolds, for which algebra structure constants play the role of spin connection coefficients. Compact group manifolds are given by ${\cal M} \simeq G/\Gamma$, where $G$ is a Lie group in one-to-one correspondence with a Lie algebra $\mathfrak{g}$, and $\Gamma$ is a lattice, a discrete subgroup ensuring compactness. We discuss the set of relevant Lie algebras, with an emphasis on 6-dimensional real solvable algebras, giving rise to a solvmanifold ${\cal M}$ (a subclass being nilmanifolds from nilpotent algebras). We finally explain how the structure constants can be variables appearing the supergravity equations to solve, and discuss how to handle this data at best in view of algebra identification and compactness conditions.\\

One way to introduce group manifolds is through the formalism of vielbeins, that we first recall. Consider a manifold ${\cal M}$ admitting a metric $g$, expressed locally in terms of coordinates $y^m$ and the 1-forms $\d y^m$, as $\d s^2 = g_{mn} (y)\, \d y^m \d y^n$. Locally, one can introduce vielbeins $e$ of coefficient $e^a{}_m (y)$, where the range of both indices is the same and given by the dimension of the manifold. Inverse vielbeins $e^{-1}$ have for coefficient $e^m{}_a$. The vielbeins are such that $g_{mn} = e^a{}_m\, \eta_{ab}\, e^b{}_n$, where $\eta$ is the local identity metric, namely $\eta_{ab}$ has only $\pm 1$ diagonal entries, depending on the signature; for us, with a $6$-dimensional Euclidian manifold ${\cal M}$, it will be $\delta_{ab}$. Vielbeins allow to define a local basis of 1-forms $e^a=e^a{}_m \d y^m$, a.k.a.~coframes, at a point $P$ on the cotangent space $T_P^* {\cal M}$, and through their inverse, of vectors $\del_a = e^m{}_a \del_m$ on the tangent space $T_P {\cal M}$. More generally, tensors can be expressed with one kind of indices or the other, $a$ or $m$, by simple multiplication of vielbein coefficients or their inverse: for example, $\nabla_a= e^m{}_a \nabla_m,\ F_{1\,a}=e^m{}_a F_{1\,m}$, etc.

Having introduced vielbeins, one can always consider the quantity $f^a{}_{bc}$, a priori coordinate dependent and locally defined, as $f^a{}_{bc}= -2 e^m{}_{[c} \del_{b]} e^a{}_m$. This is equivalently defined from the 1-forms or the vectors as
\beq
\d e^a = -\frac12 f^a{}_{bc} \, e^b \w e^c \ , \qquad 2\,\del_{[b} \del_{c]} = f^a{}_{bc} \, \del_a \ . \label{MCbracket}
\eeq
The quantity $f^a{}_{bc}$ is interesting as we now explain. Let us first consider the spin connection coefficient $\omega^a_{bc}$ defined as $e^n{}_b e^a{}_m \nabla_n V^m = \nabla_b V^a = \del_b V^a + \omega^a_{bc}\, V^c$. When choosing the Levi-Civita connection, then the spin connection, and thus the Riemann tensor, the Ricci tensor and scalar, can be expressed in terms of $f^a{}_{bc}$ only. For instance, one gets the following relations
\bea
&& \omega^a_{bc} = \frac{1}{2} \left(f^a{}_{bc} + \eta^{ad}\eta_{ce} f^e{}_{db} + \eta^{ad}\eta_{be} f^e{}_{dc}  \right) \ ,\quad  f^a{}_{bc} = 2 \omega^a_{[bc]}\ ,\\
&& {\cal R} = 2 \eta^{ab} \del_a f^c{}_{bc} - \eta^{cd} f^a{}_{ac} f^b{}_{bd} - \frac{1}{4} \left( 2 \eta^{cd} f^a{}_{bc} f^b{}_{ad} + \eta_{ad} \eta^{be} \eta^{cg} f^a{}_{bc} f^d{}_{eg} \right) \ , \label{RNotconstant}\\
&& {\cal R}^a{}_{[bcd]} = \del_{[b} f^a{}_{cd]} + f^e{}_{[cd} f^a{}_{b]e} \ , \label{BIRiemannNotconstant}
\eea
and we refer e.g.~to \cite[Sec.1, App.A]{Andriot:2013xca} for more details, while \eqref{BIRiemannNotconstant} is discussed in \cite[(3.5)]{Andriot:2014uda}. All geometric quantities, relevant to our purposes, can thus be expressed solely in terms of $f^a{}_{bc}$, which will be important in the following.\\

As we will see, for group manifolds, the $f^a{}_{bc}$ are constant, which makes them special. A first hint of this specificity can be found in \eqref{MCbracket}: we see that the antisymmetrisation (or commutator) of the vectors can be understood as a Lie bracket, provided the $f^a{}_{bc}$ appearing there is constant. In that case, $f^a{}_{bc}$ would correspond to the structure constant of a Lie algebra $\mathfrak{g}$. To be sure of having a Lie algebra, the Jacobi identity on $f^a{}_{bc}$ should also be satisfied. The latter turns out to be given by the Riemann Bianchi identity \eqref{BIRiemann}, ${\cal R}^a{}_{[bcd]}=0$, once the $f^a{}_{bc}$ are constant, as can be seen from \eqref{BIRiemannNotconstant}. In short, having constant $f^a{}_{bc}$ make those correspond to structure constants of a Lie algebra $\mathfrak{g}$, verifying the standard relations
\beq
\left[ \del_b , \del_c \right]= f^a{}_{bc}\, \del_a \ , \qquad f^e{}_{[cd} f^a{}_{b]e} = 0 \ . \label{algebravectors}
\eeq
Note that the $f^a{}_{bc}$ introduced above are real, so we will restrict to real Lie algebras. All this holds locally so far, and at a given point $P$ of ${\cal M}$, we read from the vectors in \eqref{algebravectors} the identification of the tangent space $T_P {\cal M} \simeq \mathfrak{g}$, where the algebra is given by the local constants $f^a{}_{bc}$. But one may want to have the same constants everywhere on the manifold ${\cal M}$, i.e.~have all the introduced quantities to be globally defined. This is possible if the manifold is a Lie group $G$ (we recall that Lie groups are differentiable manifolds admitting metrics), ${\cal M} \simeq G$, where $G$ is obtained by the exponential map from a certain $\mathfrak{g}$. Indeed, it is known that at the identity point $\mathbb{1}$ of $G$, one has $T_{\mathbb{1}} G \simeq \mathfrak{g}$, matching with $T_{\mathbb{1}} {\cal M}$. The exponential map and the left invariance of the group elements then allows to move to any point $P$ of $G$, and to still have $T_P G \simeq \mathfrak{g}$. Therefore the basis of vectors is globally defined, as well as the $f^a{}_{bc}$. Similarly, the 1-forms $e^a$ are globally defined, and at each point $P$, one has $T_P^* G \simeq \mathfrak{g}^*$; the first equation in \eqref{MCbracket} is then known as {\sl the Maurer-Cartan equation}, which is key on group manifolds. In short, {\sl we will consider ${\cal M}$ given by a Lie group $G$. Its $f^a{}_{bc}$, related to the spin connection coefficients, are the structure constants of its underlying (real) Lie algebra $\mathfrak{g}$.} As stressed, all relevant geometric quantities, namely the local ones appearing in equations, will be expressed only in terms of these structure constants, and will be constant themselves.

To be more precise, we should be talking of isomorphism classes of Lie algebras rather of one Lie algebra $\mathfrak{g}$, but we will omit this distinction in the following for simplicity. Related to this point, we will consider the unique Lie group $G$ obtained by exponential map which is connected and simply-connected. The group $G$ considered here is then in one-to-one correspondence with the algebra. This $G$ is also the universal cover of the possible groups obtained by exponential map from the (class) $\mathfrak{g}$. Briefly, let us just recall the example $\mathfrak{so}(3) \simeq \mathfrak{su}(2)$ and $SO(3)=SU(2)/\mathbb{Z}_2$. Note that having at first the universal cover is not so much a restriction, because further quotients by discrete subgroups will be considered, either as a lattice or with orientifolds, etc. More on algebraic, and geometric, aspects of $\mathfrak{g}$ and $G$ can be found in \cite{Andriot:2010ju}.

Finally, we will allow for a discrete subgroup $\Gamma$, in such a way that what is eventually considered is ${\cal M} \simeq G/\Gamma$. $\Gamma$ plays a role at the global level: its discrete identifications allow to make ${\cal M}$ compact when $G$ is not, in which case $\Gamma$ is known as a lattice. This extra lattice action does not change the $f^a{}_{bc}$ and the previous (local) expressions, but for the $e^a$ to be globally defined (as they should), they need to be invariant under the lattice. Then, the resulting group manifold ${\cal M} \simeq G/\Gamma$ is parallelizable, since each basis direction is globally defined. A $6$-dimensional flat torus is a first example of such a group manifold: $T^6 \simeq U(1)^6 \simeq (\mathbb{R}/\mathbb{Z})^6$. There, $f^a{}_{bc}=0$, i.e.~the algebra is abelian: it is $6\, \mathfrak{u}(1)$. {\sl Specifying the Lie algebra $\mathfrak{g}$ and the lattice $\Gamma$ is equivalent to define the group manifold ${\cal M}$.} As we will discuss, not every group $G$ admits a lattice. Therefore specifying a real Lie algebra $\mathfrak{g}$ is not sufficient to our purposes, because the compactness of ${\cal M}$ is not guaranteed, and still has to be checked. Explicit examples of group manifolds and their lattice are discussed in Appendix \ref{ap:groupman}.\\

Having motivated and introduced group manifolds, let us now be more concrete and discuss the Lie algebras $\mathfrak{g}$ that will be encountered. Focusing on $d=4$, we will be interested in $6$-dimensional ${\cal M}$, therefore on $6$-dimensional real Lie algebras. An account on those can be found in \cite[Sec.2.1, 2.2]{Andriot:2022yyj}, and we review a few elements here. We start by Levi's decomposition, which states that any (real, finite dimensional) Lie algebra $\mathfrak{g}$ is a semi-direct sum of a semi-simple algebra $\mathfrak{t}$ (appearing e.g.~in gauge groups of particle physics) and of the largest solvable ideal $\mathfrak{s}$ (a.k.a.~the radical of $\mathfrak{g}$), meaning
\beq
\mathfrak{g} = \mathfrak{t} \niplus \mathfrak{s} \ .
\eeq
A Lie algebra is then either solvable (it is then its own radical), or it is not; most of the focus will be on the former. A particular class of solvable algebras are nilpotent ones: one has the following inclusions
\beq
{\rm abelian} \ \subset \ {\rm nilpotent} \ \subset \ {\rm solvable} \ \subset \ {\rm Lie} \ .
\eeq
The precise definitions of nilpotent and solvable have to do with series of ideals and can be found e.g.~in \cite[Sec.2.1]{Andriot:2022yyj}. Beyond abelian algebras $\mathfrak{u}(1) \oplus \dots \oplus \mathfrak{u}(1)$, the real Lie algebras of lowest dimension can first be found in dimension 3: we give those below for illustration, in terms of their non-zero structure constants in some basis
\begin{gather}
\label{alg3d}
\begin{aligned}
{\rm Heis}_3 \ \ \text{(nilpotent)}:\quad & f^1{}_{23} = 1  \\
\mathfrak{g}_{3.5}^0 = \mathfrak{iso}(2) \ \ \text{(solvable)}:\quad & f^1{}_{23} = 1 \ ,\ f^2{}_{13} = -1  \\
\mathfrak{g}_{3.4}^{-1} = \mathfrak{iso}(1,1) \ \ \text{(solvable)}:\quad & f^1{}_{23} = 1 \ ,\ f^2{}_{13} = 1 \ \Leftrightarrow \  f^1{}_{13} = 1 \ ,\ f^2{}_{23} = -1 \\
\mathfrak{so}(3) = \mathfrak{su}(2) \ \ \text{(simple)}:\quad & f^1{}_{23} = 1 \ ,\ f^2{}_{31} = 1 \ ,\ f^3{}_{12} = 1  \\
\mathfrak{so}(2,1) = \mathfrak{sl}(2,\mathbb{R}) \ \ \text{(simple)}:\quad & f^1{}_{23} = 1 \ ,\ f^2{}_{31} = 1 \ ,\ f^3{}_{12} = -1  \\
\end{aligned}
\end{gather} 
where the ``solvable'' examples are non-nilpotent.

Another point, important in the following, is unimodularity (also known as unipotence) of the Lie algebra: this condition is given by
\beq
\forall b\ , \quad \sum_a f^a{}_{ab}= 0 \ . \label{unimodularity}
\eeq
All semi-simple and nilpotent algebras are unimodular. This condition is necessary to ensure compactness of the group manifold ${\cal M}$: more precisely, for the connected and simply-connected $G$ to admit a lattice, $\mathfrak{g}$ must be unimodular. Indeed, considering the exterior derivative of a globally defined 5-form, $\d (e^{a_1}\w \dots \w e^{a_5})$, where $a_1,\dots,a_5 \neq b$, one has $\d (e^{a_1}\w \dots \w e^{a_5}) \propto {\rm vol}_6\, \sum_a f^a{}_{ab}$, and $\int_{{\cal M}} \d (e^{a_1}\w \dots \w e^{a_5})=0$ for a compact manifold (without boundary), hence the condition \eqref{unimodularity}. In the following, {\sl we are thus interested in $6$-dimensional real unimodular Lie algebras}. 

Focusing on such algebras, let us first discuss those which are non-solvable: there exists only 16 of them. They can be found, with some of their properties, in \cite[Tab.1]{Andriot:2022yyj}. Among those, only 6 actually allow for a compact ${\cal M}$; the 10 others contain $\mathfrak{so}(2,1)$ or $\mathfrak{so}(3,1)$, preventing the corresponding group to admit a lattice. Among the previous 6, the only semi-simple algebra is $\mathfrak{so}(3) \oplus \mathfrak{so}(3)$. It gives the compact group manifold $SU(2) \times SU(2)$, on which the first type IIA supergravity de Sitter solution was found \cite{Caviezel:2008tf}.

We now turn to $6$-dimensional (real unimodular) solvable Lie algebras. Since the first non-abelian solvable algebras appear in dimension 3, 6-dimensional solvable algebras, decomposable or not, must be of the following form
\beq
\text{6d solvable:}\qquad  6\, \mathfrak{u}(1) \ ,\ \mathfrak{s}_3 \oplus 3\, \mathfrak{u}(1) \ ,\ \mathfrak{s}_3 \oplus \tilde{\mathfrak{s}}_3\ ,\ \mathfrak{s}_4 \oplus 2\, \mathfrak{u}(1)\ ,\ \mathfrak{s}_5 \oplus \mathfrak{u}(1)\ ,\ \mathfrak{s}_6 \ ,\label{solvable}
\eeq
where $2\, \mathfrak{u}(1)$ stands for $\mathfrak{u}(1) \oplus \mathfrak{u}(1)$, etc., and $\mathfrak{s}_n$ stands for an $n$-dimensional indecomposable solvable algebra. There is 1 algebra $6\, \mathfrak{u}(1)$, 3 algebras $\mathfrak{s}_3 \oplus 3\, \mathfrak{u}(1) $ and 6 algebras $\mathfrak{s}_3 \oplus \tilde{\mathfrak{s}}_3$. Details and properties of these first 10 solvable algebras, built from 3-dimensional ones, can be found in \cite[Tab.2]{Andriot:2022yyj}. They all provide a compact ${\cal M}$.

The amount of the other (real) solvable algebras in \eqref{solvable}, based on $\mathfrak{s}_4,\, \mathfrak{s}_5,\, \mathfrak{s}_6$, is more delicate: for those, we follow \cite[App.A]{Bock:2009}. We find that there are 5 real $\mathfrak{s}_4$ that give a $G$ admitting a lattice, which is more restrictive than being unimodular; 1 of them is nilpotent. Turning to $\mathfrak{s}_5$, it becomes important to mention the following point: solvable algebras are typically classified by their nilradical $\mathfrak{n}$, which is the (unique) maximal nilpotent ideal of $\mathfrak{g}$. One has that ${\rm dim}\, \mathfrak{n} \geq \tfrac12\, {\rm dim}\, \mathfrak{g}$ \cite{Mubarakzyanov:1963}. Overall, we read from \cite[App.A]{Bock:2009} that there are 39 indecomposable solvable 5-dimensional real Lie algebras $\mathfrak{s}_5$, 6 of them being nilpotent. Among the 33 non-nilpotent ones, 19 are unimodular. We finally turn to $\mathfrak{s}_6$, 6-dimensional indecomposable real solvable algebras. We read from \cite[App.A]{Bock:2009} that there are 24 nilpotent ones (${\rm dim}\, \mathfrak{n}=6$) \cite{Morozov:1958}, 100 with ${\rm dim}\, \mathfrak{n}=5$ (99 in \cite{Mubarakzyanov:1963bis}, 1 in \cite{Turkowski:1990}) and 40 with ${\rm dim}\, \mathfrak{n}=4$ \cite{Turkowski:1990},\footnote{The only 2 real 6-dimensional solvable algebras with ${\rm dim}\, \mathfrak{n}=3$ are decomposable.} giving a total of 164 $\mathfrak{s}_6$. From \cite[App.A]{Bock:2009}, we count 60 with ${\rm dim}\, \mathfrak{n}=5$ and 13 with ${\rm dim}\, \mathfrak{n}=4$ that are unimodular, and even less admit a lattice. For completeness, let us report that later work \cite{Shabanskaya:2011, Shabanskaya:2013} revisited the old work \cite{Mubarakzyanov:1963bis} on the case ${\rm dim}\, \mathfrak{n}=5$, proposing various corrections and additions. Similarly, possible restrictions on continuous parameters appearing in some algebras were discussed in \cite[App.A]{Andriot:2022yyj}. A more complete and recent list of solvable algebras can also be found in \cite{Snobl:2014}. Precise amounts, and details of some algebras, may thus vary a little.

Overall, using above counts, this provides a maximum of $6+10+5+25+97 =143$ 6-dimensional real Lie algebras that stand a chance to provide a compact group manifold ${\cal M}$, either because we know they allow for a lattice or because they are unimodular. While this is not comparable to the typical Calabi-Yau datasets, it is still a large enough number to look and find a diversity of (de Sitter) solutions of 10d type II supergravities. 

We recall that choosing or identifying a Lie algebra among those above is not sufficient to our purposes, since we need to make sure the resulting group manifold is compact. When a nilpotent, resp.~solvable group admits a lattice, namely it can give a compact ${\cal M}$, this group manifold is called a nilmanifold, resp.~solvmanifold. This is actually always possible from a nilpotent algebra, using a theorem from \cite{Malcev:1951} (see below). But this is not systematic, and also not fully determined, for solvable ones, which complicates our studies. For more detail on compactness and the existence of a lattice for solvable groups, we refer to \cite{Bock:2009, Andriot:2010ju, Grana:2013ila}.\\

Having detailed the Lie algebras of interest to our group manifolds, let us now discuss how this information is encoded and treated in a solution. When looking for supergravity solutions on a group manifold, we solve (local) equations, where the geometric data only appears through the structure constants. For example, in the Einstein equation appears the Ricci tensor and scalar, given on a group manifold (with unimodular algebra)\footnote{The more general expression in terms of (non-constant, non unimodular) $f^a{}_{bc}$ can be found in \cite[(2.20)]{Andriot:2014uda}.} by
\begin{gather}
\label{Riccitensor}
\begin{aligned}
2\, {\cal R}_{cd} & = - f^b{}_{ac} f^a{}_{bd} - \eta^{bg} \eta_{ah} f^h{}_{gc} f^a{}_{bd} + \frac{1}{2} \eta^{ah} \eta^{bj} \eta_{ci} \eta_{dg} f^i{}_{aj} f^g{}_{hb} \ ,\\
2\, {\cal R} & = - \eta^{cd} f^b{}_{ac} f^a{}_{bd} - \frac{1}{2} \eta^{ab} \eta^{eh} \eta_{cd} f^c{}_{ae} f^d{}_{bh} \ ,
\end{aligned}
\end{gather}
There are then two strategies: a first one is to fix the algebra and the basis, thus the structure constants, and look for solutions with those. This was pursued e.g.~in \cite{Grana:2006kf, Danielsson:2011au}. While this method has advantages as we will see, the main drawback is that this initial fixing could be restrictive. A second strategy is to leave the $f^a{}_{bc}$ free as variables, allowing a complete freedom on the algebra and the basis, except for them to respect the Jacobi identities and orientifold projection conditions. This enhances the chances to find solutions.\footnote{One way to understand this point is as follows: starting in a basis of $e^a$ with many possibly non-zero $f^a{}_{bc}$ and the metric $\eta_{ab}$, one can change basis to one for ${e^a}'$ generating less ${f^a{}_{bc}}'$, e.g.~corresponding to a simple presentation of the algebra, as given in known tables. The cost is then that the change of basis would generate a new metric $g_{ab}$ with more entries, in particular non-diagonal ones, capturing the rest of the initial information. There is conservation of the number of variables, or degrees of freedom. In the first strategy where one fixes the basis, one typically has few $f^a{}_{bc}$, and the metric $\eta_{ab}$, therefore not allowing for many free variables. This reduces the space where a solution could be found.} Having obtained one, we get a concrete set of structure constants, as e.g.~in \cite{Andriot:2022way}. The drawback is that we are then left to identify the algebra, with a priori no guarantee that the corresponding group will admit a lattice, i.e.~that the manifold is compact. To solve this problem, a method and tools have been developed in \cite{Andriot:2022yyj}.

Given a concrete set of structure constants (obeying the Jacobi identities), identifying the corresponding Lie algebra means find it among the algebras detailed above. Those are listed in tables in e.g.~\cite{Bock:2009, Snobl:2014} in a certain basis, which is most likely not the same as the one in which we have our solution. As explained, $\mathfrak{g}$ is rather an isomorphism class of algebras, meaning in particular that a change of basis can map one set of structure constants to another, equivalent one. An example of this is given for $\mathfrak{g}_{3.4}^{-1}$ in \eqref{alg3d}, and some more involved analytical changes of basis are presented in \cite[App.C]{Andriot:2020wpp} and \cite{Andriot:2022yyj}. This mathematical problem of identifying an algebra from its structure constants given in an arbitrary basis is well-known. Building on \cite{Snobl:2014}, we discussed it in \cite[Sec.2.3]{Andriot:2022yyj} and provided numerical tools to handle it in our case: essentially, one makes use of various (basis) invariants of the algebra. Eventually this allowed us to identify all algebras of solutions obtained in this way.

A further difficulty has to do with the lattice action and compactness. If (by chance) the existence of a lattice is established for a given $G$, this is typically done in one specific basis, where one can write explicitly the globally defined 1-forms $e^a$. In that basis, one infers quantization constraints on the structure constants, that we will come back to. Most likely, a change of basis from the one where we found the solution is necessary, which adds some complication (see in particular \cite[App.A]{Andriot:2022way} on preserving sources volume forms). This will play a role for the classicality study, discussed in Section \ref{sec:dSsolclassical}.

Last but not least, let us underline that {\sl structure constants are not necessarily integer}, contrary to a common belief. The confusion might be due to a standard result valid for semi-simple Lie algebras: for those, there exist the Chevalley basis where structure constants values are integers. In physics, one often uses these algebras in such basis, together with normalised generators, which prevents from rescaling quantities away from this situation. A rescaling of the generators, here the vectors $\del_a$ or even the 1-forms $e^a$, usually changing the metric, can however change the structure constant values. In addition, it is important to note that {\sl there exists no analogue to the Chevalley basis for solvable algebras, which therefore do not necessarily have integer structure constants}. Then, not only having integer structure constants depends on the choice of basis (involving possible rescalings), but it is not even guaranteed for solvable algebras. For some nilpotent algebras though, such as Heis${}_3$, the structure constants can be brought to an integer; it was also proven \cite{Malcev:1951} that structure constants of nilpotent algebras can always be made rational in a certain basis, which in turns guarantees the existence of a lattice. We see here an interesting relation between the existence of a lattice and quantization (meaning discretization) of structure constants: this is actually the closest we have to an integer condition for general solvable algebras. Indeed, in concrete examples, requiring the existence of a lattice, i.e.~discrete identifications which ensure compactness, sets some quantization constraints on structure constants. We give explicit examples of group manifolds and their lattice quantization conditions in Appendix \ref{ap:groupman}. The solution on a solvmanifold, to be discussed in Section \ref{sec:dSsolclassical}, will only require the product of some structure constants to be related to an integer, but not the individual $f^a{}_{bc}$.

This ends our presentation of group manifolds, ${\cal M} \simeq G/\Gamma$, that will be used as compactification manifolds in our supergravity de Sitter solutions, as we now present.

\subsubsection{Solution ansatz, gauged supergravity and dimensional reduction}\label{sec:groupmandimred}

We first discuss in this section the 10d ansatz for de Sitter solutions, and summarize the mathematical problem corresponding to find some. We then explain that the former is also a compactification ansatz, giving through a consistent truncation a 4d effective theory, that can be used as well to find solutions. Such 4d theories are often identified as gauged supergravities. We finally recall that those are not necessarily low energy effective theories, but they may still be used to study perturbative (in)stability of a solution.\\

We focus on solutions of 10d type IIA/B supergravities on a 4d de Sitter spacetime, times a 6d compact group manifold ${\cal M}$. As motivated at the beginning of Section \ref{sec:dSsol}, we include $D_p/O_p$ sources, and no further ingredient. Orientifolds are then mandatory due to Maldacena-Nu\~nez no-go theorem \ref{nogo1}. Furthermore, we follow the ansatz described in Section \ref{sec:10dansatz}, considering smeared sources, with constant warp factor and dilaton. We use notations of that section. In particular, the equations to solve are listed there: these are \eqref{Einstein6d} - \eqref{BI}, together with the Riemann BI, while the orientifold involution also needs to be taken into account.

The restriction to ${\cal M}$ being a group manifold, introduced and reviewed in Section \ref{sec:groupman}, brings further notations and simplifications. We use the vielbein formalism introduced there, with which the 6d metric gets written $g_{mn} = e^a{}_m\, \delta_{ab}\, e^b{}_n$, and one gets on $T^* {\cal M}$ the basis of globally defined 1-forms $e^a = e^a{}_m\,\d y^m$. All forms encountered will then be expanded on that basis: this means that the supergravity fluxes $H,\, F_p$ will have the corresponding components $H_{abc},\, F_{q\, a_1 \dots a_q}$, and the volume forms will also be expressed in terms of $e^a$. In addition, the $e^a$, acted upon by the exterior derivative $\d$, obey the Maurer-Cartan equation \eqref{MCbracket} that we repeat here for convenience
\beq
\d e^a = -\frac12 f^a{}_{bc} \, e^b \w e^c \ , \label{MC}
\eeq
where the $f^a{}_{bc}$ are the structure constants of the corresponding Lie algebra $\mathfrak{g}$. This allows to express the left-hand side of the supergravity form equations, namely the flux e.o.m. \eqref{eomfluxes} and the BI \eqref{BI}, only in terms of the structure constants, the flux components and their derivatives. For example, the $H$-flux BI, $\d H=0$, becomes
\beq
\del_{[a}H_{bcd]} - \frac{3}{2} f^e{}_{[ab} H_{cd]e} = 0 \ .
\eeq
Similarly, tensors components get expressed with indices $a$. The ones not mentioned so far, entering the supergravity equations, are the 6d Ricci tensor ${\cal R}_{ab}$ and the 6d energy momentum tensor for sources $T_{ab}$. The former was given in \eqref{Riccitensor} for a group manifold with unimodular algebra (which will the case here): this ${\cal R}_{ab}$ is only expressed in terms of $f^a{}_{bc}$. The same holds for the Ricci scalar ${\cal R}_6$ \eqref{Riccitensor}, appearing in equations. Finally $T_{ab}$ only depends on $T_{10}^s$ for each source $s$ as can be seen from \eqref{Tmn}: we get
\beq
T_{ab} =  \sum_{{\rm sources}} \delta_a^{a_{||}} \delta_b^{b_{||}}\, \delta_{a_{||} b_{||}}\, \frac{T_{10}^s}{p+1} \ ,\label{Tab}
\eeq
where the notation of parallel directions, inferred from the generic notations of \eqref{Tmn}, is detailed below; see also \cite{Andriot:2017jhf} for a derivation of \eqref{Tab}. We conclude that the only variables entering the supergravity equations to solve, \eqref{Einstein6d} - \eqref{BI}, are
\beq
{\rm Variables:}\qquad {\cal R}_4 \, ,\ f^a{}_{bc}\, ,\ H_{abc}\, ,\ g_s\, F_{q\, a_1 \dots a_q} \, ,\ g_s\, T_{10}^s \ , \label{variables}
\eeq
together with derivative of flux components, and transverse volume forms that still need to be specified. In addition, the orientifold involution and the Riemann BI \eqref{BIRiemann} should be taken into account. We will come back to the former, while the Riemann BI on de Sitter times a group manifold becomes the Jacobi identity on the structure constants \eqref{algebravectors}, obeyed given a Lie algebra $\mathfrak{g}$.\\

So far, we only rewrote conveniently the problem of finding a solution of interest, using a formalism adapted to group manifolds, without going beyond the ansatz of Section \ref{sec:10dansatz}. We now consider two further ansatz restrictions. First, we work in a basis where
\beq
\forall b\ ,\quad f^a{}_{ab}=0 \quad \text{without sum on}\ a \ . \label{faab=0}
\eeq
This is typically an appropriate basis to have globally defined 1-forms, and more easily identify the lattice action \cite{Andriot:2010ju} (see also Section \ref{sec:groupman}). In addition, the algebra is then automatically unimodular \eqref{unimodularity}, as required for compactness of ${\cal M}$. In turn, the restriction \eqref{faab=0} may exclude certain Lie algebras.

The second restriction in our ansatz is a crucial one: we consider that the flux components $H_{abc}\, ,\ F_{q\, a_1 \dots a_q}$ are constant. Any change of basis to be performed on the $e^a$ will also be constant, thus preserving this property. Given that any other variable entering the equations is constant, due to the group manifold and the smeared approximation, this is a natural working assumption. It has several major benefits. First, the flux e.o.m.~and BI simplify. All equations to solve, namely \eqref{Einstein6d} - \eqref{BI} and the Jacobi identities, then become algebraic, forming a system of multivariate quadratic polynomial equations. All their variables, listed in \eqref{variables}, are constant. Second, the orientifold involution then imposes constraints, the orientifold projection conditions, as described in Appendix \ref{ap:Op}: they set to zero some of the variables, an important simplification, that we will come back to.

To summarize, we look for solutions of type IIA/B supergravities with $D_p/O_p$ on a 10d spacetime which is the direct product dS${}_4 \, \times \, {\cal M}$, a 4d de Sitter spacetime times a 6d group manifold. Using the ansatz of Section \ref{sec:10dansatz} where sources are smeared, together with two further restrictions (\eqref{faab=0} and constant flux components), the problem simplifies. {\sl It amounts to solving a system of quadratic polynomial equations, obtained from \eqref{Einstein6d} - \eqref{BI} and the Jacobi identities of a Lie algebra $\mathfrak{g}$. The variables are constant and listed in \eqref{variables}.} As discussed in Section \ref{sec:groupman}, the compactness of ${\cal M}$ will remain to be checked. We are also left to discuss the orientifold involution, that we now turn to.\\

The ansatz of Section \ref{sec:10dansatz} specified that the set of directions parallel or transverse to the $D_p/O_p$ should be globally identified in the 10d spacetime, allowing to define the projection to these directions. Given the (external) space-filling assumption for the sources, and the focus on group manifolds, the global identification of directions boils down to be done among the globally defined directions of ${\cal M}$. In other words, for each source, the parallel and transverse directions appear globally as follows in the 6d metric
\beq
\d s^2_6 = \delta_{ab}\, e^a e^b = \delta_{a_{||}b_{||}}\, e^{a_{||}} e^{b_{||}} + \delta_{a_{\bot}b_{\bot}}\, e^{a_{\bot}} e^{b_{\bot}}  \ . \label{metricparbot}
\eeq
Note that no assumption is made on the coordinate dependence of the 1-forms, so this split is not necessarily a direct product. As an example, consider an $O_6$ along internal dimensions $a=1,2,3$, thus transverse to $a=4,5,6$. With the identity metric, this allows to define the volume forms discussed in \eqref{volumeforms}. One has ${\rm vol}_6 = e^1 \w e^2 \w e^3 \w e^4 \w e^5 \w e^6$, also denoted $e^{123456}$, where the ordering reflects the orientation. In the $O_6$ example, one has ${\rm vol}_{||_s} = e^{123}$, where by convention, we always choose the natural ordering, and as a consequence of \eqref{volumeforms}, ${\rm vol}_{\bot_s} = e^{456}$; the orientation may give the transverse volume form a sign in some cases. Since these volume forms do not include further factors, they disappear from the flux BI \eqref{BI} when projecting on form components. So they do not bring any further variable.

In general, we work in a 6d basis adequate to the parallel and transverse directions to sources. For example, flux components can be written as in \eqref{F(n)}, meaning here on the group manifold
\beq
F^{(n)}_{q\, a_{1||} \dots a_{n||} a_{n+1 \bot} \dots a_{q \bot}} \ .\label{fluxcompo}
\eeq
Since the field components are constant, the orientifold involution becomes constraining orientifold projection conditions, as explained in Appendix \ref{ap:Op}. Then, only some of the flux components \eqref{fluxcompo} survive the orientifold projection conditions: those specified in \eqref{Opflux}. Looking at the other variables \eqref{variables}, we also see constraints on the structure constants: only the following ones survive an orientifold projection
\beq
f^{a_{||}}{}_{b_{||}c_{||}} \, , \ f^{a_{||}}{}_{b_{\bot}c_{\bot}} \, ,\ f^{a_{\bot}}{}_{b_{\bot}c_{||}} \ , \label{Opfabc}
\eeq
as can be seen from preserving the Maurer-Cartan equation \eqref{MC} with globally defined $e^a$. In short, the orientifold projection conditions select some of the variables listed in \eqref{variables}.

Finally, note that some of the $D_p/O_p$ sources can be parallel to each other, while being optionally located at different points in their transverse space. This is here identified by the fact they have the same parallel and transverse directions, or the same volume forms. Such parallel sources are said to be in the same set, labeled by $I$. If there is more than one set $I=1$, then the sources can be said to be intersecting: set $I=1$ intersects set $I=2$, etc. Because the volume forms in one set are the same, ${\rm vol}_{||_s},\, {\rm vol}_{\bot_s}$ can be traded for ${\rm vol}_{||_I},\, {\rm vol}_{\bot_I}$. In equations, only the combined contributions appear, namely $T_{10}^I = \sum_{s \in I} T_{10}^s$. These variables include the number of (parallel smeared) sources in the set $I$, $N_{D_p/O_p}^I$, times their charges, making them proportional to $2^{p-5}N_{O_p}^I-N_{D_p}^I$. A localized version, not considered here, would trade the number of sources for the sum of various $\delta$-function localizing them.

{\sl In short, the problem of finding a de Sitter solution of type IIA/B supergravities becomes, with this ansatz, mathematically well-defined, as summarized above: it amounts to solve a system of quadratic polynomial equations, whose constant variables are given in the list \eqref{variables}, simplified by the orientifold projection conditions.} The latter select some flux components \eqref{Opflux}, some structure constants \eqref{Opfabc}, and gathers parallel source contributions into those of sets $g_s\, T_{10}^I$. Once a solution to this finite list of equations is found,\footnote{The system of equations is obtained from \eqref{Einstein6d}-\eqref{BI} and the Jacobi identities. Our ansatz together with the orientifold projection however simplify the system, since the list of variables gets reduced. Some of the flux BI are usually automatically satisfied. In addition, as shown in \cite[Sec.3.2]{Andriot:2019wrs} for sources of single dimensionality $p=4,5,6$, the off-diagonal internal Einstein equations under an orientifold projection, namely components ${}_{a_{||_I}b_{\bot_I}}$ for a set $I$ with an $O_p$, identically vanish and are thus trivially satisfied.} we recall that compactness of ${\cal M}$ needs to be checked, a point discussed in Section \ref{sec:groupman}.

The mathematical problem essentially depends on the orientifolds present, because of their various projection conditions. This led to a classification of possible solutions, proposed in \cite{Andriot:2022way, Andriot:2022yyj}, depending on $O_p$ and $D_p$, that we will discuss and exemplify in Section \ref{sec:dSsolexistence}. In those works, numerical tools were developed, and made available, to provide the system of polynomial equations for each class of solutions, namely for each possible $D_p/O_p$ source configuration, and solve it.\\

There exists another way to get to the same system of equations and find solutions, at least when considering only equations of motion and leaving aside Bianchi identities: a specific dimensional reduction to a 4d effective theory. Dimensional reductions were introduced and discussed in Section \ref{sec:dimred}. With the example of \eqref{equiv}, we saw how a 4d effective theory of scalar fields with a scalar potential $V$ could, by extremizing $V$, give the same equations as some 10d equations of motion. To achieve this, an important point was that the 10d solution ansatz equally served as the compactification ansatz to perform the dimensional reduction. Here, the same can be achieved with a compactification on a group manifold,  considering precisely the ansatz described above for the 10d solution, meaning constant background supergravity fields on the $e^a$ basis, selected by orientifold projection, etc. Given background fields, one should also select a finite set of fluctuations, a truncation, which eventually correspond to 4d scalar fields, as explained in Section \ref{sec:dimred}. For each class of solutions of \cite{Andriot:2022way, Andriot:2022yyj}, this was systematically done in \cite{Andriot:2022bnb}, considering left-invariant fields (i.e.~fluctuations expanded on a basis of $e^a$ with 6d-independent coefficients), leading to a 4d effective theory.\footnote{This dimensional reduction carries some subtleties discussed in \cite{Andriot:2022bnb}. It requires to adapt the 10d supergravity Chern-Simons terms and the definition of the RR fluxes, to incorporate properly the presence of background fluxes. The resulting 4d scalar potentials \eqref{potIIAB}, including especially supergravity gauge potentials corresponding to 4d axions, are non-trivial and may be of broader interest.} As an illustration, we give the resulting scalar potential, where one can see most of the 4d axions contributions; not all field dependence is made explicit here, and we refer to \cite{Andriot:2022bnb} for more details on notations.
\bea
V_{{\rm IIA}} = \frac{M_p^2}{2} \frac{e^{2\phi}}{vol_6}\Bigg[ && \hspace{-0.22in} - {\cal R}_6 + \frac12 |H|^2 - e^{\phi} \sum_{p=4,6,8} \frac{T_{10}^{(p)}}{p+1} \nn\\
&& \hspace{-0.22in} + \frac{e^{2\phi}}{2} \Big( F_0^2 + |F_2 + F_0 B|^2 + |F_4 + C_1\w H + F_2 \w B + \frac12 F_0\, B\w B|^2 \nn\\
&& \hspace{-0.22in} + |F_6 + C_3 \w H + F_4 \w B + C_1 \w H \w B + \frac12 F_2 \w B \w B + \frac16 F_0\, B\w B\w B|^2 \Big) \Bigg] \nn\\
V_{{\rm IIB}} = \frac{M_p^2}{2} \frac{e^{2\phi}}{vol_6}\Bigg[ && \hspace{-0.22in} - {\cal R}_6 + \frac12 |H|^2 - e^{\phi} \sum_{p=3,5,7,9} \frac{T_{10}^{(p)}}{p+1} \label{potIIAB}\\
&& \hspace{-0.22in} + \frac{e^{2\phi}}{2} \Big( |F_1|^2 + |F_3 - C_0\w H + F_1 \w B |^2 \nn\\
&& \hspace{-0.22in} + |F_5 - C_2 \w H + F_3 \w B - C_0 \w H \w B + \frac12 F_1 \w B \w B |^2 \Big) \Bigg] \nn
\eea
It was then showed that for each solution class, the equations obtained by extremizing the scalar potential are in one-to-one correspondence with the 10d equations of motion \cite[Sec.4]{Andriot:2022bnb}. The truncation picked then qualifies as a {\sl consistent truncation} (see Section \ref{sec:dimred}), for 4d extrema solutions. This has the consequence that {\sl solutions can be looked for using the 4d theory instead}, keeping in mind that Bianchi and Jacobi identities remain to be checked. This method has been used in the literature.

It was not a new result that compactifications on group manifolds to maximally symmetric spacetimes give rise, through specific dimensional reductions, to some 4d effective theories and allow for consistent truncation; the novelty of \cite{Andriot:2022bnb} is rather the systematic and completeness of the claim, in presence of orientifolds. Such a reduction was first known under the name of a Scherk-Schwarz truncation \cite{Scherk:1979zr}, and was studied in the realm of {\sl gauged supergravities} \cite{Samtleben:2008pe, Trigiante:2016mnt} (see Section \ref{sec:dimred}). These supergravity theories, here in 4d, depend on an object called the embedding tensor. The components of the latter are typically constant, and may be identified with the above variables \eqref{variables} such as $F_{q\, a_1 \dots a_q} \, , \ f^a{}_{bc}$, etc. As expected their components generate the various terms in the scalar potential. The precise mapping between the 10d type II compactification and the 4d gauged supergravity ingredients had been established explicitly in various concrete settings, see e.g.~\cite{Andrianopoli:2005jv, Villadoro:2005cu, DallAgata:2009wsi, Dibitetto:2010rg, Dibitetto:2011gm}, in which case searching solutions via gauged supergravity is equally relevant. The 10d BI may even be captured by extra constraints that the embedding tensor needs to obey. However, one should also keep in mind that some choices for the embedding tensor may not find an obvious 10d supergravity compactification realisation: an example could be a non-compact group manifold, or more involved, the appearance of non-geometric fluxes discussed in Section \ref{sec:listnongeo}. We will come back in Section \ref{sec:dSsolexistence} to de Sitter solutions obtained in gauged supergravities.

Let us recall that group manifolds are parallelizable, like a flat torus. As a consequence, supersymmetry is not broken by the geometry, and one can in principle have a maximal 4d (gauged) supergravity, i.e.~${\cal N}=8$, by compactifying on such ${\cal M}$. Each set of parallel sources $D_p/O_p$ however breaks supersymmetry by half, leading non-maximal (gauged) supergravities.

Another common dimensional reduction is that on manifolds or backgrounds admitting an SU(3)$\times$ SU(3) structure, generalizing the Calabi-Yau compactifications (see Section \ref{sec:dimred}). Those lead to a 4d ${\cal N}=2$ supergravity, and to one with ${\cal N}=1$ when including an orientifold \cite{Grana:2005jc, Koerber:2010bx}; those theories may also be identified as a gauged supergravity. Solutions on group manifolds can admit an SU(3)$\times$ SU(3) structure, then allowing for the formulation of the 4d effective theory with this formalism. In that case, it has also been argued that dimensional reduction can correspond to a consistent truncation (see e.g.~\cite{Grana:2005ny, Kashani-Poor:2006ofe, Cassani:2009ck, Andriot:2018tmb} and references therein). Then this formalism was used to look for de Sitter solutions from a type II supergravity compactification, using a 4d ${\cal N}=1$ effective theory \cite{Caviezel:2009tu, Danielsson:2011au}. However, it does not appear possible to express any 4d ${\cal N}=1$ theory, obtained by a consistent truncation on a group manifold from 10d type II supergravities, with this formalism of SU(3)$\times$ SU(3) structure, as pointed-out in \cite[Sec.6]{Andriot:2022bnb}.

To summarize, we have explained that the above solution ansatz of 10d type II supergravities on a group manifold can be used to perform a dimensional reduction to a 4d effective theory.\footnote{The 4d effective theory obtained in this way does not question the compactness of ${\cal M}$; the truncation typically removes any explicit 6d dependence, and the 6d volume integral is formally absorbed into $M_p^2$ without computing its value (in particular whether it is infinite). The 4d theory depends on $f^a{}_{bc}$, which only indicate the group, but not whether there exists a lattice (see Section \ref{sec:groupman}). This is how 4d gauged supergravities can be obtained by consistent truncations on non-compact manifolds. As mentioned for the 10d solution ansatz, the compactness remains to be checked. In some cases, the lattice action may however change the 4d theory by removing some modes: it then provides a further truncation \cite{Grana:2013ila}. The 4d effective theory mentioned so far thus does not take this into account, and could be further altered when considering the lattice. See also \cite{Andriot:2020vlg} for related comments.} This ansatz is then known to allow for a consistent truncation, meaning that extremizing the resulting 4d scalar potential gives as well a 10d solution, up to few extra conditions such as the Bianchi and Jacobi identities. {\sl This provides an alternative to look for de Sitter solutions, using the 4d theory.} That method has been used with 4d gauged supergravities, and 4d theories built from SU(3)$\times$ SU(3) structures.\\

It is well-known that a 4d effective theory obtained through a consistent truncation is not necessarily a low energy effective theory, as explained in Section \ref{sec:dimred}. In other words, using such a 4d effective theory may not be suited to study low energy physics around e.g.~a de Sitter extremum, a possible disappointment for cosmological applications. Nevertheless, it may still match, in some cases, a low energy effective theory, as shown in the case of a nilmanifold in \cite{Andriot:2018tmb}. In addition, we explained in Section \ref{sec:dimred} that perturbative stability can be studied using a 4d effective theory of a few degrees of freedom (e.g.~those kept by the consistent truncation). Indeed, we showed that a perturbative instability found in such a theory can only get worse by adding more fields (e.g.~other light fields that were truncated out). It will turn out that all known de Sitter solutions found this way happen to be (strongly) unstable, with $\eta_V \leq -2.4$. The use of a 4d effective theory obtained through consistent truncation therefore remains adequate to study and characterise a perturbative instability.
 
Having presented the 10d solution ansatz and the mathematical problem to solve, as well as 4d theories that can equivalently be used to find solutions and study their stability, we turn to the solutions found.

\subsubsection{Existence of supergravity solutions}\label{sec:dSsolexistence}

Solutions of 10d type II supergravities on a 4d de Sitter spacetime times a 6d group manifold, obeying the ansatz described in Section \ref{sec:groupmandimred} (including that of Section \ref{sec:10dansatz}), do exist. They were found numerically in \cite{Caviezel:2008tf, Flauger:2008ad, Caviezel:2009tu, Danielsson:2010bc, Danielsson:2011au, Roupec:2018mbn, Andriot:2020wpp, Andriot:2021rdy, Andriot:2022way, Andriot:2024cct}, using either a 10d formulation, or equivalently a 4d theory, as explained in Section \ref{sec:groupmandimred}. These solutions circumvent the existence no-go theorems discussed in Section \ref{sec:nogo10d}. The latter are useful, as they give hints of the necessary ingredients to find a solution, i.e.~favorable regions of parameter space. We recall from \eqref{summarydS}, in the case of sources $D_p/O_p$ of single dimensionality $p$ (possibly intersecting), the necessity of having $O_p$, together with negative internal curvature, and some non-zero RR flux $F_{6-p}$. This gives a first idea of the content of these de Sitter solutions, but also indicates their complexity. The necessity of all these ingredients, and the absence of supersymmetry (see Section \ref{sec:nodSQG}), points towards using numerical tools to find the solutions. We recall from Section \ref{sec:groupmandimred} that the problem of finding these solutions is nevertheless mathematically well-defined: one should solve a system of quadratic polynomial equations, depending on a finite list of (constant) variables. The compactness of the group manifold needs in addition to be checked. The problem being well-defined eases its numerical implementation, and the intuition from the no-go theorems helps in the search for solutions, allowing us to eventually find some. We present these de Sitter solutions in this section, together with some of their properties, either observed or conjectured.\\

To better describe the solutions found, it is useful to first introduce {\sl a classification} proposed in \cite{Andriot:2022way}. It uses the ansatz of Section \ref{sec:groupmandimred}, and classifies solutions with maximally symmetric spacetimes according to their source content. A class is mostly defined by its intersecting sets containing $O_p$. This is first because $O_p$ are necessary to get de Sitter and Minkowski solutions, and also because each set with $O_p$ imposes projection conditions, which determine the supergravity fields, i.e.~the allowed solution variables. The classes are denoted $s_{p_1 \dots}$ or $m_{p_1 \dots}$. There, $s$ and $m$ stand respectively for classes with sources of single or multiple dimensionalities, while dimensionalities $p$ of sets with $O_p$ are listed as $p_1 \dots$. For example, $s_{6666}$ stands for a class of type IIA solutions with 4 intersecting sets containing $O_6$ (and possibly $D_6$), along internal directions 123, 145, 256, 346 (up to relabeling). The sources in this class can be viewed as T-dual to those of $m_{5577}$, which has 2 intersecting sets containing $O_5$ (and possibly $D_5$), and 2 with $O_7$ (and possibly $D_7$), along directions 12, 34, and 1356, 2456.

The classification is constructed under the assumption that $D_p/O_p$ sources present in the solution are such that $T_{10}^I \neq 0$: this means that $O_p$ contributions do not compensate those of $D_p$ in the same set $I$ of parallel sources. In other words, they generate a ``tadpole'', canceled by fluxes. This is reminiscent of the Maldacena-Nu\~nez no-go theorem (\ref{nogo1}), which requires that, at least for one set, $T_{10}^I>0$. The assumption of all $T_{10}^I \neq 0$ for the sources present may seem restrictive,\footnote{Requiring $T_{10}^I \neq 0$ for a set of $D_p/O_p$ makes it impossible to have $p=9$ sources, due to their BI \eqref{BI}. In addition, with the ansatz of Section \ref{sec:groupmandimred} that requires constant fluxes, one obtains $\d F_0=0$. This implies from its BI a vanishing $T_{10}^I$ for $p=8$ sources. In short, $p=8,9$ sources cannot be present in the solutions considered. Solutions may still turn out to be compatible with their presence (meaning with the $O_p$ projection), even though they cannot be accounted for explicitly. We also recall the no-go theorem (\ref{nogo2}) which states the absence of de Sitter solution with $p=8,9$ sources alone. However, the latter could a priori be present in solutions together with other $D_p/O_p$ of different dimensionalities, the details of no-go theorems having then to be checked.} but all de Sitter solutions found so far do fall in one of the 21 classes of \cite{Andriot:2022way}.\footnote{Some solutions of type IIA supergravity with $O_6$ were found in \cite{Flauger:2008ad, Danielsson:2011au} with some additional orbifold, which may not be identified as a pure (intersecting) orientifold action: only those solutions may not fit in the classification.} We show these solutions below in Table \ref{tab:dSsolclass}. Note that the number of classes is limited by the combination of three facts: having a finite number of variables, the decrease of the latter with orientifold projection conditions, and the necessity of having $T_{10}^I \neq 0$ in the RR flux BI, which in turn determines possible placements of sources, i.e.~the allowed source configurations. We refer to \cite{Andriot:2022way} for more details. We give in \eqref{s55} one explicit example of a solution class, with its allowed sources and its variables (up to the $g_s$ factors in the list of variables \eqref{variables})
\begin{gather}
\label{s55}
\begin{aligned}
\text{{\bf Class $s_{55}$ :}} \quad & \\
\text{Sources:}\quad & \ \begin{tabular}{|l|c|c|c|c|c|c|c|c|} 
\hline
&\multicolumn{6}{|c|}{} & & \\[-8pt]
\multirow{3}{*}{Sources set $I$} & \multicolumn{6}{|c|}{Internal dimension $a$} & \multirow{3}{*}{${\rm vol}_{||_I}$} & \multirow{3}{*}{${\rm vol}_{\bot_I}$}  \\[4pt]
\hhline{~------~~}
&&&&&&&&\\[-8pt]
 & 1 & 2 & 3 & 4 & 5 & 6 & & \\[4pt]
\hline
\hline
&&&&&&&&\\[-8pt]
$I=1$: $O_5$, possible $D_5$ & $\times$ & $\times$ & & & & & $e^1 \w e^2$ & $e^3 \w e^4 \w e^5 \w e^6$ \\[4pt]
\hline
&&&&&&&&\\[-8pt]
$I=2$: $O_5$, possible $D_5$ & & & $\times$ & $\times$ & & & $e^3 \w e^4$ & $e^1 \w e^2 \w e^5 \w e^6$ \\[4pt]
\hline
&&&&&&&&\\[-8pt]
$I=3$: no $O_5$, possible $D_5$ & & & & & $\times$ & $\times$ & $e^5 \w e^6 $ & $e^1 \w e^2 \w e^3 \w e^4$ \\[4pt]
\hline \end{tabular}\\
& \\ 
& \ T_{10}^1 \,, \quad T_{10}^2 \,,\quad T_{10}^3 \,,\quad {\cal R}_4 \,, \\
& \ F_{1 \ 5} \,, \quad F_{1 \ 6} \,,\\
& \ F_{3\ 135} \,, \ F_{3\ 136} \,, \ F_{3\ 235} \,, \ F_{3\ 236} \,, \ F_{3\ 145} \,, \ F_{3\ 146} \,, \ F_{3\ 245} \,, \ F_{3\ 246} \,,\\
\text{Variables:}\quad & \ F_{5 \ 12345} \,, \quad F_{5 \ 12346} \,, \\
& \ H_{125} \,, \quad H_{126} \,, \quad H_{345} \,, \quad H_{346} \,,\\
& \ f^{1}{}_{35} \,, \quad f^{1}{}_{36} \,, \quad f^{1}{}_{45} \,, \quad f^{1}{}_{46} \,, \quad f^{2}{}_{35} \,, \quad f^{2}{}_{36} \,, \quad f^{2}{}_{45} \,, \quad f^{2}{}_{46} \,, \\
& \ f^{3}{}_{15} \,, \quad f^{3}{}_{16} \,, \quad f^{3}{}_{25} \,, \quad f^{3}{}_{26} \,, \quad f^{4}{}_{15} \,, \quad f^{4}{}_{16} \,, \quad f^{4}{}_{25} \,, \quad f^{4}{}_{26} \,,\\
& \ f^{5}{}_{13} \,, \quad f^{5}{}_{23} \,, \quad f^{5}{}_{14} \,, \quad f^{5}{}_{24} \,, \quad f^{6}{}_{13} \,, \quad f^{6}{}_{23} \,, \quad f^{6}{}_{14} \,, \quad f^{6}{}_{24} \,.
\end{aligned}
\end{gather}

Having established in this way 21 classes, de Sitter solutions were searched and found in 6 of them, detailed in Table \ref{tab:dSsolclass}. We will first provide one example of such a de Sitter solution, and further comment on all known solutions.
\begin{table}[ht!]
\begin{center}
\begin{tabular}{|c||c|}
\hline
&\\[-8pt]
Solution class & De Sitter solution found \\[4pt]
\hline
&\\[-8pt]
$s_{55}$ & \cite{Andriot:2020wpp, Andriot:2021rdy, Andriot:2022way, Andriot:2024cct} \\[4pt] 
\hline
&\\[-8pt]
$s_{66}$ & \cite{Andriot:2022way}\\[4pt] 
\hline
&\\[-8pt]
$s_{6666}$ & \cite{Caviezel:2008tf, Flauger:2008ad, Danielsson:2010bc, Danielsson:2011au, Roupec:2018mbn, Andriot:2022way} \\[4pt] 
\hline
&\\[-8pt]
$m_{46}$ & \cite{Andriot:2022way} \\[4pt] 
\hline
&\\[-8pt]
$m_{55}$ & \cite{Andriot:2022way} \\[4pt] 
\hline
&\\[-8pt]
$m_{5577}$ & \cite{Caviezel:2009tu, Andriot:2022way}\\[4pt] 
\hline
\end{tabular}
\end{center}\caption{Known de Sitter solutions verifying the ansatz of Section \ref{sec:groupmandimred}, with the reference where they were found. They are listed in the solution class to which they belong, following the classification of \cite{Andriot:2022way} described in the main text. The indices in the solution class name indicate the dimensionality $p$ of each set of parallel sources containing $O_p$: this illustrates the content of the solutions found.}\label{tab:dSsolclass}
\end{table}

We give in \eqref{sol29} the de Sitter solution 29 of the class $s_{55}$, found in \cite{Andriot:2024cct}. Supergravity quantities are given in units of $2\pi l_s$. Among the $s_{55}$ variables list \eqref{s55}, we only give those that take a non-zero value. We recall that we work here in 10d string frame.
\begin{gather}
\label{sol29}
\begin{aligned}
\hspace{-0.2in} &\text{{\bf Solution $s_{55}^+ 29$:}}\quad {\cal R}_4 = 0.020309\ ,\\[8pt]
\hspace{-0.2in} &g_s T_{10}^1 = 10\ ,\ g_s T_{10}^2 = -0.079765\ ,\ g_s T_{10}^3 = -1.064125\ ,\\[6pt]
\hspace{-0.2in} &g_s F_{1\, 5} = -0.231074\ ,\ g_s F_{135} = -0.659250\ ,\ g_s F_{136} = -0.662773\ ,\, g_s F_{146} = 0.084135 \ ,\\[6pt]
\hspace{-0.2in} &g_s F_{235} = -0.635765\ ,\ g_s F_{236} = -0.320255\ ,\ g_s F_{246} = -0.120817\ ,\\[6pt]
\hspace{-0.2in} & H_{125} = -0.002972\ ,\ H_{346} = -0.181872\ ,\\[6pt]
\hspace{-0.2in} &f^1{}_{45} = 0.829116\ ,\ f^1{}_{46} = -0.837373\ ,\ f^2{}_{35} = -0.256521\ ,\ f^2{}_{45} = -0.066547\ ,\\[6pt]
\hspace{-0.2in} &f^2{}_{46} = -0.807542\ ,\ f^3{}_{15} = -0.013195\ ,\ f^3{}_{25} = 0.013682\ ,\ f^6{}_{14} = -0.553790\ .
\end{aligned}
\end{gather}
Regarding the absolute values of the variables, let us mention that those can undergo an overall rescaling, as we will detail in Section \ref{sec:dSsolclassical}. Some signs are however important, starting with ${\cal R}_4>0$ for de Sitter, and $T_{10}^3\leq 0$ for an absence of $O_5$ in that set. As expected from the no-go theorems and constraints \eqref{summarydS}, one has $T_{10} = \sum_I T_{10}^I >0$, $F_{1} \neq 0$, and ${\cal R}_6 = - 0.741775  < 0$. In addition, there exists an analytical change of basis that simplifies the set of structure constants (as discussed in Section \ref{sec:groupman}), allowing to identify the manifold algebra as $\mathfrak{g}_{3.5}^0 \oplus \mathfrak{g}_{3.5}^0$, listed in \eqref{alg3d}. The latter is known to allow for a lattice for the group, making the group manifold ${\cal M}$ compact. It can be noted that most entries in the solution are not integers; we will come back in Section \ref{sec:dSsolclassical} to the necessary quantizations of these supergravity variables, to ensure a stringy origin. A related discussion about structure constants can be found in Section \ref{sec:groupman}. We finally note that $F_5=0$, in \eqref{sol29} as in all known $s_{55}$ solutions. More generally, one has $F_5=0$ or $F_6=0$ in all de Sitter solutions of \cite{Andriot:2022way}.

All solutions found before 2020 were obtained in $s_{6666}$, namely in type IIA supergravity with $O_6$ and $F_0$, except for one found in $m_{5577}$, in type IIB with $O_5, O_7$. All these solutions, except for \cite{Flauger:2008ad}, were obtained by imposing the values of structure constants, often to rational if not integer values. In \cite{Caviezel:2008tf, Caviezel:2009tu}, solutions  were found on SU(2)$\times$SU(2), and \cite{Flauger:2008ad, Danielsson:2010bc, Danielsson:2011au, Roupec:2018mbn} found solutions on non-nilpotent solvable, or non-solvable, algebras. This represents a total of about 400 de Sitter solutions \cite{Roupec:2018mbn}. From 2020 onwards were found the 29 $s_{55}$ solutions \cite{Andriot:2020wpp, Andriot:2021rdy, Andriot:2022way, Andriot:2024cct} described above, and 32 other solutions \cite{Andriot:2022way}, bringing the overall number of solutions to slightly less than 500. A database of these last 61 solutions can be found in \cite{Andriot:2022way, Andriot:2022bnb, Andriot:2024cct}, together with numerical tools to find and study them. For all these recent solutions, the method used was rather to let the structure constants as free variables, and determine a posteriori the corresponding algebra and group manifold (see Section \ref{sec:groupman} for the different approaches to structure constants). The algebras of these recent solutions were identified in \cite{Andriot:2022yyj}.\footnote{The compactness of the corresponding manifolds, i.e.~the existence of lattices, is also determined there, except for one solution: $s_{66}^+ 1$ found on $\mathfrak{g}_{6.88}^{0,1,0}$. We now know that the latter also allows for a lattice, by private communication with Christoph Bock.} There again, no solution was found on a nilmanifold. As proven at the end of Section \ref{sec:nogo10d}, there exists no de Sitter solution on nilmanifolds in the case of parallel sources. Here even with intersecting sources, no such solution was found, and it is tempting to conjecture a general no-go theorem against de Sitter solutions on nilmanifolds.

A few elements can give hints of solution classes where solutions would be found. First, it was noticed in \cite[(7.1)]{Andriot:2017jhf} that equations would be more easily solved if sets of $D_p/O_p$, $p\geq 5$, would share $p-5$ common directions. It turns out to be the case for $s_{6666}$, where the first solutions were found, and it motivated the search for solutions in $s_{55}$ \cite{Andriot:2020wpp} which has precisely such a source configuration \eqref{s55}. One may also note that this property matches the condition to preserve supersymmetry, namely ${\cal N}=1$ in 4d, with intersecting sources (see e.g.~\cite[2.4.2]{Andriot:2022way}). This could serve as a guideline, and indeed, among all solutions found, listed in Table \ref{tab:dSsolclass}, many have source configurations preserving supersymmetry. But it is not the case for all: some in $m_{46}$ and $m_{55}$ have more $D_p$, which break it. Second, one may get an intuition of where to find solutions through T-duality, which in general maps equivalent models with seemingly different scalar potentials. It turns out that some source configurations of different classes are T-dual to each other: for instance, we mentioned that the configuration of $m_{5577}$ is T-dual to that of $s_{6666}$. This may have motivated the first search in the former, that led to the de Sitter solution in \cite{Caviezel:2009tu}. However, beyond sources, the rest of the solutions in the respective classes are not necessarily T-dual to each other. And indeed, some $f^a{}_{bc}$ in the solution of \cite{Caviezel:2009tu} would transform, under standard T-duality rules on fluxes \cite{Shelton:2005cf}, to non-geometric $Q$-flux (see Section \ref{sec:listnongeo}), which are naturally absent of solutions in $s_{6666}$. The same holds for $s_{55}$ and $m_{46}$: they can have their source configurations T-dual, but not necessarily their complete solutions \cite{Andriot:2022way}. These hints provide some understanding of the various classes in Table \ref{tab:dSsolclass}, but not of all solutions.\\

In the 15 remaining classes of \cite{Andriot:2022way}, no de Sitter solution was found. Among those, no-go theorems against de Sitter solutions were proven for 5 of them, leaving 10 classes for which it could not be assessed whether numerics failed to find a solution, due to computation complexity, or whether another no-go theorem remains to be proven. This situation, and the overview provided by such scans for de Sitter solutions on group manifolds, can lead one to conjecture such no-go theorems. We discuss an important one in the remainder of this section, with evidence in its favour, and proof attempts.

A key observation is that {\sl in all de Sitter solutions found, there were at least three intersecting sets} ($I=1,2,3$) of $D_p/O_p$ sources. While it is clear from the names of classes $s_{6666}$ and $m_{5577}$, it is a priori not obvious for $s_{55}, s_{66}, m_{55}, m_{46}$ where extra sets containing only $D_p$ are allowed but not mandatory. But for these classes, we only found de Sitter solutions with at least 3 sets. Looking for solutions in $s_{55}$ with only 2 sets (where the $O_5$ are), i.e.~with $T_{10}^3=0$, only led us to find Minkowski or anti-de Sitter solutions \cite{Andriot:2022way}; de Sitter ones were found with $T_{10}^3 < 0$. This observation hints at a no-go theorem. It was already proposed in \cite[Conjecture 1]{Andriot:2019wrs} that de Sitter solutions with only parallel sources, i.e.~1 set, do not exist, as recalled in Section \ref{sec:nogo10d}: this is illustrated by all constraints indicated in Table \ref{tab:nogosummary}, and more. It was then proposed in \cite[Conjecture 4]{Andriot:2022way} that 2 sets of sources is still not enough, meaning that
\beq
\text{{\sl De Sitter solutions require at least 3 intersecting sets of $D_p/O_p$ sources.}} \label{Conj3sets}
\eeq
Here, ``de Sitter solutions'' naturally refers to those obeying the ansatz of Section \ref{sec:groupmandimred}, namely supergravity compactifications on group manifolds. The claim may hold beyond this ansatz, to e.g.~classical compactifications, even though de Sitter solutions of \cite{Cordova:2018dbb, Cordova:2019cvf}, discussed in Section \ref{sec:listbackreact}, only possess parallel sources.

Since 3 intersecting source sets break the 32 supercharges of type II to 4, at most, the above conjectured no-go theorem also gets the following important formulation \cite{Andriot:2022way}
\beq
\text{{\sl De Sitter solutions do not exist in theories with more than 4 supercharges.}}\label{Conj4supercharges}
\eeq
Here, ``theories'' naturally refer to effective ones in $d$ dimensions obtained through the compactifications we considered (and maybe beyond them). The statement \eqref{Conj4supercharges} is powerful, because unless one considers a non-supersymmetric theory, it constrains the dimension $d$ where one can find de Sitter solutions. Indeed, we first recall that 4 supercharges corresponds in $d=4$ to having (minimal) ${\cal N}=1$ supersymmetry. The statement implies that de Sitter solutions cannot be found in a supersymmetric effective theory in dimension $d=5$ or higher, because those have at least 8 supercharges \cite{Andriot:2022xjh}. This idea was discussed in Section \ref{sec:listclassdim}, where we also recalled results from no-go theorems in higher dimensions \cite{VanRiet:2011yc, Andriot:2022xjh} that support it. These no-go theorems are applicable to our ansatz, and allow to exclude many options for de Sitter in $d\geq 5$, in agreement with \eqref{Conj4supercharges}, even though few possibilities remain.

These ideas become particularly relevant when considering gauged supergravities in various dimensions, given the relation of these theories to the compactification ansatz considered here (see Section \ref{sec:groupmandimred}). In short, finding a de Sitter solution in a $d$-dimensional gauged supergravity, $d\geq 5$, with an identified compactification origin, would falsify the conjectured no-go theorem. We however do not know of such a counter-example. Rather, (metastable) de Sitter solutions were found in gauged supergravities with extended supersymmetries, always with ingredients that make the compactification origin unclear. For example, some were found with non-compact gaugings (which can indicate non-compact extra dimensions) and/or Fayet-Iliopoulos terms (whose higher dimensional origin is unclear) \cite{Fre:2002pd, deRoo:2002jf, Cosemans:2005sj, Ogetbil:2008tk, Roest:2009tt, DallAgata:2012plb}, or with non-geometric fluxes \cite{deCarlos:2009qm, Dibitetto:2011gm} (see Section \ref{sec:listnongeo}); see also \cite{Catino:2013syn} whose higher dimensional origin is not obvious to us. Further arguments against de Sitter extrema, using swampland input, applied to ${\cal N}=2$, 4d gauged supergravities or higher dimensions can be found in \cite{Cribiori:2020use, DallAgata:2021nnr, Hebecker:2023qke, Cribiori:2023ihv}. To conclude, restricting to our compactification ansatz on group manifolds, or to corresponding gauged supergravities, we see good evidence that the statements \eqref{Conj3sets} or \eqref{Conj4supercharges} could hold true. One may consider extending their validity more generally to classical de Sitter solutions from string compactifications. It is unclear if the statements hold beyond such framework: they could still be true, as e.g.~the recent construction of a 5d de Sitter maximum using Casimir energy \cite{ValeixoBento:2025yhz} happens in a non-supersymmetric effective theory.\\

We presented some example-based evidence in favour of \eqref{Conj3sets} or \eqref{Conj4supercharges}, namely \cite[Conjecture 4]{Andriot:2022way}, supporting as well \cite[Conjecture 1]{Andriot:2019wrs}. We also mentioned related no-go theorems in higher dimensions, discussed in Section \ref{sec:listclassdim}. In the following, we report on further efforts in trying to prove such (non-)existence claims.

The problem of finding a solution was said to be mathematically well-defined. For example, for the class $s_{55}$ \eqref{s55}, there are 56 non-trivial equations (quadratic polynomials of field components) and 44 variables (including ${\cal R}_4$). Proving a de Sitter existence no-go theorem only amounts to find a combination of some of these equations that implies ${\cal R}_4 \leq 0$. In Section \ref{sec:nogo10d}, we made this explicit, combining the dilaton e.o.m., the 4d and 10d trace of Einstein equation, and a sourced flux BI, leading to the constraints \eqref{summarydS}. Efforts to prove \cite[Conjecture 1]{Andriot:2019wrs}, namely the absence of de Sitter solution with only parallel sources, made use of more equations. In \cite{Andriot:2016xvq} we added to previous equations the trace of the 6d Einstein equation along parallel directions to the sources, while in \cite{Andriot:2018ept} we did the analogue in 4d by including $\del_{\sigma} V=0$ (see Section \ref{sec:rhotau}). Combining this equation to the others led to further constraints or no-go theorems, mentioned at the end of Section \ref{sec:nogo10d}, including the absence of solution on a nilmanifold, but not to a complete exclusion of de Sitter solutions with parallel sources. Further attempts were pursued in \cite{Andriot:2019wrs} by considering individual 6d Einstein equations instead of their trace, leading again to constraints mentioned at the end of Section \ref{sec:nogo10d}. In \cite[Sec.3.3]{Andriot:2019wrs}, an explicit example was provided where all equations could be solved (as well as orientifold projection conditions and compactness of ${\cal M}$), except for the individual 6d Einstein equations, while their 6d trace was also solved. This shows the necessity of including the individual 6 Einstein equations to prove the no-go theorem, which is not easy to do formally beyond examples.

Related efforts and results exist in the gauged supergravity literature: having only 1 set of parallel sources corresponds there to having a half-maximal supergravity, e.g.~${\cal N}=4$ in 4d. For the latter theory, it is shown in \cite{Dibitetto:2010rg} that a flux compactification origin requires the gaugings to be non-semi-simple, excluding e.g.~the de Sitter solutions of \cite{deRoo:2003rm}. In addition, considering an SO(3) truncation of 4d ${\cal N}=4$ gauged supergravities, also known as STU-models, no de Sitter extremum was found in \cite{Dibitetto:2011gm} (neither in \cite{Balaguer:2024cyb} that includes open strings), after a complete scan, except for one with non-geometric fluxes. A similar result was obtained recently in \cite{Arboleya:2025jko} for the same truncation in 3d half-maximal gauged supergravities, leading to RSTU-models: a full search did not provide any de Sitter solution. While this is all in agreement with \cite[Conjecture 1]{Andriot:2019wrs}, this is not yet a complete proof given the specific truncation. Finally, \cite{Dibitetto:2019odu} considered half-maximal ${\cal N}=(1,1)$ gauged supergravities in 6d, with again an eye to a flux compactification origin with sources. A complete scan led again to no de Sitter solution (except for one that goes beyond our ansatz by including Kaluza--Klein monopoles, assuming the latter gets a 10d realisation, and for which compactness of the manifold is however not ensured); this is again in agreement with \cite[Conjecture 1]{Andriot:2019wrs}.

There is no proof either for \cite[Conjecture 4]{Andriot:2022way}, i.e.~the absence of such de Sitter solutions with 2 intersecting sets of sources. In \cite[Sec.4]{Andriot:2022way}, we provided an argument for both conjectured no-go theorems based on T-duality. Using the classification of possible solutions, we considered all possible source configurations with exactly 2 sets: either those where both sets contain $O_p$, or those where only one does and the other set contains only $D_p$. For each of these two options, all source configurations turned out to be related by T-duality: for example, the two sets of $D_7/O_7$, placed as in $s_{77}$, can be T-dualised to the two sets of $D_6/O_6$ in $s_{66}$, etc. Whether all solutions in each class can as well be T-dualised to one in the other class is less clear, because some of the supergravity fields T-dualise to non-geometric fluxes, as explained above. Nevertheless, each of the two T-duality chains contains a class where a no-go theorem against de Sitter had been found, namely $s_7$ and $s_{77}$ using no-go theorem (\ref{nogo2}). This T-dual relation to a case without de Sitter solution is again a hint of a general no-go theorem, even though this could not be made to work in details.

Finally, let us mention some unpublished attempts to prove these conjectures,\footnote{These attempts were led by Paul Marconnet and myself in 2022 and 2023.} motivated by the fact that the problem is mathematically well-defined. ``Linearisation'' methods exist to transform a system of quadratic polynomial equations into a linear one: essentially, one introduces new variables that capture the quadratic products, for example $x_1 = f^1{}_{23} H_{145}$, etc. This increases the number of variables, but rewrites the system as a linear one. Then, techniques can be used to determine whether solutions exist or not. Such methods become however quickly computationally expensive, and we did not manage to reach a conclusive answer.\\ 

Having presented the de Sitter solutions found, with their observed and conjectured properties, we now turn to the study of their perturbative stability. There again, the overview provided by the solution dataset will hint at universal properties, with attempts of proving them.

\subsubsection{Stability}\label{sec:dSsolstab}

We discuss in this section the perturbative (in)stability of de Sitter solutions of 10d type II supergravities on 6d group manifolds, obeying the ansatz of Section \ref{sec:groupmandimred}. As just presented in Section \ref{sec:dSsolexistence}, there exists a little less than 500 such solutions, falling in 6 classes of the classification of \cite{Andriot:2022way}. To study the perturbative stability, we consider a 4d effective theory, obtained by a truncation but not necessarily as a low energy approximation. We then study the 4d extremum of the scalar potential, corresponding to the de Sitter solution. As explained in Section \ref{sec:dimred} or \ref{sec:groupmandimred}, finding one field direction along which the potential is a maximum, i.e.~a tachyon, is then sufficient to conclude on a perturbative instability. For the de Sitter solutions of interest, using either the $(\rho,\tau,\sigma)$ theory \eqref{rhotausigmatheory} or the 4d (gauged) supergravities has allowed to show that {\sl all of them are perturbatively unstable}. More precisely, all these solutions were found with $\eta_V \leq -2.49$.\footnote{An account on the stability and $\eta_V$ of solutions found before 2020 is given in \cite[Sec.3]{Andriot:2018mav}, while the $\eta_V$ values for all solutions found from 2020 onwards can be found in \cite[Sec.5.2]{Andriot:2022bnb}, together with $\eta_V = -4.36757$ for $s_{55}^+ 29$ found in \cite{Andriot:2024cct}.} In type IIA, the less unstable ones were found in $s_{6666}$ with $\eta_V \simeq -2.5$ \cite{Flauger:2008ad} or $\eta_V\simeq -2.495$ \cite[Tab.1]{Andriot:2018mav}. The same highest value was found in type IIB for solution $s_{55}^+22$, $\eta_V\simeq -2.4940$, in \cite[Tab.6]{Andriot:2022bnb} (this IIA/B numerical coincidence is surprising, we comment on it in Section \ref{sec:sumclassdS}). Given this situation, it is tempting to conjecture a systematic instability for these de Sitter solutions; this was done more broadly for classical de Sitter solutions in \cite[Conjecture 2]{Andriot:2019wrs}, recalled as (\ref{nogostab1}) in Section \ref{sec:nogostab}. We rephrase the conjecture here focusing on our setup:
\beq
\text{{\sl De Sitter solutions, obeying the ansatz of Section \ref{sec:groupmandimred}, are perturbatively unstable.}} \label{conjstab}
\eeq
After few comments on such perturbative instability, we will come back in detail to this conjecture \eqref{conjstab}, discussing attempts to prove it, together with other related conjectures.\\

We start by making a few remarks on perturbative instability. First, while many de Sitter constructions from string theory have aimed at de Sitter minima, we can mention some recent ones that have also obtained a maximum, as here: those with Casimir energy, discussed in Section \ref{sec:listCas}, and non-geometric ones using specific Landau-Ginzburg models, discussed in Section \ref{sec:listnongeo}; we note that both do not have orientifolds. Let us also mention the argument of \cite{Catino:2012dq}, enforcing de Sitter extrema to be unstable at weak string coupling, for certain 4d ${\cal N}=1$ supergravities; it would be interesting to study to which string constructions this precisely applies.

Second, we recall from Section \ref{sec:dSconjsum} that having a de Sitter maximum, even a fairly unstable one with $\eta_V < -1$, is not necessarily in contradiction with cosmological observations. While it does not allow to realise a standard cosmological constant, it could still agree with a dynamical dark energy scenario for the recent universe, essentially corresponding to a hilltop quintessence model. We also noted in Section \ref{sec:dSconjsum} that constraints on single field inflation models suggest a concave scalar potential, i.e.~$\eta_V <0$, with a smaller $|\eta_V|$, while multifield models allow for more options.

Third, we recall from Section \ref{sec:swamp} that non-supersymmetric solutions, with maximally symmetric spacetimes, are believed to be unstable, either perturbatively or non-perturbatively. References on this idea for (anti-)de Sitter were given and discussed in Section \ref{sec:swamp}, and we mention here \cite{Acharya:2019mcu, Acharya:2020hsc} about Minkowski; discussions of this idea can also be found e.g.~in \cite{GarciaEtxebarria:2020xsr, Agmon:2022thq}. As de Sitter solutions are non-supersymmetric, this belief agrees with our unstable de Sitter solutions, here with a perturbative realisation. This also agrees with the RdSC \eqref{dSconjRef}, as discussed in Section \ref{sec:reftestdSconj}.\\

We now make further comments about the probability of finding a (meta)stable de Sitter solution. We recall from Section \ref{sec:dimred} that having a perturbatively unstable de Sitter extremum corresponds to having at least one negative eigenvalue in the mass matrix. The size of that matrix corresponds to the number of scalar fields, and can be large: for example, in the 6 classes where de Sitter solutions have been found (see Section \ref{sec:dSsolexistence}), the number of fields in the corresponding 4d theory is either 14 or 22 \cite[Tab.1]{Andriot:2022bnb}. Considering randomly a matrix of this size, it is clear that having all eigenvalues positive is much less likely than having at least one negative; one could infer that (meta)stable de Sitter solutions are, in terms of probability, much harder to find. This was made more precise in \cite{Marsh:2011aa}, considering 4d ${\cal N}=1$ supergravity with random potentials, leading to a random matrix for the Hessian (whose signature is the same as that of the mass matrix). It is was quantitatively shown that having all eigenvalues positive is extremely unlikely. Of course, one may object that the potential we are considering is not random, possibly favoring stability more than in a random model. Nevertheless, this suggests a naturalness to the presence of a tachyon, and a general difficulty in finding (local) minima, if any. 

This possible inherent scarcity of de Sitter minima leads us to ask about the capacities of the tools at hand to find them. Finding numerically such a minimum has been argued to be challenging, and qualified as a NP-hard problem \cite{Halverson:2018cio}. Numerical searches in \cite{Danielsson:2012by}, in particular with random scans, also led to argue that (isotropically meta)stable de Sitter solutions are rare, and require non-geometric fluxes. In \cite{Andriot:2021rdy}, we used a more guided strategy: we identified some analytical stability no-go theorems, such as conditions $C5$ or $C10$ discussed in (\ref{nogostab4}) in Section \ref{sec:nogostab}, and we directed the numerical searches precisely in regions of parameter space where such a condition was not obeyed. We then found new and different de Sitter solutions, however still unstable (except one, $s_{55}^+ 18$, to which we will come back), mostly with new kinds of tachyons. It therefore remains unclear whether it is just difficult, both inherently and numerically, to find de Sitter minima in our setup, or whether it is impossible because they do not exist. This brings us back to questioning the conjecture \eqref{conjstab}.\\

If conjecture \eqref{conjstab} would hold true, a related question is whether the tachyon is universal, meaning whether the tachyonic field direction could be identified in some unique manner for all compactifications. Most attempts to prove conjecture \eqref{conjstab} have to some extent assumed this idea of having a universal tachyon. We recall from Section \ref{sec:nogostab} two such approaches. First, as presented there in (\ref{nogostab3}), a de Sitter solution in 4d ${\cal N}=1$ supergravity was shown to be tachyonic in the case where this extremum is close a no-scale Minkowski solution \cite{Junghans:2016abx}. In addition, a large class of such de Sitter solutions was shown to admit a universal tachyon, which would align with the sgoldstino in the Minkowski limit. Looking for 10d realisations of such a 4d scenario, explicit examples were found among the known de Sitter solutions in classes $s_{6666}$ and $m_{5577}$ \cite{Junghans:2016uvg}. Second, as mentioned in (\ref{nogostab2}), it was conjectured \cite{Danielsson:2012et}, initially for $s_{6666}$, that
\begin{gather}
\label{conjstabrhotausigma}
\begin{aligned}
& \text{{\sl De Sitter solutions, obeying the ansatz of Section \ref{sec:groupmandimred}, have a tachyon}} \\
& \text{{\sl that lies among the fields $(\rho,\tau,\sigma_I)$.}}
\end{aligned}
\end{gather}
These fields were introduced in Section \ref{sec:rhotau}. As suggested in (\ref{nogostab2}), this claim on the tachyon could be broadened to other classical de Sitter solutions, but we restrict here to group manifolds and the ansatz of Section \ref{sec:groupmandimred}, a setup with which this conjecture has been tested. We recall from Section \ref{sec:dimred} that finding a tachyon among a finite set of fields is sufficient to conclude on the perturbative instability when considering more fields; the idea of this conjecture is thus to identify the minimal number of fields to find a systematic tachyon, which almost amounts to finding a universal field direction. As discussed in Section \ref{sec:nogostab}, $(\rho,\tau,\sigma_I)$ forms indeed such a minimal set of fields: a tachyon is not always found with $(\rho,\tau)$, but a tachyon has always been found among $(\rho,\tau,\sigma_I)$ whenever tested, as long as ${\cal M}$ is compact. This interpretation of the tachyonic field direction was actually shown to be compatible with the previous one in the examples considered in classes $s_{6666}$ and $m_{5577}$ \cite{Junghans:2016uvg}. For solutions found from 2020, a more systematic machinery was developed to test this $(\rho,\tau,\sigma_I)$ conjecture \eqref{conjstabrhotausigma} for all solution classes \cite{Andriot:2020wpp, Andriot:2021rdy, Andriot:2022yyj}. The corresponding $\eta_V$ could systematically  be computed for this subset of fields, and the conjecture \eqref{conjstabrhotausigma} was successfully verified in all but one example, $s_{55}^+ 18$, as discussed below. Using the tools developed, we compute for the later solution $s_{55}^+ 29$ \eqref{sol29} the value $\eta_V = -1.43763$ for the four fields $(\rho,\tau,\sigma_1,\sigma_2)$, verifying again the conjecture \eqref{conjstabrhotausigma}. For all these solutions since 2020 except $s_{55}^+ 18$, we find one and only one tachyonic field direction among $(\rho,\tau,\sigma_I)$.

Proving the conjecture \eqref{conjstabrhotausigma} of a systematic $(\rho,\tau,\sigma_I)$ tachyon would imply the conjecture \eqref{conjstab}, and would identify the relevant subset of fields, if not a universal tachyonic direction. We focus on this question in the following, before eventually coming back to the initial and more general conjecture \eqref{conjstab}. Attempts to prove \eqref{conjstabrhotausigma} were carried-out in general type II compactifications on group manifolds, without succeeding. For example in \cite[Sec.3.3]{Andriot:2018ept}, assuming parallel sources, a parameter space region was left where stability could in principle still occur. Further attempts and tools were developed in subsequent works, especially with intersecting sets of sources, as reported in Section \ref{sec:nogostab}, without reaching a complete proof. An important question to prove instability is to know the equations involved, similarly to the existence no-go theorems. Proving the presence of a tachyon among $(\rho,\tau,\sigma_I)$ requires to include the second derivatives of the potential with respect to those fields (only), and further extra equations to form a relevant combination, as presented in Section \ref{sec:nogostab}: the extremization (or equations of motion) with respect to those fields, the expression of the potential itself, and the fluxes BI. This list of equations was not sufficient in \cite{Andriot:2018ept, Andriot:2019wrs, Andriot:2021rdy} to formally prove the systematic presence of a tachyon in general compactifications. However, tachyons were found formally in \cite{Danielsson:2012et, Junghans:2016uvg, Garg:2018zdg} for the specific de Sitter solutions in $s_{6666}$ on the manifold $SU(2) \times SU(2)$. These works used further extremization conditions, in particular with respect to axions, that should correspond in 10d to the flux equations \eqref{eomfluxes}. These extra equations could therefore be necessary to a proof, but they are not easy to use generally beyond a concrete example.

More advanced attempts to prove the presence of a systematic tachyon among $(\rho,\tau,\sigma_I)$ were carried-out in \cite{Andriot:2021rdy}, focusing on the solution class $s_{55}$; we report on the methods developed in Section \ref{sec:nogostab}. This study revealed two problems with this conjecture \eqref{conjstabrhotausigma}. First, a solution was found, $s_{55}^+ 18$, without any tachyon among these fields. However, it was also noted that the group manifold ${\cal M}$ for this solution is not compact. This was actually in line with various de Sitter minima found in gauged supergravities with non-compact extra dimensions, as reported in Section \ref{sec:dSsolexistence}. Later, a tachyon was found for this solution when including further fields \cite{Andriot:2022bnb}, thus still in agreement with \eqref{conjstab}. Nevertheless, this shows a first delicate point: {\sl compactness of the manifold is an information that needs to be included}, if one wants to prove a systematic $(\rho,\tau,\sigma_I)$ tachyon as in conjecture \eqref{conjstabrhotausigma}! In other words, considering combinations of equations as discussed above cannot be enough; this certainly makes it more difficult.\footnote{A possible way to incorporate this information, briefly alluded to in \cite[Sec.4.2]{Andriot:2021rdy} with $R_4<0$, is to consider the sign of certain curvature terms. Indeed, some of them correspond to (partial) traces of the Killing form, and signs of eigenvalues of the Killing form can be related to compactness \cite[Sec.2.1]{Andriot:2022yyj}.} Second, it was shown explicitly that different solutions in $s_{55}$ can have tachyons in different field directions among $(\rho,\tau,\sigma_I)$: this means that {\sl there is no universal tachyon}, even though the field subset might still be identified.

This situation may seem at odds with respect to results obtained for specific examples in classes $s_{6666}$ and $m_{5577}$. There could be an explanation: the 4d effective theories, obtained through consistent truncation, are very different. For $s_{6666}$ and $m_{5577}$, the 4d theory can be written as an ${\cal N}=1$ supergravity in terms of $SU(3)\times SU(3)$ structures. As mentioned in \cite[Sec.6]{Andriot:2022bnb}, such a formulation may not exist for compactifications in $s_{55}$. Furthermore, the 4d effective theory for $s_{6666}$ and $m_{5577}$ possesses 14 scalar fields, while the one for $s_{55}$ has 22 \cite[Tab.1]{Andriot:2022bnb}, indicating a clear difference between these 4d theories. In that situation, results on tachyons may then differ, and more conclusive results might be obtained in the case of $s_{6666}$ and $m_{5577}$. Whether de Sitter solutions in $s_{55}$ admit or not a limit to a no-scale Minkowski solution (see e.g.~\cite[Sec.2.4]{Andriot:2020wpp}) would also be a way to distinguish their tachyonic behaviour from that of other classes, in view of above results.

A last, more speculative point to be made about conjecture \eqref{conjstabrhotausigma} is its possible relation to another conjecture involving $(\rho,\tau,\sigma_I)$: the Massless Minkowski Conjecture (MMC) \cite{Andriot:2022yyj}. We recall it here
\begin{gather}
\label{MMC}
\begin{aligned}
\text{MMC}:\quad & \text{{\sl 10d supergravity solutions compactified to 4d Minkowski always admit}} \\
& \text{{\sl a 4d massless scalar, among the fields $(\rho,\tau,\sigma_I)$.}}
\end{aligned}
\end{gather}
Restricting to the ansatz of Section \ref{sec:groupmandimred}, this was observed to hold true for a variety of Minkowski solutions found in the different compactification classes of \cite{Andriot:2022way}, and beyond it, in many examples in the literature, as discussed in \cite{Andriot:2022yyj}. We also recall that the systematic massless scalar is not necessarily a flat direction. For example it may appear in the scalar potential in an overall factor. Specificities of the MCC is also the reference to the subset of fields $(\rho,\tau,\sigma_I)$, and the fact the claim does not depend on supersymmetry (neither of the theory nor of the solution). Both points strongly suggest a possible relation to the de Sitter tachyon among $(\rho,\tau,\sigma_I)$, as conjectured in \eqref{conjstabrhotausigma}. Indeed, in the case where the Minkowski solution could be lifted, by a deformation, to a de Sitter one, then the massless mode of the MMC might become the systematic tachyon of \eqref{conjstabrhotausigma}; this idea has not been investigated. Such a relation is also reminiscent of the stability no-go theorem on nearly no-scale de Sitter solutions (\ref{nogostab3}). We refer to \cite{Andriot:2022yyj} for a more detailed discussion.\\

We finally come back to the initial conjecture \eqref{conjstab}, going beyond the fields $(\rho,\tau,\sigma_I)$. We described in Section \ref{sec:groupmandimred} how a consistent truncation to left-invariant modes on group manifolds had provided us in \cite{Andriot:2022bnb} with a 4d effective theory for each of the 21 compactification classes of \cite{Andriot:2022way}. It should be noted  that the dimensional reduction presents various subtleties, starting with the 10d supergravity Chern-Simons terms and the definition of RR fluxes in presence of background fluxes, or the presence of the $B$-field in the DBI action. The dimensional reduction eventually leads to general 4d scalar potentials \eqref{potIIAB} including 4d axions, as well as kinetic terms for all fields. We also recall that extremizing this 4d scalar potential was explicitly shown to be equivalent to the 10d equations of motion. With these tools at hand, the instability of all de Sitter solutions could be studied, and their $\eta_V$ values were computed, for the more complete set of 14 or 22 4d fields, depending on the class, instead of the 4 to 6 $(\rho,\tau,\sigma_I)$. From this study, we first verified that all the de Sitter solutions were perturbatively unstable, including $s_{55}^+ 18$ discussed above, in line with \eqref{conjstab}. We recall that they obey $\eta_V \leq -2.49$. With these extra fields, some de Sitter solutions got a second tachyon. The tachyonic field directions are commented on in \cite{Andriot:2022bnb}: while a first tachyon along $(\rho,\tau,\sigma_I)$ would now appear along the dilaton and diagonal metric components, the new fields, i.e.~axions and off-diagonal metric components, would also contribute, either together in (both) tachyon(s) or separately to the second tachyon. The role of the compactness of the manifold was also investigated. We note for instance that all solutions on compact manifolds (especially those in $s_{6666}$ and $m_{5577}$) have only one tachyon, except for $s_{55}^+ 19, 22-27$. For those, we recall the differences in the 4d effective theory, mentioned above. We refer to \cite{Andriot:2022bnb} for a more detailed discussion.\\

We hope that the material presented in this section will help understanding the seemingly systematic appearance of a strong perturbative instability for the de Sitter solutions of interest. The latter may still provide valuable cosmological solutions.

\subsubsection{Classicality}\label{sec:dSsolclassical}

We have considered for now de Sitter solutions of 10d type II supergravities on a 6d group manifold, as described by the ansatz of Section \ref{sec:groupmandimred}. As explained at the beginning of Section \ref{sec:dSsol}, these qualify as ``classical'' string backgrounds provided certain conditions are met. In this section, we report on tests of the ``classicality'' of these de Sitter supergravity solutions, to guarantee their stringy origin. We start by presenting the five required conditions, before testing them explicitly on solution $s_{55}^+29$ \eqref{sol29}. We then review related results in the literature, suggesting the absence of a parametric control on classicality for a de Sitter solution. We highlight a possible loophole to such a conclusion, in the case of an anisotropic compactification on a specific geometry. We illustrate this idea by presenting a concrete example, where scaling by a parameter in solution $s_{55}^+29$, improves classicality. While it is unclear at this stage whether a classical de Sitter solution exists, this section defines the problem, indicates difficulties to be encountered and illustrates possible resolutions to this question.\\

The two main conditions are that the $\alpha'$-corrections and the string-loop corrections are both negligible, giving a low-energy (in string spectrum, and higher derivatives) and perturbative approximation that would match a supergravity description. To be certain that this holds, one should compute these corrections. However, this is not an easy task on general backgrounds, so these conditions are often traded for simpler ones. String-loop corrections are plausibly considered negligible if the string coupling, $g_s$, is found small compared to $1$ to allow for a perturbative expansion. Similarly, $\alpha'$-corrections are thought negligible if the physical lengths encountered, $r$, are large compared to $l_s$, typically then looking at the 6d volume or (some) individual 6d lengths. These simpler requirements can be viewed as necessary conditions. To make them sufficient, one could find a free parameter $\gamma$ on which $g_s$ and $r/l_s$ depend, in such a way that taking $\gamma$ to be arbitrarily large also makes $r/l_s$ and $1/g_s$ arbitrarily large. In that case, the corrections can be made as negligible as desired, without needing to compute them in detail: one talks of a parametric control on corrections. This is to be contrasted with a numerical control, where one only finds after estimation or computation that the corrections are indeed negligible, because their numerical value is comparatively small, but this hierarchy cannot be tuned to arbitrary precision.

For the supergravity solutions on group manifolds, three other conditions should be met. First the harmonic fluxes should be quantized. This is because they can be phrased in terms of a $U(1)$-bundle, in which the corresponding harmonic forms, or the de Rham cohomology equivalence classes, are discretized (here in units of the string length to guarantee a stringy origin). Second, the amount of $D_p/O_p$ sources should also discretized, essentially as an integer. It has the important consequence that one infers an upper bound, related to the number of orientifolds, since those correspond to involution fixed points in the manifold, coming as a finite and not very large number. While counting properly sources appears as an obvious requirement, we recall that their contributions appear only through the quantities $T_{10}^I$, which include at the same time the source charges (here again in string length units) and the transverse volume, on top of their number. Third, the compactness of the group manifold, sometimes obtained through a lattice action, imposes certain discretization conditions on the structure constants. In short, we would talk of the fluxes, sources and lattice quantization conditions. Note that these three requirements all need a detailed knowledge of the group manifold geometry (harmonic forms, involution fixed points, lattice), which complicates their verification.

We should emphasize that some of these conditions differ from more standard compactifications, on Calabi-Yau or tori. In the latter, one only considers cycles in the geometry (focusing on massless modes), leading to have only harmonic forms. Any flux component should then be quantized, and every source is wrapping a cycle, while the question of compactness and structure constants does not occur. In group manifold solutions, non-harmonic forms typically appear: one should then identify the harmonic parts of the fluxes, i.e.~discard their exact and co-closed pieces following Hodge decomposition, which can be involved. Similarly, the sources often wrap non-cycles in ${\cal M}$ (e.g.~fibers), even though their (internal) world-volume, as a pull-back from ${\cal M}$, remains a cycle (e.g.~a torus): we refer to \cite[App.A]{Andriot:2024cct} for a treatment of this subtlety. Finally, the volume of each cycle in a Calabi-Yau should be large in string units, and there can be many of them. But a group manifold is parallelizable, so the number of cycles is limited. Their volume is essentially given in terms of only 6 internal lengths, $r_{a=1 \dots 6}$, that normalise the $e^a$ (lengths are in the $e^a$, since the metric does not carry length in our conventions, as e.g.~$\delta_{ab}$). The focus here is then mostly on those 6 lengths.

To summarize, testing the classicality of a supergravity solution amounts to verify five requirements, namely having large 6d lengths and a small string coupling, together with the fluxes, sources and lattice quantization conditions. The last three should in principle be verified first, to ensure the consistency of the string compactification setting itself; then one looks at the values of the lengths or volume(s) and of the string coupling in the supergravity solution, to see if they can reach the classical regime.\\

We now provide a concrete example to illustrate these requirements. We are going to test the classicality of the type IIB supergravity solution $s_{55}^+ 29$, presented in \eqref{sol29} and obtained in \cite{Andriot:2024cct}. To that end, we first need to rewrite the solution in an adequate fashion to verify the five requirements. That work was essentially done in \cite{Andriot:2020vlg} for the very similar solution $s_{55}^+ 14$, up to few $2\pi$ factors missing and corrected in \cite{Andriot:2024cct}. We rewrite the supergravity solution variables as follows
\begin{gather}
\label{var3}
\begin{aligned}
& \frac{1}{\lambda}\, g_s F_{3\, \omega_1} = \frac{1}{2\pi l_s} \times  g_s\, N_{\omega_1}\, \frac{l_s^3}{r_1r_4r_6} \ ,\\
& \frac{1}{\lambda^2} g_s T_{10}^1 = \frac{1}{(2\pi l_s)^2} \times 6 g_s\, N_{s1}\, \frac{l_s^4}{r_3r_4r_5r_6}  \ ,\\
& \frac{1}{\lambda}\, f^1{}_{64} = \frac{1}{2\pi l_s} \times 2\pi \, N_1\, \frac{r_1\, l_s}{r_4r_6} \ ,\ \text{etc.}
\end{aligned}
\end{gather}
The flux rewriting comes from the standard quantization of a $U(1)$-bundle harmonic flux on a cycle $\Sigma_{\omega}$ (with component $F_{q\, \omega}$), while that of $T_{10}^I$ comes from the smeared source contribution definition
\beq
\frac{1}{(2\pi l_s)^{q-1}} \int_{\Sigma_{\omega}} F_q = N_{\omega} \in \mathbb{Z} \ ,\qquad \frac{T_{10}^I}{p+1} = N_{sI}\, \frac{(2\pi l_s)^{7-p}}{vol_{\bot_I}} \ ,\ N_{sI}= 2^{p-5} N_{O_p}^I - N_{D_p}^I \ ,
\eeq
and $N_{sI}$ counts the number of $O_p/D_p$ sources (up to a charge difference factor) in the set $I$ of parallel sources. The lengths $r_a$ enter each 1-form $e^a$, as made explicit e.g.~in \eqref{eaexample}. The relevant volumes can be computed explicitly and are essentially products of $2\pi r_a$. The structure constants also depend on the $r_a$: the form of this dependence can be inferred from the Maurer-Cartan equation \eqref{MC}, while the rest of the $f^a{}_{bc}$ is captured by a coefficient $N_a$, as $N_1$ above, that we now turn to. 

The solution $s_{55}^+29$ was obtained on a solvmanifold, whose Lie algebra is $\mathfrak{g}_{3.5}^0 \oplus \mathfrak{g}_{3.5}^0$, depending mainly on two corresponding pairs of structure constants, $f^2{}_{35},f^3{}_{25}$ and $f^1{}_{46},f^6{}_{14}$. We recall from the end of Section \ref{sec:groupman} that structure constants of solvable algebras do not have to be integers. However, they obey a quantization condition due to the lattice action that makes the manifold compact. These lattice quantization conditions are discussed in detail in Appendix \ref{ap:groupman}. The manifold is chosen such that the only conditions to be obeyed by the $N_a$ parameters of the four structure constants are $N_6 = -1/N_1,\, N_3 = -1/N_2$, while these $N_a$ take real values. The resulting solvmanifold is topologically a torus, with a non-Ricci flat metric.

Last but not least, the parameter $\lambda>0$ appearing in \eqref{var3} corresponds to a scaling freedom that we made manifest. Indeed, under the following $\lambda$-scaling
\beq
F_{q\, a_1\dots a_q},\ H_{a_1a_2a_3},\ f^a{}_{bc} \rightarrow \frac{1}{\lambda}\, F_{q\, a_1\dots a_q},\ \frac{1}{\lambda}\, H_{a_1a_2a_3},\ \frac{1}{\lambda}\, f^a{}_{bc} \ , \qquad T_{10}^I \rightarrow \frac{1}{\lambda^2}\, T_{10}^I \ , \label{lambdascaling}
\eeq
the 10d equations are invariant. This means that the scaling provides a new solution, still de Sitter, as ${\cal R}_4 \rightarrow {\cal R}_4 / \lambda^2$. On $s_{55}^+29$, we first perform a rescaling, and quantize the resulting solution as in \eqref{var3}. We see that the supergravity variables in \eqref{var3} appear in units of $2\pi l_s$, which can also be absorbed in $\lambda$. Finally, the 6d lengths $r_a$ appear in units of $l_s$, suggesting the classicality requirement $r_a/l_s > 1$.

The five classicality requirements introduced above are now expressed as follows for this example
\begin{gather}
\label{classrequir}
\begin{aligned}
& N_{\omega},\, N_{sI} \in \mathbb{Z}\ ,\quad N_{sI} \leq 2^{p-5} N_{O_p}^I \ ,\quad N_a: \,\text{lattice quantized}\\
& g_s < 1 \ ,\quad \frac{r_a}{l_s} > 1\ (?)
\end{aligned}
\end{gather}
where we express explicitly the upper bound due to the fixed number of $O_p$, and we will come back to the question mark on the minimal value for $r_a/l_s$.

We now have all ingredients to write the solution $s_{55}^+ 29$ \eqref{sol29} in a way where we can test the five requirements of classicality, as expressed in \eqref{var3} and \eqref{classrequir}. The rewriting of the solution gives the following (best) values
\begin{gather}
\label{sol29var3}
\begin{aligned}
&\text{{\bf Solution $s_{55}^+29$:}}\\[8pt]
&g_s = 0.532758\ ,\  r_1 = 4.704542\ ,\ r_2 = 112.925701\ ,\  r_3 = 0.067605\ ,\ r_4 = 14.968801\ ,\\[6pt]
&r_5 = 172.058417\ ,\ r_6 = 0.077310\ ,\ \lambda = 1.622330\ ,\\[6pt]
&N_{s1} = 16\ ,\  N_{s2} = -67\ ,\  N_{s3} = -68\ ,\ N_{1\,  5} = -46\ ,\  N_{\omega_1} = 1\ ,\  N_{\omega_2} = -18\ ,\\[6pt]
&N_1 = 0.020207\ ,\ N_2 = -0.002592\ ,\  N_3 = -1/N_2\ ,\ N_6 = -1/N_1\ ,
\end{aligned}
\end{gather}
where $r_a$ stands for $r_a/l_s$, and we left aside few other (non-quantized) solution parameters, capturing non-harmonic fluxes or off-diagonal metric components. We see from \eqref{sol29var3} that the fluxes, sources and lattice quantization conditions are obeyed. In particular, the number of $O_5$ is fixed to be 16. This is because the 4-dimensional transverse geometry to the $O_5$ is in that example a torus, giving $2^4=16$ involution fixed points. 

We also obtain $g_s \approx 0.53 < 1$, which is not very small, and from which one may question the string loop corrections. More dramatically, one may be bothered by some $r_a/l_s$ that are not smaller than 1. The search of \cite{Andriot:2024cct} actually did not manage to do better on these values. As discussed there, it is not clear why this could not be improved: either an unknown no-go theorem prevents it or the (advanced) numerical tools face too many limitations. There are however two reasons not to be worried about $r_3/l_s$ and $r_6/l_s$. First, these 6d lengths, appearing as normalisations in the corresponding $e^a$, are not 1-cycle volumes (or circle radii). Indeed, we have $\d e^3 \neq 0$, $\d e^6 \neq 0$; there is thus no associated winding mode. Second, as we will argue in the following, they do not govern the size of $\alpha'$-corrections; in other words maybe one should not require these ratios to be greater than 1 to start with.

This example provides a first illustration of tests of classicality, with its load of technical and numerical difficulties. While the conclusion may not be completely satisfying, we will come back to this specific solution because it offers in addition some parametric control on its classicality, as motivated above. While this property was found in \cite{Andriot:2024cct}, let us first review previous results in the literature.\\

The first classicality tests of de Sitter supergravity solutions were only performed in 2018 after the dSC \eqref{dSconj} of \cite{Obied:2018sgi}, as discussed at the end of Section \ref{sec:nodSQG} (except for one prior solution example in \cite[Sec.6]{Danielsson:2011au}). Those essentially followed the strategy outlined: considering solutions with quantized fluxes, sources and structure constants, or at least minimal bounds for those (e.g.~if a quantity is an integer, then it is larger than 1), one looks whether $1/g_s$ and the 6d volume can be made large. To that end, one may look for possible scaling symmetries, where a parameter $\gamma$ can be found, which maintains the adequate quantization conditions while making $1/g_s$ and the volume arbitrarily large. This relates to the parametric control discussed above, and the ``asymptotic classical regime'' \eqref{asymptoticclass}. The conclusion of \cite{Roupec:2018mbn, Banlaki:2018ayh}, which studied the type IIA solutions of the class $s_{6666}$, was that such a scaling cannot be found, without violating the upper bound on the number of $O_6$; similar obstructions were found in \cite{Junghans:2018gdb}. While these first results are interesting and hint at an impossibility of a parametric control, and possibly an absence of classical de Sitter solutions, we can mention already one possible loophole: anisotropy.

There could be ``anisotropic'' solutions, where the 6d geometry has inhomogeneous lengths; for example, the various $r_a$ could scale in different fashions with a parameter $\gamma$, making some $r_a$ much larger than others. In that case, the volume, as the product of the $r_a$, would still be made large by the scalings. However, the scaling of the 4d potential terms, or of the 10d solution contributions, with respect to the 6d volume is then different than the one with respect to individual lengths $r_a$. The previous conclusions may then be altered. Having first investigated scalings of the volume and the dilaton, essentially in the class $s_{6666}$, \cite{Junghans:2018gdb} then turned to the scaling of an arbitrary field, for example that of one $r_a$. Upon few assumptions we will come back to, the conclusion of \cite{Junghans:2018gdb} remained the same: it was claimed to be {\sl impossible to obtain a parametric control on classicality} for de Sitter solutions.

Related results were obtained in \cite{Andriot:2019wrs} for de Sitter solutions in type IIA/B supergravity with sources of single dimensionality $p=4$, $5$ or $6$: it was then shown that $1/g_s$ and the volume could not be made arbitrarily large, unless $|{\cal R}_6|$ takes specifically small values, for example thanks to an anisotropy, i.e.~internal length hierarchies. The argument of \cite{Andriot:2019wrs} can be phrased as follows, using the $(\rho,\tau)$ theory \eqref{rhotautheory}, here for a single $p$. From $\del_{\tau}V=0$, one derives
\beq
-2 ({\cal R}_6 \times \rho) \leq 3 g_s \tau^{-1} \rho^{p-\frac{11}{2}} \left( \sum_s \frac{T_{10}^s}{p+1} \times \rho^{\frac{9-p}{2}} \right) \ .
\eeq
Considering homogeneous lengths $L/ l_s \sim \rho^{1/2}$, the argument uses the fact that the parentheses on the right is bounded from above by the number of $O_p$. The right-hand side then gets arbitrary small in the asymptotic classical limit \eqref{asymptoticclass} for $p\leq 6$. The left-hand side captures a dimensionless curvature number, and has to be positive for de Sitter. So unless this curvature number can also be arbitrarily small, a point we discuss at the end of Appendix \ref{ap:groupman} for group manifolds, one faces an obstruction. Such difficulty was already mentioned in \cite{Junghans:2018gdb, Banlaki:2018ayh} for $p=6$.\footnote{Similar reasonings with isotropic compactifications also lead to bounds on flux numbers, and can relate the limit of $\rho$ or $L$ to that of $g_s$ \cite{Banlaki:2018ayh, Andriot:2019wrs}.} A conjecture was also proposed to capture the difficulties observed. \cite[Conjecture 3]{Andriot:2019wrs} was phrased as follows: {\sl Classical de Sitter solutions cannot have at the same time a large internal volume, a small string coupling, a bounded number of orientifolds and quantized fluxes}. Note that compactness, meaning here the lattice conditions, are not explicitly mentioned.

More works studied this question, as we now briefly mention. Similar results to the above were obtained in \cite{Grimm:2019ixq} for type IIA/B compactifications on Calabi-Yau manifolds. The results also go beyond this perturbative framework, to F- or M-theory compactifications, in asymptotic limits in some fields. The conclusion is again that there is no de Sitter solution at parametric control. Similar conclusions, for an isotropic compactification, were reached recently in \cite{Tringas:2025uyg} using the $(\rho,\tau)$ theory \eqref{rhotautheory} in various dimensions $d\geq 3$, comparing as well to the question of scale separation. Difficulties were also pointed-out in the earlier work \cite{Andriot:2020vlg}, for having simultaneously parametric control on classicality and on scale separation for a de Sitter solution. Finally, the question of parametric control with weak coupling and large volume was also addressed in the case de Sitter solutions with Casimir energy (see Section \ref{sec:listCas}) in \cite[Sec.3.4.1]{Parameswaran:2024mrc}, where again a 6d anisotropy was shown to be a necessary condition.

To summarize, many works have pointed-out in some framework the impossibility of getting a classical de Sitter solution with parametric control with a large volume and a small string coupling: this means that de Sitter solutions found in 10d supergravity are unlikely to be found with an arbitrarily large volume and $1/g_s$, which would have allowed to neglect to any desired approximation the corresponding corrections, without computing them in detail. Two possible ways-out were however mentioned: first, one may give up on a parametric control, for a numerical control, and second, one may give up on a parametric control through the volume, in favor of a distinguished parametric control on each internal length, the latter implying 6d anisotropy. That second option may still suffer from an obstruction obtained in \cite{Junghans:2018gdb}, but we will present in the following an example that circumvents it.\\

Before doing so, we make a few comments. First, having a parametric control or a scaling symmetry in a de Sitter solution can be viewed via 10d supergravity fields, but also through a 4d potential. At the de Sitter 4d extremum, the scalar field values, e.g.~$\rho_0$, are fixed in terms of the potential terms, which correspond to 10d supergravity fields, e.g.~a flux integer number $N_{\omega}$. Scaling or changing the control parameter in order e.g.~to get the de Sitter solution at larger volume means changing $\rho_0$, through a change of e.g.~$N_{\omega}$. We would like to clarify that this is different from having a fixed scalar potential $V$, with $N_{\omega}$ fixed, and looking at the possible presence of extrema at large $\rho$, and further concluding on their absence in the asymptotics because of a runaway. In other words, verifying the SdSC \eqref{SdSC} is a different task than verifying whether a de Sitter extremum can be moved at large volume. In the second case only, one may allow for varying and making large flux integers $N_{\omega}$ (so-called unbounded fluxes, not tied to a tadpole condition giving them an upper bound). Finding a classical de Sitter solution at (arbitrarily) large volume, by changing in this way the 10d fields or the 4d potential, would then not invalidate the SdSC \eqref{SdSC}; the example displayed below may fall in this category.

Another possible confusion with parametric control is the distinction between classicality and scale separation. We have explained above what is meant by a classical solution, while scale separation was presented at the beginning of Section \ref{sec:quint}, and refers to an energy gap between e.g.~the cosmological constant and the Kaluza--Klein scale. The two requirements are a priori distinct, but both may or may not be achieved with parametric control. A possible source of confusion is the DGKT anti-de Sitter solution \cite{DeWolfe:2005uu, Camara:2005dc}, for which both are realised thanks to the same parameter, an unbounded flux integer. But there are also counter-examples to such a situation: the scale separated anti-de Sitter solutions belonging to the class $m_{5577}$ of \cite{Caviezel:2008ik, Petrini:2013ika} were argued to be non-classical in \cite{Cribiori:2021djm}. For de Sitter, although examples barely exist, (parametric) classicality and scale separation would tend not to be realised together \cite{Andriot:2020vlg, Tringas:2025uyg} (contrary to the claim in \cite[Sec.5]{Andriot:2019wrs}, that uses an isotropic ${\cal R}_6$): this is illustrated in \cite{Andriot:2024cct} for the solution example to be presented below. 

Finally, let us comment on numerical control. Using the expression \eqref{dimred1} for the 4d reduced Planck mass, with $M_p=1/l_p$, and introducing an average 6d length $L$, corresponding to the volume, as $\int \d^6 y \sqrt{|g_6^0|} = (2\pi L)^6$, one obtains the expression
\beq
\frac{L}{l_s} = \left( \sqrt{\pi}\, g_s\, \frac{L}{l_p} \right)^\frac{1}{4} <  10^4 \ , \quad g_s > 10^{-16} \ .
\eeq
The upper bound on $L/l_s$ comes from asking that the Kaluza--Klein (KK) scale, here naively $1/L$, is larger than the LHC energy,\footnote{This comparison to the LHC energy may be alleviated by considering different models, where only gravity propagates in extra dimensions, i.e.~the visible sector of particle physics does not. What we consider here might be viewed as a conservative bound. However, considering instead a higher energy scale from the early universe, e.g.~that of inflation, could be even more restrictive \cite{Parameswaran:2016fqr}.} i.e.~$L/l_p < 10^{15}$, and that we are in a perturbative regime with $g_s<1$. The lower bound on $g_s$ comes from requiring $L/l_s > 1$. The upper bound on $L/l_s$ may even be lowered by asking that the KK scale is higher and $g_s$ smaller; in turn this would raise the lower bound on $g_s$. The KK scale can still be made fairly high (e.g.~$10^{10}$ times the LHC scale). The limiting point here is rather that $L/l_s$ cannot be very large. This emphasizes that the low-energy approximation, and thus the classical regime, can get important bounds from observations, in conflict with a parametric control that would bring $L/l_s$ to asymptotic values. From this perspective, having only a numerical control is adequate.\\

We end the section by presenting an example of an anisotropic scaling, to control classicality. As shown in \cite[Sec.4.1]{Andriot:2024cct}, specific solutions among the class $s_{55}$, such as $s_{55}^+ 14, 29$, possess an interesting family of ``$\gamma$-scaling''. First, modulo discretization of the continuous parameter $\gamma\geq 1$, this scaling respects the five requirements \eqref{classrequir}: for example, some unbounded integers only get larger, not smaller, some lengths $r_a$ also get larger, while lattice quantization conditions remain invariant. Second, the scaling maps a solution to another one, as its sole effect on the supergravity fields is exactly a $\lambda$-scaling \eqref{lambdascaling}. For example, one gets
\beq
F_{3\, \omega_1} \rightarrow \frac{1}{\gamma} F_{3\, \omega_1}  \ ,\quad f^1{}_{64} \rightarrow \frac{1}{\gamma} f^1{}_{64} \ ,\quad T_{10}^1 \rightarrow \frac{1}{\gamma^2} T_{10}^1 \ , \quad \text{etc.} \label{gammascalingsugra}
\eeq  
Among this family of $\gamma$-scaling, one simple example is
\beq
r_{1,4,5} \rightarrow \gamma\ r_{1,4,5} \ ,\ N_{\omega_1} \rightarrow \gamma\, N_{\omega_1} \ , \ N_1 \rightarrow \gamma^{-1} N_1 \ , \ N_6 \rightarrow \gamma\, N_6 \ . \label{gammascaling}
\eeq
One can verify the result on the supergravity fields \eqref{var3} as being a $\lambda$-scaling as in \eqref{gammascalingsugra}. One also verifies from \eqref{gammascaling} that the classicality requirements are indeed met, provided $\gamma$ gets discretized to preserve flux quantization on $N_{\omega_1}$. This scaling can then be seen as that of an unbounded flux number, here of an harmonic part of $F_3$. The scaling also acts on the continuous parameters $N_1, N_6$, while respecting the lattice quantization condition, namely here the toroidal topology. This is possible thanks to having a solvmanifold, instead of a nilmanifold, as explained in Appendix \ref{ap:groupman}. We also explain there, and in \cite[Sec.4.3]{Andriot:2024cct}, how the Ricci scalar and its scaling behaviour differs between nil- and solvmanifolds, offering a loophole to the no-go theorem of \cite{Junghans:2018gdb}, that otherwise obstructs parametric control on classicality.

An important observation is that the scaling is anisotropic: only some $r_a$ grow. More general versions in the family of $\gamma$-scaling leave only $r_3, r_6$ invariant. The compactification is then anisotropic, as advocated above. In a 4d potential, one should thus consider the scaling of each term by looking at the dependence on the individual lengths (possibly identified with diagonal metric components), rather than looking at that of an average length $L$ or volume. However, \cite{Andriot:2024cct} failed to obtain a realisation of the scaling on the 4d fields and potential, preventing us from a clear picture there.

Regarding corrections, it should be noted that $g_s$ is left invariant by this scaling. One is then stuck with the value $g_s \approx 0.53$, which may or may not be satisfying. One could talk of (at best) a numerical control on string loop corrections. However, one gets a parametric control on $\alpha'$-corrections as we now explain. The scaling on the supergravity fields \eqref{gammascalingsugra} results in a $\lambda$-scaling, and diminishes each derivative contribution. One then creates a tunable hierarchy with higher-derivative terms which appear as $\alpha'$-corrections, provided those are expressed in terms of the standard supergravity fields; the Riemann tensor, if built from structure constants, should scale in the same way. In other words, the scaling of the various quantities precisely combines in such a way as to make $\alpha'$-corrections negligible; the possibly worrisome lengths, $r_3, r_6$, that are small and do not scale, never appear alone in those corrections, but always combined with other quantities that scale down the corrections. 

This $\gamma$-scaling is an example where an anisotropic scaling, combined with an unbounded flux and a specific geometry, allows for some parametric control on classicality (except for $g_s$). While a more detailed study of this solution and its corrections is lacking for now, it offers some hope of finding a classical de Sitter solution. We recall however that (no parametric) scale separation may appear as an independent issue. As we will discuss in Section \ref{sec:backreact}, the backreaction of sources, ignored so far, may also be problematic.

\subsection{Backreaction and localized sources}\label{sec:backreact}

In Section \ref{sec:dSsol}, we have so far considered supergravity solutions with ``smeared sources''. Our main ansatz for 10d supergravity solutions with $O_p/D_p$ sources was introduced in Section \ref{sec:10dansatz}, and completed in Section \ref{sec:groupmandimred} when focusing on 6d group manifolds. This ansatz fulfils the ``smeared approximation'', defined and discussed around  \eqref{smearing}. It implies that the warp factor and the dilaton are constant. The warp factor is meant to capture sources backreaction on the 10d metric: the smearing can then be viewed as considering an averaged, or integrated, solution. The same interpretation holds for the $\delta$-function localizing sources in their transverse dimensions: those are traded for their integral in this approximation. We refer to Section \ref{sec:10dansatz} for more detail. In this section, we now examine the possibility of going beyond this smeared approximation and having solutions with backreacted and localized sources. At first sight, those would appear as true and complete supergravity solutions. There are some loopholes to such a claim, namely the string corrections and the orientifold hole, that we will come back to. In this section, we investigate whether such solutions can found, keeping an eye towards the de Sitter solutions on group manifolds discussed in Section \ref{sec:dSsolgroupman}.

To provide further motivation, let us mention few related works. A first question to be asked, of relevance to the supergravity de Sitter solutions discussed previously, is the following: given a smeared solution, does a localized version exist?  As we will argue, there are several examples with parallel $O_p/D_p$ sources, i.e.~transverse to the same directions, or equivalently in one set $I$, for which the smeared solution can be localized (see e.g.~\cite{Baines:2020dmu} with parallel $O_6$ and $F_0$). However, early works on this localization question \cite{Blaback:2010sj, Junghans:2013xza} showed that a local version of the solution does not always exist. Another important question is whether, and how, the 4d effective theory gets modified by a non-trivial warp factor: early works discussing this point include \cite{Giddings:2005ff, Shiu:2008ry, Douglas:2008jx, Frey:2008xw, Martucci:2009sf, Martucci:2014ska}; a recent one is \cite[App.B.2]{Andriot:2023fss}.

Other relevant works are related to the search for de Sitter solutions. In \cite{Burgess:2011rv, Gautason:2013zw}, it was shown that the 4d curvature or cosmological constant can be related and expressed in terms of boundary conditions of certain supergravity fields close to an $O_p/D_p$ source. Probing this ``near-brane'' behaviour of fields requires the source to be localized and backreacted; in turn this backreaction could play a role in the value of the cosmological constant, an important point when looking for de Sitter solutions. Also, new mathematical tools were recently used in \cite{Luca:2022inb, DeLuca:2023kjj} to describe warped compactifications, especially due to $O_p/D_p$ sources, in presence of a cosmological constant, with various applications (e.g.~scale separation and Laplacian spectrum). Finally, let us also point-out some constraints on the existence of de Sitter solutions, proposed in \cite{Das:2019vnx} (see also \cite{Faruk:2024usu}) in a warped and localized setting.\footnote{Such results are reminiscent, even though different, of a no-go theorem proposed against supergravity de Sitter solutions with localized sources in \cite{Dasgupta:2014pma}. The latter was argued to be too strong in \cite[App.A]{Junghans:2016uvg}.} Backreacted solutions may thus carry important contributions regarding de Sitter.

With these motivations in mind, we first review here some basics of the backreaction, and then present some difficulties, in particular the orientifold hole and having intersecting sources. Those are encountered when trying to localize the de Sitter supergravity solutions on group manifolds. More work exist on the backreaction and boundary conditions close to localized sources, for different supergravity de Sitter solutions: for this we refer to Section \ref{sec:listbackreact}, even though some of the questions tackled here are also relevant to those.\\

Let us start by recalling the $p$-brane solution of supergravity in asymptotic Minkowski spacetime (eventually corresponding to the $D_p$-brane in 10d string theory). Early and review references include \cite{Duff:1993ye, Youm:1997hw, Johnson:2000ch}, and we follow here \cite{Andriot:2019hay}. The $p$-brane solution in 10d string frame admits the following metric
\beq
\d s_{10}^2 = H^{-\frac{1}{2}}\, \left( \eta_{\mu\nu} \d x^{\mu} \d x^{\nu} + \delta_{m_{||}n_{||}} \d y^{m_{||}} \d y^{n_{||}} \right) + H^{\frac{1}{2}}\, \delta_{m_{\bot}n_{\bot}} \d y^{m_{\bot}} \d y^{n_{\bot}} \ , \label{10dmetricwarp}
\eeq
where we split the $p+1$ world-volume direction into the 4d and parallel 6d ones, anticipating compactification, in agreement with the 10d metric \eqref{10dmetricmaxsym}. We identify the warp factor as $e^A = H^{-\frac{1}{4}}$, where this terminology sometimes also refers to $H$ itself; $e^A$ or $H$ depend only on the transverse coordinates $y^{m_{\bot}}$. The solution also has a varying dilaton $e^{\phi} = e^{\phi_0}\, H^{-\frac{p-3}{4}}$, with a constant $\phi_0$, and an electric gauge potential $C_{p+1}$ that depends on $H$. The equation of motion for the latter, or the Bianchi identity for its magnetic flux, leads to the following equation on the warp factor
\beq
\delta^{m_{\bot}n_{\bot}} \del_{m_{\bot}} \del_{n_{\bot}} H = \frac{Q}{\sqrt{g_{\bot}}} \, \delta (\overrightarrow{y}_{\!\!\bot}) \ , \label{warpeq}
\eeq
for a $p$-brane located at $y^{m_{\bot}}=0$ in its transverse space, denoted collectively $\overrightarrow{y}_{\!\!\bot}=\overrightarrow{0}$. Anticipating on a compactification setting, we introduced the (unwarped) volume factor of that transverse space, $\sqrt{g_{\bot}}$, which is simply $1$ here in this solution since $g_{m_{\bot}n_{\bot}}=\delta_{m_{\bot}n_{\bot}}$. One normalises $\int_{\bot} \d^{\bot} y\, \delta (\overrightarrow{y}_{\!\!\bot}) = \int_{\bot} {\rm vol}_{\bot}\, \frac{1}{\sqrt{g_{\bot}}}\, \delta (\overrightarrow{y}_{\!\!\bot}) = 1$. Finally, $Q$ is the $p$-brane charge, that gets adjusted to the string theory value when matching the $D_p$-brane, namely $Q_{D_p}= -2\kappa_{10}^2 T_p g_s = -(2\pi l_s)^{7-p} g_s$. Overall, the right-hand side of \eqref{warpeq} is nothing but $g_s \frac{T_{10}^s}{p+1}$ in our notations \eqref{T10s}. From equation \eqref{warpeq}, we read that the solution is $H = Q\, G + H_0$, where $G$ is the Green's function for the Laplacian in the transverse flat space, and $H_0$ a constant. For $n$ transverse dimensions, and a transverse radius $|\overrightarrow{y}_{\!\!\bot}|$, we recall that
\beq
n\geq 3:\ G \sim -\frac{1}{|\overrightarrow{y}_{\!\!\bot}|^{n-2}} \ ,\quad n=2:\ G \sim \ln |\overrightarrow{y}_{\!\!\bot}| \ ,\quad n=1:\ G \sim |\overrightarrow{y}_{\!\!\bot}| \ ,\label{Greenflat}
\eeq
up to positive factors and possible additional constants. This makes the definition of co-dimension 2 and 1 branes delicate ($D_7$ and $D_8$ in 10d), given their asymptotics, while objects with co-dimension 3 and higher have a more straightforward interpretation. For the latter, we clearly see that away from the source, $H$ weakens, and is eventually just the constant $H_0$ in the asymptotics $|\overrightarrow{y}_{\!\!\bot}| \rightarrow \infty$. This is why having a constant warp factor, as in the smeared approximation, can be understood has having no backreaction: this corresponds here as being in the asymptotic Minkowski spacetime, where the metric is undeformed by the presence of an object. In turn, this gives a constant dilaton.

We turn to the case of interest: the 6d manifold is compact. Considering parallel $O_p/D_p$ sources, located at points $\overrightarrow{y}_{\!\!i}$ in their common transverse compact space, we can write a 10d string frame metric similar to \eqref{10dmetricwarp}: the unwarped parallel and transverse metrics can be traded for more general ones, but we keep the inverse powers of $H$ along the different directions, namely
\beq
\d s_{10}^2 = H^{-\frac{1}{2}}\, \left( \d \tilde{s}_4^2 + \d \tilde{s}_{6||}^2 \right) + H^{\frac{1}{2}}\, \d \tilde{s}_{6\bot}^2 \ . \label{4+6dmetricwarp}
\eeq
In typical solutions, the type IIA/B (magnetic) flux BI \eqref{BI} gives rise to the following equation, similarly to \eqref{warpeq}
\beq
\Delta_{\bot} H + C = \frac{1}{\sqrt{g_{\bot}}} \sum_i Q_i \, \delta(\overrightarrow{y}_{\!\!\bot}-\overrightarrow{y}_{\!\!i}) \ , \label{DeltaH}
\eeq
where the Laplacian $\Delta_{\bot}$ and the determinant $g_{\bot}$ refer to the unwarped transverse metric $\d \tilde{s}_{6\bot}^2$. Explicit examples of such solutions in compactifications to 4d Minkowski can be found in \cite{Giddings:2001yu, Grana:2006kf, Andriot:2016ufg}. The constant $C$ comes from extra fluxes in the BI ($H\w F_{6-p}$) or from the connection term on the flux (or structure constant term for a group manifold). With a compact transverse space being a manifold without boundary, and with a globally defined function $H$, integrating \eqref{DeltaH} over it gives $C \times \int_{\bot} {\rm vol}_{\bot} = \sum_i Q_i$.\footnote{Having $C \neq 0$ corresponds to what is known as having a tadpole and canceling it by extra fluxes, while $C=0$ indicates a cancelation of source charges. In other words, on a compact space, having one source alone is not possible: one needs either another source of opposite charge, or some extra flux.} This integration corresponds precisely to the smeared approximation \eqref{smearing}. Indeed, taking a constant $H= H_0$ and $\delta(\cdot) \rightarrow 1$ in \eqref{DeltaH} amounts precisely to considering this integrated equation, up to the adequate volume normalisation. This integration, or smearing, gives a solution with non-localized sources. On the other hand, the localized/backreacted solution is given again by $H =\sum_i Q_i\, G + H_0$, where this time $G$ is the generalized Green's function for the Laplacian on the compact transverse space. Finding the latter can be difficult, but the case of a warp factor on a transverse torus has been treated in \cite{Andriot:2019hay}.\\

From this brief recap, there are two interesting lessons for the question of localizing a smeared solution. First, in the case of parallel $O_p/D_p$ sources, we read from the metric that the only difference between smeared (unwarped) and backreacted (warped) is a multiplication by a power of the warp factor. This leads to the suspicion, verified in some supersymmetric type II Minkowski solutions \cite{Grana:2006kf, Blaback:2010sj}, that a local solution can be obtained from a smeared one by the simple operation on the coframe 1-forms $e^{a_{||}} \rightarrow e^A \, e^{a_{||}}$, $e^{a_{\bot}} \rightarrow e^{-A} \, e^{a_{\bot}}$, together with $\del_{a_{||}} A=0$. Such a rule could be very useful for group manifold compactifications. This should go together with $g_s \rightarrow g_s\, e^{A(p-3)}$ for $e^{\phi}$, while one also needs to add some pure derivatives of $A$, e.g.~as a component in the sourced flux $F_{8-p}$. To check this idea in general, one should start from a warped ansatz, and compare the 10d equations to the unwarped ones: the only difference should be \eqref{DeltaH}. This was carried-out to some extend in \cite{Andriot:2016xvq}; see also the treatment of a class of Minkowski solutions in \cite{Andriot:2016ufg}. This simple rule offers hope of an easy localization of smeared solutions, at least on group manifolds. A different ansatz for warped de Sitter solutions with parallel sources, in particular one where the dilaton is independent of the warp factor, was however proposed in \cite{Cordova:2018dbb}, as commented on in Section \ref{sec:listbackreact}; this proposal followed warped solutions studied on Minkowski \cite{Macpherson:2016xwk} and anti-de Sitter \cite{Cordova:2018eba}. The standard warp factor dependence presented above may therefore not be the only option.

\begin{figure}[ht!]
\begin{center}
\includegraphics[width=0.6\textwidth]{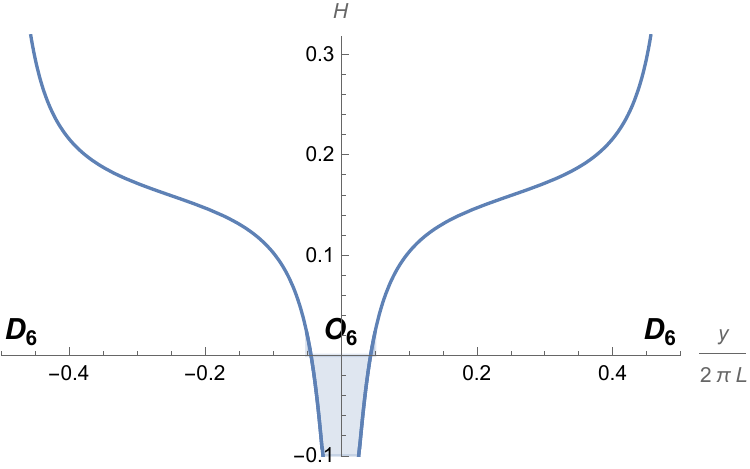}
\caption{Warp factor $H$ (normalised) along a transverse periodic coordinate $\sigma=\frac{y}{2\pi L}$, for an $O_6$ located at $\sigma=0$ and 2 $D_6$ located at $\sigma=\pm \frac12$, with a transverse $T^3$. The shaded region around $\sigma=0$ corresponds to the orientifold hole, where $H<0$. The figure is adapted from \cite[Fig.2a]{Andriot:2019hay}.}\label{fig:warp}
\end{center}
\end{figure}

A second lesson is the backreaction of orientifold planes. The previous solutions apply as well to $O_p$, with the charge $Q_{O_p}= -2^{p-5} Q_{D_p} > 0$. For one of those, one then gets $H_{O_p}=H_0+ Q_{O_p} G$, i.e.~for $p\leq 6$ in 10d
\beq
H_{O_p} = H_0 - 2^{p-5}\, c_0 \, g_s \, \frac{(2\pi l_s)^{7-p}}{|\overrightarrow{y}_{\!\!\bot}|^{7-p}} \ , \label{HOp}
\eeq
where $c_0$ is the positive missing numerical factor in \eqref{Greenflat}. From the metric \eqref{10dmetricwarp} or \eqref{4+6dmetricwarp}, it is understood that $H>0$, and with an asymptotic Minkowski spacetime, one would take $H_0>0$. However, it is clear from \eqref{HOp} that there exist a finite radius $r_H$ away from the source at which $H_{O_p}=0$. Closer to the $O_p$, one gets $H_{O_p}<0$ (see Figure \ref{fig:warp}), which gives an ill-defined (complex!) metric. We may refer to the location of this specific radius $r_H$ as a horizon, or as the size of the orientifold hole. For the asymptotic Minkowski solution, a naive estimate gives this size $r_H$ to be of order $g_s^{1/(7-p)}\, l_s$. It is well-known that the backreaction close to the $O_p$ is problematic due to the hole, but it is generally believed to be an artefact of supergravity, and thought to be cured by string theory corrections when getting close to the source, in agreement with the effect being relevant at distance of order $g_s^{1/(7-p)}\, l_s$. Another argument is the existence of an M-theory uplift of $O_6$ as a smooth Atiyah-Hitchin geometry. In the case when the latter is not possible, due to a Romans mass $F_0 \neq 0$, it has been argued \cite{Saracco:2012wc} that there is actually no problematic region around a single $O_6$ (see however \cite{Baines:2020dmu}); the case of multiple intersecting $O_6$ could be different though, and will be discussed below.

However, when turning to solutions on compact manifolds, the size of the hole can be different. We discussed this matter already when referring to the singular bulk problem \cite{Gao:2020xqh} in Section \ref{sec:listnonpert} and \ref{sec:listcrit}: this problem is that the size of the hole is large, meaning comparable to the overall size of the manifold. In that case, the supergravity description cannot be trusted in a large part of the space, which is problematic for the overall consistency. In a compact space, when getting very close to a source, one may approximate locally the metric by a (warped) flat space. As a consequence, the behaviour of $H$ close to the source should be the same as the asymptotic Minkowski solution, namely given by the Green's functions \eqref{Greenflat}. A non-trivial analytical check of this behaviour was done in \cite[App.A]{Andriot:2019hay}, for the warp factor on a torus and using generalized Green's functions. However, there is no asymptotic limit in a compact space: going away from the source, the warp factor changes and eventually adapts to become globally defined when going around the manifold. As a consequence, the same problematic behaviour is present close to an $O_p$ on a compact manifold, but the size of the hole may differ on a compact space, possibly being large and leading to the singular bulk problem. We will come back to this point for de Sitter solutions. Note that $D_p$ do not face such a hole: the different sign of $Q_{D_p}$ prevents it, but it makes $H$ diverge. We illustrate the $O_p$ and $D_p$ backreactions in Figure \ref{fig:warp}. The divergence of $H$ near the brane may also appear as a pathology in the metric, which also calls for string corrections close to $D_p$. From this perspective, finding solutions of supergravity with localized sources may even be disputed since string theory is in any case necessary to get a complete description. Here we would aim at having backreacted solutions everywhere except in a small stringy region around sources; we will see whether this is possible.

We now come back to the smeared de Sitter solutions of type II supergravities found on 6d group manifolds, discussed in Section \ref{sec:dSsolgroupman}. When trying to apply the above knowledge to localize these smeared solutions, we face an immediate issue. As emphasized in Section \ref{sec:dSsolexistence}, the de Sitter solutions found have (at least 3) intersecting sets of sources. In other words, the $O_p/D_p$ are not parallel, so the simple metric \eqref{4+6dmetricwarp} does not apply. We now need to find a metric, or warp factors, which capture backreactions of each (or all) set of parallel sources, together with their intersection; or at least such a solution on a substantial part of the compact manifold, up to small regions requiring a stringy description. This is actually a longstanding supergravity problem \cite{Lu:1997mi, Youm:1999ti, Cvetic:2000cj, Arapoglu:2003ah}, which is not related to de Sitter: it can already be discussed in 10d asymptotic Minkowski, where typical attempts require a partial smearing of sources. A review can be found in \cite{Smith:2002wn}. We should mention very specific anti-de Sitter examples, where localized solutions were achieved \cite{Assel:2011xz, Assel:2012cj, Rota:2015aoa}, but in general, the problem of backreaction of intersecting sources remains unsolved, as recently revisited in \cite{Bardzell:2024anh, Lin:2025ucv}. This stands as an obstacle in localizing the de Sitter solutions of interest. We should recall that examples of intersecting brane configurations, in non-backreacted (probe brane) regimes, are plenty in the literature, either in particle physics models or in a holographic framework. The question here is therefore not so much the existence of such configurations, but rather their description in a supergravity regime.

In this situation of multiple intersecting $O_p/D_p$ sources, the ansatz \eqref{10dmetricwarp} or \eqref{4+6dmetricwarp} for the metric is not suited. Generalizations of it have been considered, following the rule of the inverse powers of warp factor along parallel and transverse dimensions for each source set, without necessarily providing satisfying solutions, as just discussed. In our situation with a compactification, all $O_p/D_p$ are space-filling in 4d, so they share the 4 external dimensions in their world-volume. As a consequence, an alternative or minimal ansatz is the 10d string frame metric \eqref{10dmetricmaxsym}, that we repeat here for convenience
\beq
\d s_{10}^2 = e^{2A}\, \d \tilde{s}_4^2 + \d s_{6}^2   = e^{2A(y)} \, \tilde{g}_{\mu \nu}(x) \d x^{\mu} \d x^{\nu} + g_{mn}(y) \d y^m \d y^n \ .
\eeq
There, a common warp factor dependence, $e^{2A}$, is only considered along 4d. The internal metric $g_{mn}(y)$ may still capture backreaction in the form of warp factor(s), but this does not need to be specified. Many works, including those to be discussed in the following, have thus used \eqref{10dmetricmaxsym} as the backreacted 10d metric. The resulting 10d supergravity equations are listed in Appendix \ref{ap:warpeq}.

Recently, a different approach to the question of backreaction was considered, in the context of anti-de Sitter solutions with parametric control on classicality and scale separation. Those refer to the DGKT solution \cite{DeWolfe:2005uu, Camara:2005dc} that we discussed already in Section \ref{sec:dSsolclassical}. A simple instance of this solution is found in type IIA supergravity with $O_6$-planes on a torus orbifold. The $O_6$ can be understood as intersecting, and one may wonder about their backreaction. The idea developed in \cite{Junghans:2020acz, Marchesano:2020qvg} is to make use of the control parameter $N$ present in the smeared solution to discuss the backreaction. As mentioned, this solution admits a parameter $N$, related to an unbounded flux number, which allows at the same time to ensure classicality (in particular a small $g_s$ and large 6d radii) and scale separation of the anti-de Sitter solution. As can be seen in \eqref{HOp}, the non-constant (non-smeared) part of the warp factor for the $O_6$ depends on $g_s\times l_s/L$, where $L$ is a typical radius in the transverse torus. And it turns out, in the DGKT solution, that $g_s\times l_s/L \sim N^{-1}$, meaning that the parameter $N$ can adjust at will $g_s\times l_s/L$ to be small, as in a classical limit. Then this non-smeared piece of the warp factor can be made parametrically small (as long as we are not too close to the $O_6$ location, and in particular, away from the hole). It is precisely this mechanism which is used in \cite{Junghans:2020acz}: the warp factor $e^{A}$, and all fields accordingly, are written as an expansion in the control parameter $N$. The zeroth order is the smeared solution, and the first order is a backreaction correction with a non-trivial warp factor. It should be noted that at this expansion level, the intersection itself is not captured. This explains the possibility of solving the equations using above Green's function, as for parallel sources. Many more examples of backreacted anti-de Sitter solutions have since then been discussed, including e.g.~\cite{Cribiori:2021djm, Marchesano:2022rpr}. Given the corrections are by definition small here, one does not expect major qualitative changes to the solution. One may even wonder the size of these (classical, supergravity) corrections, compared to e.g.~$\alpha'$-corrections, as discussed in \cite{Junghans:2020acz, Andriot:2023fss}.\\

Inspired by this approach, and given the difficulties to proceed otherwise with intersecting sources, one could consider mimicking it to get a first evaluation of the backreaction of intersecting $O_p/D_p$ sources in de Sitter solutions. The key ingredient that allowed to account for backreaction in the anti-de Sitter solutions was the free parameter, that provided classicality and scale separation in those examples. Interestingly, as reviewed around \eqref{gammascaling}, we have one example of a smeared supergravity de Sitter solution on a group manifold, with intersecting sources, where we have one such parameter at hand. This solution is $s_{55}^+29$ of \cite{Andriot:2024cct}, and the free parameter $\gamma$ allows for a (partial) control on classicality of the solution, ensuring in particular a control on $\alpha'$-corrections. Would it then be possible to expand the warp factor, the dilaton and all supergravity fields in perturbative powers of $\gamma$, to get first order corrections from the backreaction, analogously to the above? This has not been attempted explicitly.

However, a general discussion of the backreaction in de Sitter solutions was carried-out in \cite[Sec.4]{Junghans:2023lpo}. This backreaction was related to the question of the size of the orientifold hole, and the singular bulk problem. There is a reason to relate the two, as we now explain. When being close to an $O_p$, $p\leq 6$ in 10d, we argued above that we can write the warp factor, up to an overall normalisation, as
\beq
H_{O_p} \simeq 1 - \left(\frac{r_H}{r_\bot}\right)^{7-p} \ ,
\eeq
with $r_\bot$ the transverse radius coordinate, and $r_H$ a constant, which is nothing but the size of the hole. If we introduce again $L$ the typical radius of transverse dimensions, the ratio $r_H/L$ allows to compare the size of the hole to that of the (transverse) space in which the hole lives, and measures the singular problem. At the same time, we argued that the first corrections from the backreaction to the smeared solutions came from the non-constant term in $H_{O_p}$, while staying away from the hole. The size of these corrections is thus governed by $r_H/L$, which should correspond to the free parameter in the expansion: making it small allows to have a perturbative expansion. Therefore, we see that having a parameter making backreaction corrections small, as in the anti-de Sitter example, is directly related to having tunably small hole that avoids the singular bulk problem. In \cite{Junghans:2023lpo} a more precise estimate is proposed for the non-constant part of the backreaction, that is
\beq
\left(\frac{r_H}{L}\right)^{7-p} \sim g_s\, (2\pi l_s)^{7-p}\, \frac{{\cal D}^2}{vol_{\bot}} \ ,
\eeq
where ${\cal D}$ is the diameter of the transverse space, that is its largest length. Considering the solution $s_{55}^+29$ with the $\gamma$-scaling \eqref{gammascaling}, and applying the above to the set $I=1$ of $O_5$ along direction 12, we get\footnote{I am indebted to D.~Junghans for related private exchanges.}
\beq
g_s\, (2\pi l_s)^{7-p}\, \frac{{\cal D}^2}{vol_{\bot}} \sim g_s\, l_s^2\, \frac{(r_5)^2}{r_3\, r_4\,r_5\,r_6} \sim \gamma^0 \ .
\eeq
The same result is obtained for the set $I=2$ along 34. In this example, and using the formula from \cite{Junghans:2023lpo}, the conclusion is that the backreaction estimate does not scale with the free parameter $\gamma$. As a consequence, one cannot perform the desired perturbative expansion of the warp factor in the adequate parameter $\gamma$, and equivalently, the size of the $O_5$ hole cannot be made arbitrarily small thanks to $\gamma$, therefore not avoiding the singular bulk problem. While this brief analysis of the backreaction in this de Sitter solution would require a more careful and dedicated study, it provides a first hint that the perturbative method used for DGKT does not apply. It also emphasizes the importance of the singular bulk problem, and more generally underlines the difficulties to be encountered with sources backreaction in de Sitter supergravity solutions.

\subsection{Summary - what should be done?}\label{sec:sumclassdS}

In Section \ref{sec:dSsol}, we have presented attempts at getting de Sitter solutions in the classical regime of string theory, where string loops and $\alpha'$-corrections can be neglected, and we have discussed properties of such solutions. We have focused on the main approach to this problem, that is, searching and studying solutions of 10d type IIA/B supergravities with a 4d de Sitter spacetime and 6 extra dimensions gathered as a group manifold. While this approach was briefly mentioned in Section \ref{sec:listclassgroup}, other options for classical de Sitter were discussed in Section \ref{sec:listbackreact}, \ref{sec:listclassingr} and \ref{sec:listclassdim}.

De Sitter solutions on group manifolds obey an ansatz described in Section \ref{sec:10dansatz} and  \ref{sec:groupmandimred}, where all supergravity variables are constant, and in particular $O_p/D_p$ sources are smeared. This ansatz allows for a compactification with a consistent truncation, that gives specific 4d effective theories, sometimes corresponding to gauged supergravities. Those provide an alternative (4d) way of looking for solutions, and give the tools to study their perturbative stability. Group manifolds and their appearance in supergravity solutions are reviewed in Section \ref{sec:groupman}: they are built as ${\cal M} \simeq G/\Gamma$, where $G$ is a Lie group and $\Gamma$ a lattice making the manifold ${\cal M}$ compact. Relevant Lie algebras are reviewed as well, with a focus on nilpotent and solvable ones. Those give rise to nilmanifolds and solvmanifolds, which geometrically are essentially toroidal fibrations; some lattices are discussed in Appendix \ref{ap:groupman}. A little less than 500 supergravity de Sitter solutions on group manifolds have been found. They fall in 6 compactification classes among 21 as presented in Section \ref{sec:dSsolexistence}. They are all found to be perturbatively unstable as detailed in Section \ref{sec:dSsolstab}. Last but not least, whether these supergravity solutions actually have a string theory origin in the classical regime is a delicate and technical question introduced in Section \ref{sec:dSsolclassical}. Most tests and results on this classicality have for now concluded negatively, with one possible counter-example where a partial parametric control on classicality was put forward. In that example and in general, it seems that anisotropy in the 6d geometry is necessary to classicality. {\sl Up to that example and up to now, it is therefore unclear that any classical de Sitter solutions actually exist.}

In Section \ref{sec:backreact}, we went beyond the smeared approximation to discuss the $O_p/D_p$ sources backreaction. We present there what is expected for a non-constant warp factor to capture this backreaction, and have sources localized in their transverse dimensions. The supergravity equations with a non-trivial warp factor are given in Appendix \ref{ap:warpeq}. As we explain, having intersecting sources in the de Sitter solutions, instead of only parallel ones, makes it much harder to describe their backreaction. An additional problem is the control over the orientifold hole, related to the singular bulk problem: the larger it is, the less one can describe the solution within supergravity. This point is in addition possibly related to that of having or not a parametric control on classicality. Overall, describing the backreaction in de Sitter solutions is one of the main challenges to tackle, that recent works, mentioned in Section \ref{sec:backreact}, have revived. Backreaction could also play a non-negligible role in actually getting a de Sitter solution.\\

Other challenges have been highlighted. One of them is to prove the conjectured statements \eqref{Conj3sets} or \eqref{Conj4supercharges}, saying that de Sitter solutions need at least three intersecting sets of $O_p/D_p$, or that they do not exist in theories with more than 4 supercharges. Attempts to prove these claims been discussed in Section \ref{sec:dSsolexistence}. This problem of finding or not a solution is mathematically well-defined, at least within the ansatz of group manifolds, making this challenge of proving the claim appealing. In addition, if we require the theory to be supersymmetric, and we insist on obtaining a de Sitter solution in this way, then the claim has two crucial consequences. First, the spacetime dimension cannot be more than $d=4$ (supersymmetric theories in higher dimensions have more supercharges): this gives an unexpected explanation to the dimensionality of the observed universe, complementary to \cite{Brandenberger:1988aj} (see also \cite{Brandenberger:2024ddi} and references therein). A second consequence is that we automatically get necessary ingredients to build (realistic) particle physics models. Indeed, some of those have precisely been built as intersecting brane models with at least 3 sets, as reviewed in \cite{Angelantonj:2002ct, Blumenhagen:2005mu, Blumenhagen:2006ci, Honecker:2016gyz} (see also \cite{Marchesano:2024gul}), even though these constructions exists so far mostly on Minkowski times a toroidal orbifold. Through the claim, one also gets a maximum of ${\cal N}=1$ supersymmetry, a necessity to get chirality for particle physics in 4d. It is remarkable that focusing on a cosmological question automatically places us in a setting adequate to construct viable particle physics. This hints at the idea that incorporating matter from the start may actually be necessary to our searches for de Sitter solutions; this point was already made for quintessence models, e.g.~in Section \ref{sec:phantom} (see also Section \ref{sec:topicsnotcovered}).

The apparent importance of having intersecting sources, most of the time containing some $D_p$-branes, brings back the question of open string degrees of freedom. Those have been completely ignored so far. However, as mentioned in Appendix \ref{ap:actionsugra}, they should obey their own equations of motion, that could possibly be identified to so-called calibration conditions. Those could provide extra constraints, e.g.~on background fluxes, which would certainly be interesting to check on de Sitter solution. A difficulty is the absence of supersymmetry in the solution, while calibration conditions are usually worked-out in a supersymmetric context. We may note though that the source configurations are by themselves often supersymmetric (meaning without considering background fields), as recalled in Section \ref{sec:dSsolexistence}. If open string equations of motion, or calibration conditions, or extremization of a 4d potential for open string moduli (see e.g.~\cite{Balaguer:2024cyb}), are not satisfied, this would indicate some possible (previously unseen) dynamics, or instability, in the system; and satisfying them could turn out to be incompatible with having a de Sitter solution. Once again, including matter, here in the form of open string degrees of freedom and related constraints, could be decisive with respect to de Sitter and cosmology.

Another challenge is to understand and prove the perturbative instability, systematically observed in the supergravity de Sitter solutions on group manifolds, as stated in \eqref{conjstab}. To prove such a stability no-go theorem, various strategies have been presented in Section \ref{sec:nogostab} and \ref{sec:dSsolstab}, together with partial results. One is the proposal \eqref{conjstabrhotausigma} of a systematic tachyon among the fields $(\rho,\tau,\sigma_I)$, which however requires, for some compactification class at least, a non-trivial specification of the compactness of the group manifold. In that discussion and for other points, we noticed that the specific compactification class, and resulting 4d theory, may matter in getting conclusive results. For example, some classes lead to having 14 scalar fields in 4d, while others have 22, hence probably giving different behaviours with respect to the instability. Another surprising observation is that the highest values of $\eta_V$ obtained on those de Sitter solutions seem to match in IIA and IIB: $\eta_V \simeq -2.494$, as reported in Section \ref{sec:dSsolstab}. It is surprising because in each theory, these specific solutions and values have been obtained with different tools, and the classes to which they belong, $s_{6666}$ and $s_{55}$, are not T-dual nor related in any obvious way. While this could be a coincidence, it may also offer another hint at a systematic tachyon.

Last but not least, a major challenge remains the numerics. Numerical tools have been developed for each task, namely finding supergravity solutions, testing their classicality, and determining their (in)stability, sometimes in combined ways. As mentioned in Section \ref{sec:dSsolclassical}, we clearly faced limitations with the numerical tools in trying to find solutions while obeying constraints (e.g.~to be classical), in spite of the many different numerical strategies and methods existing in the literature. As mentioned, the problem of finding a supergravity solution (and sometimes even a classical one, or a perturbatively stable one) is mathematically well-defined given the ansatz. But its numerical implementation remains challenging, as it involves a large and constrained parameter space, with tens of variables, some real and some integer, with bounds to be obeyed, etc. The difficulty in finding a desired solution at a specific location in a large parameter space is related to the difficulty in finding a de Sitter solution at numerical control: for the latter one needs very specific field values, corresponding to a particular location, since there is no freedom to extend to asymptotics. Similarly, if there was the possibility of a parametric control on classicality, the solution loci in parameter space would have more than 0 dimension, possibly enlarging chances to be found by a numerical search. This is why a solution at numerical control is probably even more difficult to find, unless ingredients are from the start designed to that purpose (e.g.~in another context, considering Riemann-flat manifolds). In any case, progress on the numerical tools would help determining whether the solutions found so far share the observed common features because they suffer a (numerical) lamppost effect, or because unknown no-go theorems are at work.

\newpage

\begin{appendix}

\section{Type II supergravities}\label{ap:sugra}

In this appendix, we give our conventions for $10$-dimensional (10d) type IIA and IIB supergravities, consistently with \cite[App.A]{Andriot:2016xvq}. We start with notations for differential forms and tensors, then give the 10d action and equations, and finally detail the target-space involution of orientifolds. We also provide the equations in 4+6 dimensions, when having a warp factor.

\subsection{Differential forms and tensors}\label{ap:diffforms}

A differential $p$-form $A_p$ is denoted in our conventions as
\beq
A_p = \frac{1}{p!} A_{m_1\dots m_p} \d x^{m_1} \w \dots \w \d x^{m_p} \ .
\eeq
In a $d$-dimensional spacetime, with metric $\d s_d^2 = g_{mn} \d x^m \d x^n$, the volume form is
\beq
{\rm vol}_d = \sqrt{|g_d|}\, \d x^0 \w \dots \w \d x^{d-1} = \frac{\sqrt{|g_d|}}{d!} \epsilon_{m_1 \dots m_d}  \d x^{m_1} \w \dots \w \d x^{m_d} = \d^d x \sqrt{|g_d|} \ ,
\eeq
where $|g_d|$ denotes the absolute value of the metric determinant, and the antisymmetric Levi-Civita symbol obeys $ \epsilon_{0 \dots d-1} = 1 $. The Hodge star of a $p$-form, $p\leq d$, is given by
\beq
*_d (\d x^{m_1} \w \dots \w \d x^{m_p}) = \frac{\sqrt{|g_d|}}{p! (d-p)!} \, g^{m_1 n_1} \dots g^{m_p n_p} \, \epsilon_{n_1 \dots n_p n_{p+1} \dots n_d} \, \d x^{n_{p+1}} \w \dots \w \d x^{n_d} \ .
\eeq
From this definition, one gets $*_d *_d A_p = s \, (-1)^{p(d-p)} A_p$ for a metric signature $s$. The contraction of two $p$-forms (where indices are lifted by the inverse metric), and the square of $A_p$, are given by
\beq
A_p \w *_d B_p = B_p \w *_d A_p = {\rm vol}_d \, \frac{1}{p!} A_{m_1\dots m_p} B^{m_1\dots m_p} \ ,\qquad A_p \w *_d A_p = {\rm vol}_d \ |A_p|^2 \ .
\eeq
For completeness, we recall, for a metric $g_{mn}$ with Levi-Civita connection, the connection coefficients (or Christoffel symbols), the covariant derivatives, the Riemann and Ricci tensors, and the Ricci scalar
\begin{gather}
\label{tensors}
\begin{aligned}
& \Gamma_{np}^m = \frac12 g^{mq} (\del_n g_{qp} + \del_p g_{qn} - \del_q g_{np}) \ ,\\
& \nabla_m V^n = \del_m V^n + \Gamma_{mp}^n V^p \, ,\quad \nabla_m V_n = \del_m V_n - \Gamma_{mn}^p V_p \ ,\\
& {\cal R}^n{}_{rms} = \del_m \Gamma^n_{sr} - \del_s \Gamma^n_{mr} + \Gamma^q_{sr} \Gamma^n_{mq} - \Gamma^q_{mr} \Gamma^n_{sq} \ , \\
& {\cal R}_{rs} = {\cal R}^n{}_{rns} \, ,\quad {\cal R}= g^{mn} {\cal R}_{mn} \ .
\end{aligned}
\end{gather}

\subsection{10d action}\label{ap:actionsugra}

Type IIA and type IIB supergravities are 10d theories of action ${\cal S}$. The latter contains a bulk part, together with the world-volume actions of $D_p$-branes and $O_p$-plane orientifolds, collectively called sources. As motivated in the main text, we do not include here further ingredients (e.g.~$N\!S_5$-branes). We only discuss in the following the bosonic part of ${\cal S}$, because fermions will vanish in the solutions considered; see e.g.~\cite[Sec.8.5.3]{Dudas:2025ubq} for fermionic terms. One has
\beq
{\cal S} = {\cal S}_{{\rm NSNS}} + {\cal S}_{{\rm RR}} + {\cal S}_{{\rm CS}} + \sum_{{\rm sources}} \left( {\cal S}_{{\rm DBI}} + {\cal S}_{{\rm WZ}} \right) \ .
\eeq
We detail in the following each term in the 10d string frame, which refers to the 10d metric used and the dependence on the dilaton. A standard Einstein-Hilbert term would be restored by a change of metric to the 10d Einstein frame. The action depends on the dimensionful constant ${\kappa}_{10}$, related to the 10d Planck mass. It can be expressed in terms of the fundamental string length $l_s$ as $2 {\kappa}_{10}^2 = (2\pi)^7 (\alpha')^4$ with $\alpha' = l_s^2$.\\

The Neveu-Schwarz-Neveu-Schwarz (NSNS) term is universal and depends on fundamental fields that are the metric $g_{MN}$, the $2$-form $B$-field and the dilaton (scalar) field $\phi$. One has
\beq
{\cal S}_{{\rm NSNS}} = \frac{1}{2 {\kappa}_{10}^2} \int \d^{10} x \sqrt{|g_{10}|}\, e^{-2\phi} \left( {\cal R}_{10} + 4 (\del \phi)^2 - \frac{1}{2} |H|^2 \right) \ , \label{SNSNS}
\eeq
with ${\cal R}_{10}$ the 10d Ricci scalar, the kinetic term $(\del \phi)^2 = \del_M \phi \,\del^M \phi$, and the $3$-form $H$-flux being the field-strength $H = \d B$ for the $U(1)$ potential $B$.

The Ramond-Ramond (RR) sector depends on the $U(1)$ gauge potentials $q$-form $C_q$, with $q=1,3$ in IIA and $q=0,2,4$ in IIB. Their field strengths are the RR fluxes: these are $q$-forms $F_{q}$ generically given by $F_q = \d C_{q-1} - H\w C_{q-3} +\delta_{{\rm IIA}}\, e^B|_q \, F_0$. More explicitly, one has
\begin{gather}
\label{RRflux}
\begin{aligned}
{\rm IIA:}\quad & F_0\, ,\ F_2= \d C_1 + B\, F_0\, ,\ F_4^{10}= \d C_3 - H\w C_1 + \frac12 B\w B\, F_0 \ ,\\
{\rm IIB:}\quad & F_1= \d C_0\, ,\ F_3= \d C_2 - H\, C_0\, ,\ F_5^{10}= \d C_4 - H\w C_2 \ ,
\end{aligned}
\end{gather}
where the labels ${^{10}}$ are introduced for later convenience. The scalar quantity $F_0$ is called the Romans mass; including it promotes type IIA to massive type IIA supergravity. The action terms are the standard Maxwell ones
\begin{gather}
\label{SRR}
\begin{aligned}
&{\cal S}_{{\rm RR}\, {\rm IIA}} = \frac{1}{2 {\kappa}_{10}^2} \int \d^{10} x \sqrt{|g_{10}|} \left( - \frac{1}{2} |F_0|^2 - \frac{1}{2} |F_2|^2 - \frac{1}{2} |F_4^{10}|^2  \right) \ , \\
&{\cal S}_{{\rm RR}\, {\rm IIB}} = \frac{1}{2 {\kappa}_{10}^2} \int \d^{10} x \sqrt{|g_{10}|} \left( - \frac{1}{2} |F_1|^2 - \frac{1}{2} |F_3|^2 - \frac{1}{4} |F_5^{10}|^2  \right) \ ,
\end{aligned}
\end{gather}
together with the Chern-Simons (CS) terms
\begin{gather}
\label{SCS}
\begin{aligned}
& S_{{\rm CS}\, {\rm IIA}} = -\frac{1}{4 \kappa_{10}^2} \int B\w \d C_3 \w \d C_3 \quad \text{(for $F_0=0$)} \ ,\\
& S_{{\rm CS}\, {\rm IIB}} = -\frac{1}{4 \kappa_{10}^2} \int B\w \d C_4 \w \d C_2 \ .
\end{aligned}
\end{gather}
The CS terms are given here up to a total derivative, for $F_0=0$. For $F_0 \neq0$, the IIA term gets more complicated, but its expression is not necessary here: it only contributes to the gauge potentials equations of motion (e.o.m.), that will be derived differently. Also, the CS terms are topological, i.e.~they do not depend on the metric, and they do not depend on the dilaton either. Finally, let us point-out that subtleties occur when considering these CS terms on a compact space with background fluxes: a detailed account can be found e.g.~in \cite{Andriot:2022bnb}. The latter can be used to derive an effective scalar potential in 4d, \eqref{potIIAB}, for the gauge potential degrees of freedom.

In type IIB supergravity, one must consider in addition the following on-shell constraint on the $5$-form flux: it has to be anti-self dual
\beq
F_5^{10} = - *_{10} F_5^{10} \ . \label{F510ASD}
\eeq
This makes its Maxwell action a pseudo-action, because the constraint makes the square of this form vanish on-shell: $|F_5^{10}|^2=0$.\\

We turn to the source contributions, namely $D_p$ and $O_p$ of even, resp.~odd, dimensionality $p$ for IIA, resp.~IIB, with $0 \leq p \leq 9$. The Dirac-Born-Infeld (DBI) and Wess-Zumino (WZ) action terms are defined on the $p+1$-world-volume $\Sigma_{p+1}$ of a $D_p$-brane, and by extension of an $O_p$-plane. The pull-back from the 10d spacetime to $\Sigma_{p+1}$ is denoted $\iota^*[\cdot]$. The action terms (for $\iota^*[B]=0$) are given by
\begin{gather}
\begin{aligned}
& {\cal S}_{{\rm DBI}} = - c_p\, T_p \int_{\Sigma_{p+1}} \d^{p+1} \xi \ e^{-\phi}\, \sqrt{|\iota^*[g_{10}]+ {\cal F} |} \ ,\\
&  {\cal S}_{{\rm WZ}} =  \ c_p\, \mu_p \int_{\Sigma_{p+1}}  \sum_q\, \iota^*[C_q]\w e^{{\cal F}} \bigg|_{p+1} \ ,
\end{aligned}
\end{gather}
with the brane tension $T_p=(2\pi l_s)^{-p}\, l_s^{-1} $, the charge $\mu_p$, and the notation $c_p=1$ for $D_p$, $c_p= -2^{p-5}$ for $O_p$. We do not include here $\alpha'$-corrections to these actions, as discussed e.g.~in \cite{Dasgupta:1999ss}. For BPS sources, as considered here, one has $\mu_p =T_p$. The brane world-volume $2$-form flux ${\cal F}$ obeys $\d {\cal F}=0$; for $O_p$ there is no such degree of freedom and one sets ${\cal F}=0$. The above actions also depend on the pull-back of the $B$-field, but we have not included it for simplicity. The $B$-field is odd under the orientifold involution (discussed in Appendix \ref{ap:Op}) so its pull-back to the world-volume must vanish. It could however be non-zero on $D_p$-branes; in most solutions to be considered, it will nevertheless vanish. A Freed-Witten anomaly can be due to a non-vanishing $H$-flux along the brane world-volume (see e.g.~\cite{Camara:2005dc, Christensen:2025ktc}); having $\iota^*[B]=0$ could then avoid this. An off-shell dependence of the action on the $B$-field could still contribute to its equation of motion, as e.g.~in \cite[(5.3)]{Koerber:2007hd}. We ignore this contribution here, as it only occurs in very specific cases, and refer to \cite[App.A]{Andriot:2016xvq} and \cite[Foot.5]{Andriot:2022bnb} for more details.

$D_p$-branes world-volume bosonic degrees of freedom are the gauge potential associated to ${\cal F}$, and the transverse scalar fields. These correspond to the open string degrees of freedom. They have their own equations of motion, which will be ignored here, but those could provide further interesting constraints (see Section \ref{sec:sumclassdS}). Some of these equations could be equivalent to so-called calibration conditions, usually discussed in a supersymmetric context, because the latter are meant to minimize the brane energy. We refer to \cite[App.B]{Andriot:2016xvq} for an account on these conditions, see also \cite{Menet:2025nbf} for recent discussions in a non-supersymmetric context. From now on, we consider as well ${\cal F}=0$.

In order to derive the (closed string, i.e.~bulk supergravity fields) equations of motion, the source actions need to be promoted to 10d ones. This is done by introducing a bulk $(9-p)$-form $\delta^{\bot}_{9-p}$ that localizes any world-volume form, $\iota^*[A_{p+1}]$, into the source transverse dimensions
\beq
\int_{\Sigma_{p+1}} \iota^*[A_{p+1}] \ \equiv\ \int A_{p+1} \w \delta^{\bot}_{9-p} \ \equiv\ \frac{1}{-c_p T_p\ 2 \kappa_{10}^2} \, \int A_{p+1} \w {\rm vol}_{\bot_s}\ \frac{T_{10}^s}{p+1} \ .
\eeq
For future convenience, we introduced a quantity $T_{10}^s$ related to the component of the form $\delta^{\bot}_{9-p}$, where $s$ is labeling the particular source $s$ considered. We also introduced, formally, the $(9-p)$-form ${\rm vol}_{\bot_s}$ corresponding to the volume form of transverse dimensions. Neither $\delta^{\bot}_{9-p}$ nor $T_{10}^s\, {\rm vol}_{\bot_s}$ are metric dependent. In the case where the source embedding in the bulk is straightforward, i.e.~when bulk parallel and transverse directions to the source are clearly identified, then one can understand $\delta^{\bot}_{9-p}= \d^{\bot}x\ \delta(\bot)$, i.e.~a $\delta$-function localisation in transverse dimensions with $\int_{\bot} \delta^{\bot}_{9-p}=1$. The pull-back is then nothing but a projection to the parallel dimensions, $P[\cdot]$, and one has
\beq
\frac{T_{10}^s}{p+1} = -c_p T_p \, 2 \kappa_{10}^2\, \sqrt{\frac{|P[g_{10}]|}{|g_{10}|}} \, \delta(\bot) \ . \label{T10s}
\eeq
This gives $T_{10}^s<0$ for $D_p$ and $T_{10}^s>0$ for $O_p$ (or more precisely the integral of $T_{10}^s$). These quantities finally allow us to promote $D_p$ and $O_p$ source world-volume actions to 10d ones, as
\begin{gather}
\label{Ssource10d}
\begin{aligned}
& {\cal S}_{{\rm DBI}} = - c_p\, T_p \int e^{-\phi}\, {\rm vol}_{p+1} \w \delta^{\bot}_{9-p} = \frac{1}{2 \kappa_{10}^2} \,  \int e^{-\phi}\, {\rm vol}_{p+1} \w {\rm vol}_{\bot_s}\ \frac{T_{10}^s}{p+1}  \ ,\\
&  {\cal S}_{{\rm WZ}} =  \ c_p\, T_p \int  C_{p+1} \w \delta^{\bot}_{9-p} =-\frac{1}{2 \kappa_{10}^2} \,  \int  C_{p+1} \w {\rm vol}_{\bot_s}\ \frac{T_{10}^s}{p+1} \ ,
\end{aligned}
\end{gather}
where we recall the restrictions $B={\cal F}=0$ and $\mu_p=T_p$. In addition, ${\rm vol}_{p+1}$ is the bulk volume form of parallel directions to the source, and we take as a definition ${\rm vol}_{p+1}\w {\rm vol}_{\bot_s} = {\rm vol}_{10}$.

Note that an anti-$D_p$ brane, $\bar{D}_p$, and an anti-orientifold, $\bar{O}_p$, have an opposite charge $-\mu_p$, modifying the above WZ term to the opposite. Unless specified, we will not consider such sources in the following.

This ends the presentation of our conventions for 10d type II supergravities actions. Detailed relations of these conventions to others in the literature can be found in \cite[App.A]{Andriot:2016xvq}. We now turn to the 10d equations.

\subsection{10d equations}\label{ap:sugraeom}

From the above action, one can derive 10d equations of motion (e.o.m.); to those one should add various Bianchi identities (BI) that need to be satisfied. We start with the metric and the dilaton. For those, the topological terms CS and WZ do not contribute. Regarding the DBI term, it is convenient to introduce an energy momentum tensor for sources (in string frame), as follows
\beq
T_{MN} = -\frac{4 \kappa_{10}^2\, e^{\phi}}{\sqrt{|g_{10}|}} \sum_{{\rm sources}} \frac{\delta S_{{\rm DBI}}}{\delta g^{MN}} = \sum_{{\rm sources}} P[g_{MN}]\, \frac{T_{10}^s}{p+1} \ .
\eeq
This definition gives the convenient trace
\beq
T_{10} \equiv g^{MN} T_{MN} = \sum_{{\rm sources}} T_{10}^s \ .
\eeq
These quantities also appear in the dilaton e.o.m.~since
\beq
-\frac{2 \kappa_{10}^2\, e^{\phi}}{\sqrt{|g_{10}|}} \sum_{{\rm sources}} \frac{\delta S_{{\rm DBI}}}{\delta \phi} = \sum_{{\rm sources}} \frac{T_{10}^s}{p+1} \ .
\eeq
From all this, we deduce the dilaton equation of motion
\beq
2 {\cal R}_{10} - |H|^2 + 8\Delta \phi - 8 (\del \phi)^2 + e^{\phi} \sum_{{\rm sources}} \frac{T_{10}^s}{p+1} = 0 \ ,\label{10ddileom}
\eeq
and the 10d trace-reversed Einstein equation in type IIA and type IIB
\begin{gather}
\label{Einstein10d}
\begin{aligned}
{\cal R}_{MN} & = \frac{1}{4} H_{MPQ}H_N{}^{PQ} + \frac{e^{2\phi}}{2} \left( F_{2\, MP}F_{2\, N}{}^P + \frac{1}{3!} F_{4\, MPQR}^{10} F_{4\, N}^{10\ PQR} \right) + \frac{e^{\phi}}{2} T_{MN} - 2 \nabla_M \del_N \phi \\
& +\frac{g_{MN}}{16} \left(- 2 |H|^2 + e^{2\phi} (|F_0|^2 - |F_2|^2 - 3 |F_4^{10}|^2 ) - e^{\phi} T_{10} - 4 \Delta \phi + 8 (\del \phi)^2  \right) \ ,\\
{\cal R}_{MN} &= \frac{1}{4} H_{MPQ}H_N{}^{PQ} + \frac{e^{2\phi}}{2} \left( F_{1\, M}F_{1\, N} + \frac{1}{2} F_{3\, MPQ}F_{3\, N}{}^{PQ} + \frac{1}{2\cdot 4!} F_{5\, MPQRS}^{10}F_{5\, N}^{10\ PQRS} \right) \\
&  + \frac{e^{\phi}}{2} T_{MN} - 2 \nabla_M \del_N \phi + \frac{g_{MN}}{16} \left(- 2 |H|^2 -2 e^{2\phi} |F_3|^2 - e^{\phi} T_{10} - 4 \Delta \phi + 8 (\del \phi)^2  \right) \ ,
\end{aligned}
\end{gather}
with the 10d Laplacian $\Delta \phi = g^{MN} \nabla_M \del_N \phi = \frac{1}{\sqrt{|g_{10}|}} \del_M (\sqrt{|g_{10}|} g^{MN} \del_N \phi )$. We refer to \cite[App.A]{Andriot:2016xvq} for the details of their derivation. The 10d trace of Einstein equation is given by
\begin{gather}
\label{10dtraceE}
\begin{aligned}
& 4 {\cal R}_{10} - |H|^2 - \frac{e^{2\phi}}{2} \left(5 |F_0|^2 + 4 |F_1|^2 + 3 |F_2|^2 + 2 |F_3|^2 + |F_4^{10}|^2 \right)\\
& + \frac{e^{\phi}}{2} T_{10} +18 \Delta \phi -20 (\del \phi)^2 = 0 \ ,
\end{aligned}
\end{gather}
picking the right RR fluxes for each theory. We finally recall that the Riemann tensor is subject to a Bianchi identity
\beq
{\cal R}^M{}_{[NPQ]}=0 \ , \label{BIRiemann}
\eeq
which is trivial when given an explicit dependence on a globally defined metric.

The gauge potentials equations of motion can be derived from the above action. It is however simpler to obtain them using the democratic formalism \cite{Bergshoeff:2001pv}, that introduces a pseudo-action for all RR fluxes and discards the CS term. This is especially helpful when $F_0 \neq 0$. This formalism allows in addition to derive the Bianchi identities (BI), in presence of a (magnetic) source. We refer to \cite[App.A]{Andriot:2016xvq} for a detailed account of this derivation, and give here the results. The equation of motion for $C_{q-1}$ is given generically by $\d(*_{10} F_q) + H\w *_{10} F_{q+2} =  (-1)^{q} 2 \kappa_{10}^2 \sum_{\text{$q-2$-sources}} c_{q-2} \mu_{q-2}\, \delta^{\bot}_{11-q}$, where the sources refer to $D_p$ and $O_p$ with $0\leq p\leq 9$. More explicitly, with above conventions, the e.o.m.~are
\begin{gather}
\label{IIFeom}
\begin{aligned}
\text{IIA:} &\quad \d(*_{10} F_2) + H\w *_{10} F_{4}^{10} = - \sum_{\text{$0$-sources}}  T_{10}^s\, {\rm vol}_{\bot_s} \ ,\\
&\quad \d(*_{10} F_4^{10}) + H\w F_{4}^{10} = - \sum_{\text{$2$-sources}} \frac{T_{10}^s}{3}\, {\rm vol}_{\bot_s}\ ,\\
\text{IIB:} &\quad \d(*_{10} F_1) + H\w *_{10} F_{3} = 0 \ ,\\
&\quad \d(*_{10} F_3) + H\w *_{10} F_{5}^{10} =  \sum_{\text{$1$-sources}} \frac{T_{10}^s}{2} \, {\rm vol}_{\bot_s} \ ,\\
&\quad \d(*_{10} F_5^{10}) + H\w F_{3} = \sum_{\text{$3$-sources}} \frac{T_{10}^s}{4} \, {\rm vol}_{\bot_s}\ .
\end{aligned}
\end{gather}
The RR BI are generically given by $\d F_{q} - H \w F_{q-2} = (-1)^{q+[(q+1)/2]}  2 \kappa_{10}^2 \sum_{\text{$8-q$-sources}} c_{8-q} \mu_{8-q}\, \delta^{\bot}_{q+1}$, where $[\cdot]$ denotes the integer part, and more explicitly
\begin{gather}
\label{IIBI}
\begin{aligned}
\text{IIA:} &\quad \d F_{0} = - \sum_{\text{$8$-sources}} \frac{T_{10}^s}{9}\, {\rm vol}_{\bot_s} \ ,\\
&\quad \d F_{2} - H \w F_{0} =  \sum_{\text{$6$-sources}} \frac{T_{10}^s}{7}\, {\rm vol}_{\bot_s} \ ,\\
&\quad \d F_{4}^{10} - H \w F_{2} = - \sum_{\text{$4$-sources}} \frac{T_{10}^s}{5}\, {\rm vol}_{\bot_s} \ ,\\
\text{IIB:} &\quad \d F_{1} = - \sum_{\text{$7$-sources}} \frac{T_{10}^s}{8}\, {\rm vol}_{\bot_s} \ ,\\
&\quad \d F_{3} - H \w F_{1} =  \sum_{\text{$5$-sources}} \frac{T_{10}^s}{6}\, {\rm vol}_{\bot_s} \ ,\\
&\quad \d F_{5}^{10} - H \w F_{3} = - \sum_{\text{$3$-sources}} \frac{T_{10}^s}{4}\, {\rm vol}_{\bot_s} \ .
\end{aligned}
\end{gather}
To the above, let us add the condition on $p=9$ sources: the $D_9$ and $O_9$ contributions have to cancel, which amounts to
\beq
\sum_{\text{$9$-sources}} T_{10}^s = 0 \ .
\eeq
Justifications for this condition can be found e.g.~around \cite[(2.3)]{Andriot:2022xjh}.

Finally, we recall that the above source contributions are due to the WZ term. Therefore, contributions from $\bar{D}_p$ or $\bar{O}_p$ would come with a minus sign in the above equations (keeping the same definition for $\mu_p$ and $T_p$). But their contribution to the Einstein or dilaton equations, due to the DBI term, would be the same as for $D_p$ and $O_p$.

We finally turn to the $B$-field, whose equation of motion can as well be derived with the CS term or with the democratic formalism, making use of the RR e.o.m. With the restrictions previously discussed (no $N\!S_5$-brane, ignored source contributions), we obtain the following BI and e.o.m.
\begin{gather}
\label{BeomBI}
\begin{aligned}
&\quad \d H = 0 \ , \phantom{\frac12}\\
\text{IIA:} &\quad \d (e^{-2 \phi} *_{10} H) - F_0 \w *_{10} F_2 - F_2 \w *_{10} F_4^{10} - \frac{1}{2} F_4^{10} \w F_4^{10} = 0 \ ,\\
\text{IIB:} &\quad \d (e^{-2 \phi} *_{10} H) - F_{1} \w *_{10} F_{3} - F_{3} \w *_{10} F_{5}^{10}  = 0 \ .
\end{aligned}
\end{gather}
This ends the list of 10d equations to solve.

\subsection{Orientifold involution and projection}\label{ap:Op}

Since we allow for orientifold $O_p$-planes, the solutions will need to respect their target-space involution $\sigma$. For an $O_p$ transverse to a line along $x^{\bot}$, this involution may act as $x^{\bot} \rightarrow - x^{\bot}$. The $O_p$ is then located at the fixed point of the involution, namely $x^{\bot}=0$. The supergravity fields also inherit a certain action of this involution. In the remainder of this appendix, we restrict to space-filling $O_p$, i.e.~the $p+1$-dimensional world-volume is along the 4d spacetime, leaving us with $3 \leq p \leq 9$. We also restrict to internal (6d) flux-forms $H, F_q$, allowing for $0 \leq q \leq 6$. This restrictive ansatz is motivated in Section \ref{sec:10dansatz}. The involution then acts as follows on supergravity fields \cite[Sec.3.1]{Koerber:2007hd}
\begin{gather}
\label{sigmaOp}
\begin{aligned}
& \sigma\big(g_{MN}\, \d x^M \d x^N\big) = g_{MN}\, \d x^M \d x^N \ ,\quad \sigma(\phi) =\phi \ ,\quad \sigma(H)=-H \ ,\\
& \sigma(F_q) =\pm (-1)^{\frac{q(q-1)}{2}}\ F_q \ ,\quad \pm\ {\rm for}\ {}^{p=3,6,7}_{p=4,5,8,9} \ .
\end{aligned}
\end{gather}
The involution is expressed here on fluxes, but its action on gauge potentials can easily be deduced, given that $\d$ is invariant: for example, $\sigma(B) = - B$.\\

Let us now consider that the directions parallel, e.g.~$\d x^{||}$, and transverse, $\d x^{\bot}$, to the $O_p$ are globally identified in the internal space. The involution on the above tensors then gives constraints on their components. For example, considering $\sigma\big(g_{|| \bot} \d x^{||} \d x^{\bot} \big)$ gives $\sigma(g_{|| \bot} ) = - g_{|| \bot} $. This implies that the latter component has to be an odd function with respect to the $O_p$ locus. If in addition we restrict to constant components, we conclude on $g_{|| \bot} =0 $: one can then talk of orientifold projection conditions.

Fluxes can be decomposed in terms of their components with $n$ legs parallel to the $O_p$, and we introduce accordingly an upper index ${}^{(n)}$ \cite{Andriot:2016xvq}, for instance with the component
\beq
F_{q\ m_{1||} \dots m_{n||} m_{n+1 \bot} \dots m_{q \bot}}^{(n)} \ . \label{F(n)}
\eeq
For an internal (6d) form $F_q$ and a (4d) space-filling $O_p$, one has $0\leq n \leq p-3$. Restricting further to constant components, we obtain orientifold projection conditions: generically, only the following flux components survive \cite{Andriot:2018ept, Andriot:2022way}
\beq
F_{6-p}^{(0)} \ ,\ F_{8-p}^{(1)} \ ,\ F_{10-p}^{(2)} \ ,\ F_{12-p}^{(3)} \ ,\ H^{(0)} \ ,\ H^{(2)} \ , \label{fluxlistconstant}
\eeq
and more explicitly
\begin{gather}
\label{Opflux}
\begin{aligned}
O_3:\quad & F_{3}^{(0)} \ ,\ H^{(0)} \ ,\\
O_4:\quad & F_{2}^{(0)} \ ,\ F_{4}^{(1)} \ ,\ H^{(0)} \ ,\\
O_5:\quad & F_{1}^{(0)} \ ,\ F_{3}^{(1)} \ ,\ F_{5}^{(2)} \ ,\ H^{(0)} \ ,\ H^{(2)} \ ,\\
O_6:\quad & F_{0}^{(0)} \ ,\ F_{2}^{(1)} \ ,\ F_{4}^{(2)} \ ,\ F_{6}^{(3)} \ ,\ H^{(0)} \ ,\ H^{(2)} \ ,\\
O_7:\quad & F_{1}^{(1)} \ ,\ F_{3}^{(2)} \ ,\ F_{5}^{(3)} \ ,\ H^{(2)} \ ,\\
O_8:\quad & F_{2}^{(2)} \ ,\ F_{4}^{(3)} \ ,\ H^{(2)} \ ,\\
O_9:\quad & F_{3}^{(3)} \ .
\end{aligned}
\end{gather}
This gives for $O_5$ the contribution $(*_6 F_5)^{(0)}$ and for $O_7$ a $(*_6 F_5)^{(1)}$. The list \eqref{Opflux} will be useful when looking for solutions with constant flux components.

\subsection{Warped (4+6)d equations}\label{ap:warpeq}

In this appendix, we start from the 10d type IIA/B supergravity equations given in Appendix \ref{ap:sugraeom}, and rewrite them using the compactification ansatz described at the beginning of Section \ref{sec:10dansatz}. This ansatz splits the 10d spacetime into a maximally symmetric 4d spacetime, times a 6d manifold, together with a warp factor $e^A$, as indicated by the metric \eqref{10dmetricmaxsym}. We give here the resulting (4+6)d equations, with an explicit dependence on $A$ and the dilaton $\phi$, recalling that both only depend on the 6d coordinates $y^m$. These are the equations to be obtained before applying the smeared approximation \eqref{smearing} discussed in section \ref{sec:10dansatz}. These equations are useful when looking for solutions with backreacted and localized sources, as discussed in Section \ref{sec:backreact}. We follow partly \cite[App.A.1]{Andriot:2023fss}.

We start from the 10d dilaton e.o.m.~\eqref{10ddileom}. Using \cite[(2.28)]{Andriot:2017oaz} for the rewriting of ${\cal R}_{10}$, and \cite[(C.5)]{Andriot:2016xvq} for the dilaton terms, we obtain
\bea
\hspace{-0.35in} && 0 = 2 e^{-2A} \tilde{{\cal R}}_4 + 2 {\cal R}_6 - |H|^2 + e^{\phi} \sum_{{\rm sources}}  \frac{T_{10}^s}{p+1} \label{warpdil} \\
\hspace{-0.35in}&& \phantom{=0} - 2 e^{-4A} (\del e^{2A} )^2 - 8 e^{-2A} \Delta_6 e^{2A} +2 e^{-4\phi} (\del e^{2\phi})^2 - 4  e^{2\phi} \Delta_6 e^{-2\phi} -8 e^{2\phi - 2A} \del_m e^{2A} \del^m e^{-2\phi} \nn
\eea
The Laplacian on the 6d manifold with respect to $g_{mn}$ is denoted $\Delta_6$.

We turn to the Einstein equation, given in \eqref{Einstein10d}. From there, we first obtain the 4d Einstein equation, proportional to its trace. The latter is given as follows, using the first relations in \cite[App. C]{Andriot:2016xvq}
\begin{gather}
\label{warp4dEinstein}
\begin{aligned}
0 = &\, 2e^{-2A} \tilde{{\cal R}}_4  +|H|^2 + \frac{e^{2\phi}}{2} \sum_{q=0}^6 (q-1) |F_q|^2   +\frac{e^{\phi}}{2}  \sum_{{\rm sources}}  \frac{p-7}{p+1}\, T_{10}^s \\
&  -4 e^{-2A} \Delta_6\, e^{2A} -4 e^{-4A} (\del e^{2A})^2 - e^{2\phi} \Delta_6\, e^{-2\phi}  - 6 e^{2\phi-2A}  \del_{m} e^{2A} \del^{m} e^{-2\phi}  \ .
\end{aligned}
\end{gather}
From \eqref{Einstein10d}, we also derive the 6d Einstein equation, using again \cite[(C.5)]{Andriot:2016xvq} for the dilaton, and \cite[(2.23c)]{Andriot:2017oaz} for the Ricci tensor
\begin{gather}
\label{warp6dEinstein}
\begin{aligned}
0= &\ {\cal R}_{mn}  -\frac{1}{4} H_{mpq}H_n^{\ \ pq} - \frac{e^{\phi}}{2}T_{mn} \\
& - \frac{e^{2\phi}}{2}\left(F_{2\ mp}F_{2\ n}^{\ \ \ \ p} +\frac{1}{3!} F_{4\ mpqr}F_{4\ n}^{\ \ \ pqr} \right) \\
& - \frac{e^{2\phi}}{2} \left( F_{1\, m}F_{1\, n} + \frac{1}{2} F_{3\, mpq}F_{3\, n}{}^{pq} + \frac{1}{2\cdot 4!} F_{5\, mpqrs}F_{5\, n}^{\ \ pqrs} - \frac{1}{2} (*_6 F_{5})_m (*_6 F_{5})_n  \right) \\
& - \frac{g_{mn}}{4} \Bigg(-\frac{e^{\phi}}{4} T_{10} -\frac{1}{2} |H|^2 + \frac{e^{2\phi}}{4} \bigg(\sum_{q=0}^4 (1-q) |F_q|^2 + 3|F_6|^2 \bigg) +\frac{1}{2} e^{2\phi} \Delta_6\, e^{-2\phi} \\
&+ e^{2\phi-2A} \del_{p} e^{2A} \del^{p} e^{-2\phi} \Bigg)
- 2 e^{-2A} \nabla_n \del_m e^{2A} + e^{-4A} \del_m e^{2A} \del_n e^{2A} + 2\nabla_m \del_n{\phi}
\end{aligned}
\end{gather}
where even/odd RR fluxes should be considered in IIA/B. A rewriting of the three above equations in terms of $e^A$ and $e^{\phi}$, and not their squares, can be found in \cite[App.A.1]{Andriot:2023fss}.

Finally, the gauge potential equations of motion of Appendix \ref{ap:sugraeom} become
\begin{gather}
\label{eAeom}
\begin{aligned}
\text{IIA:} &\quad e^{-4A} \d (e^{-2 \phi + 4A} *_{6} H) - F_0 \w *_{6} F_2 - F_2 \w *_{6} F_4 - F_4 \w *_{6} F_6 = 0 \ ,\\
&\quad e^{-4A} \d(e^{4A} *_{6} F_2) + H\w *_{6} F_{4} = 0 \ ,\\
&\quad e^{-4A} \d(e^{4A} *_{6} F_4) + H\w *_{6} F_{6} = 0\ ,\\
&\quad \d(e^{4A} *_{6} F_6) =0 \ ,\\[0.4cm]
\text{IIB:}  &\quad e^{-4A} \d (e^{-2 \phi + 4A} *_{6} H) - F_{1} \w *_{6} F_{3} - F_{3} \w *_{6} F_{5}  = 0 \ ,\\
&\quad e^{-4A} \d(e^{4A} *_{6} F_1) + H\w *_{6} F_{3} = 0 \ ,\\
&\quad e^{-4A} \d(e^{4A} *_{6} F_3) + H\w *_{6} F_{5} =  0 \ ,\\
&\quad \d(e^{4A} *_{6} F_5)  = 0\ .
\end{aligned}
\end{gather}
while the flux BI are the same as \eqref{BI}.

We do not look at the Riemann BI \eqref{BIRiemann}, but the interested reader can find the necessary expressions for the components of the Riemann tensor in \cite[(2.23)]{Andriot:2017oaz}.

\newpage

\section{Classical scalar field cosmology}\label{ap:cosmo}

In this appendix, we introduce the standard formalism of cosmology in presence of scalar fields, useful to inflation or quintessence models; the latter are studied in Section \ref{sec:quint}. We do so at the classical level: we do not discuss fluctuations beyond a background solution, nor quantum corrections. We consider a 4d spacetime; generalization to $d$ dimensions can be found in \cite[App.A.1]{Andriot:2024jsh}.\\

We start with Einstein gravity in presence of a perfect fluid. The 4d action is given by
\beq
{\cal S}= \int \d^4 x \sqrt{|g_4|} \left(\frac{M_p^2}{2} {\cal R}_4 - {\cal L} \right) \ ,
\eeq
where $M_p=(8\pi G)^{-1/2}$ is the 4d reduced Planck mass, and we set $\bar{h}=c=1$. ${\cal L}$ is the (opposite of the) Lagrangian density for the perfect fluid; ${\cal S}_{{\rm p.f.}} = \int \d^4 x \sqrt{|g_4|} ( - {\cal L} )$ denotes the corresponding action. The Einstein equation is given by
\beq
{\cal R}_{\mu\nu} - \frac{1}{2} g_{\mu\nu} {\cal R}_4 = \frac{1}{M_p^2}\, T_{\mu\nu} \ , \label{Einstein}
\eeq
where the energy momentum tensor is defined as
\beq
T_{\mu\nu} = - \frac{2}{\sqrt{|g_4|}} \frac{\delta {\cal S}_{{\rm p.f.}}}{\delta g^{\mu\nu}} = - g_{\mu\nu}\, {\cal L} + 2 \frac{\delta {\cal L}}{\delta g^{\mu\nu}} \ .
\eeq
For a relativistic perfect fluid, the latter is usually expressed as $T^{\mu\nu}= (\rho + p) u^{\mu} u^{\nu} + p\, g^{\mu\nu}$ in terms of the 4-velocity $\{u^{\mu}\}$, the energy density $\rho$ and the pressure $p$. We now specify this form further. We take as metric signature $(-,+,+,+)$. We take the time $t$ to be defined such that $g_{00}=-1$ and the metric is block diagonal between time and space, namely $g_{0i}=0$. Then, in rest frame, the energy momentum tensor of a perfect fluid takes the form $T_{00}=\rho$, $T_{0i}=0$, $T_{ij}=g_{ij}\, p$. This gives
\beq
\rho = {\cal L} + 2 \frac{\delta {\cal L}}{\delta g^{00}} \ ,\qquad g_{ij}\, p= - g_{ij}\, {\cal L} +2 \frac{\delta {\cal L}}{\delta g^{ij}} \ . \label{rhop}
\eeq
Let us apply these formulas to various cases of interest, with the notation $\del_0 \varphi =\del_t \varphi =\dot{\varphi}$:
\begin{itemize}
  \item Consider scalar fields $\varphi^k$ which only depend on time. Their kinetic energy is given by ${\cal L}_{{\rm kin}} = \frac{1}{2} g_{kl}\, \del_{\mu}\varphi^k \del^{\mu}\varphi^l = - \frac{1}{2} g_{kl}\, \dot{\varphi}^k \dot{\varphi}^l $, where $g_{kl}(\varphi)$ is the field space metric. One gets $\rho_{{\rm kin}} = \frac{1}{2} g_{kl}\, \dot{\varphi}^k \dot{\varphi}^l $ and $p_{{\rm kin}} = \rho_{{\rm kin}}$.
  \item We now consider a scalar potential: ${\cal L}_V = V(\varphi)$. One gets $\rho_V = V = - p_V$.
  \item We turn to non-relativistic matter, for which we take a Lagrangian density inversely proportional to the space volume: ${\cal L}_m = {\cal L}_{m0}/ \sqrt{{\rm det} g_{ij}}$. We obtain $\rho_m = {\cal L}_m$ and $p_m=0$.
  \item We finally consider radiation: as an example, one can take ${\cal L}_r = \frac{1}{2} F_{\mu\nu} F^{\mu\nu}$. The above block diagonal assumption is not easily compatible with having a non-zero electric field in rest frame: we thus take $F_{0i}=0$. Taking for simplicity a diagonal metric, one concludes that $\rho_r= \frac{1}{2} F_{ij} F^{ij} = 3\, p_r$.
\end{itemize}
In all these examples, we read in the equation of state $p = w\, \rho $ that the parameter $w$ is constant.\\

We now consider the following general action for scalar field cosmology
\beq
{\cal S}= \int \d^4 x \sqrt{|g_4|} \left(\frac{M_p^2}{2} {\cal R}_4 - \frac{1}{2} g_{kl} \del_{\mu}\varphi^k \del^{\mu}\varphi^l -V (\varphi) - {\cal L}_m - {\cal L}_r \right) \ , \label{Scosmo}
\eeq
with the same definitions as above, in particular the dependence $\varphi^k(t)$, and ${\cal L}_m$, ${\cal L}_r$ standing for (non-relativistic) matter, (relativistic) radiation. Assuming this universe content to have perfect fluid behaviours, we can derive the same Einstein equation as above with a total energy density and pressure given by the sums on each constituent $n$ of $\rho_n, p_n$. We also introduce the equation of state parameter $w_n$ for each constituent with $p_n = w_n\, \rho_n$.

In addition, we specialize to solutions whose metric is given by Friedmann-Lema\^itre-Robertson-Walker (FLRW) with spatial curvature $k$ (where $k=0, 1$ or $-1$, for flat, closed or open universe respectively):
\beq
\d s_4^2 = - \d t^2 + a^2(t) \left( \frac{\d r^2}{1 - k\, r^2} + r^2\, \d \Omega_2^2 \right) \ . \label{FLRW}
\eeq
We take the scale factor $a(t)>0$, except possibly at the time origin. One defines the Hubble parameter as $H=\dot{a}/a$. We then derive the equations of motion from \eqref{Scosmo}: the Einstein equation \eqref{Einstein} gives the two Friedmann equations $F_1 = F_2 =0$, to which one should add each scalar field equation of motion $E^k=0$. For the latter, we need to specify whether the scalar fields couple (classically) to the matter and radiation, namely if ${\cal L}_m$ and ${\cal L}_r$ depend on $\varphi^k$. For now, we assume it is not the case: the scalars are only minimally coupled to gravity. We then have
\begin{gather}
\label{F1F2Ei}
\begin{aligned}
&F_1 = 3 H^2 - \frac{1}{M_p^2} \sum_n \rho_n \ ,\qquad F_2 = \dot{H} + \frac{1}{2 M_p^2} \sum_n (1 + w_n) \rho_n \ ,\\
&E^k = \ddot{\varphi}^k + \Gamma_{lp}^k\, \dot{\varphi}^l \dot{\varphi}^p + 3H \dot{\varphi}^k + g^{kl} \del_{\varphi^l} V  \ ,
\end{aligned}
\end{gather}
where $\Gamma_{lp}^k$ is the Christoffel symbol for the field space metric $g_{kl}(\varphi)$. The constituents $n$ in the sums are the same as above, together with the spatial curvature contribution that is read directly from the equations: we list all those in Table \ref{tab:rhow}. The $\rho_n$ and $w_n$ are obtained from the results above, combined with the FLRW metric that obeys the assumptions. We introduce the total quantities for the scalar fields, namely $\rho_{\varphi} = \rho_{{\rm kin}} + \rho_V$ and $p_{\varphi} = p_{{\rm kin}} + p_V = \rho_{{\rm kin}} -V$, which define $w_{\varphi}$. In addition, we normalise without loss of generality $a(t_0)=1$, while all quantities at $t_0$ are denoted with an index ${}_0$. Unless mentioned otherwise, $t_0$ will be the time today, meaning the age of the universe.
\begin{table}[H]
\begin{center}
\begin{tabular}{|l|c||c|c|}
\hline
&&&\\[-8pt]
Constituent & $n$ & $\rho_n$ & $w_n$ \\[4pt]
\hline
&&&\\[-8pt]
radiation & $r$ & $\rho_{r0}\, a^{-4}$ & $\frac{1}{3}$ \\[4pt]
\hhline{----}
&&&\\[-8pt]
matter & $m$ & $\rho_{m0}\, a^{-3}$ & $0$ \\[4pt]
\hhline{----}
&&&\\[-8pt]
curvature & $k$ & $-3k\, M_p^2\, a^{-2}$ & $-\frac{1}{3}$ \\[4pt]
\hline
&&&\\[-8pt]
scalar & $\varphi$ & $\frac{1}{2} g_{kl}\, \dot{\varphi}^k \dot{\varphi}^l  + V$ & $\frac{\frac{1}{2} g_{kl}\, \dot{\varphi}^k \dot{\varphi}^l  - V}{\frac{1}{2} g_{kl}\, \dot{\varphi}^k \dot{\varphi}^l  + V}$ \\[6pt]
\hhline{----}
\end{tabular}
\end{center}\caption{Energy density $\rho_n$ and equation of state parameter $w_n$ of each universe constituent $n$ entering \eqref{F1F2Ei}.}\label{tab:rhow}
\end{table}
\noindent The equations in \eqref{F1F2Ei} are related by the following relation
\beq
\dot{F}_1 = - \dot{\varphi}^l g_{lk} \, E^k + 6 H\, F_2 \ .
\eeq
This allows to solve two instead of three equations when looking for cosmological solutions.

For the first three components in Table \ref{tab:rhow}, one has $\rho_n = \rho_{n0}\, a^{-3(1+w_n)}$. From this, it is straightforward to verify that the following continuity equation holds
\beq
\dot{\rho}_n = -3 (1+w_n) H \rho_n \ . \label{continuityeq}
\eeq
This also holds on-shell for $\rho_{\varphi}$, therefore for all constituents of Table \ref{tab:rhow}, as one can verify from the equalities $\dot{\rho}_{\varphi} = \dot{\varphi}^l g_{lk} (E^k - 3 H \dot{\varphi}^k ) = \dot{\varphi}^l g_{lk} E^k - 3 H (1+w_{\varphi}) \rho_{\varphi} $.\\

We now mention few properties of the solutions of interest. We will restrict ourselves to $V>0$ and $g_{kl}$ definite positive. It is then straightforward to see that $-1 \leq w_{\varphi} \leq 1 $; quintessence models typically admit solutions with $w_{\varphi}$ varying between these bounds. This property will be crucial in view of recent cosmological observations \cite{DES:2024jxu, DESI:2024mwx, DESI:2025zgx}. This should be contrasted with a cosmological constant, that gives the constant value $w_{\varphi}=-1$, since it corresponds to a potential extremum with no kinetic energy.

In addition, we will consider solutions describing an expanding universe at any finite time, namely $\dot{a} >0 $: this gives $H >0$. We will also have $\rho_{r0} \geq0$, $\rho_{m0} \geq0$: except for a closed universe, one then has $(1+w_n) \rho_n \geq 0$, and the sum of those will be non-zero in our solutions. As a consequence, we get from $F_2=0$ that $\dot{H}<0$, i.e.~$H$ decreases with time.\\

We end this appendix with few useful definitions. Since $H \neq 0$, we can introduce the energy density parameters for each constituent: $\Omega_n = \frac{\rho_n}{3 M_p^2\, H^2}$. The first Friedmann equation gets rewritten as
\beq
\frac{F_1}{3 H^2} = 1 - \sum_n \Omega_n \ . 
\eeq
Except for a closed universe, we get $\Omega_n \geq 0$. $F_1=0$ then indicates that the $\Omega_n$ give the proportion of each constituent in the universe. For example, we will use the following fiducial values for today's universe (compatible with, or close to, most observational data)
\beq
\Omega_{r0} = 0.0001 \ ,\ \Omega_{m0} = 0.3149 \ ,\ \Omega_{\varphi0} = 0.6850 \ ,\ \Omega_{k0}=0 \ .\label{Onfiducial}
\eeq

Let us also introduce the effective equation of state parameter $w_{{\rm eff}}= \sum_n p_n / \sum_n \rho_n$. One verifies that $w_{{\rm eff}}= \sum_n w_n \Omega_n$. In addition, one has
\beq
\frac{F_1 +2 F_2}{3 H^2} = \frac{2}{3 H^2} \frac{\ddot{a}}{a} + \frac{1}{3} + w_{{\rm eff}} \ .\label{awe}
\eeq 
Another equivalent parameter is $\epsilon = - \dot{H}/H^2$. By definition, one has
\beq
\frac{\ddot{a}}{a} = H^2 (1 - \epsilon) \ . \label{aeps}
\eeq 
Using equations of motion, it can also be written as $\epsilon = 3/2\, (1 + w_{{\rm eff}})$. From \eqref{awe} or \eqref{aeps}, we conclude on the following condition for acceleration
\beq
{\rm Acceleration:}\quad\quad \ddot{a}>0 \ \Leftrightarrow \ w_{{\rm eff}} < - \frac{1}{3} \ \Leftrightarrow \ \epsilon < 1 \ .\label{acccond}
\eeq

A last useful definition is that of the cosmological event horizon. We recall the distance measured by an observer at a time $t_0$ between two points $A,B$ (located at $r_A,r_B$ in comoving coordinates in \eqref{FLRW}): $a(t_0) \int_A^B \d r / \sqrt{1-k\, r^2}$. The event horizon measures at a time $t_0$ the distance that will be traveled by light until the end of time (denoted $t=\infty$) if emitted at $t_0$. It is given by the same formula, where the integrand gets replaced using that for light, $\d s =0$. We get
\beq
d_e = a(t_0) \int_{t_0}^{\infty} \frac{\d t}{a(t)} \ . \label{horizon}
\eeq
If $d_e$ is finite, this means there is a horizon, beyond which the universe is causally disconnected (at $t_0$). If $d_e$ is infinite, there is no such horizon. As an example, let us take the pure de Sitter solution, with cosmological constant $\Lambda$ and $a(t_0)=1$: it is given by $a(t)=e^{H_0\, (t-t_0)}$, where $H_0 =\sqrt{\Lambda / 3}$. In that case, it is straightforward to verify the existence of an event horizon, with $d_e = 1/H_0$.

\newpage

\section{Group manifolds and their lattice}\label{ap:groupman}

As reviewed in Section \ref{sec:groupman}, group manifolds are built as a Lie group, most of the time quotiented by a discrete subgroup, the lattice, which makes the manifold compact. In turn the lattice imposes quantization conditions on the structure constants of the corresponding Lie algebra. In this appendix, we discuss these lattice quantization conditions for some Lie algebras in the list \eqref{alg3d}. These conditions are important to guarantee compactness of the manifold, and they stand among the requirements to be satisfied by a supergravity solution, to eventually claim the latter to be classical. This is explained in Section \ref{sec:dSsolclassical}, which motivates and guides the following presentation.\\

We start with the 3-dimensional solvable algebra $\mathfrak{g}_{3.5}^0$. The 6d Lie algebra underlying the de Sitter solution $s_{55}^+ 29$ \eqref{sol29} is made of two copies of the latter, $\mathfrak{g}_{3.5}^0 \oplus \mathfrak{g}_{3.5}^0$, and the 6d manifold of this solution is then a direct product of two corresponding 3d solvmanifolds. In \eqref{sol29}, the solution exhibits many structure constants, but a change of basis allows to keep only two pairs of non-zero structure constants, $f^2{}_{35},f^3{}_{25}$ and $f^1{}_{64},f^6{}_{14}$, taking the same value as in the initial basis \cite{Andriot:2020vlg, Andriot:2024cct}. Each pair corresponds to one copy of the 3d solvable algebra, $\mathfrak{g}_{3.5}^0$, on which we focus. For the pair $f^1{}_{64},f^6{}_{14}$, we can write the Maurer-Cartan equations \eqref{MC} as follows
\beq
\d e^1 = - N_1\, \frac{r_1}{r_4r_6} \, e^6 \w e^4 \ ,\quad \d e^6 =  - N_6\, \frac{r_6}{r_1r_4} \, e^1 \w e^4  \ ,\quad \d e^4 = 0 \ ,\quad N_6 < 0\ ,\ N_1>0 \ ,\label{MCsol29}
\eeq
where we use the parametrisation \eqref{var3} of the structure constants. We recall that each $r_a$ is a length associated to $e^a$, while $N_a$ is a free parameter. We now consider 3 coordinates $y^m \in [ 0, 2\pi]$, with periodic identification $y^m \sim y^m + 2 \pi$. A solution to the Maurer-Cartan equations \eqref{MCsol29} is given by
\beq
\hspace{-0.25in}
\left(\!\! \begin{array}{c} \phantom{\sqrt{\left| \frac{N_1}{N_6} \right|}} \hspace{-0.4in}e^6 \\ \phantom{\sqrt{\left| \frac{N_1}{N_6} \right|}} \hspace{-0.4in}e^1 \\ \,e^4 \end{array} \!\!\right) = \left( \begin{array}{rrc} r_6\, \cos(\sqrt{|N_1 N_6|}\, y^4)  & \ \ - r_6\, \sqrt{\left| \frac{N_6}{N_1} \right|}\,  \sin(\sqrt{|N_1 N_6|}\, y^4)  &  \\ r_1\, \sqrt{\left| \frac{N_1}{N_6} \right|}\, \sin(\sqrt{|N_1 N_6|}\, y^4)  & r_1 \, \cos(\sqrt{|N_1 N_6|}\, y^4)  & \\ & & r_4 \end{array} \right) \left(\!\! \begin{array}{c} \phantom{\sqrt{\left| \frac{N_1}{N_6} \right|}} \hspace{-0.4in}\d y^6 \\ \phantom{\sqrt{\left| \frac{N_1}{N_6} \right|}} \hspace{-0.4in}\d y^1 \\ \,\d y^4 \end{array} \!\!\right) \label{eaexample}
\eeq
where the square matrix is nothing but the vielbein with coefficient $e^a{}_m$. From there, when considering $y^4 +2\pi$, various conditions combined with extra coordinate identifications are possible in order to leave the 1-forms $e^a$ invariant. The simplest choice, to be followed for solution $s_{55}^+ 29$, is to require
\beq
\sqrt{|N_1 N_6|} \in \mathbb{N}^* \ .
\eeq
This choice requires no extra coordinate identification beyond the periodic ones above. Therefore, the manifold is topologically simply a torus, here a priori with a non-Ricci flat metric $\d s^2 = \delta_{ab} e^a e^b$. Other choices of conditions and coordinate identifications are possible, leading to non-trivial lattice actions, giving manifolds whose cover is still a torus. More details on these lattices can be found in \cite{Andriot:2010ju, Grana:2013ila, Andriot:2020vlg} while further instances of these geometries can be found in \cite{Andriot:2015sia, Acharya:2019mcu, Acharya:2020hsc}. A a result, what was considered in \cite{Andriot:2024cct} for solution $s_{55}^+ 29$ was
\beq
N_6 = - \frac{1}{N_1} \ .
\eeq
It is crucial to note that there is no further quantization conditions on the structure constants. In particular, {\sl $N_1$ and $N_6$ do not have to be integers}, as we explained at the end of Section \ref{sec:groupman}: this is specific to solvable algebras. A rescaling of $r_1/r_6$ can bring the algebra to a basis where $N_1=-N_6=1$, but in general and in an arbitrary basis, they do not have to be integer.

We now turn to the 3d nilpotent Lie algebra Heis${}_3$ \eqref{alg3d}, that gives a nilmanifold; we follow the same presentation as above to offer a comparison. The Maurer-Cartan equations \eqref{MC} are given by
\beq
\d e^2 = 0 \ ,\quad \d e^3 = 0 \ ,\quad \d e^1 = - N \, \frac{r_1}{r_2 r_3}\, e^2 \w e^3 \ .
\eeq
We introduce again 3 coordinates $z^m \in [ 0, 1]$, where we now take for convenience the periodic identification to $z^m \sim z^m + 1$. A solution to the Maurer-Cartan equation is given by
\beq
e^2 = r_2 \, \d z^2 \ ,\quad e^3 = r_3 \, \d z^3 \ ,\quad e^1 = r_1 ( \d z^1 + N \, z^3 \, \d z^2 ) \ .
\eeq
Having globally defined 1-forms $e^a$ requires a lattice action, corresponding to the following coordinate identification, on top of the previous periodic ones
\beq
z^3 \sim z^3 + 1 \ ,\quad z^1 \sim z^1 - N \, z^2 \ .
\eeq
This indicates that the circle along $z^1$ gets twisted as a fiber over the $T^2$ base along $z^2, z^3$, hence the name of twisted torus for this nilmanifold. This extra coordinate identification also needs to be well-defined when $z^2 \sim z^2 + 1$. To that end, one needs $z^1 \sim z^1 + N$, which can be achieved with $z^1$ periodic identifications, if and only if
\beq
N \in \mathbb{Z}^* \ .
\eeq
This consistency requirement is one way to see that the parameter $N$ needs to be an integer, for a nilmanifold. Another way is to recognise the fibration as a $U(1)$-bundle. A detailed discussion of this condition can be found in \cite[App.A]{Andriot:2016rdd}. Finally, to be analogous to the previous discussion for the solvmanifold with \eqref{eaexample}, one can require for compactness that the vielbein matrix, evaluated at the period $z^3 =1$, is integer (up to the $r_a$), to allow for a compensation by an extra coordinate identification. This imposes again $N \in \mathbb{Z}^*$, and makes the $e^a$ globally defined.

So there is an important difference between structure constants, when parameterized as in \eqref{var3} with $N_a$: the latter needs to be integer for nilmanifolds, but not necessarily for solvmanifolds. This is read from the lattice quantization conditions. Looking at the curvature, or the 3d Ricci scalar, we obtain for the nilmanifold, respectively the solvmanifold
\beq
{\cal R}_{\text{Nil}_3} = - \frac{1}{2} (f^1{}_{23})^2 \ ,\qquad  {\cal R}_{\text{Solv}_3} = - \frac{1}{2} (f^6{}_{14} + f^1{}_{64})^2 \ .
\eeq
Considering for simplicity all lengths to be of the same order, $r_a = L$, related to the volume, together with the above lattice quantization conditions, we obtain
\beq
-2 {\cal R}_{\text{Nil}_3} \times L^2 =  N^2 \geq 1 \ ,\qquad  - 2 {\cal R}_{\text{Solv}_3} \times L^2 = \left(N_6 - \frac{1}{N_6}\right)^2 \geq 0 \ .
\eeq
We see that the nilmanifold curvature (times $-L^2$) gets bounded from below due to the integer condition, while the solvmanifold does not. This relates to one loophole in the no-go theorem of \cite{Junghans:2018gdb}, as we explain in detail in \cite[Sec.4.3]{Andriot:2024cct} and mention in Section \ref{sec:dSsolclassical}. Circumventing this no-go theorem is possible thanks to having a continuous parameter, here $N_6 \neq 1$, and to the explicit dependence on the lengths $r_a$ (not displayed here) that are anisotropic \cite[Sec.4.3]{Andriot:2024cct}.

For completeness, let us add a word on $\mathfrak{g}_{3.4}^{-1}$. There, the pair of structure constants is such that $f^1{}_{64}  f^6{}_{14} >0$, while this product was negative for $\mathfrak{g}_{3.5}^0$. As a result, the $\cos, \sin$ of \eqref{eaexample} essentially get replaced by $\cosh, \sinh$ for $\mathfrak{g}_{3.4}^{-1}$. The quantization condition still require entries in the (vielbein) matrix, analogous to that in \eqref{eaexample}, to take integer values (up to the $r_a$). This is much more complicated to achieve with $\cosh, \sinh$ functions, and the ratio $N_6/N_1$. We refer to \cite{Andriot:2010ju} for detailed lattice conditions. In \cite{Andriot:2020vlg, Andriot:2024cct}, we failed to satisfy them numerically for concrete solutions having this underlying Lie algebra. In other words, when solvable Lie algebras are known to allow for a lattice (as studied in \cite{Bock:2009}), this holds for some structure constant values, but not necessarily for others, such as those of a given supergravity solution.

\end{appendix}

\newpage

\addcontentsline{toc}{section}{References}
\bibliographystyle{JHEP}
\bibliography{Refs}

@article{Andriot:2010ju,
      author         = "Andriot, David and Goi, Enrico and Minasian, Ruben and
                        Petrini, Michela",
      title          = "{Supersymmetry breaking branes on solvmanifolds and de
                        Sitter vacua in string theory}",
      journal        = "JHEP",
      volume         = "1105",
      pages          = "028",
      doi            = "10.1007/JHEP05(2011)028",
      year           = "2011",
      eprint         = "1003.3774",
      archivePrefix  = "arXiv",
      primaryClass   = "hep-th",
      reportNumber   = "IPHT-T10-022",
      SLACcitation   = "%%CITATION = ARXIV:1003.3774;%%",
}

@article{Andriot:2012vb,
    author = "Andriot, David and Larfors, Magdalena and Lust, Dieter and Patalong, Peter",
    title = "{(Non-)commutative closed string on T-dual toroidal backgrounds}",
    eprint = "1211.6437",
    archivePrefix = "arXiv",
    primaryClass = "hep-th",
    reportNumber = "LMU-ASC-82-12, CERN-PH-TH-2012-325, MPP-2012-152, CERN-PH-TH-2012-235",
    doi = "10.1007/JHEP06(2013)021",
    journal = "JHEP",
    volume = "06",
    pages = "021",
    year = "2013"
}

@article{Andriot:2013xca,
    author = "Andriot, David and Betz, Andr{\'e}",
    title = "{$\beta$-supergravity: a ten-dimensional theory with non-geometric fluxes, and its geometric framework}",
    eprint = "1306.4381",
    archivePrefix = "arXiv",
    primaryClass = "hep-th",
    doi = "10.1007/JHEP12(2013)083",
    journal = "JHEP",
    volume = "12",
    pages = "083",
    year = "2013"
}

@article{Andriot:2014uda,
    author = "Andriot, David and Betz, Andr{\'e}",
    title = "{NS-branes, source corrected Bianchi identities, and more on backgrounds with non-geometric fluxes}",
    eprint = "1402.5972",
    archivePrefix = "arXiv",
    primaryClass = "hep-th",
    doi = "10.1007/JHEP07(2014)059",
    journal = "JHEP",
    volume = "07",
    pages = "059",
    year = "2014"
}

@article{Andriot:2015sia,
      author         = "Andriot, David",
      title          = "{New supersymmetric vacua on solvmanifolds}",
      journal        = "JHEP",
      volume         = "02",
      year           = "2016",
      pages          = "112",
      doi            = "10.1007/JHEP02(2016)112",
      eprint         = "1507.00014",
      archivePrefix  = "arXiv",
      primaryClass   = "hep-th",
      SLACcitation   = "%%CITATION = ARXIV:1507.00014;%%"
}

@article{Andriot:2015aza,
      author         = "Andriot, David",
      title          = "{A no-go theorem for monodromy inflation}",
      journal        = "JCAP",
      volume         = "1603",
      year           = "2016",
      number         = "03",
      pages          = "025",
      doi            = "10.1088/1475-7516/2016/03/025",
      eprint         = "1510.02005",
      archivePrefix  = "arXiv",
      primaryClass   = "hep-th",
      SLACcitation   = "%%CITATION = ARXIV:1510.02005;%%"
}

@article{Andriot:2016rdd,
    author = "Andriot, David and Cacciapaglia, Giacomo and Deandrea, Aldo and Deutschmann, Nicolas and Tsimpis, Dimitrios",
    title = "{Towards Kaluza-Klein Dark Matter on Nilmanifolds}",
    eprint = "1603.02289",
    archivePrefix = "arXiv",
    primaryClass = "hep-th",
    doi = "10.1007/JHEP06(2016)169",
    journal = "JHEP",
    volume = "06",
    pages = "169",
    year = "2016"
}

@article{Andriot:2016xvq,
      author         = "Andriot, David and Blåbäck, Johan",
      title          = "{Refining the boundaries of the classical de Sitter
                        landscape}",
      journal        = "JHEP",
      volume         = "03",
      year           = "2017",
      pages          = "102",
      doi            = "10.1007/JHEP03(2017)102, 10.1007/JHEP03(2018)083",
      note           = "[Erratum: JHEP03,083(2018)]",
      eprint         = "1609.00385",
      archivePrefix  = "arXiv",
      primaryClass   = "hep-th",
      reportNumber   = "IPHT-T16-080",
      SLACcitation   = "%%CITATION = ARXIV:1609.00385;%%"
}

@article{Andriot:2016ufg,
      author         = "Andriot, David and Blåbäck, Johan and Van Riet, Thomas",
      title          = "{Minkowski flux vacua of type II supergravities}",
      journal        = "Phys. Rev. Lett.",
      volume         = "118",
      year           = "2017",
      number         = "1",
      pages          = "011603",
      doi            = "10.1103/PhysRevLett.118.011603,
                        10.1103/PhysRevLett.120.169901",
      note           = "[Erratum: Phys. Rev. Lett.120,no.16,169901(2018)]",
      eprint         = "1609.00729",
      archivePrefix  = "arXiv",
      primaryClass   = "hep-th",
      reportNumber   = "IPHT-T16-081",
      SLACcitation   = "%%CITATION = ARXIV:1609.00729;%%"
}

@article{Andriot:2017oaz,
    author = "Andriot, David and Lucena G{\'o}mez, Gustavo",
    title = "{Signatures of extra dimensions in gravitational waves}",
    eprint = "1704.07392",
    archivePrefix = "arXiv",
    primaryClass = "hep-th",
    doi = "10.1088/1475-7516/2017/06/048",
    journal = "JCAP",
    volume = "06",
    pages = "048",
    year = "2017",
    note = "[Erratum: JCAP 05, E01 (2019)]"
}

@article{Andriot:2017jhf,
      author         = "Andriot, David",
      title          = "{On classical de Sitter and Minkowski solutions with
                        intersecting branes}",
      journal        = "JHEP",
      volume         = "03",
      year           = "2018",
      pages          = "054",
      doi            = "10.1007/JHEP03(2018)054",
      eprint         = "1710.08886",
      archivePrefix  = "arXiv",
      primaryClass   = "hep-th",
      reportNumber   = "CERN-TH-2017-222",
      SLACcitation   = "%%CITATION = ARXIV:1710.08886;%%"
}

@article{Andriot:2018tmb,
      author         = "Andriot, David and Tsimpis, Dimitrios",
      title          = "{Laplacian spectrum on a nilmanifold, truncations and
                        effective theories}",
      journal        = "JHEP",
      volume         = "09",
      year           = "2018",
      pages          = "096",
      doi            = "10.1007/JHEP09(2018)096",
      eprint         = "1806.05156",
      archivePrefix  = "arXiv",
      primaryClass   = "hep-th",
      SLACcitation   = "%%CITATION = ARXIV:1806.05156;%%"
}

@article{Andriot:2018wzk,
      author         = "Andriot, David",
      title          = "{On the de Sitter swampland criterion}",
      journal        = "Phys. Lett.",
      volume         = "B785",
      year           = "2018",
      pages          = "570-573",
      doi            = "10.1016/j.physletb.2018.09.022",
      eprint         = "1806.10999",
      archivePrefix  = "arXiv",
      primaryClass   = "hep-th",
      reportNumber   = "CERN-TH-2018-148",
      SLACcitation   = "%%CITATION = ARXIV:1806.10999;%%"
}

@article{Andriot:2018ept,
      author         = "Andriot, David",
      title          = "{New constraints on classical de Sitter: flirting with
                        the swampland}",
      journal        = "Fortsch. Phys.",
      volume         = "67",
      year           = "2019",
      number         = "1-2",
      pages          = "1800103",
      doi            = "10.1002/prop.201800103",
      eprint         = "1807.09698",
      archivePrefix  = "arXiv",
      primaryClass   = "hep-th",
      reportNumber   = "CERN-TH-2018-173",
      SLACcitation   = "%%CITATION = ARXIV:1807.09698;%%"
}

@article{Andriot:2018mav,
      author         = "Andriot, David and Roupec, Christoph",
      title          = "{Further refining the de Sitter swampland conjecture}",
      journal        = "Fortsch. Phys.",
      volume         = "67",
      year           = "2019",
      number         = "1-2",
      pages          = "1800105",
      doi            = "10.1002/prop.201800105",
      eprint         = "1811.08889",
      archivePrefix  = "arXiv",
      primaryClass   = "hep-th",
      reportNumber   = "CERN-TH-2018-251",
      SLACcitation   = "%%CITATION = ARXIV:1811.08889;%%"
}

@article{Andriot:2019wrs,
      author         = "Andriot, David",
      title          = "{Open problems on classical de Sitter solutions}",
      journal        = "Fortsch. Phys.",
      volume         = "67",
      year           = "2019",
      number         = "7",
      pages          = "1900026",
      doi            = "10.1002/prop.201900026",
      eprint         = "1902.10093",
      archivePrefix  = "arXiv",
      primaryClass   = "hep-th",
      reportNumber   = "CERN-TH-2019-019",
      SLACcitation   = "%%CITATION = ARXIV:1902.10093;%%"
}

@article{Andriot:2019hay,
    author = "Andriot, David and Tsimpis, Dimitrios",
    title = "{Gravitational waves in warped compactifications}",
    eprint = "1911.01444",
    archivePrefix = "arXiv",
    primaryClass = "hep-th",
    doi = "10.1007/JHEP06(2020)100",
    journal = "JHEP",
    volume = "06",
    pages = "100",
    year = "2020"
}

@article{Andriot:2020lea,
    author = "Andriot, David and Cribiori, Niccolò and Erkinger, David",
    title = "{The web of swampland conjectures and the TCC bound}",
    eprint = "2004.00030",
    archivePrefix = "arXiv",
    primaryClass = "hep-th",
    doi = "10.1007/JHEP07(2020)162",
    journal = "JHEP",
    volume = "07",
    pages = "162",
    year = "2020"
}

@article{Andriot:2020wpp,
    author = "Andriot, David and Marconnet, Paul and Wrase, Timm",
    title = "{New de Sitter solutions of 10d type IIB supergravity}",
    eprint = "2005.12930",
    archivePrefix = "arXiv",
    primaryClass = "hep-th",
    doi = "10.1007/JHEP08(2020)076",
    journal = "JHEP",
    volume = "08",
    pages = "076",
    year = "2020"
}

@article{Andriot:2020vlg,
    author = "Andriot, David and Marconnet, Paul and Wrase, Timm",
    title = "{Intricacies of classical de Sitter string backgrounds}",
    eprint = "2006.01848",
    archivePrefix = "arXiv",
    primaryClass = "hep-th",
    doi = "10.1016/j.physletb.2020.136015",
    journal = "Phys. Lett. B",
    volume = "812",
    pages = "136015",
    year = "2021"
}

@article{Andriot:2021rdy,
    author = "Andriot, David",
    title = "{Tachyonic de Sitter solutions of 10d type II supergravities}",
    eprint = "2101.06251",
    archivePrefix = "arXiv",
    primaryClass = "hep-th",
    doi = "10.1002/prop.202100063",
    month = "1",
    year = "2021"
}

@article{Andriot:2021gwv,
    author = "Andriot, David and Marconnet, Paul and Tsimpis, Dimitrios",
    title = "{Warp factor and the gravitational wave spectrum}",
    eprint = "2103.09240",
    archivePrefix = "arXiv",
    primaryClass = "hep-th",
    doi = "10.1088/1475-7516/2021/07/040",
    journal = "JCAP",
    volume = "07",
    pages = "040",
    year = "2021"
}

@article{Andriot:2022way,
    author = "Andriot, David and Horer, Ludwig and Marconnet, Paul",
    title = "{Charting the landscape of (anti-) de Sitter and Minkowski solutions of 10d supergravities}",
    eprint = "2201.04152",
    archivePrefix = "arXiv",
    primaryClass = "hep-th",
    doi = "10.1007/JHEP06(2022)131",
    journal = "JHEP",
    volume = "06",
    pages = "131",
    year = "2022"
}

@article{Andriot:2022yyj,
    author = "Andriot, David and Horer, Ludwig and Marconnet, Paul",
    title = "{Exploring the landscape of (anti-) de Sitter and Minkowski solutions: group manifolds, stability and scale separation}",
    eprint = "2204.05327",
    archivePrefix = "arXiv",
    primaryClass = "hep-th",
    doi = "10.1007/JHEP08(2022)109",
    journal = "JHEP",
    volume = "08",
    pages = "109",
    year = "2022",
    note = "[Erratum: JHEP 09, 184 (2022)]"
}

@article{Andriot:2022xjh,
    author = "Andriot, David and Horer, Ludwig",
    title = "{(Quasi-) de Sitter solutions across dimensions and the TCC bound}",
    eprint = "2208.14462",
    archivePrefix = "arXiv",
    primaryClass = "hep-th",
    doi = "10.1007/JHEP01(2023)020",
    journal = "JHEP",
    volume = "01",
    pages = "020",
    year = "2023"
}

@article{Andriot:2022bnb,
    author = "Andriot, David and Marconnet, Paul and Rajaguru, Muthusamy and Wrase, Timm",
    title = "{Automated consistent truncations and stability of flux compactifications}",
    eprint = "2209.08015",
    archivePrefix = "arXiv",
    primaryClass = "hep-th",
    doi = "10.1007/JHEP12(2022)026",
    journal = "JHEP",
    volume = "12",
    pages = "026",
    year = "2022",
    note = "[Addendum: JHEP 04, 044 (2023)]"
}

@article{Andriot:2022brg,
    author = "Andriot, David and Horer, Ludwig and Tringas, George",
    title = "{Negative scalar potentials and the swampland: an Anti-Trans-Planckian Censorship Conjecture}",
    eprint = "2212.04517",
    archivePrefix = "arXiv",
    primaryClass = "hep-th",
    doi = "10.1007/JHEP04(2023)139",
    journal = "JHEP",
    volume = "04",
    pages = "139",
    year = "2023"
}

@article{Andriot:2023isc,
    author = "Andriot, David",
    title = "{Bumping into the species scale with the scalar potential}",
    eprint = "2305.07480",
    archivePrefix = "arXiv",
    primaryClass = "hep-th",
    doi = "10.1002/prop.202300139",
    journal = "Fortsch. Phys.",
    volume = "71",
    pages = "2300139",
    year = "2023"
}

@article{Andriot:2023wvg,
    author = "Andriot, David and Tsimpis, Dimitrios and Wrase, Timm",
    title = "{Accelerated expansion of an open universe and string theory realizations}",
    eprint = "2309.03938",
    archivePrefix = "arXiv",
    primaryClass = "hep-th",
    doi = "10.1103/PhysRevD.108.123515",
    journal = "Phys. Rev. D",
    volume = "108",
    number = "12",
    pages = "123515",
    year = "2023"
}

@article{Andriot:2023fss,
    author = "Andriot, David and Tringas, George",
    title = "{Extensions of a scale-separated AdS$_{4}$ solution and their mass spectrum}",
    eprint = "2310.06115",
    archivePrefix = "arXiv",
    primaryClass = "hep-th",
    doi = "10.1007/JHEP01(2024)008",
    journal = "JHEP",
    volume = "01",
    pages = "008",
    year = "2024"
}

@article{Andriot:2024cct,
    author = "Andriot, David and Ruehle, Fabian",
    title = "{On classical de Sitter solutions and parametric control}",
    eprint = "2403.07065",
    archivePrefix = "arXiv",
    primaryClass = "hep-th",
    doi = "10.1007/JHEP06(2024)101",
    journal = "JHEP",
    volume = "06",
    pages = "101",
    year = "2024"
}

@article{Andriot:2024jsh,
    author = "Andriot, David and Parameswaran, Susha and Tsimpis, Dimitrios and Wrase, Timm and Zavala, Ivonne",
    title = "{Exponential quintessence: curved, steep and stringy?}",
    eprint = "2405.09323",
    archivePrefix = "arXiv",
    primaryClass = "hep-th",
    doi = "10.1007/JHEP08(2024)117",
    journal = "JHEP",
    volume = "08",
    pages = "117",
    year = "2024"
}

@article{Andriot:2024sif,
    author = "Andriot, David",
    title = "{Quintessence: An Analytical Study, With Theoretical and Observational Applications}",
    eprint = "2410.17182",
    archivePrefix = "arXiv",
    primaryClass = "hep-th",
    doi = "10.1002/prop.70007",
    journal = "Fortsch. Phys.",
    volume = "73",
    number = "6",
    pages = "e70007",
    year = "2025"
}

@article{Andriot:2025gyr,
    author = "Andriot, David and Rajaguru, Muthusamy and Tringas, George",
    title = "{Single versus multifield scalar potentials from string theory}",
    eprint = "2501.17775",
    archivePrefix = "arXiv",
    primaryClass = "hep-th",
    doi = "10.1007/JHEP05(2025)046",
    journal = "JHEP",
    volume = "05",
    pages = "046",
    year = "2025"
}

@article{Andriot:2025cyi,
    author = "Andriot, David and Cribiori, Niccol{\`o} and Van Riet, Thomas",
    title = "{Scale separation, rolling solutions, and entropy bounds}",
    eprint = "2504.08634",
    archivePrefix = "arXiv",
    primaryClass = "hep-th",
    doi = "10.1103/5rkw-5qfk",
    journal = "Phys. Rev. D",
    volume = "112",
    number = "2",
    pages = "026028",
    year = "2025"
}

@article{Andriot:2025los,
    author = "Andriot, David",
    title = "{Phantom matters}",
    eprint = "2505.10410",
    archivePrefix = "arXiv",
    primaryClass = "hep-th",
    doi = "10.1016/j.dark.2025.102000",
    journal = "Phys. Dark Univ.",
    volume = "49",
    pages = "102000",
    year = "2025"
}

@article{Bergshoeff:2001pv,
    author = "Bergshoeff, Eric and Kallosh, Renata and Ortin, Tomas and Roest, Diederik and Van Proeyen, Antoine",
    title = "{New formulations of D = 10 supersymmetry and D8 - O8 domain walls}",
    eprint = "hep-th/0103233",
    archivePrefix = "arXiv",
    reportNumber = "UG-00-15, SU-ITP-01-09, IFT-UAM-CSIC-00-35, KUL-TF-01-06",
    doi = "10.1088/0264-9381/18/17/303",
    journal = "Class. Quant. Grav.",
    volume = "18",
    pages = "3359--3382",
    year = "2001"
}

@article{Koerber:2007hd,
    author = "Koerber, Paul and Tsimpis, Dimitrios",
    title = "{Supersymmetric sources, integrability and generalized-structure compactifications}",
    eprint = "0706.1244",
    archivePrefix = "arXiv",
    primaryClass = "hep-th",
    reportNumber = "MPP-2007-66, LMU-ASC-37-07",
    doi = "10.1088/1126-6708/2007/08/082",
    journal = "JHEP",
    volume = "08",
    pages = "082",
    year = "2007"
}

@article{Camara:2005dc,
    author = "Camara, Pablo G. and Font, A. and Ibanez, L. E.",
    title = "{Fluxes, moduli fixing and MSSM-like vacua in a simple IIA orientifold}",
    eprint = "hep-th/0506066",
    archivePrefix = "arXiv",
    reportNumber = "IFT-UAM-CSIC-05-28",
    doi = "10.1088/1126-6708/2005/09/013",
    journal = "JHEP",
    volume = "09",
    pages = "013",
    year = "2005"
}

@article{Dasgupta:1999ss,
    author = "Dasgupta, Keshav and Rajesh, Govindan and Sethi, Savdeep",
    title = "{M theory, orientifolds and G - flux}",
    eprint = "hep-th/9908088",
    archivePrefix = "arXiv",
    reportNumber = "IASSNS-HEP-99-75, NSF-ITP-99-095",
    doi = "10.1088/1126-6708/1999/08/023",
    journal = "JHEP",
    volume = "08",
    pages = "023",
    year = "1999"
}

@article{Burgess:2011rv,
    author = "Burgess, C. P. and Maharana, Anshuman and van Nierop, L. and Nizami, A. A. and Quevedo, F.",
    title = "{On Brane Back-Reaction and de Sitter Solutions in Higher-Dimensional Supergravity}",
    eprint = "1109.0532",
    archivePrefix = "arXiv",
    primaryClass = "hep-th",
    reportNumber = "DAMTP-2011-66",
    doi = "10.1007/JHEP04(2012)018",
    journal = "JHEP",
    volume = "04",
    pages = "018",
    year = "2012"
}

@article{Shiu:2011zt,
    author = "Shiu, Gary and Sumitomo, Yoske",
    title = "{Stability Constraints on Classical de Sitter Vacua}",
    eprint = "1107.2925",
    archivePrefix = "arXiv",
    primaryClass = "hep-th",
    reportNumber = "MAD-TH-11-06, TIFR-TH-11-31",
    doi = "10.1007/JHEP09(2011)052",
    journal = "JHEP",
    volume = "09",
    pages = "052",
    year = "2011"
}

@article{Hertzberg:2007wc,
    author = "Hertzberg, Mark P. and Kachru, Shamit and Taylor, Washington and Tegmark, Max",
    title = "{Inflationary Constraints on Type IIA String Theory}",
    eprint = "0711.2512",
    archivePrefix = "arXiv",
    primaryClass = "hep-th",
    reportNumber = "MIT-CTP-3905, SLAC-PUB-12999",
    doi = "10.1088/1126-6708/2007/12/095",
    journal = "JHEP",
    volume = "12",
    pages = "095",
    year = "2007"
}

@article{Silverstein:2007ac,
    author = "Silverstein, Eva",
    title = "{Simple de Sitter Solutions}",
    eprint = "0712.1196",
    archivePrefix = "arXiv",
    primaryClass = "hep-th",
    reportNumber = "SLAC-PUB-13016, SITP-07-20",
    doi = "10.1103/PhysRevD.77.106006",
    journal = "Phys. Rev. D",
    volume = "77",
    pages = "106006",
    year = "2008"
}

@article{VanRiet:2011yc,
    author = "Van Riet, Thomas",
    title = "{On classical de Sitter solutions in higher dimensions}",
    eprint = "1111.3154",
    archivePrefix = "arXiv",
    primaryClass = "hep-th",
    doi = "10.1088/0264-9381/29/5/055001",
    journal = "Class. Quant. Grav.",
    volume = "29",
    pages = "055001",
    year = "2012"
}

@article{Danielsson:2012et,
    author = "Danielsson, Ulf H. and Shiu, Gary and Van Riet, Thomas and Wrase, Timm",
    title = "{A note on obstinate tachyons in classical dS solutions}",
    eprint = "1212.5178",
    archivePrefix = "arXiv",
    primaryClass = "hep-th",
    reportNumber = "UUITP-26-12, MAD-TH-12-10, SU-ITP-12-44",
    doi = "10.1007/JHEP03(2013)138",
    journal = "JHEP",
    volume = "03",
    pages = "138",
    year = "2013"
}

@article{Marconnet:2022fmx,
    author = "Marconnet, Paul and Tsimpis, Dimitrios",
    title = "{Universal accelerating cosmologies from 10d supergravity}",
    eprint = "2210.10813",
    archivePrefix = "arXiv",
    primaryClass = "hep-th",
    doi = "10.1007/JHEP01(2023)033",
    journal = "JHEP",
    volume = "01",
    pages = "033",
    year = "2023"
}

@article{Tsimpis:2020ysl,
    author = "Tsimpis, Dimitrios",
    title = {{Consistent truncation on Calabi-Yau and Nearly-K{\"a}hler manifolds}},
    eprint = "2002.09359",
    archivePrefix = "arXiv",
    primaryClass = "hep-th",
    doi = "10.22323/1.376.0122",
    journal = "PoS",
    volume = "CORFU2019",
    pages = "122",
    year = "2020"
}

@article{Lin:2024eqq,
    author = "Lin, Jieming and Skrzypek, Torben and Stelle, K. S.",
    title = "{Compactification on Calabi-Yau threefolds: consistent truncation to pure supergravity}",
    eprint = "2412.00186",
    archivePrefix = "arXiv",
    primaryClass = "hep-th",
    reportNumber = "Imperial/TP/2024/KS/01, DESY-24-179",
    doi = "10.1007/JHEP03(2025)200",
    journal = "JHEP",
    volume = "03",
    pages = "200",
    year = "2025"
}

@article{Trigiante:2016mnt,
    author = "Trigiante, Mario",
    title = "{Gauged Supergravities}",
    eprint = "1609.09745",
    archivePrefix = "arXiv",
    primaryClass = "hep-th",
    doi = "10.1016/j.physrep.2017.03.001",
    journal = "Phys. Rept.",
    volume = "680",
    pages = "1--175",
    year = "2017"
}

@article{Grana:2005jc,
    author = "Grana, Mariana",
    title = "{Flux compactifications in string theory: A Comprehensive review}",
    eprint = "hep-th/0509003",
    archivePrefix = "arXiv",
    reportNumber = "LPTENS-05-26, CPHT-RR-049-0805",
    doi = "10.1016/j.physrep.2005.10.008",
    journal = "Phys. Rept.",
    volume = "423",
    pages = "91--158",
    year = "2006"
}

@article{Koerber:2010bx,
    author = "Koerber, Paul",
    title = "{Lectures on Generalized Complex Geometry for Physicists}",
    eprint = "1006.1536",
    archivePrefix = "arXiv",
    primaryClass = "hep-th",
    reportNumber = "KUL-TF-10-04",
    doi = "10.1002/prop.201000083",
    journal = "Fortsch. Phys.",
    volume = "59",
    pages = "169--242",
    year = "2011"
}

@article{Smith:2024ejf,
    author = "Smith, George R. and Tennyson, David and Waldram, Daniel",
    title = "{All-orders moduli for type II flux backgrounds}",
    eprint = "2409.03847",
    archivePrefix = "arXiv",
    primaryClass = "hep-th",
    reportNumber = "Imperial/TP/24/DW/1; MI-HET-828",
    month = "9",
    year = "2024"
}

@inproceedings{Gibbons:1984kp,
    author = "Gibbons, G. W.",
    title = "{ASPECTS OF SUPERGRAVITY THEORIES}",
    booktitle = "{XV GIFT Seminar on Supersymmetry and Supergravity}",
    reportNumber = "Print-85-0061 (CAMBRIDGE)",
    month = "6",
    year = "1984"
}

@article{deWit:1986mwo,
    author = "de Wit, B. and Smit, D. J. and Hari Dass, N. D.",
    title = "{Residual Supersymmetry of Compactified D=10 Supergravity}",
    reportNumber = "NIKHEF-H-86-15",
    doi = "10.1016/0550-3213(87)90267-7",
    journal = "Nucl. Phys. B",
    volume = "283",
    pages = "165",
    year = "1987"
}

@article{Maldacena:2000mw,
    author = "Maldacena, Juan Martin and Nunez, Carlos",
    editor = "Duff, Michael J. and Liu, J. T. and Lu, J.",
    title = "{Supergravity description of field theories on curved manifolds and a no go theorem}",
    eprint = "hep-th/0007018",
    archivePrefix = "arXiv",
    doi = "10.1142/S0217751X01003937",
    journal = "Int. J. Mod. Phys. A",
    volume = "16",
    pages = "822--855",
    year = "2001"
}

@inproceedings{Townsend:2003qv,
    author = "Townsend, Paul K.",
    title = "{Cosmic acceleration and M-theory}",
    booktitle = "{14th International Congress on Mathematical Physics}",
    eprint = "hep-th/0308149",
    archivePrefix = "arXiv",
    reportNumber = "DAMTP-2003-79",
    doi = "10.1142/9789812704016_0067",
    pages = "655--662",
    year = "2006"
}

@article{Wrase:2010ew,
    author = "Wrase, Timm and Zagermann, Marco",
    editor = "Anagnostopoulos, Konstantinos and Zoupanos, George",
    title = "{On Classical de Sitter Vacua in String Theory}",
    eprint = "1003.0029",
    archivePrefix = "arXiv",
    primaryClass = "hep-th",
    doi = "10.1002/prop.201000053",
    journal = "Fortsch. Phys.",
    volume = "58",
    pages = "906--910",
    year = "2010"
}

@article{Blaback:2010sj,
    author = "Blaback, Johan and Danielsson, Ulf H. and Junghans, Daniel and Van Riet, Thomas and Wrase, Timm and Zagermann, Marco",
    title = "{Smeared versus localised sources in flux compactifications}",
    eprint = "1009.1877",
    archivePrefix = "arXiv",
    primaryClass = "hep-th",
    reportNumber = "UUITP-28-10, ITP-UH-15-10",
    doi = "10.1007/JHEP12(2010)043",
    journal = "JHEP",
    volume = "12",
    pages = "043",
    year = "2010"
}

@article{Cribiori:2019clo,
    author = "Cribiori, Niccol{\`o} and Junghans, Daniel",
    title = "{No classical (anti-)de Sitter solutions with O8-planes}",
    eprint = "1902.08209",
    archivePrefix = "arXiv",
    primaryClass = "hep-th",
    doi = "10.1016/j.physletb.2019.04.030",
    journal = "Phys. Lett. B",
    volume = "793",
    pages = "54--58",
    year = "2019"
}

@article{Haque:2008jz,
    author = "Haque, Sheikh Shajidul and Shiu, Gary and Underwood, Bret and Van Riet, Thomas",
    title = "{Minimal simple de Sitter solutions}",
    eprint = "0810.5328",
    archivePrefix = "arXiv",
    primaryClass = "hep-th",
    reportNumber = "SLAC-PUB-14712, MAD-TH-2008-13, SU-ITP-08-27, UG-FT-241-08, CAFPE-111-08",
    doi = "10.1103/PhysRevD.79.086005",
    journal = "Phys. Rev. D",
    volume = "79",
    pages = "086005",
    year = "2009"
}

@article{Caviezel:2008tf,
    author = "Caviezel, Claudio and Koerber, Paul and Kors, Simon and Lust, Dieter and Wrase, Timm and Zagermann, Marco",
    title = "{On the Cosmology of Type IIA Compactifications on SU(3)-structure Manifolds}",
    eprint = "0812.3551",
    archivePrefix = "arXiv",
    primaryClass = "hep-th",
    reportNumber = "MPP-2008-165, LMU-ASC-64-08",
    doi = "10.1088/1126-6708/2009/04/010",
    journal = "JHEP",
    volume = "04",
    pages = "010",
    year = "2009"
}

@article{Flauger:2008ad,
    author = "Flauger, Raphael and Paban, Sonia and Robbins, Daniel and Wrase, Timm",
    title = "{Searching for slow-roll moduli inflation in massive type IIA supergravity with metric fluxes}",
    eprint = "0812.3886",
    archivePrefix = "arXiv",
    primaryClass = "hep-th",
    reportNumber = "UTTG-09-08, MPI-2008-170",
    doi = "10.1103/PhysRevD.79.086011",
    journal = "Phys. Rev. D",
    volume = "79",
    pages = "086011",
    year = "2009"
}

@article{Danielsson:2009ff,
    author = "Danielsson, Ulf H. and Haque, Sheikh Shajidul and Shiu, Gary and Van Riet, Thomas",
    title = "{Towards Classical de Sitter Solutions in String Theory}",
    eprint = "0907.2041",
    archivePrefix = "arXiv",
    primaryClass = "hep-th",
    reportNumber = "UUITP-18-09, MAD-TH-09-06",
    doi = "10.1088/1126-6708/2009/09/114",
    journal = "JHEP",
    volume = "09",
    pages = "114",
    year = "2009"
}

@article{Caviezel:2009tu,
    author = "Caviezel, Claudio and Wrase, Timm and Zagermann, Marco",
    title = "{Moduli Stabilization and Cosmology of Type IIB on SU(2)-Structure Orientifolds}",
    eprint = "0912.3287",
    archivePrefix = "arXiv",
    primaryClass = "hep-th",
    reportNumber = "ITP-UH-21-09, MPP-2009-212",
    doi = "10.1007/JHEP04(2010)011",
    journal = "JHEP",
    volume = "04",
    pages = "011",
    year = "2010"
}

@article{Obied:2018sgi,
    author = "Obied, Georges and Ooguri, Hirosi and Spodyneiko, Lev and Vafa, Cumrun",
    title = "{De Sitter Space and the Swampland}",
    eprint = "1806.08362",
    archivePrefix = "arXiv",
    primaryClass = "hep-th",
    reportNumber = "CALT-TH-2018-020, IPMU18-0100",
    month = "6",
    year = "2018"
}

@article{Covi:2008ea,
    author = "Covi, Laura and Gomez-Reino, Marta and Gross, Christian and Louis, Jan and Palma, Gonzalo A. and Scrucca, Claudio A.",
    title = "{de Sitter vacua in no-scale supergravities and Calabi-Yau string models}",
    eprint = "0804.1073",
    archivePrefix = "arXiv",
    primaryClass = "hep-th",
    reportNumber = "DESY-08-038, CERN-PH-TH-2008-066, ZMP-HH-08-6",
    doi = "10.1088/1126-6708/2008/06/057",
    journal = "JHEP",
    volume = "06",
    pages = "057",
    year = "2008"
}

@article{Kallosh:2014oja,
    author = "Kallosh, Renata and Linde, Andrei and Vercnocke, Bert and Wrase, Timm",
    title = "{Analytic Classes of Metastable de Sitter Vacua}",
    eprint = "1406.4866",
    archivePrefix = "arXiv",
    primaryClass = "hep-th",
    reportNumber = "SU-ITP-14-18",
    doi = "10.1007/JHEP10(2014)011",
    journal = "JHEP",
    volume = "10",
    pages = "011",
    year = "2014"
}

@article{Junghans:2016uvg,
    author = "Junghans, Daniel",
    title = "{Tachyons in Classical de Sitter Vacua}",
    eprint = "1603.08939",
    archivePrefix = "arXiv",
    primaryClass = "hep-th",
    doi = "10.1007/JHEP06(2016)132",
    journal = "JHEP",
    volume = "06",
    pages = "132",
    year = "2016"
}

@article{Junghans:2016abx,
    author = "Junghans, Daniel and Zagermann, Marco",
    title = "{A Universal Tachyon in Nearly No-scale de Sitter Compactifications}",
    eprint = "1612.06847",
    archivePrefix = "arXiv",
    primaryClass = "hep-th",
    doi = "10.1007/JHEP07(2018)078",
    journal = "JHEP",
    volume = "07",
    pages = "078",
    year = "2018"
}

@article{Danielsson:2011au,
    author = "Danielsson, Ulf H. and Haque, Sheikh S. and Koerber, Paul and Shiu, Gary and Van Riet, Thomas and Wrase, Timm",
    title = "{De Sitter hunting in a classical landscape}",
    eprint = "1103.4858",
    archivePrefix = "arXiv",
    primaryClass = "hep-th",
    reportNumber = "UUITP-05-11, MAD-TH-10-04",
    doi = "10.1002/prop.201100047",
    journal = "Fortsch. Phys.",
    volume = "59",
    pages = "897--933",
    year = "2011"
}

@article{Danielsson:2012by,
    author = "Danielsson, Ulf and Dibitetto, Giuseppe",
    title = "{On the distribution of stable de Sitter vacua}",
    eprint = "1212.4984",
    archivePrefix = "arXiv",
    primaryClass = "hep-th",
    reportNumber = "UUITP-27-12",
    doi = "10.1007/JHEP03(2013)018",
    journal = "JHEP",
    volume = "03",
    pages = "018",
    year = "2013"
}

@article{Kallosh:2018nrk,
    author = "Kallosh, Renata and Wrase, Timm",
    title = "{dS Supergravity from 10d}",
    eprint = "1808.09427",
    archivePrefix = "arXiv",
    primaryClass = "hep-th",
    doi = "10.1002/prop.201800071",
    journal = "Fortsch. Phys.",
    volume = "67",
    number = "1-2",
    pages = "1800071",
    year = "2019"
}

@article{Parameswaran:2024mrc,
    author = "Parameswaran, Susha and Serra, Marco",
    title = "{On (A)dS solutions from Scherk-Schwarz orbifolds}",
    eprint = "2407.16781",
    archivePrefix = "arXiv",
    primaryClass = "hep-th",
    doi = "10.1007/JHEP10(2024)039",
    journal = "JHEP",
    volume = "10",
    pages = "039",
    year = "2024"
}

@article{Denef:2008wq,
    author = "Denef, Frederik",
    editor = "Bachas, Costas and Baulieu, Laurent and Douglas, Michael and Kiritsis, Elias and Rabinovici, Eliezer and Vanhove, Pierre and Windey, Paul and Cugliandolo, Leticia F.",
    title = "{Lectures on constructing string vacua}",
    eprint = "0803.1194",
    archivePrefix = "arXiv",
    primaryClass = "hep-th",
    doi = "10.1016/S0924-8099(08)80029-7",
    journal = "Les Houches",
    volume = "87",
    pages = "483--610",
    year = "2008"
}

@article{Vafa:2005ui,
    author = "Vafa, Cumrun",
    title = "{The String landscape and the swampland}",
    eprint = "hep-th/0509212",
    archivePrefix = "arXiv",
    reportNumber = "HUTP-05-A043",
    month = "9",
    year = "2005"
}

@article{Palti:2019pca,
    author = "Palti, Eran",
    title = "{The Swampland: Introduction and Review}",
    eprint = "1903.06239",
    archivePrefix = "arXiv",
    primaryClass = "hep-th",
    reportNumber = "MPP-2019-53",
    doi = "10.1002/prop.201900037",
    journal = "Fortsch. Phys.",
    volume = "67",
    number = "6",
    pages = "1900037",
    year = "2019"
}

@article{vanBeest:2021lhn,
    author = "van Beest, Marieke and Calder\'on-Infante, Jos\'e and Mirfendereski, Delaram and Valenzuela, Irene",
    title = "{Lectures on the Swampland Program in String Compactifications}",
    eprint = "2102.01111",
    archivePrefix = "arXiv",
    primaryClass = "hep-th",
    doi = "10.1016/j.physrep.2022.09.002",
    journal = "Phys. Rept.",
    volume = "989",
    pages = "1--50",
    year = "2022"
}

@article{Agmon:2022thq,
    author = "Agmon, Nathan Benjamin and Bedroya, Alek and Kang, Monica Jinwoo and Vafa, Cumrun",
    title = "{Lectures on the string landscape and the Swampland}",
    eprint = "2212.06187",
    archivePrefix = "arXiv",
    primaryClass = "hep-th",
    month = "12",
    year = "2022"
}

@article{ValeixoBento:2025yhz,
    author = "Valeixo Bento, Bruno and Montero, Miguel",
    title = "{An M-theory dS maximum from Casimir energies on Riemann-flat manifolds}",
    eprint = "2507.02037",
    archivePrefix = "arXiv",
    primaryClass = "hep-th",
    reportNumber = "IFT-025-070",
    month = "7",
    year = "2025"
}

@article{Bena:2023sks,
    author = "Bena, Iosif and Gra{\~n}a, Mariana and Van Riet, Thomas",
    title = "{Trustworthy de Sitter compactifications of string theory: a comprehensive review}",
    eprint = "2303.17680",
    archivePrefix = "arXiv",
    primaryClass = "hep-th",
    month = "3",
    year = "2023"
}

@article{Kachru:2003aw,
    author = "Kachru, Shamit and Kallosh, Renata and Linde, Andrei D. and Trivedi, Sandip P.",
    title = "{De Sitter vacua in string theory}",
    eprint = "hep-th/0301240",
    archivePrefix = "arXiv",
    reportNumber = "SLAC-PUB-9630, SU-ITP-03-01, TIFR-TH-03-03",
    doi = "10.1103/PhysRevD.68.046005",
    journal = "Phys. Rev. D",
    volume = "68",
    pages = "046005",
    year = "2003"
}

@article{Danielsson:2018ztv,
    author = "Danielsson, Ulf H. and Van Riet, Thomas",
    title = "{What if string theory has no de Sitter vacua?}",
    eprint = "1804.01120",
    archivePrefix = "arXiv",
    primaryClass = "hep-th",
    reportNumber = "UUITP-09/18, UUITP-09-18",
    doi = "10.1142/S0218271818300070",
    journal = "Int. J. Mod. Phys. D",
    volume = "27",
    number = "12",
    pages = "1830007",
    year = "2018"
}

@article{Brennan:2017rbf,
    author = "Brennan, T. Daniel and Carta, Federico and Vafa, Cumrun",
    title = "{The String Landscape, the Swampland, and the Missing Corner}",
    eprint = "1711.00864",
    archivePrefix = "arXiv",
    primaryClass = "hep-th",
    reportNumber = "IFT-UAM-CSIC-17-105",
    doi = "10.22323/1.305.0015",
    journal = "PoS",
    volume = "TASI2017",
    pages = "015",
    year = "2017"
}

@book{DallAgata:2021uvl,
    author = "Dall{\textquoteright}Agata, Gianguido and Zagermann, Marco",
    title = "{Supergravity: From First Principles to Modern Applications}",
    doi = "10.1007/978-3-662-63980-1",
    isbn = "978-3-662-63978-8, 978-3-662-63980-1",
    series = "Lecture Notes in Physics",
    volume = "991",
    month = "7",
    year = "2021",
    publisher = "Springer"
}

@article{Ooguri:2016pdq,
    author = "Ooguri, Hirosi and Vafa, Cumrun",
    title = "{Non-supersymmetric AdS and the Swampland}",
    eprint = "1610.01533",
    archivePrefix = "arXiv",
    primaryClass = "hep-th",
    reportNumber = "CALT-TH-2016-027, IPMU16-0139",
    doi = "10.4310/ATMP.2017.v21.n7.a8",
    journal = "Adv. Theor. Math. Phys.",
    volume = "21",
    pages = "1787--1801",
    year = "2017"
}

@article{Anninos:2011ui,
    author = "Anninos, Dionysios and Hartman, Thomas and Strominger, Andrew",
    title = "{Higher Spin Realization of the dS/CFT Correspondence}",
    eprint = "1108.5735",
    archivePrefix = "arXiv",
    primaryClass = "hep-th",
    doi = "10.1088/1361-6382/34/1/015009",
    journal = "Class. Quant. Grav.",
    volume = "34",
    number = "1",
    pages = "015009",
    year = "2017"
}

@article{Anninos:2017eib,
    author = "Anninos, Dionysios and Denef, Frederik and Monten, Ruben and Sun, Zimo",
    title = "{Higher Spin de Sitter Hilbert Space}",
    eprint = "1711.10037",
    archivePrefix = "arXiv",
    primaryClass = "hep-th",
    doi = "10.1007/JHEP10(2019)071",
    journal = "JHEP",
    volume = "10",
    pages = "071",
    year = "2019",
    note = "[Erratum: JHEP 06, 085 (2024)]"
}

@article{Collier:2025lux,
    author = {Collier, Scott and Eberhardt, Lorenz and M{\"u}hlmann, Beatrix},
    title = "{A microscopic realization of dS$_3$}",
    eprint = "2501.01486",
    archivePrefix = "arXiv",
    primaryClass = "hep-th",
    doi = "10.21468/SciPostPhys.18.4.131",
    journal = "SciPost Phys.",
    volume = "18",
    number = "4",
    pages = "131",
    year = "2025"
}

@article{Castro:2023dxp,
    author = "Castro, Alejandra and Coman, Ioana and Fliss, Jackson R. and Zukowski, Claire",
    title = "{Keeping matter in the loop in dS$_{3}$ quantum gravity}",
    eprint = "2302.12281",
    archivePrefix = "arXiv",
    primaryClass = "hep-th",
    doi = "10.1007/JHEP07(2023)120",
    journal = "JHEP",
    volume = "07",
    pages = "120",
    year = "2023",
    note = "[Erratum: JHEP 09, 004 (2024)]"
}

@inproceedings{Witten:2001kn,
    author = "Witten, Edward",
    title = "{Quantum gravity in de Sitter space}",
    booktitle = "{Strings 2001: International Conference}",
    eprint = "hep-th/0106109",
    archivePrefix = "arXiv",
    month = "6",
    year = "2001"
}

@article{Hull:1998vg,
    author = "Hull, C. M.",
    title = "{Timelike T duality, de Sitter space, large N gauge theories and topological field theory}",
    eprint = "hep-th/9806146",
    archivePrefix = "arXiv",
    reportNumber = "QMW-PH-98-28",
    doi = "10.1088/1126-6708/1998/07/021",
    journal = "JHEP",
    volume = "07",
    pages = "021",
    year = "1998"
}

@article{Strominger:2001pn,
    author = "Strominger, Andrew",
    title = "{The dS / CFT correspondence}",
    eprint = "hep-th/0106113",
    archivePrefix = "arXiv",
    doi = "10.1088/1126-6708/2001/10/034",
    journal = "JHEP",
    volume = "10",
    pages = "034",
    year = "2001"
}

@article{Balasubramanian:2001nb,
    author = "Balasubramanian, Vijay and de Boer, Jan and Minic, Djordje",
    title = "{Mass, entropy and holography in asymptotically de Sitter spaces}",
    eprint = "hep-th/0110108",
    archivePrefix = "arXiv",
    reportNumber = "VPI-IPPAP-01-01, UPR-964-T",
    doi = "10.1103/PhysRevD.65.123508",
    journal = "Phys. Rev. D",
    volume = "65",
    pages = "123508",
    year = "2002"
}

@article{Bedroya:2025ltj,
    author = "Bedroya, Alek and Steinhardt, Paul J.",
    title = "{Holography vs. Scale Separation}",
    eprint = "2509.25313",
    archivePrefix = "arXiv",
    primaryClass = "hep-th",
    month = "9",
    year = "2025"
}

@article{Banks:2000fe,
    author = "Banks, Tom",
    editor = "Duff, Michael J. and Liu, J. T. and Lu, J.",
    title = "{Cosmological breaking of supersymmetry?}",
    eprint = "hep-th/0007146",
    archivePrefix = "arXiv",
    reportNumber = "RUNHETC-2000-24, SCIPP-00-23",
    doi = "10.1142/S0217751X01003998",
    journal = "Int. J. Mod. Phys. A",
    volume = "16",
    pages = "910--921",
    year = "2001"
}

@article{Banks:2001yp,
    author = "Banks, Tom and Fischler, W.",
    title = "{M theory observables for cosmological space-times}",
    eprint = "hep-th/0102077",
    archivePrefix = "arXiv",
    reportNumber = "RUNHETC-2001-5, SCIPP-01-2, UTTG-02-01",
    month = "2",
    year = "2001"
}

@article{Banks:2002wr,
    author = "Banks, T. and Fischler, W. and Paban, S.",
    title = "{Recurrent nightmares? Measurement theory in de Sitter space}",
    eprint = "hep-th/0210160",
    archivePrefix = "arXiv",
    reportNumber = "RUNHETC-2002-35, SCIPP-02-25, UTTG-11-02",
    doi = "10.1088/1126-6708/2002/12/062",
    journal = "JHEP",
    volume = "12",
    pages = "062",
    year = "2002"
}

@article{Roupec:2018mbn,
    author = "Roupec, Christoph and Wrase, Timm",
    title = "{de Sitter Extrema and the Swampland}",
    eprint = "1807.09538",
    archivePrefix = "arXiv",
    primaryClass = "hep-th",
    doi = "10.1002/prop.201800082",
    journal = "Fortsch. Phys.",
    volume = "67",
    number = "1-2",
    pages = "1800082",
    year = "2019"
}

@article{Junghans:2018gdb,
    author = "Junghans, Daniel",
    title = "{Weakly Coupled de Sitter Vacua with Fluxes and the Swampland}",
    eprint = "1811.06990",
    archivePrefix = "arXiv",
    primaryClass = "hep-th",
    doi = "10.1007/JHEP03(2019)150",
    journal = "JHEP",
    volume = "03",
    pages = "150",
    year = "2019"
}

@article{Banlaki:2018ayh,
    author = "Banlaki, Andreas and Chowdhury, Abhishek and Roupec, Christoph and Wrase, Timm",
    title = "{Scaling limits of dS vacua and the swampland}",
    eprint = "1811.07880",
    archivePrefix = "arXiv",
    primaryClass = "hep-th",
    doi = "10.1007/JHEP03(2019)065",
    journal = "JHEP",
    volume = "03",
    pages = "065",
    year = "2019"
}

@article{Garg:2018reu,
    author = "Garg, Sumit K. and Krishnan, Chethan",
    title = "{Bounds on Slow Roll and the de Sitter Swampland}",
    eprint = "1807.05193",
    archivePrefix = "arXiv",
    primaryClass = "hep-th",
    doi = "10.1007/JHEP11(2019)075",
    journal = "JHEP",
    volume = "11",
    pages = "075",
    year = "2019"
}

@article{Ooguri:2018wrx,
    author = "Ooguri, Hirosi and Palti, Eran and Shiu, Gary and Vafa, Cumrun",
    title = "{Distance and de Sitter Conjectures on the Swampland}",
    eprint = "1810.05506",
    archivePrefix = "arXiv",
    primaryClass = "hep-th",
    doi = "10.1016/j.physletb.2018.11.018",
    journal = "Phys. Lett. B",
    volume = "788",
    pages = "180--184",
    year = "2019"
}

@article{Rudelius:2019cfh,
    author = "Rudelius, Tom",
    title = "{Conditions for (No) Eternal Inflation}",
    eprint = "1905.05198",
    archivePrefix = "arXiv",
    primaryClass = "hep-th",
    doi = "10.1088/1475-7516/2019/08/009",
    journal = "JCAP",
    volume = "08",
    pages = "009",
    year = "2019"
}

@article{Bedroya:2019snp,
    author = "Bedroya, Alek and Vafa, Cumrun",
    title = "{Trans-Planckian Censorship and the Swampland}",
    eprint = "1909.11063",
    archivePrefix = "arXiv",
    primaryClass = "hep-th",
    doi = "10.1007/JHEP09(2020)123",
    journal = "JHEP",
    volume = "09",
    pages = "123",
    year = "2020"
}

@article{Rudelius:2021oaz,
    author = "Rudelius, Tom",
    title = "{Dimensional reduction and (Anti) de Sitter bounds}",
    eprint = "2101.11617",
    archivePrefix = "arXiv",
    primaryClass = "hep-th",
    doi = "10.1007/JHEP08(2021)041",
    journal = "JHEP",
    volume = "08",
    pages = "041",
    year = "2021"
}

@article{Vafa:2025nst,
    author = "Vafa, Cumrun",
    title = "{On the origin and fate of our universe}",
    eprint = "2501.00966",
    archivePrefix = "arXiv",
    primaryClass = "hep-th",
    doi = "10.1007/s10714-025-03353-w",
    journal = "Gen. Rel. Grav.",
    volume = "57",
    number = "1",
    pages = "19",
    year = "2025"
}

@article{Denef:2018etk,
    author = "Denef, Frederik and Hebecker, Arthur and Wrase, Timm",
    title = "{de Sitter swampland conjecture and the Higgs potential}",
    eprint = "1807.06581",
    archivePrefix = "arXiv",
    primaryClass = "hep-th",
    doi = "10.1103/PhysRevD.98.086004",
    journal = "Phys. Rev. D",
    volume = "98",
    number = "8",
    pages = "086004",
    year = "2018"
}

@article{Murayama:2018lie,
    author = "Murayama, Hitoshi and Yamazaki, Masahito and Yanagida, Tsutomu T.",
    title = "{Do We Live in the Swampland?}",
    eprint = "1809.00478",
    archivePrefix = "arXiv",
    primaryClass = "hep-th",
    reportNumber = "IPMU-18-0143",
    doi = "10.1007/JHEP12(2018)032",
    journal = "JHEP",
    volume = "12",
    pages = "032",
    year = "2018"
}

@article{Choi:2018rze,
    author = "Choi, Kiwoon and Chway, Dongjin and Shin, Chang Sub",
    title = "{The dS swampland conjecture with the electroweak symmetry and QCD chiral symmetry breaking}",
    eprint = "1809.01475",
    archivePrefix = "arXiv",
    primaryClass = "hep-th",
    doi = "10.1007/JHEP11(2018)142",
    journal = "JHEP",
    volume = "11",
    pages = "142",
    year = "2018"
}

@article{Conlon:2018eyr,
    author = "Conlon, Joseph P.",
    title = "{The de Sitter swampland conjecture and supersymmetric AdS vacua}",
    eprint = "1808.05040",
    archivePrefix = "arXiv",
    primaryClass = "hep-th",
    doi = "10.1142/S0217751X18501786",
    journal = "Int. J. Mod. Phys. A",
    volume = "33",
    number = "29",
    pages = "1850178",
    year = "2018"
}

@article{Cicoli:2018kdo,
    author = "Cicoli, Michele and De Alwis, Senarath and Maharana, Anshuman and Muia, Francesco and Quevedo, Fernando",
    title = "{De Sitter vs Quintessence in String Theory}",
    eprint = "1808.08967",
    archivePrefix = "arXiv",
    primaryClass = "hep-th",
    doi = "10.1002/prop.201800079",
    journal = "Fortsch. Phys.",
    volume = "67",
    number = "1-2",
    pages = "1800079",
    year = "2019"
}

@article{Hebecker:2018vxz,
    author = "Hebecker, Arthur and Wrase, Timm",
    title = "{The Asymptotic dS Swampland Conjecture - a Simplified Derivation and a Potential Loophole}",
    eprint = "1810.08182",
    archivePrefix = "arXiv",
    primaryClass = "hep-th",
    doi = "10.1002/prop.201800097",
    journal = "Fortsch. Phys.",
    volume = "67",
    number = "1-2",
    pages = "1800097",
    year = "2019"
}

@article{Blaback:2013fca,
    author = {Bl{\r{a}}b{\"a}ck, Johan and Danielsson, Ulf and Dibitetto, Giuseppe},
    title = "{Accelerated Universes from type IIA Compactifications}",
    eprint = "1310.8300",
    archivePrefix = "arXiv",
    primaryClass = "hep-th",
    reportNumber = "UUITP-16-13",
    doi = "10.1088/1475-7516/2014/03/003",
    journal = "JCAP",
    volume = "03",
    pages = "003",
    year = "2014"
}

@article{Dine:1985he,
    author = "Dine, Michael and Seiberg, Nathan",
    title = "{Is the Superstring Weakly Coupled?}",
    reportNumber = "Print-85-0524 (IAS,PRINCETON)",
    doi = "10.1016/0370-2693(85)90927-X",
    journal = "Phys. Lett. B",
    volume = "162",
    pages = "299--302",
    year = "1985"
}

@article{Bedroya:2025ris,
    author = "Bedroya, Alek and Lee, Hayden and Steinhardt, Paul",
    title = "{A species scale-driven breakdown of effective field theory in time-dependent string backgrounds}",
    eprint = "2504.13260",
    archivePrefix = "arXiv",
    primaryClass = "hep-th",
    month = "4",
    year = "2025"
}

@article{Rudelius:2021azq,
    author = "Rudelius, Tom",
    title = "{Asymptotic observables and the swampland}",
    eprint = "2106.09026",
    archivePrefix = "arXiv",
    primaryClass = "hep-th",
    doi = "10.1103/PhysRevD.104.126023",
    journal = "Phys. Rev. D",
    volume = "104",
    number = "12",
    pages = "126023",
    year = "2021"
}

@article{Shiu:2023fhb,
    author = "Shiu, Gary and Tonioni, Flavio and Tran, Hung V.",
    title = "{Late-time attractors and cosmic acceleration}",
    eprint = "2306.07327",
    archivePrefix = "arXiv",
    primaryClass = "hep-th",
    doi = "10.1103/PhysRevD.108.063528",
    journal = "Phys. Rev. D",
    volume = "108",
    number = "6",
    pages = "063528",
    year = "2023"
}

@article{Calderon-Infante:2022nxb,
    author = "Calder\'on-Infante, Jos\'e and Ruiz, Ignacio and Valenzuela, Irene",
    title = "{Asymptotic accelerated expansion in string theory and the Swampland}",
    eprint = "2209.11821",
    archivePrefix = "arXiv",
    primaryClass = "hep-th",
    reportNumber = "CERN-TH-2022-153, IFT-UAM/CSIC-22-110",
    doi = "10.1007/JHEP06(2023)129",
    journal = "JHEP",
    volume = "06",
    pages = "129",
    year = "2023"
}

@article{Ooguri:2006in,
    author = "Ooguri, Hirosi and Vafa, Cumrun",
    title = "{On the Geometry of the String Landscape and the Swampland}",
    eprint = "hep-th/0605264",
    archivePrefix = "arXiv",
    reportNumber = "CALT-68-2600, HUTP-06-A017",
    doi = "10.1016/j.nuclphysb.2006.10.033",
    journal = "Nucl. Phys. B",
    volume = "766",
    pages = "21--33",
    year = "2007"
}

@article{Klaewer:2016kiy,
    author = "Klaewer, Daniel and Palti, Eran",
    title = "{Super-Planckian Spatial Field Variations and Quantum Gravity}",
    eprint = "1610.00010",
    archivePrefix = "arXiv",
    primaryClass = "hep-th",
    doi = "10.1007/JHEP01(2017)088",
    journal = "JHEP",
    volume = "01",
    pages = "088",
    year = "2017"
}

@article{Baume:2016psm,
    author = "Baume, Florent and Palti, Eran",
    title = "{Backreacted Axion Field Ranges in String Theory}",
    eprint = "1602.06517",
    archivePrefix = "arXiv",
    primaryClass = "hep-th",
    doi = "10.1007/JHEP08(2016)043",
    journal = "JHEP",
    volume = "08",
    pages = "043",
    year = "2016"
}

@article{Lee:2019wij,
    author = "Lee, Seung-Joo and Lerche, Wolfgang and Weigand, Timo",
    title = "{Emergent strings from infinite distance limits}",
    eprint = "1910.01135",
    archivePrefix = "arXiv",
    primaryClass = "hep-th",
    reportNumber = "CERN-TH-2019-159",
    doi = "10.1007/JHEP02(2022)190",
    journal = "JHEP",
    volume = "02",
    pages = "190",
    year = "2022"
}

@article{Etheredge:2022opl,
    author = "Etheredge, Muldrow and Heidenreich, Ben and Kaya, Sami and Qiu, Yue and Rudelius, Tom",
    title = "{Sharpening the Distance Conjecture in diverse dimensions}",
    eprint = "2206.04063",
    archivePrefix = "arXiv",
    primaryClass = "hep-th",
    reportNumber = "ACFI-T22-07",
    doi = "10.1007/JHEP12(2022)114",
    journal = "JHEP",
    volume = "12",
    pages = "114",
    year = "2022"
}

@article{Dvali:2007hz,
    author = "Dvali, Gia",
    title = "{Black Holes and Large N Species Solution to the Hierarchy Problem}",
    eprint = "0706.2050",
    archivePrefix = "arXiv",
    primaryClass = "hep-th",
    doi = "10.1002/prop.201000009",
    journal = "Fortsch. Phys.",
    volume = "58",
    pages = "528--536",
    year = "2010"
}

@article{Dvali:2007wp,
    author = "Dvali, Gia and Redi, Michele",
    title = "{Black Hole Bound on the Number of Species and Quantum Gravity at LHC}",
    eprint = "0710.4344",
    archivePrefix = "arXiv",
    primaryClass = "hep-th",
    doi = "10.1103/PhysRevD.77.045027",
    journal = "Phys. Rev. D",
    volume = "77",
    pages = "045027",
    year = "2008"
}

@article{Monnee:2025ynn,
    author = "Monnee, Jeroen and Weigand, Timo and Wiesner, Max",
    title = "{Physics and Geometry of Complex Structure Limits in Type IIB Calabi-Yau Compactifications}",
    eprint = "2509.07056",
    archivePrefix = "arXiv",
    primaryClass = "hep-th",
    month = "9",
    year = "2025"
}

@article{vandeHeisteeg:2023ubh,
    author = "van de Heisteeg, Damian and Vafa, Cumrun and Wiesner, Max",
    title = "{Bounds on Species Scale and the Distance Conjecture}",
    eprint = "2303.13580",
    archivePrefix = "arXiv",
    primaryClass = "hep-th",
    doi = "10.1002/prop.202300143",
    journal = "Fortsch. Phys.",
    volume = "71",
    number = "10-11",
    pages = "2300143",
    year = "2023"
}

@article{Castellano:2023stg,
    author = "Castellano, Alberto and Ruiz, Ignacio and Valenzuela, Irene",
    title = "{Universal Pattern in Quantum Gravity at Infinite Distance}",
    eprint = "2311.01501",
    archivePrefix = "arXiv",
    primaryClass = "hep-th",
    reportNumber = "CERN-TH-2023-203",
    doi = "10.1103/PhysRevLett.132.181601",
    journal = "Phys. Rev. Lett.",
    volume = "132",
    number = "18",
    pages = "181601",
    year = "2024"
}

@article{vandeHeisteeg:2023uxj,
    author = "van de Heisteeg, Damian and Vafa, Cumrun and Wiesner, Max and Wu, David H.",
    title = "{Bounds on field range for slowly varying positive potentials}",
    eprint = "2305.07701",
    archivePrefix = "arXiv",
    primaryClass = "hep-th",
    doi = "10.1007/JHEP02(2024)175",
    journal = "JHEP",
    volume = "02",
    pages = "175",
    year = "2024"
}

@article{Silverstein:2008sg,
    author = "Silverstein, Eva and Westphal, Alexander",
    title = "{Monodromy in the CMB: Gravity Waves and String Inflation}",
    eprint = "0803.3085",
    archivePrefix = "arXiv",
    primaryClass = "hep-th",
    reportNumber = "SU-ITP-08-07, SLAC-PUB-13183",
    doi = "10.1103/PhysRevD.78.106003",
    journal = "Phys. Rev. D",
    volume = "78",
    pages = "106003",
    year = "2008"
}

@article{Palti:2017elp,
    author = "Palti, Eran",
    title = "{The Weak Gravity Conjecture and Scalar Fields}",
    eprint = "1705.04328",
    archivePrefix = "arXiv",
    primaryClass = "hep-th",
    doi = "10.1007/JHEP08(2017)034",
    journal = "JHEP",
    volume = "08",
    pages = "034",
    year = "2017"
}

@article{Bedroya:2022tbh,
    author = "Bedroya, Alek",
    title = "{Holographic origin of TCC and the distance conjecture}",
    eprint = "2211.09128",
    archivePrefix = "arXiv",
    primaryClass = "hep-th",
    doi = "10.1007/JHEP06(2024)016",
    journal = "JHEP",
    volume = "06",
    pages = "016",
    year = "2024"
}

@article{Bedroya:2024zta,
    author = "Bedroya, Alek and Lu, Qianshu and Steinhardt, Paul J.",
    title = "{TCC in the interior of moduli space and its implications for the string landscape and cosmology}",
    eprint = "2407.08793",
    archivePrefix = "arXiv",
    primaryClass = "hep-th",
    doi = "10.1007/JHEP08(2025)007",
    journal = "JHEP",
    volume = "08",
    pages = "007",
    year = "2025"
}

@article{Hellerman:2001yi,
    author = "Hellerman, Simeon and Kaloper, Nemanja and Susskind, Leonard",
    title = "{String theory and quintessence}",
    eprint = "hep-th/0104180",
    archivePrefix = "arXiv",
    reportNumber = "SU-ITP-01-25",
    doi = "10.1088/1126-6708/2001/06/003",
    journal = "JHEP",
    volume = "06",
    pages = "003",
    year = "2001"
}

@article{Fischler:2001yj,
    author = "Fischler, W. and Kashani-Poor, A. and McNees, R. and Paban, S.",
    title = "{The Acceleration of the universe, a challenge for string theory}",
    eprint = "hep-th/0104181",
    archivePrefix = "arXiv",
    reportNumber = "UTTG-07-01",
    doi = "10.1088/1126-6708/2001/07/003",
    journal = "JHEP",
    volume = "07",
    pages = "003",
    year = "2001"
}

@article{Hebecker:2023qke,
    author = "Hebecker, Arthur and Schreyer, Simon and Venken, Victoria",
    title = "{No asymptotic acceleration without higher-dimensional de Sitter vacua}",
    eprint = "2306.17213",
    archivePrefix = "arXiv",
    primaryClass = "hep-th",
    doi = "10.1007/JHEP11(2023)173",
    journal = "JHEP",
    volume = "11",
    pages = "173",
    year = "2023"
}

@article{Friedrich:2025aec,
    author = "Friedrich, Bjoern and Hebecker, Arthur and Schiller, Daniel",
    title = "{Localized Gravity, de Sitter, and the Horizon Criterion}",
    eprint = "2505.07934",
    archivePrefix = "arXiv",
    primaryClass = "hep-th",
    month = "5",
    year = "2025"
}

@article{Rudelius:2022gbz,
    author = "Rudelius, Tom",
    title = "{Asymptotic scalar field cosmology in string theory}",
    eprint = "2208.08989",
    archivePrefix = "arXiv",
    primaryClass = "hep-th",
    doi = "10.1007/JHEP10(2022)018",
    journal = "JHEP",
    volume = "10",
    pages = "018",
    year = "2022"
}

@article{Casas:2024oak,
    author = "Casas, Gonzalo F. and Ruiz, Ignacio",
    title = "{Cosmology of light towers and swampland constraints}",
    eprint = "2409.08317",
    archivePrefix = "arXiv",
    primaryClass = "hep-th",
    reportNumber = "IFT-UAM/CSIC-24-129",
    doi = "10.1007/JHEP12(2024)193",
    journal = "JHEP",
    volume = "12",
    pages = "193",
    year = "2024"
}

@article{Lust:2019zwm,
    author = {L{\"u}st, Dieter and Palti, Eran and Vafa, Cumrun},
    title = "{AdS and the Swampland}",
    eprint = "1906.05225",
    archivePrefix = "arXiv",
    primaryClass = "hep-th",
    doi = "10.1016/j.physletb.2019.134867",
    journal = "Phys. Lett. B",
    volume = "797",
    pages = "134867",
    year = "2019"
}

@article{Burgess:2020nec,
    author = "Burgess, C. P. and de Alwis, S. P. and Quevedo, F.",
    title = "{Cosmological Trans-Planckian Conjectures are not Effective}",
    eprint = "2011.03069",
    archivePrefix = "arXiv",
    primaryClass = "hep-th",
    doi = "10.1088/1475-7516/2021/05/037",
    journal = "JCAP",
    volume = "05",
    pages = "037",
    year = "2021"
}

@article{Guo:2025mlb,
    author = "Guo, Xu and Pang, Yi and Sezgin, Ergin",
    title = "{4D de Sitter from 6D gauged supergravity with Green-Schwarz counterterm}",
    eprint = "2510.11794",
    archivePrefix = "arXiv",
    primaryClass = "hep-th",
    reportNumber = "MI-HET-867 and USTC-ICTS/PCFT-25-39",
    month = "10",
    year = "2025"
}

@article{Agrawal:2018rcg,
    author = "Agrawal, Prateek and Obied, Georges",
    title = "{Dark Energy and the Refined de Sitter Conjecture}",
    eprint = "1811.00554",
    archivePrefix = "arXiv",
    primaryClass = "hep-ph",
    doi = "10.1007/JHEP06(2019)103",
    journal = "JHEP",
    volume = "06",
    pages = "103",
    year = "2019"
}

@article{Shlivko:2024llw,
    author = "Shlivko, David and Steinhardt, Paul J.",
    title = "{Assessing observational constraints on dark energy}",
    eprint = "2405.03933",
    archivePrefix = "arXiv",
    primaryClass = "astro-ph.CO",
    doi = "10.1016/j.physletb.2024.138826",
    journal = "Phys. Lett. B",
    volume = "855",
    pages = "138826",
    year = "2024"
}

@article{Bayat:2025xfr,
    author = "Bayat, Zahra and Hertzberg, Mark P.",
    title = "{Examining quintessence models with DESI data}",
    eprint = "2505.18937",
    archivePrefix = "arXiv",
    primaryClass = "astro-ph.CO",
    doi = "10.1088/1475-7516/2025/08/065",
    journal = "JCAP",
    volume = "08",
    pages = "065",
    year = "2025"
}

@article{Planck:2018jri,
    author = "Akrami, Y. and others",
    collaboration = "Planck",
    title = "{Planck 2018 results. X. Constraints on inflation}",
    eprint = "1807.06211",
    archivePrefix = "arXiv",
    primaryClass = "astro-ph.CO",
    doi = "10.1051/0004-6361/201833887",
    journal = "Astron. Astrophys.",
    volume = "641",
    pages = "A10",
    year = "2020"
}

@article{Brown:2017osf,
    author = "Brown, Adam R.",
    title = "{Hyperbolic Inflation}",
    eprint = "1705.03023",
    archivePrefix = "arXiv",
    primaryClass = "hep-th",
    doi = "10.1103/PhysRevLett.121.251601",
    journal = "Phys. Rev. Lett.",
    volume = "121",
    number = "25",
    pages = "251601",
    year = "2018"
}

@article{Garcia-Saenz:2018ifx,
    author = "Garcia-Saenz, Sebastian and Renaux-Petel, S{\'e}bastien and Ronayne, John",
    title = "{Primordial fluctuations and non-Gaussianities in sidetracked inflation}",
    eprint = "1804.11279",
    archivePrefix = "arXiv",
    primaryClass = "astro-ph.CO",
    doi = "10.1088/1475-7516/2018/07/057",
    journal = "JCAP",
    volume = "07",
    pages = "057",
    year = "2018"
}

@article{Achucarro:2018vey,
    author = "Ach{\'u}carro, Ana and Palma, Gonzalo A.",
    title = "{The string swampland constraints require multi-field inflation}",
    eprint = "1807.04390",
    archivePrefix = "arXiv",
    primaryClass = "hep-th",
    doi = "10.1088/1475-7516/2019/02/041",
    journal = "JCAP",
    volume = "02",
    pages = "041",
    year = "2019"
}

@article{Bjorkmo:2019aev,
    author = "Bjorkmo, Theodor and Marsh, M. C. David",
    title = "{Hyperinflation generalised: from its attractor mechanism to its tension with the {\textquoteleft}swampland conditions{\textquoteright}}",
    eprint = "1901.08603",
    archivePrefix = "arXiv",
    primaryClass = "hep-th",
    doi = "10.1007/JHEP04(2019)172",
    journal = "JHEP",
    volume = "04",
    pages = "172",
    year = "2019"
}

@article{Bjorkmo:2019fls,
    author = "Bjorkmo, Theodor",
    title = "{Rapid-Turn Inflationary Attractors}",
    eprint = "1902.10529",
    archivePrefix = "arXiv",
    primaryClass = "hep-th",
    doi = "10.1103/PhysRevLett.122.251301",
    journal = "Phys. Rev. Lett.",
    volume = "122",
    number = "25",
    pages = "251301",
    year = "2019"
}

@article{Cicoli:2020noz,
    author = "Cicoli, Michele and Dibitetto, Giuseppe and Pedro, Francisco G.",
    title = "{Out of the Swampland with Multifield Quintessence?}",
    eprint = "2007.11011",
    archivePrefix = "arXiv",
    primaryClass = "hep-th",
    doi = "10.1007/JHEP10(2020)035",
    journal = "JHEP",
    volume = "10",
    pages = "035",
    year = "2020"
}

@article{Akrami:2020zfz,
    author = "Akrami, Yashar and Sasaki, Misao and Solomon, Adam R. and Vardanyan, Valeri",
    title = "{Multi-field dark energy: Cosmic acceleration on a steep potential}",
    eprint = "2008.13660",
    archivePrefix = "arXiv",
    primaryClass = "astro-ph.CO",
    doi = "10.1016/j.physletb.2021.136427",
    journal = "Phys. Lett. B",
    volume = "819",
    pages = "136427",
    year = "2021"
}

@article{Aragam:2020uqi,
    author = "Aragam, Vikas and Paban, Sonia and Rosati, Robert",
    title = "{The Multi-Field, Rapid-Turn Inflationary Solution}",
    eprint = "2010.15933",
    archivePrefix = "arXiv",
    primaryClass = "hep-th",
    reportNumber = "UTTG-10-2020",
    doi = "10.1007/JHEP03(2021)009",
    journal = "JHEP",
    volume = "03",
    pages = "009",
    year = "2021"
}

@article{Freigang:2023ogu,
    author = "Freigang, Julian and Lust, Dieter and Nian, Guo-En and Scalisi, Marco",
    title = "{Cosmic acceleration and turns in the Swampland}",
    eprint = "2306.17217",
    archivePrefix = "arXiv",
    primaryClass = "hep-th",
    reportNumber = "LMU-ASC 22/23, MPP-2023-135",
    doi = "10.1088/1475-7516/2023/11/080",
    journal = "JCAP",
    volume = "11",
    pages = "080",
    year = "2023"
}

@inproceedings{Greene:1996cy,
    author = "Greene, Brian R.",
    title = "{String theory on Calabi-Yau manifolds}",
    booktitle = "{Theoretical Advanced Study Institute in Elementary Particle Physics (TASI 96): Fields, Strings, and Duality}",
    eprint = "hep-th/9702155",
    archivePrefix = "arXiv",
    reportNumber = "CU-TP-812",
    pages = "543--726",
    month = "6",
    year = "1996"
}

@article{Becker:2006ks,
    author = "Becker, Katrin and Becker, Melanie and Vafa, Cumrun and Walcher, Johannes",
    title = "{Moduli Stabilization in Non-Geometric Backgrounds}",
    eprint = "hep-th/0611001",
    archivePrefix = "arXiv",
    reportNumber = "HUTP-06-A044",
    doi = "10.1016/j.nuclphysb.2007.01.034",
    journal = "Nucl. Phys. B",
    volume = "770",
    pages = "1--46",
    year = "2007"
}

@article{Baykara:2024vss,
    author = "Baykara, Zihni Kaan and Tarazi, Houri-Christina and Vafa, Cumrun",
    title = "{The Quasicrystalline String Landscape}",
    eprint = "2406.00129",
    archivePrefix = "arXiv",
    primaryClass = "hep-th",
    month = "5",
    year = "2024"
}

@article{Baykara:2024tjr,
    author = "Baykara, Zihni Kaan and Tarazi, Houri-Christina and Vafa, Cumrun",
    title = "{New Non-Supersymmetric Tachyon-Free Strings}",
    eprint = "2406.00185",
    archivePrefix = "arXiv",
    primaryClass = "hep-th",
    month = "5",
    year = "2024"
}

@article{ValeixoBento:2025qih,
    author = "Valeixo Bento, Bruno and Montero, Miguel",
    title = "{de Sitter no-go's for Riemann-flat manifolds and a link to semidefinite optimisation}",
    eprint = "2510.18945",
    archivePrefix = "arXiv",
    primaryClass = "hep-th",
    reportNumber = "IFT-25-115",
    month = "10",
    year = "2025"
}

@article{Danielsson:2010bc,
    author = "Danielsson, Ulf H. and Koerber, Paul and Van Riet, Thomas",
    title = "{Universal de Sitter solutions at tree-level}",
    eprint = "1003.3590",
    archivePrefix = "arXiv",
    primaryClass = "hep-th",
    reportNumber = "KUL-TF-10-03, UUITP-07-10",
    doi = "10.1007/JHEP05(2010)090",
    journal = "JHEP",
    volume = "05",
    pages = "090",
    year = "2010"
}

@article{Cordova:2018dbb,
    author = "C{\'o}rdova, Clay and De Luca, G. Bruno and Tomasiello, Alessandro",
    title = "{Classical de Sitter Solutions of 10-Dimensional Supergravity}",
    eprint = "1812.04147",
    archivePrefix = "arXiv",
    primaryClass = "hep-th",
    reportNumber = "CALT-TH-2018-053",
    doi = "10.1103/PhysRevLett.122.091601",
    journal = "Phys. Rev. Lett.",
    volume = "122",
    number = "9",
    pages = "091601",
    year = "2019"
}

@article{Cordova:2019cvf,
    author = "C{\'o}rdova, Clay and De Luca, G. Bruno and Tomasiello, Alessandro",
    title = "{New de Sitter Solutions in Ten Dimensions and Orientifold Singularities}",
    eprint = "1911.04498",
    archivePrefix = "arXiv",
    primaryClass = "hep-th",
    doi = "10.1007/JHEP08(2020)093",
    journal = "JHEP",
    volume = "08",
    pages = "093",
    year = "2020"
}

@article{Kim:2020ysx,
    author = "Kim, Nakwoo",
    title = "{Towards an explicit construction of de Sitter solutions in classical supergravity}",
    eprint = "2004.05885",
    archivePrefix = "arXiv",
    primaryClass = "hep-th",
    doi = "10.1007/JHEP10(2020)057",
    journal = "JHEP",
    volume = "10",
    pages = "057",
    year = "2020"
}

@article{Bena:2020qpa,
    author = "Bena, Iosif and De Luca, G. Bruno and Gra{\~n}a, Mariana and Lo Monaco, Gabriele",
    title = "{Oh, wait, O8 de Sitter may be unstable!}",
    eprint = "2010.05936",
    archivePrefix = "arXiv",
    primaryClass = "hep-th",
    doi = "10.1007/JHEP03(2021)168",
    journal = "JHEP",
    volume = "03",
    pages = "168",
    year = "2021"
}

@article{Horer:2024hgy,
    author = "Horer, Ludwig and Junghans, Daniel",
    title = "{Almost classical de Sitter?}",
    eprint = "2406.01690",
    archivePrefix = "arXiv",
    primaryClass = "hep-th",
    doi = "10.1007/JHEP09(2024)038",
    journal = "JHEP",
    volume = "09",
    pages = "038",
    year = "2024"
}

@article{Burgess:2024jkx,
    author = "Burgess, C. P. and Muia, F. and Quevedo, F.",
    title = "{4D de Sitter from String Theory via 6D Supergravity}",
    eprint = "2408.03852",
    archivePrefix = "arXiv",
    primaryClass = "hep-th",
    reportNumber = "CERN-TH-2024-107",
    month = "8",
    year = "2024"
}

@article{Blaback:2018hdo,
    author = {Bl{\r{a}}b{\"a}ck, Johan and Danielsson, Ulf and Dibitetto, Giuseppe},
    title = "{A new light on the darkest corner of the landscape}",
    eprint = "1810.11365",
    archivePrefix = "arXiv",
    primaryClass = "hep-th",
    reportNumber = "ROM2F/2018/06, UUITP-50/18",
    month = "10",
    year = "2018"
}

@article{Villadoro:2007yq,
    author = "Villadoro, Giovanni and Zwirner, Fabio",
    title = "{Beyond Twisted Tori}",
    eprint = "0706.3049",
    archivePrefix = "arXiv",
    primaryClass = "hep-th",
    reportNumber = "DFPD-07-TH-09",
    doi = "10.1016/j.physletb.2007.07.002",
    journal = "Phys. Lett. B",
    volume = "652",
    pages = "118--123",
    year = "2007"
}

@article{Blaback:2019zig,
    author = {Bl{\r{a}}b{\"a}ck, Johan and Danielsson, Ulf and Dibitetto, Giuseppe and Giri, Suvendu},
    title = "{Constructing stable de Sitter in M-theory from higher curvature corrections}",
    eprint = "1902.04053",
    archivePrefix = "arXiv",
    primaryClass = "hep-th",
    reportNumber = "ROM2F/2019/02, UUITP-6/19",
    doi = "10.1007/JHEP09(2019)042",
    journal = "JHEP",
    volume = "09",
    pages = "042",
    year = "2019"
}

@article{Cribiori:2023ihv,
    author = "Cribiori, Niccol{\`o} and Montella, Carmine",
    title = "{Quantum gravity constraints on scale separation and de Sitter in five dimensions}",
    eprint = "2303.04162",
    archivePrefix = "arXiv",
    primaryClass = "hep-th",
    reportNumber = "MPP-2023-45",
    doi = "10.1007/JHEP05(2023)178",
    journal = "JHEP",
    volume = "05",
    pages = "178",
    year = "2023"
}

@article{Farakos:2020idt,
    author = "Farakos, Fotis and Tringas, George and Van Riet, Thomas",
    title = "{Classical de Sitter solutions in three dimensions without tachyons?}",
    eprint = "2007.12084",
    archivePrefix = "arXiv",
    primaryClass = "hep-th",
    doi = "10.1140/epjc/s10052-020-08525-3",
    journal = "Eur. Phys. J. C",
    volume = "80",
    number = "10",
    pages = "947",
    year = "2020"
}

@article{Emelin:2021gzx,
    author = "Emelin, Maxim and Farakos, Fotis and Tringas, George",
    title = "{Three-dimensional flux vacua from IIB on co-calibrated G2 orientifolds}",
    eprint = "2103.03282",
    archivePrefix = "arXiv",
    primaryClass = "hep-th",
    doi = "10.1140/epjc/s10052-021-09261-y",
    journal = "Eur. Phys. J. C",
    volume = "81",
    number = "5",
    pages = "456",
    year = "2021"
}

@article{Farakos:2025shl,
    author = "Farakos, Fotis and Tringas, George and Van Riet, Thomas",
    title = "{Warped $\hbox {G}_2$-throats in IIA and uplift dSillusions}",
    eprint = "2505.21104",
    archivePrefix = "arXiv",
    primaryClass = "hep-th",
    doi = "10.1140/epjc/s10052-025-14769-8",
    journal = "Eur. Phys. J. C",
    volume = "85",
    number = "9",
    pages = "1018",
    year = "2025"
}

@article{Tringas:2025uyg,
    author = "Tringas, George and Wrase, Timm",
    title = "{Scale separation from O-planes}",
    eprint = "2504.15436",
    archivePrefix = "arXiv",
    primaryClass = "hep-th",
    doi = "10.1007/JHEP08(2025)073",
    journal = "JHEP",
    volume = "08",
    pages = "073",
    year = "2025"
}

@article{Aharony:2010af,
    author = "Aharony, Ofer and Jafferis, Daniel and Tomasiello, Alessandro and Zaffaroni, Alberto",
    title = "{Massive type IIA string theory cannot be strongly coupled}",
    eprint = "1007.2451",
    archivePrefix = "arXiv",
    primaryClass = "hep-th",
    doi = "10.1007/JHEP11(2010)047",
    journal = "JHEP",
    volume = "11",
    pages = "047",
    year = "2010"
}

@article{Heckman:2018mxl,
    author = "Heckman, Jonathan J. and Lawrie, Craig and Lin, Ling and Zoccarato, Gianluca",
    title = "{F-theory and Dark Energy}",
    eprint = "1811.01959",
    archivePrefix = "arXiv",
    primaryClass = "hep-th",
    reportNumber = "UPR-1294-T",
    doi = "10.1002/prop.201900057",
    journal = "Fortsch. Phys.",
    volume = "67",
    number = "10",
    pages = "1900057",
    year = "2019"
}

@article{Heckman:2019dsj,
    author = "Heckman, Jonathan J. and Lawrie, Craig and Lin, Ling and Sakstein, Jeremy and Zoccarato, Gianluca",
    title = "{Pixelated Dark Energy}",
    eprint = "1901.10489",
    archivePrefix = "arXiv",
    primaryClass = "hep-th",
    doi = "10.1002/prop.201900071",
    journal = "Fortsch. Phys.",
    volume = "67",
    number = "11",
    pages = "1900071",
    year = "2019"
}

@article{Grimm:2019ixq,
    author = "Grimm, Thomas W. and Li, Chongchuo and Valenzuela, Irene",
    title = "{Asymptotic Flux Compactifications and the Swampland}",
    eprint = "1910.09549",
    archivePrefix = "arXiv",
    primaryClass = "hep-th",
    doi = "10.1007/JHEP06(2020)009",
    journal = "JHEP",
    volume = "06",
    pages = "009",
    year = "2020",
    note = "[Erratum: JHEP 01, 007 (2021)]"
}

@article{Green:2011cn,
    author = "Green, Stephen R. and Martinec, Emil J. and Quigley, Callum and Sethi, Savdeep",
    title = "{Constraints on String Cosmology}",
    eprint = "1110.0545",
    archivePrefix = "arXiv",
    primaryClass = "hep-th",
    reportNumber = "EFI-11-25",
    doi = "10.1088/0264-9381/29/7/075006",
    journal = "Class. Quant. Grav.",
    volume = "29",
    pages = "075006",
    year = "2012"
}

@article{Gautason:2012tb,
    author = "Gautason, Fridrik Freyr and Junghans, Daniel and Zagermann, Marco",
    title = "{On Cosmological Constants from alpha'-Corrections}",
    eprint = "1204.0807",
    archivePrefix = "arXiv",
    primaryClass = "hep-th",
    reportNumber = "ITP-UH-08-12",
    doi = "10.1007/JHEP06(2012)029",
    journal = "JHEP",
    volume = "06",
    pages = "029",
    year = "2012"
}

@article{Kutasov:2015eba,
    author = "Kutasov, David and Maxfield, Travis and Melnikov, Ilarion and Sethi, Savdeep",
    title = "{Constraining de Sitter Space in String Theory}",
    eprint = "1504.00056",
    archivePrefix = "arXiv",
    primaryClass = "hep-th",
    doi = "10.1103/PhysRevLett.115.071305",
    journal = "Phys. Rev. Lett.",
    volume = "115",
    number = "7",
    pages = "071305",
    year = "2015"
}

@article{Quigley:2015jia,
    author = "Quigley, Callum",
    title = "{Gaugino Condensation and the Cosmological Constant}",
    eprint = "1504.00652",
    archivePrefix = "arXiv",
    primaryClass = "hep-th",
    doi = "10.1007/JHEP06(2015)104",
    journal = "JHEP",
    volume = "06",
    pages = "104",
    year = "2015"
}

@article{Leedom:2022zdm,
    author = "Leedom, Jacob M. and Righi, Nicole and Westphal, Alexander",
    title = "{Heterotic de Sitter beyond modular symmetry}",
    eprint = "2212.03876",
    archivePrefix = "arXiv",
    primaryClass = "hep-th",
    reportNumber = "DESY 22-173, KCL-PH-TH/2022-54",
    doi = "10.1007/JHEP02(2023)209",
    journal = "JHEP",
    volume = "02",
    pages = "209",
    year = "2023"
}

@article{Alvarez-Garcia:2024vnr,
    author = "{\'A}lvarez-Garc{\'\i}a, Rafael and Knei{\ss}l, Christian and Leedom, Jacob M. and Righi, Nicole",
    title = "{Open Strings and Heterotic Instantons}",
    eprint = "2407.20319",
    archivePrefix = "arXiv",
    primaryClass = "hep-th",
    reportNumber = "DESY 24-114, KCL-PH-TH/2024-44, MPP-2024-157, ZMP-HH/24-15",
    month = "7",
    year = "2024"
}

@article{Basile:2021krk,
    author = "Basile, Ivano and Platania, Alessia",
    title = "{String tension between de Sitter vacua and curvature corrections}",
    eprint = "2103.06276",
    archivePrefix = "arXiv",
    primaryClass = "hep-th",
    doi = "10.1103/PhysRevD.104.L121901",
    journal = "Phys. Rev. D",
    volume = "104",
    number = "12",
    pages = "L121901",
    year = "2021"
}

@article{Basile:2020mpt,
    author = "Basile, Ivano and Lanza, Stefano",
    title = "{de Sitter in non-supersymmetric string theories: no-go theorems and brane-worlds}",
    eprint = "2007.13757",
    archivePrefix = "arXiv",
    primaryClass = "hep-th",
    doi = "10.1007/JHEP10(2020)108",
    journal = "JHEP",
    volume = "10",
    pages = "108",
    year = "2020"
}

@article{Florakis:2016ani,
    author = "Florakis, Ioannis and Rizos, John",
    title = "{Chiral Heterotic Strings with Positive Cosmological Constant}",
    eprint = "1608.04582",
    archivePrefix = "arXiv",
    primaryClass = "hep-th",
    reportNumber = "CERN-TH-2016-179",
    doi = "10.1016/j.nuclphysb.2016.09.018",
    journal = "Nucl. Phys. B",
    volume = "913",
    pages = "495--533",
    year = "2016"
}

@article{Leone:2025mwo,
    author = "Leone, Giorgio and Raucci, Salvatore",
    title = "{Aspects of strings without spacetime supersymmetry}",
    eprint = "2509.24703",
    archivePrefix = "arXiv",
    primaryClass = "hep-th",
    reportNumber = "IFT-UAM/CSIC-25-100",
    month = "9",
    year = "2025"
}

@article{Alvarez-Gaume:1986ghj,
    author = "Alvarez-Gaume, Luis and Ginsparg, Paul H. and Moore, Gregory W. and Vafa, C.",
    title = "{An O(16) x O(16) Heterotic String}",
    reportNumber = "HUTP-86/A013",
    doi = "10.1016/0370-2693(86)91524-8",
    journal = "Phys. Lett. B",
    volume = "171",
    pages = "155--162",
    year = "1986"
}

@article{Dixon:1986iz,
    author = "Dixon, Lance J. and Harvey, Jeffrey A.",
    editor = "Schellekens, B.",
    title = "{String Theories in Ten-Dimensions Without Space-Time Supersymmetry}",
    reportNumber = "PRINT-86-0244 (PRINCETON)",
    doi = "10.1016/0550-3213(86)90619-X",
    journal = "Nucl. Phys. B",
    volume = "274",
    pages = "93--105",
    year = "1986"
}

@article{Sugimoto:1999tx,
    author = "Sugimoto, Shigeki",
    title = "{Anomaly cancellations in type I D-9 - anti-D-9 system and the USp(32) string theory}",
    eprint = "hep-th/9905159",
    archivePrefix = "arXiv",
    reportNumber = "YITP-99-25",
    doi = "10.1143/PTP.102.685",
    journal = "Prog. Theor. Phys.",
    volume = "102",
    pages = "685--699",
    year = "1999"
}

@inproceedings{Sagnotti:1995ga,
    author = "Sagnotti, Augusto",
    title = "{Some properties of open string theories}",
    booktitle = "{International Workshop on Supersymmetry and Unification of Fundamental Interactions (SUSY 95)}",
    eprint = "hep-th/9509080",
    archivePrefix = "arXiv",
    reportNumber = "ROM2F-95-18",
    pages = "473--484",
    month = "9",
    year = "1995"
}

@article{Sagnotti:1996qj,
    author = "Sagnotti, Augusto",
    editor = "Lust, D. and Otto, H. J. and Weigt, G.",
    title = "{Surprises in open string perturbation theory}",
    eprint = "hep-th/9702093",
    archivePrefix = "arXiv",
    reportNumber = "ROM2F-97-4",
    doi = "10.1016/S0920-5632(97)00344-7",
    journal = "Nucl. Phys. B Proc. Suppl.",
    volume = "56",
    pages = "332--343",
    year = "1997"
}

@article{Fraiman:2023cpa,
    author = "Fraiman, Bernardo and Gra{\~n}a, Mariana and Parra De Freitas, H{\'e}ctor and Sethi, Savdeep",
    title = "{Non-supersymmetric heterotic strings on a circle}",
    eprint = "2307.13745",
    archivePrefix = "arXiv",
    primaryClass = "hep-th",
    doi = "10.1007/JHEP12(2024)082",
    journal = "JHEP",
    volume = "12",
    pages = "082",
    year = "2024"
}

@article{DeLuca:2021pej,
    author = "De Luca, G. Bruno and Silverstein, Eva and Torroba, Gonzalo",
    title = "{Hyperbolic compactification of M-theory and de Sitter quantum gravity}",
    eprint = "2104.13380",
    archivePrefix = "arXiv",
    primaryClass = "hep-th",
    doi = "10.21468/SciPostPhys.12.3.083",
    journal = "SciPost Phys.",
    volume = "12",
    number = "3",
    pages = "083",
    year = "2022"
}

@article{DallAgata:2025jii,
    author = "Dall'Agata, Gianguido and Zwirner, Fabio",
    title = "{Supersymmetry-breaking compactifications on Riemann-flat manifolds}",
    eprint = "2507.02339",
    archivePrefix = "arXiv",
    primaryClass = "hep-th",
    month = "7",
    year = "2025"
}

@article{Cicoli:2023opf,
    author = "Cicoli, Michele and Conlon, Joseph P. and Maharana, Anshuman and Parameswaran, Susha and Quevedo, Fernando and Zavala, Ivonne",
    title = "{String cosmology: From the early universe to today}",
    eprint = "2303.04819",
    archivePrefix = "arXiv",
    primaryClass = "hep-th",
    doi = "10.1016/j.physrep.2024.01.002",
    journal = "Phys. Rept.",
    volume = "1059",
    pages = "1--155",
    year = "2024"
}

@article{Giddings:2001yu,
    author = "Giddings, Steven B. and Kachru, Shamit and Polchinski, Joseph",
    title = "{Hierarchies from fluxes in string compactifications}",
    eprint = "hep-th/0105097",
    archivePrefix = "arXiv",
    reportNumber = "SLAC-PUB-8807, NSF-ITP-01-37, SU-ITP-01-16",
    doi = "10.1103/PhysRevD.66.106006",
    journal = "Phys. Rev. D",
    volume = "66",
    pages = "106006",
    year = "2002"
}

@article{Balasubramanian:2005zx,
    author = "Balasubramanian, Vijay and Berglund, Per and Conlon, Joseph P. and Quevedo, Fernando",
    title = "{Systematics of moduli stabilisation in Calabi-Yau flux compactifications}",
    eprint = "hep-th/0502058",
    archivePrefix = "arXiv",
    reportNumber = "DAMTP-2005-10, UNH-05-01, UPR-1109-T",
    doi = "10.1088/1126-6708/2005/03/007",
    journal = "JHEP",
    volume = "03",
    pages = "007",
    year = "2005"
}

@article{Bena:2009xk,
    author = "Bena, Iosif and Grana, Mariana and Halmagyi, Nick",
    title = "{On the Existence of Meta-stable Vacua in Klebanov-Strassler}",
    eprint = "0912.3519",
    archivePrefix = "arXiv",
    primaryClass = "hep-th",
    reportNumber = "IPHT-T09-237",
    doi = "10.1007/JHEP09(2010)087",
    journal = "JHEP",
    volume = "09",
    pages = "087",
    year = "2010"
}

@article{Kachru:2018aqn,
    author = "Kachru, Shamit and Trivedi, Sandip P.",
    title = "{A comment on effective field theories of flux vacua}",
    eprint = "1808.08971",
    archivePrefix = "arXiv",
    primaryClass = "hep-th",
    doi = "10.1002/prop.201800086",
    journal = "Fortsch. Phys.",
    volume = "67",
    number = "1-2",
    pages = "1800086",
    year = "2019"
}

@article{Gao:2020xqh,
    author = "Gao, Xin and Hebecker, Arthur and Junghans, Daniel",
    title = "{Control issues of KKLT}",
    eprint = "2009.03914",
    archivePrefix = "arXiv",
    primaryClass = "hep-th",
    doi = "10.1002/prop.202000089",
    journal = "Fortsch. Phys.",
    volume = "68",
    pages = "2000089",
    year = "2020"
}

@article{Carta:2019rhx,
    author = "Carta, Federico and Moritz, Jakob and Westphal, Alexander",
    title = "{Gaugino condensation and small uplifts in KKLT}",
    eprint = "1902.01412",
    archivePrefix = "arXiv",
    primaryClass = "hep-th",
    reportNumber = "DESY 19-012, DESY-19-012",
    doi = "10.1007/JHEP08(2019)141",
    journal = "JHEP",
    volume = "08",
    pages = "141",
    year = "2019"
}

@article{Carta:2021lqg,
    author = "Carta, Federico and Moritz, Jakob",
    title = "{Resolving spacetime singularities in flux compactifications {\&} KKLT}",
    eprint = "2101.05281",
    archivePrefix = "arXiv",
    primaryClass = "hep-th",
    doi = "10.1007/JHEP08(2021)093",
    journal = "JHEP",
    volume = "08",
    pages = "093",
    year = "2021"
}

@article{McAllister:2024lnt,
    author = "McAllister, Liam and Moritz, Jakob and Nally, Richard and Schachner, Andreas",
    title = "{Candidate de Sitter vacua}",
    eprint = "2406.13751",
    archivePrefix = "arXiv",
    primaryClass = "hep-th",
    reportNumber = "CERN-TH-2024-090",
    doi = "10.1103/PhysRevD.111.086015",
    journal = "Phys. Rev. D",
    volume = "111",
    number = "8",
    pages = "086015",
    year = "2025"
}

@article{Demirtas:2021nlu,
    author = "Demirtas, Mehmet and Kim, Manki and McAllister, Liam and Moritz, Jakob and Rios-Tascon, Andres",
    title = "{Small cosmological constants in string theory}",
    eprint = "2107.09064",
    archivePrefix = "arXiv",
    primaryClass = "hep-th",
    doi = "10.1007/JHEP12(2021)136",
    journal = "JHEP",
    volume = "12",
    pages = "136",
    year = "2021"
}

@article{Sethi:2017phn,
    author = "Sethi, Savdeep",
    title = "{Supersymmetry Breaking by Fluxes}",
    eprint = "1709.03554",
    archivePrefix = "arXiv",
    primaryClass = "hep-th",
    reportNumber = "EFI-17-5",
    doi = "10.1007/JHEP10(2018)022",
    journal = "JHEP",
    volume = "10",
    pages = "022",
    year = "2018"
}

@article{Kim:2023sfs,
    author = "Kim, Manki",
    title = {{On string one-loop correction to the Einstein-Hilbert term and its implications on the K{\"a}hler potential}},
    eprint = "2302.12117",
    archivePrefix = "arXiv",
    primaryClass = "hep-th",
    reportNumber = "MIT-CTP/5531",
    doi = "10.1007/JHEP07(2023)044",
    journal = "JHEP",
    volume = "07",
    pages = "044",
    year = "2023"
}

@article{Kim:2023eut,
    author = "Kim, Manki",
    title = "{On one-loop corrected dilaton action in string theory}",
    eprint = "2305.08263",
    archivePrefix = "arXiv",
    primaryClass = "hep-th",
    reportNumber = "MIT-CTP/5552",
    doi = "10.4310/ATMP.2023.v27.n7.a2",
    journal = "Adv. Theor. Math. Phys.",
    volume = "27",
    number = "7",
    pages = "1965--2044",
    year = "2023"
}

@article{Cvetic:2024wsj,
    author = "Cveti{\v{c}}, Mirjam and Wiesner, Max",
    title = "{Nonperturbative resolution of strong coupling singularities in 4D N=1 heterotic M-theory}",
    eprint = "2408.12458",
    archivePrefix = "arXiv",
    primaryClass = "hep-th",
    doi = "10.1103/PhysRevD.110.106008",
    journal = "Phys. Rev. D",
    volume = "110",
    number = "10",
    pages = "106008",
    year = "2024"
}

@article{Crino:2020qwk,
    author = "Crin{\`o}, Chiara and Quevedo, Fernando and Valandro, Roberto",
    title = "{On de Sitter String Vacua from Anti-D3-Branes in the Large Volume Scenario}",
    eprint = "2010.15903",
    archivePrefix = "arXiv",
    primaryClass = "hep-th",
    doi = "10.1007/JHEP03(2021)258",
    journal = "JHEP",
    volume = "03",
    pages = "258",
    year = "2021"
}

@article{Junghans:2022exo,
    author = "Junghans, Daniel",
    title = "{LVS de Sitter vacua are probably in the swampland}",
    eprint = "2201.03572",
    archivePrefix = "arXiv",
    primaryClass = "hep-th",
    doi = "10.1016/j.nuclphysb.2023.116179",
    journal = "Nucl. Phys. B",
    volume = "990",
    pages = "116179",
    year = "2023"
}

@article{Junghans:2022kxg,
    author = "Junghans, Daniel",
    title = "{Topological constraints in the LARGE-volume scenario}",
    eprint = "2205.02856",
    archivePrefix = "arXiv",
    primaryClass = "hep-th",
    doi = "10.1007/JHEP08(2022)226",
    journal = "JHEP",
    volume = "08",
    pages = "226",
    year = "2022"
}

@article{ValeixoBento:2023nbv,
    author = "Valeixo Bento, Bruno and Chakraborty, Dibya and Parameswaran, Susha and Zavala, Ivonne",
    title = "{De Sitter vacua {\textemdash} when are {\textquoteleft}subleading corrections{\textquoteright} really subleading?}",
    eprint = "2306.07332",
    archivePrefix = "arXiv",
    primaryClass = "hep-th",
    doi = "10.1007/JHEP11(2023)075",
    journal = "JHEP",
    volume = "11",
    pages = "075",
    year = "2023"
}

@article{Gao:2022fdi,
    author = "Gao, Xin and Hebecker, Arthur and Schreyer, Simon and Venken, Victoria",
    title = "{The LVS parametric tadpole constraint}",
    eprint = "2202.04087",
    archivePrefix = "arXiv",
    primaryClass = "hep-th",
    doi = "10.1007/JHEP07(2022)056",
    journal = "JHEP",
    volume = "07",
    pages = "056",
    year = "2022"
}

@article{Conlon:2005ki,
    author = "Conlon, Joseph P. and Quevedo, Fernando and Suruliz, Kerim",
    title = "{Large-volume flux compactifications: Moduli spectrum and D3/D7 soft supersymmetry breaking}",
    eprint = "hep-th/0505076",
    archivePrefix = "arXiv",
    reportNumber = "DAMTP-2005-48",
    doi = "10.1088/1126-6708/2005/08/007",
    journal = "JHEP",
    volume = "08",
    pages = "007",
    year = "2005"
}

@article{Gallego:2017dvd,
    author = "Gallego, Diego and Marsh, M. C. David and Vercnocke, Bert and Wrase, Timm",
    title = "{A New Class of de Sitter Vacua in Type IIB Large Volume Compactifications}",
    eprint = "1707.01095",
    archivePrefix = "arXiv",
    primaryClass = "hep-th",
    doi = "10.1007/JHEP10(2017)193",
    journal = "JHEP",
    volume = "10",
    pages = "193",
    year = "2017"
}

@article{Rummel:2011cd,
    author = "Rummel, Markus and Westphal, Alexander",
    title = "{A sufficient condition for de Sitter vacua in type IIB string theory}",
    eprint = "1107.2115",
    archivePrefix = "arXiv",
    primaryClass = "hep-th",
    reportNumber = "DESY-11-121",
    doi = "10.1007/JHEP01(2012)020",
    journal = "JHEP",
    volume = "01",
    pages = "020",
    year = "2012"
}

@article{Bernardo:2020lar,
    author = "Bernardo, Heliudson and Brahma, Suddhasattwa and Dasgupta, Keshav and Tatar, Radu",
    title = "{Crisis on Infinite Earths: Short-lived de Sitter Vacua in the String Theory Landscape}",
    eprint = "2009.04504",
    archivePrefix = "arXiv",
    primaryClass = "hep-th",
    doi = "10.1007/JHEP04(2021)037",
    journal = "JHEP",
    volume = "04",
    pages = "037",
    year = "2021"
}

@article{Cicoli:2024yqh,
    author = "Cicoli, Michele and Cunillera, Francesc and Padilla, Antonio and Pedro, Francisco G.",
    title = "{From inflation to quintessence: a history of the universe in string theory}",
    eprint = "2407.03405",
    archivePrefix = "arXiv",
    primaryClass = "hep-th",
    doi = "10.1007/JHEP10(2024)141",
    journal = "JHEP",
    volume = "10",
    pages = "141",
    year = "2024"
}

@article{deCarlos:2009fq,
    author = "de Carlos, Beatriz and Guarino, Adolfo and Moreno, Jesus M.",
    title = "{Flux moduli stabilisation, Supergravity algebras and no-go theorems}",
    eprint = "0907.5580",
    archivePrefix = "arXiv",
    primaryClass = "hep-th",
    reportNumber = "IFT-UAM-CSIC-09-36",
    doi = "10.1007/JHEP01(2010)012",
    journal = "JHEP",
    volume = "01",
    pages = "012",
    year = "2010"
}

@article{deCarlos:2009qm,
    author = "de Carlos, Beatriz and Guarino, Adolfo and Moreno, Jesus M.",
    title = "{Complete classification of Minkowski vacua in generalised flux models}",
    eprint = "0911.2876",
    archivePrefix = "arXiv",
    primaryClass = "hep-th",
    reportNumber = "IFT-UAM-CSIC-09-51",
    doi = "10.1007/JHEP02(2010)076",
    journal = "JHEP",
    volume = "02",
    pages = "076",
    year = "2010"
}

@article{Dibitetto:2010rg,
    author = "Dibitetto, Giuseppe and Linares, Roman and Roest, Diederik",
    title = "{Flux Compactifications, Gauge Algebras and De Sitter}",
    eprint = "1001.3982",
    archivePrefix = "arXiv",
    primaryClass = "hep-th",
    doi = "10.1016/j.physletb.2010.03.074",
    journal = "Phys. Lett. B",
    volume = "688",
    pages = "96--100",
    year = "2010"
}

@article{Dibitetto:2011gm,
    author = "Dibitetto, Giuseppe and Guarino, Adolfo and Roest, Diederik",
    title = "{Charting the landscape of N=4 flux compactifications}",
    eprint = "1102.0239",
    archivePrefix = "arXiv",
    primaryClass = "hep-th",
    doi = "10.1007/JHEP03(2011)137",
    journal = "JHEP",
    volume = "03",
    pages = "137",
    year = "2011"
}

@article{Blaback:2013ht,
    author = {Bl{\r{a}}b{\"a}ck, Johan and Danielsson, Ulf and Dibitetto, Giuseppe},
    title = "{Fully stable dS vacua from generalised fluxes}",
    eprint = "1301.7073",
    archivePrefix = "arXiv",
    primaryClass = "hep-th",
    doi = "10.1007/JHEP08(2013)054",
    journal = "JHEP",
    volume = "08",
    pages = "054",
    year = "2013"
}

@article{Plauschinn:2020ram,
    author = "Plauschinn, Erik",
    title = "{Moduli Stabilization with Non-Geometric Fluxes {\textemdash} Comments on Tadpole Contributions and de-Sitter Vacua}",
    eprint = "2011.08227",
    archivePrefix = "arXiv",
    primaryClass = "hep-th",
    doi = "10.1002/prop.202100003",
    journal = "Fortsch. Phys.",
    volume = "69",
    number = "3",
    pages = "2100003",
    year = "2021"
}

@article{Gao:2018ayp,
    author = "Gao, Xin and Shukla, Pramod and Sun, Rui",
    title = "{On Missing Bianchi Identities in Cohomology Formulation}",
    eprint = "1805.05748",
    archivePrefix = "arXiv",
    primaryClass = "hep-th",
    reportNumber = "IFT-UAM/CSIC-18-038, IFT-UAM-CSIC-18-038",
    doi = "10.1140/epjc/s10052-019-7291-5",
    journal = "Eur. Phys. J. C",
    volume = "79",
    number = "9",
    pages = "781",
    year = "2019"
}

@article{Shukla:2022srx,
    author = "Shukla, Pramod",
    title = "{On stable type IIA de-Sitter vacua with geometric flux}",
    eprint = "2202.12840",
    archivePrefix = "arXiv",
    primaryClass = "hep-th",
    doi = "10.1140/epjc/s10052-023-11361-w",
    journal = "Eur. Phys. J. C",
    volume = "83",
    number = "3",
    pages = "196",
    year = "2023"
}

@article{Plauschinn:2018wbo,
    author = "Plauschinn, Erik",
    title = "{Non-geometric backgrounds in string theory}",
    eprint = "1811.11203",
    archivePrefix = "arXiv",
    primaryClass = "hep-th",
    reportNumber = "LMU-ASC 79/18",
    doi = "10.1016/j.physrep.2018.12.002",
    journal = "Phys. Rept.",
    volume = "798",
    pages = "1--122",
    year = "2019"
}

@article{Becker:2007dn,
    author = "Becker, Katrin and Becker, Melanie and Walcher, Johannes",
    title = "{Runaway in the Landscape}",
    eprint = "0706.0514",
    archivePrefix = "arXiv",
    primaryClass = "hep-th",
    doi = "10.1103/PhysRevD.76.106002",
    journal = "Phys. Rev. D",
    volume = "76",
    pages = "106002",
    year = "2007"
}

@article{Ishiguro:2021csu,
    author = "Ishiguro, Keiya and Otsuka, Hajime",
    title = "{Sharpening the boundaries between flux landscape and swampland by tadpole charge}",
    eprint = "2104.15030",
    archivePrefix = "arXiv",
    primaryClass = "hep-th",
    reportNumber = "KEK-TH-2322",
    doi = "10.1007/JHEP12(2021)017",
    journal = "JHEP",
    volume = "12",
    pages = "017",
    year = "2021"
}

@article{Bardzell:2022jfh,
    author = "Bardzell, Jacob and Gonzalo, Eduardo and Rajaguru, Muthusamy and Smith, Danielle and Wrase, Timm",
    title = "{Type IIB flux compactifications with h$^{1,1}$ = 0}",
    eprint = "2203.15818",
    archivePrefix = "arXiv",
    primaryClass = "hep-th",
    doi = "10.1007/JHEP06(2022)166",
    journal = "JHEP",
    volume = "06",
    pages = "166",
    year = "2022"
}

@article{Chen:2025rkb,
    author = "Chen, Shi and van de Heisteeg, Damian and Vafa, Cumrun",
    title = "{Symmetries and M-theory-like vacua in four dimensions}",
    eprint = "2503.16599",
    archivePrefix = "arXiv",
    primaryClass = "hep-th",
    doi = "10.1007/JHEP07(2025)258",
    journal = "JHEP",
    volume = "07",
    pages = "258",
    year = "2025"
}

@article{Cremonini:2023suw,
    author = "Cremonini, Sera and Gonzalo, Eduardo and Rajaguru, Muthusamy and Tang, Yuezhang and Wrase, Timm",
    title = "{On asymptotic dark energy in string theory}",
    eprint = "2306.15714",
    archivePrefix = "arXiv",
    primaryClass = "hep-th",
    doi = "10.1007/JHEP09(2023)075",
    journal = "JHEP",
    volume = "09",
    pages = "075",
    year = "2023"
}

@article{Silverstein:2001xn,
    author = "Silverstein, Eva",
    title = "{(A)dS backgrounds from asymmetric orientifolds}",
    eprint = "hep-th/0106209",
    archivePrefix = "arXiv",
    reportNumber = "SLAC-PUB-8869",
    journal = "Clay Mat. Proc.",
    volume = "1",
    pages = "179",
    year = "2002"
}

@inproceedings{Maloney:2002rr,
    author = "Maloney, Alexander and Silverstein, Eva and Strominger, Andrew",
    title = "{De Sitter space in noncritical string theory}",
    booktitle = "{Workshop on Conference on the Future of Theoretical Physics and Cosmology in Honor of Steven Hawking's 60th Birthday}",
    eprint = "hep-th/0205316",
    archivePrefix = "arXiv",
    reportNumber = "SLAC-PUB-9228, HUTP-02-A019",
    pages = "570--591",
    month = "5",
    year = "2002"
}

@article{Dodelson:2013iba,
    author = "Dodelson, Matthew and Dong, Xi and Silverstein, Eva and Torroba, Gonzalo",
    title = "{New solutions with accelerated expansion in string theory}",
    eprint = "1310.5297",
    archivePrefix = "arXiv",
    primaryClass = "hep-th",
    reportNumber = "SLAC-PUB-15800, SU-ITP-13-20",
    doi = "10.1007/JHEP12(2014)050",
    journal = "JHEP",
    volume = "12",
    pages = "050",
    year = "2014"
}

@article{Harribey:2018xvs,
    author = "Harribey, Sabine and Tsimpis, Dimitrios",
    title = "{One-loop bosonic string and De Sitter space}",
    eprint = "1810.02236",
    archivePrefix = "arXiv",
    primaryClass = "hep-th",
    doi = "10.1016/j.nuclphysb.2019.114768",
    journal = "Nucl. Phys. B",
    volume = "948",
    pages = "114768",
    year = "2019"
}

@article{Junghans:2023lpo,
    author = "Junghans, Daniel",
    title = "{de Sitter-eating O-planes in supercritical string theory}",
    eprint = "2308.00026",
    archivePrefix = "arXiv",
    primaryClass = "hep-th",
    doi = "10.1007/JHEP12(2023)196",
    journal = "JHEP",
    volume = "12",
    pages = "196",
    year = "2023"
}

@article{Chakravarty:2024pec,
    author = "Chakravarty, Joydeep and Dasgupta, Keshav",
    title = "{What if string theory has a de Sitter excited state?}",
    eprint = "2404.11680",
    archivePrefix = "arXiv",
    primaryClass = "hep-th",
    doi = "10.1007/JHEP10(2024)065",
    journal = "JHEP",
    volume = "10",
    pages = "065",
    year = "2024"
}

@article{Bernardo:2021rul,
    author = "Bernardo, Heliudson and Brahma, Suddhasattwa and Dasgupta, Keshav and Faruk, Mir-Mehedi and Tatar, Radu",
    title = "{de Sitter Space as a Glauber-Sudarshan State: II}",
    eprint = "2108.08365",
    archivePrefix = "arXiv",
    primaryClass = "hep-th",
    doi = "10.1002/prop.202100131",
    journal = "Fortsch. Phys.",
    volume = "69",
    number = "11-12",
    pages = "2100131",
    year = "2021"
}

@article{Brahma:2020tak,
    author = "Brahma, Suddhasattwa and Dasgupta, Keshav and Tatar, Radu",
    title = "{de Sitter Space as a Glauber-Sudarshan State}",
    eprint = "2007.11611",
    archivePrefix = "arXiv",
    primaryClass = "hep-th",
    doi = "10.1007/JHEP02(2021)104",
    journal = "JHEP",
    volume = "02",
    pages = "104",
    year = "2021"
}

@article{Dasgupta:2019gcd,
    author = "Dasgupta, Keshav and Emelin, Maxim and Faruk, Mir Mehedi and Tatar, Radu",
    title = "{de Sitter vacua in the string landscape}",
    eprint = "1908.05288",
    archivePrefix = "arXiv",
    primaryClass = "hep-th",
    doi = "10.1016/j.nuclphysb.2021.115463",
    journal = "Nucl. Phys. B",
    volume = "969",
    pages = "115463",
    year = "2021"
}

@article{Banerjee:2018qey,
    author = "Banerjee, Souvik and Danielsson, Ulf and Dibitetto, Giuseppe and Giri, Suvendu and Schillo, Marjorie",
    title = "{Emergent de Sitter Cosmology from Decaying Anti{\textendash}de Sitter Space}",
    eprint = "1807.01570",
    archivePrefix = "arXiv",
    primaryClass = "hep-th",
    reportNumber = "UUITP-27-18",
    doi = "10.1103/PhysRevLett.121.261301",
    journal = "Phys. Rev. Lett.",
    volume = "121",
    number = "26",
    pages = "261301",
    year = "2018"
}

@article{Banerjee:2019fzz,
    author = "Banerjee, Souvik and Danielsson, Ulf and Dibitetto, Giuseppe and Giri, Suvendu and Schillo, Marjorie",
    title = "{de Sitter Cosmology on an expanding bubble}",
    eprint = "1907.04268",
    archivePrefix = "arXiv",
    primaryClass = "hep-th",
    reportNumber = "UUITP-26/19",
    doi = "10.1007/JHEP10(2019)164",
    journal = "JHEP",
    volume = "10",
    pages = "164",
    year = "2019"
}

@article{Randall:1999vf,
    author = "Randall, Lisa and Sundrum, Raman",
    title = "{An Alternative to compactification}",
    eprint = "hep-th/9906064",
    archivePrefix = "arXiv",
    reportNumber = "MIT-CTP-2874, PUPT-1867, BUHEP-99-13",
    doi = "10.1103/PhysRevLett.83.4690",
    journal = "Phys. Rev. Lett.",
    volume = "83",
    pages = "4690--4693",
    year = "1999"
}

@article{Banerjee:2022ree,
    author = "Banerjee, Souvik and Danielsson, Ulf and Giri, Suvendu",
    title = "{Features of a dark energy model in string theory}",
    eprint = "2212.14004",
    archivePrefix = "arXiv",
    primaryClass = "hep-th",
    reportNumber = "UUITP-63/22",
    doi = "10.1103/PhysRevD.108.126009",
    journal = "Phys. Rev. D",
    volume = "108",
    number = "12",
    pages = "126009",
    year = "2023"
}

@article{Banerjee:2023uto,
    author = "Banerjee, Souvik and Danielsson, Ulf and Zemsch, Maximilian",
    title = "{The dark bubbleography}",
    eprint = "2311.16242",
    archivePrefix = "arXiv",
    primaryClass = "hep-th",
    doi = "10.1007/JHEP02(2024)102",
    journal = "JHEP",
    volume = "02",
    pages = "102",
    year = "2024"
}

@article{Banerjee:2020wov,
    author = "Banerjee, Souvik and Danielsson, Ulf and Giri, Suvendu",
    title = "{Bubble needs strings}",
    eprint = "2009.01597",
    archivePrefix = "arXiv",
    primaryClass = "hep-th",
    reportNumber = "UUITP-33/20",
    doi = "10.1007/JHEP03(2021)250",
    journal = "JHEP",
    volume = "21",
    pages = "250",
    year = "2020"
}

@article{DES:2024jxu,
    author = "Abbott, T. M. C. and others",
    collaboration = "DES",
    title = "{The Dark Energy Survey: Cosmology Results with {\ensuremath{\sim}}1500 New High-redshift Type Ia Supernovae Using the Full 5 yr Data Set}",
    eprint = "2401.02929",
    archivePrefix = "arXiv",
    primaryClass = "astro-ph.CO",
    reportNumber = "FERMILAB-PUB-23-0821-PPD, DES-2023-805",
    doi = "10.3847/2041-8213/ad6f9f",
    journal = "Astrophys. J. Lett.",
    volume = "973",
    number = "1",
    pages = "L14",
    year = "2024"
}

@article{DESI:2024mwx,
    author = "Adame, A. G. and others",
    collaboration = "DESI",
    title = "{DESI 2024 VI: cosmological constraints from the measurements of baryon acoustic oscillations}",
    eprint = "2404.03002",
    archivePrefix = "arXiv",
    primaryClass = "astro-ph.CO",
    reportNumber = "FERMILAB-PUB-24-0154-PPD",
    doi = "10.1088/1475-7516/2025/02/021",
    journal = "JCAP",
    volume = "02",
    pages = "021",
    year = "2025"
}

@article{DESI:2025zgx,
    author = "Abdul Karim, M. and others",
    collaboration = "DESI",
    title = "{DESI DR2 results. II. Measurements of baryon acoustic oscillations and cosmological constraints}",
    eprint = "2503.14738",
    archivePrefix = "arXiv",
    primaryClass = "astro-ph.CO",
    reportNumber = "FERMILAB-PUB-25-0169-PPD",
    doi = "10.1103/tr6y-kpc6",
    journal = "Phys. Rev. D",
    volume = "112",
    number = "8",
    pages = "083515",
    year = "2025"
}

@article{Ratra:1987rm,
    author = "Ratra, Bharat and Peebles, P. J. E.",
    title = "{Cosmological Consequences of a Rolling Homogeneous Scalar Field}",
    reportNumber = "PUPT-1072",
    doi = "10.1103/PhysRevD.37.3406",
    journal = "Phys. Rev. D",
    volume = "37",
    pages = "3406",
    year = "1988"
}

@article{Peebles:1987ek,
    author = "Peebles, P. J. E. and Ratra, Bharat",
    title = "{Cosmology with a Time Variable Cosmological Constant}",
    reportNumber = "PUPT-1069",
    doi = "10.1086/185100",
    journal = "Astrophys. J. Lett.",
    volume = "325",
    pages = "L17",
    year = "1988"
}

@article{Wetterich:1994bg,
    author = "Wetterich, Christof",
    title = "{The Cosmon model for an asymptotically vanishing time dependent cosmological 'constant'}",
    eprint = "hep-th/9408025",
    archivePrefix = "arXiv",
    reportNumber = "HD-THEP-94-16",
    journal = "Astron. Astrophys.",
    volume = "301",
    pages = "321--328",
    year = "1995"
}

@article{Caldwell:1997ii,
    author = "Caldwell, R. R. and Dave, Rahul and Steinhardt, Paul J.",
    title = "{Cosmological imprint of an energy component with general equation of state}",
    eprint = "astro-ph/9708069",
    archivePrefix = "arXiv",
    doi = "10.1103/PhysRevLett.80.1582",
    journal = "Phys. Rev. Lett.",
    volume = "80",
    pages = "1582--1585",
    year = "1998"
}

@article{Caldwell:2000wt,
    author = "Caldwell, R. R.",
    editor = "Cline, D. B.",
    title = "{An introduction to quintessence}",
    doi = "10.1590/S0103-97332000000200002",
    journal = "Braz. J. Phys.",
    volume = "30",
    pages = "215--229",
    year = "2000"
}

@article{Tsujikawa:2013fta,
    author = "Tsujikawa, Shinji",
    title = "{Quintessence: A Review}",
    eprint = "1304.1961",
    archivePrefix = "arXiv",
    primaryClass = "gr-qc",
    doi = "10.1088/0264-9381/30/21/214003",
    journal = "Class. Quant. Grav.",
    volume = "30",
    pages = "214003",
    year = "2013"
}

@book{Blumenhagen:2013fgp,
    author = {Blumenhagen, Ralph and L{\"u}st, Dieter and Theisen, Stefan},
    title = "{Basic concepts of string theory}",
    doi = "10.1007/978-3-642-29497-6",
    isbn = "978-3-642-29496-9",
    publisher = "Springer",
    address = "Heidelberg, Germany",
    series = "Theoretical and Mathematical Physics",
    year = "2013"
}

@article{SupernovaSearchTeam:1998fmf,
    author = "Riess, Adam G. and others",
    collaboration = "Supernova Search Team",
    title = "{Observational evidence from supernovae for an accelerating universe and a cosmological constant}",
    eprint = "astro-ph/9805201",
    archivePrefix = "arXiv",
    doi = "10.1086/300499",
    journal = "Astron. J.",
    volume = "116",
    pages = "1009--1038",
    year = "1998"
}

@article{SupernovaCosmologyProject:1998vns,
    author = "Perlmutter, S. and others",
    collaboration = "Supernova Cosmology Project",
    title = "{Measurements of $\Omega$ and $\Lambda$ from 42 High Redshift Supernovae}",
    eprint = "astro-ph/9812133",
    archivePrefix = "arXiv",
    reportNumber = "LBNL-41801, LBL-41801",
    doi = "10.1086/307221",
    journal = "Astrophys. J.",
    volume = "517",
    pages = "565--586",
    year = "1999"
}

@article{Planck:2018vyg,
    author = "Aghanim, N. and others",
    collaboration = "Planck",
    title = "{Planck 2018 results. VI. Cosmological parameters}",
    eprint = "1807.06209",
    archivePrefix = "arXiv",
    primaryClass = "astro-ph.CO",
    doi = "10.1051/0004-6361/201833910",
    journal = "Astron. Astrophys.",
    volume = "641",
    pages = "A6",
    year = "2020",
    note = "[Erratum: Astron.Astrophys. 652, C4 (2021)]"
}

@article{Chevallier:2000qy,
    author = "Chevallier, Michel and Polarski, David",
    title = "{Accelerating universes with scaling dark matter}",
    eprint = "gr-qc/0009008",
    archivePrefix = "arXiv",
    doi = "10.1142/S0218271801000822",
    journal = "Int. J. Mod. Phys. D",
    volume = "10",
    pages = "213--224",
    year = "2001"
}

@article{Linder:2002et,
    author = "Linder, Eric V.",
    title = "{Exploring the expansion history of the universe}",
    eprint = "astro-ph/0208512",
    archivePrefix = "arXiv",
    doi = "10.1103/PhysRevLett.90.091301",
    journal = "Phys. Rev. Lett.",
    volume = "90",
    pages = "091301",
    year = "2003"
}

@article{Shiu:2023nph,
    author = "Shiu, Gary and Tonioni, Flavio and Tran, Hung V.",
    title = "{Accelerating universe at the end of time}",
    eprint = "2303.03418",
    archivePrefix = "arXiv",
    primaryClass = "hep-th",
    doi = "10.1103/PhysRevD.108.063527",
    journal = "Phys. Rev. D",
    volume = "108",
    number = "6",
    pages = "063527",
    year = "2023"
}

@article{Shiu:2024sbe,
    author = "Shiu, Gary and Tonioni, Flavio and Tran, Hung V.",
    title = "{Analytic bounds on late-time axion-scalar cosmologies}",
    eprint = "2406.17030",
    archivePrefix = "arXiv",
    primaryClass = "hep-th",
    doi = "10.1007/JHEP09(2024)158",
    journal = "JHEP",
    volume = "09",
    pages = "158",
    year = "2024"
}

@article{Shiu:2025ycw,
    author = "Shiu, Gary and Tonioni, Flavio and Tran, Hung V.",
    title = "{Long-lived SEC violation via DM/DE couplings}",
    eprint = "2506.19914",
    archivePrefix = "arXiv",
    primaryClass = "hep-th",
    month = "6",
    year = "2025"
}

@article{Marconnet:2025vhj,
    author = "Marconnet, Paul and Tsimpis, Dimitrios",
    title = "{Universal cosmologies}",
    eprint = "2505.03449",
    archivePrefix = "arXiv",
    primaryClass = "hep-th",
    doi = "10.1007/JHEP08(2025)089",
    journal = "JHEP",
    volume = "08",
    pages = "089",
    year = "2025"
}

@article{Brinkmann:2022oxy,
    author = "Brinkmann, Max and Cicoli, Michele and Dibitetto, Giuseppe and Pedro, Francisco G.",
    title = "{Stringy multifield quintessence and the Swampland}",
    eprint = "2206.10649",
    archivePrefix = "arXiv",
    primaryClass = "hep-th",
    doi = "10.1007/JHEP11(2022)044",
    journal = "JHEP",
    volume = "11",
    pages = "044",
    year = "2022"
}

@article{Cicoli:2021fsd,
    author = "Cicoli, Michele and Cunillera, Francesc and Padilla, Antonio and Pedro, Francisco G.",
    title = "{Quintessence and the Swampland: The Parametrically Controlled Regime of Moduli Space}",
    eprint = "2112.10779",
    archivePrefix = "arXiv",
    primaryClass = "hep-th",
    doi = "10.1002/prop.202200009",
    journal = "Fortsch. Phys.",
    volume = "70",
    number = "4",
    pages = "2200009",
    year = "2022"
}

@article{Cicoli:2021skd,
    author = "Cicoli, Michele and Cunillera, Francesc and Padilla, Antonio and Pedro, Francisco G.",
    title = "{Quintessence and the Swampland: The Numerically Controlled Regime of Moduli Space}",
    eprint = "2112.10783",
    archivePrefix = "arXiv",
    primaryClass = "hep-th",
    doi = "10.1002/prop.202200008",
    journal = "Fortsch. Phys.",
    volume = "70",
    number = "4",
    pages = "2200008",
    year = "2022"
}

@article{Grimm:2025cpq,
    author = "Grimm, Thomas W. and van de Heisteeg, Damian and Revello, Filippo",
    title = "{Axion-Scalar Systems and Dynamical Distances}",
    eprint = "2510.12879",
    archivePrefix = "arXiv",
    primaryClass = "hep-th",
    month = "10",
    year = "2025"
}

@article{Licciardello:2025fhx,
    author = "Licciardello, Daniele and Rahimy, Saba and Zavala, Ivonne",
    title = "{Extending the Dynamical Systems Toolkit: Coupled Fields in Multiscalar Dark Energy}",
    eprint = "2509.02539",
    archivePrefix = "arXiv",
    primaryClass = "hep-th",
    month = "9",
    year = "2025"
}

@article{Gordillo-Ruiz:2025xcg,
    author = "Gordillo-Ruiz, Hansel and Hernandez-Segura, Miguel and Portillo-Castillo, Ignacio and Ramos-Sanchez, Saul and Zavala, Ivonne",
    title = "{Rolling with modular symmetry: quintessence and de Sitter in heterotic orbifolds}",
    eprint = "2509.22781",
    archivePrefix = "arXiv",
    primaryClass = "hep-th",
    month = "9",
    year = "2025"
}

@article{Anchordoqui:2025fgz,
    author = "Anchordoqui, Luis A. and Antoniadis, Ignatios and Lust, Dieter",
    title = "{S-dual quintessence, the Swampland, and the DESI DR2 results}",
    eprint = "2503.19428",
    archivePrefix = "arXiv",
    primaryClass = "hep-th",
    reportNumber = "MPP-2025-45, LMU-ASC 06/25",
    doi = "10.1016/j.physletb.2025.139632",
    journal = "Phys. Lett. B",
    volume = "868",
    pages = "139632",
    year = "2025"
}

@article{Anchordoqui:2025epz,
    author = "Anchordoqui, Luis A. and Antoniadis, Ignatios and Cribiori, Niccol{\`o} and Hasar, Arda and Lust, Dieter and Masias, Joaquin and Scalisi, Marco",
    title = "{Bulk/boundary modular quintessence and DESI}",
    eprint = "2506.02731",
    archivePrefix = "arXiv",
    primaryClass = "hep-th",
    reportNumber = "MPP-2025-113, LMU-ASC 15/25",
    doi = "10.1007/JHEP09(2025)128",
    journal = "JHEP",
    volume = "09",
    pages = "128",
    year = "2025"
}

@article{Parameswaran:2010ec,
    author = "Parameswaran, Susha L. and Ramos-Sanchez, Saul and Zavala, Ivonne",
    title = "{On Moduli Stabilisation and de Sitter Vacua in MSSM Heterotic Orbifolds}",
    eprint = "1009.3931",
    archivePrefix = "arXiv",
    primaryClass = "hep-th",
    reportNumber = "DESY-10-148",
    doi = "10.1007/JHEP01(2011)071",
    journal = "JHEP",
    volume = "01",
    pages = "071",
    year = "2011"
}

@article{Seo:2024qzf,
    author = "Seo, Min-Seok",
    title = "{Asymptotic Behavior of Saxion{\textendash}Axion System in Stringy Quintessence Model}",
    eprint = "2403.07307",
    archivePrefix = "arXiv",
    primaryClass = "hep-th",
    doi = "10.1002/prop.202400112",
    journal = "Fortsch. Phys.",
    volume = "72",
    number = "11",
    pages = "2400112",
    year = "2024"
}

@article{Andersson:2006du,
    author = "Andersson, Lars and Heinzle, J. Mark",
    title = "{Eternal acceleration from M-theory}",
    eprint = "hep-th/0602102",
    archivePrefix = "arXiv",
    doi = "10.4310/ATMP.2007.v11.n3.a2",
    journal = "Adv. Theor. Math. Phys.",
    volume = "11",
    number = "3",
    pages = "371--398",
    year = "2007"
}

@article{Boya:2002mv,
    author = "Boya, L. J. and Per, M. A. and Segui, A. J.",
    title = "{Graphical and kinematical approach to cosmological horizons}",
    eprint = "gr-qc/0203074",
    archivePrefix = "arXiv",
    reportNumber = "DFTUZ-02-02",
    doi = "10.1103/PhysRevD.66.064009",
    journal = "Phys. Rev. D",
    volume = "66",
    pages = "064009",
    year = "2002"
}

@article{Townsend:2003fx,
    author = "Townsend, Paul K. and Wohlfarth, Mattias N. R.",
    title = "{Accelerating cosmologies from compactification}",
    eprint = "hep-th/0303097",
    archivePrefix = "arXiv",
    reportNumber = "DAMTP-2003-23",
    doi = "10.1103/PhysRevLett.91.061302",
    journal = "Phys. Rev. Lett.",
    volume = "91",
    pages = "061302",
    year = "2003"
}

@article{Ohta:2003pu,
    author = "Ohta, Nobuyoshi",
    title = "{Accelerating cosmologies from S-branes}",
    eprint = "hep-th/0303238",
    archivePrefix = "arXiv",
    reportNumber = "OU-HET-434, USTC-ICTS-03-4",
    doi = "10.1103/PhysRevLett.91.061303",
    journal = "Phys. Rev. Lett.",
    volume = "91",
    pages = "061303",
    year = "2003"
}

@article{Emparan:2003gg,
    author = "Emparan, Roberto and Garriga, Jaume",
    title = "{A Note on accelerating cosmologies from compactifications and S branes}",
    eprint = "hep-th/0304124",
    archivePrefix = "arXiv",
    doi = "10.1088/1126-6708/2003/05/028",
    journal = "JHEP",
    volume = "05",
    pages = "028",
    year = "2003"
}

@article{Coudarchet:2023mfs,
    author = "Coudarchet, Thibaut",
    title = "{Hiding the extra dimensions: A review on scale separation in string theory}",
    eprint = "2311.12105",
    archivePrefix = "arXiv",
    primaryClass = "hep-th",
    doi = "10.1016/j.physrep.2024.02.003",
    journal = "Phys. Rept.",
    volume = "1064",
    pages = "1--28",
    year = "2024"
}

@article{Uzan:2010pm,
    author = "Uzan, Jean-Philippe",
    title = "{Varying Constants, Gravitation and Cosmology}",
    eprint = "1009.5514",
    archivePrefix = "arXiv",
    primaryClass = "astro-ph.CO",
    doi = "10.12942/lrr-2011-2",
    journal = "Living Rev. Rel.",
    volume = "14",
    pages = "2",
    year = "2011"
}

@article{Uzan:2024ded,
    author = "Uzan, Jean-Philippe",
    title = "{Fundamental constants: from measurement to the universe, a window on gravitation and cosmology}",
    eprint = "2410.07281",
    archivePrefix = "arXiv",
    primaryClass = "astro-ph.CO",
    doi = "10.1007/s41114-025-00059-y",
    journal = "Living Rev. Rel.",
    volume = "28",
    number = "1",
    pages = "6",
    year = "2025"
}

@inproceedings{Dasgupta:2025ypg,
    author = "Dasgupta, Keshav and Brahma, Suddhasattwa and Kulinich, Bohdan and Maji, Archana and Ramadevi, Pichai and Tatar, Radu",
    title = "{Transient de Sitter and Quasi de Sitter States in SO(32) and E{\_}8 x E{\_}8 Heterotic String Theories}",
    eprint = "2511.03798",
    archivePrefix = "arXiv",
    primaryClass = "hep-th",
    month = "11",
    year = "2025",
    booktitle = "."
}

@article{Dudas:2025ubq,
    author = "Dudas, E. and Mourad, J. and Sagnotti, A.",
    title = "{Supersymmetry Breaking with Fields, Strings and Branes}",
    eprint = "2511.04367",
    archivePrefix = "arXiv",
    primaryClass = "hep-th",
    month = "11",
    year = "2025"
}

@article{vandenHoogen:1999qq,
    author = "van den Hoogen, R. J. and Coley, Alan A. and Wands, David",
    title = "{Scaling solutions in Robertson-Walker space-times}",
    eprint = "gr-qc/9901014",
    archivePrefix = "arXiv",
    doi = "10.1088/0264-9381/16/6/317",
    journal = "Class. Quant. Grav.",
    volume = "16",
    pages = "1843--1851",
    year = "1999"
}

@article{Gosenca:2015qha,
    author = "Gosenca, Mateja and Coles, Peter",
    title = "{Dynamical Analysis of Scalar Field Cosmologies with Spatial Curvature}",
    eprint = "1502.04020",
    archivePrefix = "arXiv",
    primaryClass = "gr-qc",
    doi = "10.21105/astro.1502.04020",
    journal = "Open J. Astrophys.",
    volume = "1",
    number = "1",
    pages = "1",
    year = "2016"
}

@article{SavasArapoglu:2017pyh,
    author = {Sava{\c{s}} Arapo{\u{g}}lu, A. and Emrah Y{\"u}kselci, A.},
    title = "{Dynamical System Analysis of Quintessence Models with Exponential Potential - Revisited}",
    eprint = "1711.03824",
    archivePrefix = "arXiv",
    primaryClass = "gr-qc",
    doi = "10.1142/S021773231950069X",
    journal = "Mod. Phys. Lett. A",
    volume = "34",
    number = "09",
    pages = "1950069",
    year = "2019"
}

@article{Bahamonde:2017ize,
    author = {Bahamonde, Sebastian and B{\"o}hmer, Christian G. and Carloni, Sante and Copeland, Edmund J. and Fang, Wei and Tamanini, Nicola},
    title = "{Dynamical systems applied to cosmology: dark energy and modified gravity}",
    eprint = "1712.03107",
    archivePrefix = "arXiv",
    primaryClass = "gr-qc",
    doi = "10.1016/j.physrep.2018.09.001",
    journal = "Phys. Rept.",
    volume = "775-777",
    pages = "1--122",
    year = "2018"
}

@article{Copeland:1997et,
    author = "Copeland, Edmund J. and Liddle, Andrew R and Wands, David",
    title = "{Exponential potentials and cosmological scaling solutions}",
    eprint = "gr-qc/9711068",
    archivePrefix = "arXiv",
    reportNumber = "SUSX-TH-97-022, SUSSEX-AST-97-11-1, PU-RCG-97-20",
    doi = "10.1103/PhysRevD.57.4686",
    journal = "Phys. Rev. D",
    volume = "57",
    pages = "4686--4690",
    year = "1998"
}

@article{Ferreira:1997hj,
    author = "Ferreira, Pedro G. and Joyce, Michael",
    title = "{Cosmology with a primordial scaling field}",
    eprint = "astro-ph/9711102",
    archivePrefix = "arXiv",
    reportNumber = "CFPA-97-TH-20",
    doi = "10.1103/PhysRevD.58.023503",
    journal = "Phys. Rev. D",
    volume = "58",
    pages = "023503",
    year = "1998"
}

@article{Freivogel:2005vv,
    author = "Freivogel, Ben and Kleban, Matthew and Rodriguez Martinez, Maria and Susskind, Leonard",
    title = "{Observational consequences of a landscape}",
    eprint = "hep-th/0505232",
    archivePrefix = "arXiv",
    reportNumber = "SU-ITP-05-19",
    doi = "10.1088/1126-6708/2006/03/039",
    journal = "JHEP",
    volume = "03",
    pages = "039",
    year = "2006"
}

@article{Buniy:2006ed,
    author = "Buniy, Roman V. and Hsu, Stephen D. H. and Zee, A.",
    title = "{Does string theory predict an open universe?}",
    eprint = "hep-th/0610231",
    archivePrefix = "arXiv",
    doi = "10.1016/j.physletb.2008.01.007",
    journal = "Phys. Lett. B",
    volume = "660",
    pages = "382--385",
    year = "2008"
}

@article{Horn:2017kmv,
    author = "Horn, Bart",
    title = "{Positive curvature and scalar field tunneling in the landscape}",
    eprint = "1707.03851",
    archivePrefix = "arXiv",
    primaryClass = "hep-th",
    doi = "10.1103/PhysRevD.99.025010",
    journal = "Phys. Rev. D",
    volume = "99",
    number = "2",
    pages = "025010",
    year = "2019"
}

@article{Cespedes:2020xpn,
    author = "Cespedes, Sebastian and de Alwis, Senarath P. and Muia, Francesco and Quevedo, Fernando",
    title = "{Lorentzian vacuum transitions: Open or closed universes?}",
    eprint = "2011.13936",
    archivePrefix = "arXiv",
    primaryClass = "hep-th",
    doi = "10.1103/PhysRevD.104.026013",
    journal = "Phys. Rev. D",
    volume = "104",
    number = "2",
    pages = "026013",
    year = "2021"
}

\end{document}